\documentclass[12pt]{article}
\usepackage[dvips]{graphicx}
\usepackage{siunitx}
\usepackage{booktabs}
\usepackage[x11names,rgb]{xcolor}
\usepackage{longtable}

\usepackage{tikz}
\usetikzlibrary{arrows,snakes,backgrounds,matrix, shapes,calc, positioning}

\usetikzlibrary{arrows}
\pgfarrowsdeclarecombine{twotriang}{twotriang}%
{latex}{latex}{latex}{latex}

\usepackage{amsfonts}
\usepackage{amsmath}
\usepackage{amssymb}
\usepackage{amsthm}
\usepackage{placeins}
\usepackage{geometry}
\usepackage{adjustbox}
\usepackage{url}
\usepackage{subcaption}
\usepackage{threeparttable}

\usepackage[pdftex]{hyperref}

\usetikzlibrary{positioning}
\usepackage{authblk}

\usepackage[symbol]{footmisc}

\begin{document}

\tikzstyle{block} = [rectangle, draw,
    text width=6em, text centered, rounded corners, minimum height=4em]
\tikzstyle{line} = [draw, -latex']
\tikzstyle{input} = [coordinate]

\title{\huge Effectiveness of iNTS vaccination in Sub-Saharan Africa}

\author[a,$\star$]{Daniele Cassese\footnote{The authors are grateful to Allan Saul for precious support to the research and Dan Yamin for helpful comments and suggestions. The authors wish to thank the S-AFRIVAC (www.s-afrivac.org) and the Vacc-iNTS (www.vacc-intsproject.eu) Consortia. 
This study has been carried out with the financial support from Regione Toscana, project S- AFRIVAC "Sviluppo di un vaccine contro la Salmonella non tifoide invasiva in Africa", under the call FAS-Salute 2014, PAR FAS 2007-2013 Program (Project No 4042.16092014.066000010; CUP B62F13000870007). This project has also received funding from the European Union's Horizon 2020 research and innovation programme under the Vacc-iNTS project (grant agreement No 815439). The sponsor had no role in the manuscript submission.}
}
\author[b]{Nicola Dimitri}
\author[c]{Gianluca Breghi}
\author[c,d,$\star$]{Tiziana Spadafina}

\affil[a]{University of Cambridge, Cambridge, CB2 3AP}
\affil[b]{Department of Economics, University of Siena, Siena, 53100}
\affil[c]{Fondazione Achille Sclavo ONLUS, Siena 53100}
\affil[d]{Sclavo Vaccines Association E.T.S., Siena 53100}
\affil[$\star$]{To whom correspondence should be addressed. E-mails: dc554@cam.ac.uk,spadafina@sclavo.org}
\date{}

\date{}

\maketitle

\begin{abstract}
{Invasive non-Typhoidal Salmonella (iNTS) is one of the leading causes of blood stream infections in Sub-Saharan Africa, especially among children. iNTS can be difficult to diagnose, particularly in areas where malaria is endemic and difficult to treat, partly because of the emergence of antibiotic resistance. We developed a mathematical model to evaluate the impact of a vaccine for iNTS in 49 countries of sub-Saharan Africa. Without vaccination we estimate 9.2 million new iNTS cases among children below 5 years old in these 49 countries from 2022 to 2038, 6.2 million of which between 2028 and 2038. The introduction of a $85\%$ ($95\%$) efficacy vaccine in 2028 would prevent 2.6 (2.9) million of these new infections. We provide the country-specific impact of a iNTS vaccine considering the different age structures and vaccine coverage levels.}
\end{abstract}

\section{Introduction}\label{sec1}

Non-typhoidal \emph{Salmonella} invasive (iNTS) disease is an emerging neglected infectious disease that causes a serious global burden of morbidity and mortality. The first global estimates of iNTS disease, produced as part of the Global Burden of Disease (GBD) 2017, showed that iNTS disease affects more than half a million people  [535,000 (409,000-705,000) cases] with an average Case Fatality Rates (CFR) of about $15\%$ \cite{Stanaway2019, Marchello22} and 4.26 million (2.38-7.38) DALYs.
Most recently, iNTS disease has been globally  associated to 87,100 [53,800-131,000] deaths due to bloodstream infection, with an age mortality rate of 1.2 years  \cite{GBD2019}. INTS is among the 10 top pathogens in the estimation of the years of life lost (YLL) burden \cite{GBD2019}. Data on iNTS incidence and the prevalence of complications and case-fatality ratio (CFR) of iNTS \cite{Marchello22} has been systematically reviewed on behalf of the Vacc-iNTS Consortium \cite{Marchello21}. They have calculated a global incidence ($95\%$ CI) of 44.8 (31.5-60.5) per 100,000 persons per year, with Africa significantly higher [51.0 (36.3-68.0)] and estimated that approximately $15\%$ of patients with iNTS disease die, similarly to  \cite{Stanaway2019}.  
Approximately $70\%$ of currently reported iNTS cases are observed in sub-Saharan Africa (sSA) \cite{Stanaway2019, Marchello21}, where it is among the leading cause of community-acquired bloodstream infections and is associated with increasing antibiotic resistance\cite{VanPuyvelde2019, Kingsley2009, Park2021,Akullian2018, Gordon2008}. In sSA, iNTS disease spreads mostly among children under 5 years of age, with comordibities as malaria, anaemia, malnutrition, and HIV infection being prominent risk factors \cite{Stanaway2019, Marchello21} together with young age. In adults, HIV infection is by far the most important risk factor. These infections usually present as a febrile illness, frequently without gastrointestinal symptoms in both young adults and children, leading to severe, extra-intestinal, invasive bacteremia. iNTS disease has been reported across Africa, demonstrating it is a widespread threat throughout the continent  \cite{Marchello21, Uche2017, Marks2017, GBD2019} with indication that the disease is endemic in much of the Region.

Over the past years, it has emerged that the serovars \emph{Enteritidis} and \emph{Typhimurium} are the ones most commonly associated with iNTS in Africa, causing more than $90\%$ of cases \cite{Marchello21, Uche2017, Crump15}  characterized by genome degradation and appear to be adapting to an invasive lifestyle. 
Recent estimates have been showing the increasing emergence of multidrug-resistant (MDR) \emph{S. Typhimurium} and \emph{S. Enteritidis} strains especially in sSA \cite{Feasey15, Kalonji15, Kariuki19, VanPuyvelde2019}, which compromise the clinical treatment of iNTS disease in settings where diagnosis, surveillance programmes and affordable medicines are often scarce \cite{Kariuki15}. \emph{Salmonella} spp. have been included in the World Health Organization (WHO) antibiotic-resistant high priority pathogens list and show a concerning increase in MDR \cite{Feasey15, Tacconelli2018}. 
iNTS causes major disease and socioeconomic burden in resource-poor communities, particularly in children, elderly people, and people with HIV infection in sSA, and no vaccine is currently available \cite{Tennant2016, Baliban2020,Piccini2020, GBD2019}. Medical need, difficult diagnosis and increasing AMR make the investigation of the sources and transmission pathways of iNTS disease a crucial point to implement effective preventive and control measures and definitely strongly advocate for rapid development of an effective vaccine.

Several iNTS vaccines are currently under development, some of which are bivalent and targeting the two serovars leading causes of iNTS,  and include live attenuated, protein-polysaccharide vaccines, multiple antigen presenting system complexes and Outer Membrane Vesicles-based vaccines. The ideal iNTS vaccine should be poor reactogenic, cross-protective against multiple serovars and at affordable production and delivery costs. Among them the generalised modules for membrane antigens (GMMA) of \emph{S.} enterica serovars \emph{Enteritidis} and \emph{Typhimurium} expressed on outer membrane vesicles. The development of this candidate has been carried out also in the context of the S-AFRIVAC project (supported by the Tuscany Region) and is advancing under the H2020 Vacc-iNTS and EDCTP2 PEDVAC- iNTS projects.

To evaluate the impact of the introduction of a vaccine against iNTS on the African burden we developed a mathematical model for the transmission of iNTS disease. %%%(\emph{Methods} and \href{run:./iNTS_SI.pdf}{\emph{SI Appendix}}) 
We simulate the transmission dynamic for each country separately, using country-specific population pyramids and comorbidity data, as well as vaccine coverage rates.
The effect of the introduction of  the iNTS	vaccine has been evaluated comparing  two different scenarios: the status quo, without any intervention, and a vaccination scenario where we consider a catch-up campaign followed by a routine campaign. The catch-up campaign starts in 2028 and lasts for one year, during which children between 9 months and 5 years of age are vaccinated, while the routine vaccination campaign lasts 9 years, during which children are vaccinated upon reaching 9 months of age. The epidemiological trajectories are projected up to 2038, and we consider two levels of vaccine efficacy, $85\%$ and $95\%$.

\section{Results}\label{sec2}

Our model estimates 478,000 iNTS cases for the year 2021 in all sSA, and under the status quo scenario the cumulated number of cases estimated from 2021 to 2038 is 9,733,000 (6,242,000 of which in the period 2028-2038). The 10 years Routine + Catch-Up (RCU) campaign between 2028 and 2038 could prevent between 2,605,000 and 2,981,000 cases when the vaccine efficacy is between 85$\%$ and 95$\%$  (Fig. \ref{fig:perc} and Supplementary Fig. 9).
This means that the model estimates a reduction of iNTS cases between 41.7$\%$ and 47.8$\%$ among children below 5 years old in all sSA. If we assume a CFR of $15\%$ the vaccine would avert between 391,000 and 447,000 deaths over the 10 years considered. Based on the model projections, the 10-year vaccination campaign would reduce the burden of the disease by 34,187,000-39,117,000 DALYs for all sSA  (Table \ref{dalys}).%(\href{run:./iNTS_SI.pdf}{\emph{SI Appendix}}).

There are considerable differences in cases prevented across countries, due to their differences in the prevalence of comorbidities among children below 5 years old and EPI vaccination coverage levels (Fig. \ref{fig:casesprev} and Supplementary Figs. 4-8). 
For countries where EPI coverage levels are below $70\%$, the reduction in iNTS cases reaches at most $34\%$-$38\%$ ($85\%$ and $95\%$ efficacy respectively). For countries that have a coverage at least of $95\%$, the model predicts that the reduction in cases can go above  $50\%$-$60\%$ ($85\%$ and $95\%$ efficacy respectively). Countries where the prevalence of comorbidities is higher (especially Malaria among children) present smaller reduction in cases given their coverage levels (Supplementary Fig. 8). %(\href{run:./iINTS_SI.pdf}{\emph{SI Appendix}, Fig. S8}). 

\renewcommand{\thefigure}{\textbf{Fig. 1 Vaccine efficacy as percentage of prevented infections per country}}
\begin{figure}[tbhp]
\centering
\includegraphics[width=0.6\linewidth]{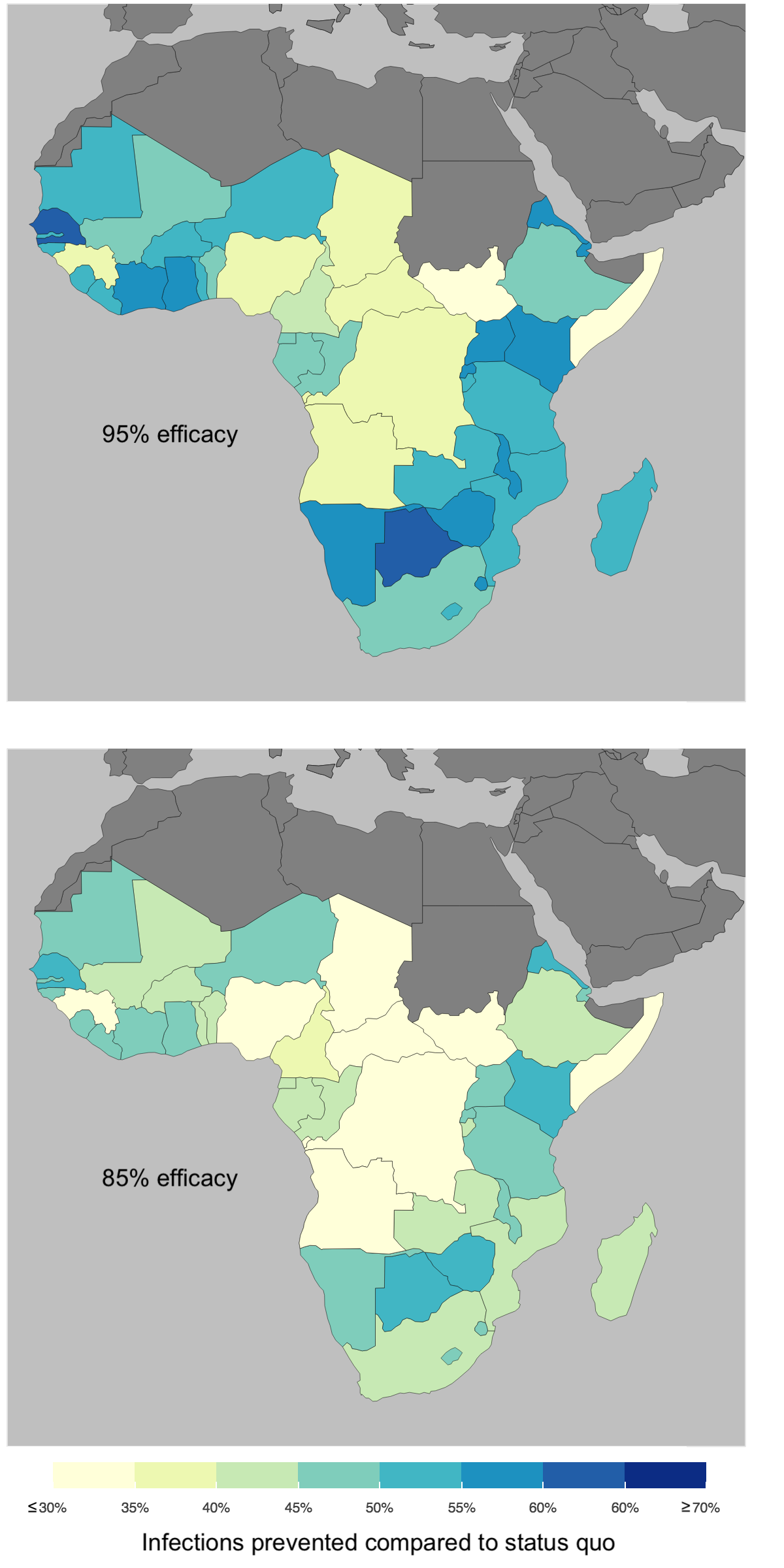}
\caption{Infections prevented between 2028 and 2038, one year Catch-Up folllowed by 9 years Routine. Vaccine assumed with 95$\%$ efficacy (\emph{Top}) and 85$\%$ efficacy (\emph{Bottom}). Differences across countries mostly due to routine vaccination coverage (see  Fig. \ref{fig:percountry}) and prevalence of comorbidities.}
\label{fig:casesprev}
\end{figure}

Nigeria and Ethiopia account for the highest number of cases prevented among children below 5 years old over the considered period [362,000 - 404,000] (35$\%$-38$\%$) and [186,000-210,000] (42$\%$-48$\%$) respectively, being among the most populous countries together with the Democratic Republic of the Congo (DRC) which accounts for [133,000-150,000] (32$\%$-36$\%$) cases prevented. Both Nigeria and DRC present low levels of coverage (65$\%$ and 66$\%$ respectively, against 80$\%$ for Ethiopia) and this explains the relatively lower reduction in cases following vaccination. Assuming a coverage of 90$\%$ for the routine vaccination the number of prevented cases would substantially increase: our model predicts [487,000-556,000] (47$\%$-53$\%$) cases prevented for Nigeria, [179,000-206,000] (44$\%$-50$\%$) for DRC and [209,000-238,000] (47$\%$-54$\%$) for Ethiopia. Tanzania and Uganda are less populous than the three countries mentioned, but thanks to higher coverage levels (91$\%$ and 99$\%$ respectively) they count for a large share of infections prevented, at [167,000-191,000] (47$\%$-54$\%$) and [147,000-174,000] (49$\%$-58$\%$). Our results show that increasing EPI vaccination coverages across countries would induce a considerable reduction in the burden of iNTS: by assuming an EPI coverage of at least 90$\%$ in each country the aggregate number of cases prevented over the period goes up to 2,996,000-3,373,000, (47$\%$-54$\%$).

\renewcommand{\thefigure}{\textbf{Fig. 2 Cumulative number of cases prevented per country}}
\begin{figure}[t]
\centering
\includegraphics[width=1\textwidth]{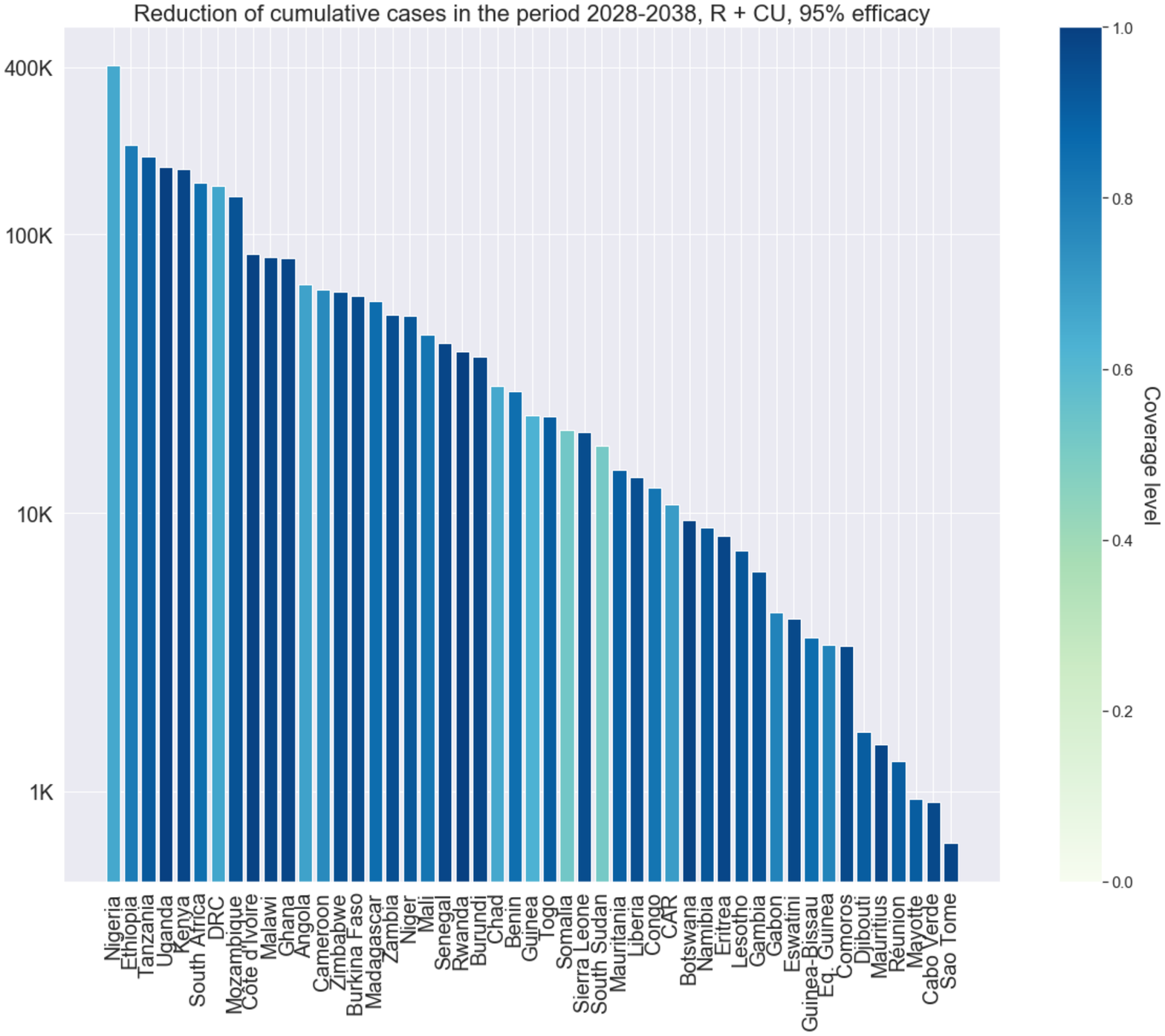}
\caption{Bars show the number of cases prevented considering a one year catch-up followed by a routine vaccination over the period 2028-2038. Vaccine efficacy is  $95\%$. The $y$-axis is in logarithmic scale. The colour of each bar corresponds to the routine coverage level in the country, as captured by the colormap on the righthand side of  the graph. }\label{fig:percountry}
\end{figure}

\clearpage
\renewcommand{\thetable}{\textbf{Table 1 Per-country cumulative cases prevented, deaths and DALYs averted}}
\newpage
{\tiny
\begin{longtable}{lcccclll}
\caption{All number are in thousands. The lower bound of cases prevented, deaths averted and DALYs averted correspond to 85$\%$ vaccine efficacy, the upper bound to 95$\%$ vaccine efficacy.}\\
\toprule
                         Country &   \begin{tabular}{@{}c@{}}Status quo \\ cases\end{tabular} & \begin{tabular}{@{}c@{}}Status quo \\ deaths\end{tabular} &  \begin{tabular}{@{}c@{}}Status quo \\DALYs\end{tabular} & \begin{tabular}{@{}c@{}}Cases\\Prevented\end{tabular} & \begin{tabular}{@{}c@{}}Deaths\\averted\end{tabular} & \begin{tabular}{@{}c@{}}DALYs\\ averted\end{tabular} \\
\midrule
                                  Angola &           193.409 &             43.847 &              2538 &     (65.795, 73.659) &     (9.869, 11.049) &         (863, 967) \\
                           Benin &            64.274 &             14.879 &               844 &     (26.886, 30.499) &      (4.033, 4.575) &         (353, 401) \\
                        Botswana &            17.168 &              4.194 &               225 &       (8.924, 10.42) &      (1.339, 1.563) &         (117, 137) \\
                    Burkina Faso &           133.042 &             30.748 &              1746 &     (56.911, 67.085) &     (8.537, 10.063) &         (747, 880) \\
                         Burundi &            78.375 &             18.243 &              1028 &      (34.077, 40.63) &      (5.112, 6.094) &         (447, 533) \\
                      Cabo Verde &             1.671 &              0.426 &                22 &       (0.872, 1.005) &      (0.131, 0.151) &           (11, 13) \\
                        Cameroon &           161.300 &             37.731 &              2117 &     (62.883, 70.574) &     (9.432, 10.586) &         (825, 926) \\
        Central African Republic &            30.435 &              7.133 &               399 &     (10.566, 11.844) &      (1.585, 1.777) &         (139, 155) \\
                            Chad &            83.618 &             19.419 &              1097 &     (28.265, 31.601) &        (4.24, 4.74) &         (371, 415) \\
                         Comoros &             6.872 &              1.634 &                90 &       (3.158, 3.706) &      (0.474, 0.556) &           (41, 49) \\
                           Congo &            29.459 &              6.806 &               386 &      (12.098, 13.68) &      (1.815, 2.052) &         (158, 179) \\
                   C\^ote d'Ivoire &           164.364 &             38.204 &              2157 &     (80.289, 94.726) &    (12.043, 14.209) &       (1053, 1243) \\
Democratic Republic of the Congo &           453.291 &            104.500 &              5948 &   (147.597, 165.409) &     (22.14, 24.811) &       (1937, 2170) \\
                        Djibouti &             3.215 &              0.808 &                42 &       (1.585, 1.797) &       (0.238, 0.27) &           (21, 24) \\
               Equatorial Guinea &             8.241 &              1.912 &               108 &       (3.306, 3.715) &      (0.496, 0.557) &           (43, 48) \\
                         Eritrea &            15.418 &              3.636 &               202 &       (7.884, 9.183) &      (1.183, 1.377) &         (103, 120) \\
                        Eswatini &             8.058 &              1.960 &               105 &       (3.979, 4.628) &      (0.597, 0.694) &           (52, 60) \\
                        Ethiopia &           486.369 &            116.275 &              6383 &   (206.646, 232.389) &    (30.997, 34.858) &       (2712, 3050) \\
                           Gabon &            10.591 &              2.570 &               139 &       (4.327, 4.857) &      (0.649, 0.729) &           (57, 64) \\
                          Gambia &            12.756 &              2.985 &               167 &       (5.919, 6.834) &      (0.888, 1.025) &           (78, 90) \\
                           Ghana &           158.806 &             37.742 &              2084 &     (77.991, 91.232) &    (11.699, 13.685) &       (1024, 1197) \\
                          Guinea &            68.041 &             15.883 &               893 &     (22.118, 24.703) &      (3.318, 3.705) &         (291, 325) \\
                   Guinea-Bissau &             7.643 &              1.811 &               100 &       (3.508, 3.958) &      (0.526, 0.594) &           (46, 52) \\
                           Kenya &           318.218 &             75.191 &              4176 &    (164.178, 190.72) &    (24.627, 28.608) &       (2154, 2503) \\
                         Lesotho &            15.160 &              3.753 &               199 &       (7.024, 8.083) &      (1.054, 1.212) &          (92, 106) \\
                         Liberia &            28.154 &              6.582 &               370 &     (12.829, 14.912) &      (1.924, 2.237) &         (169, 196) \\
                      Madagascar &           127.544 &             29.717 &              1674 &     (56.444, 63.903) &      (8.467, 9.585) &         (741, 839) \\
                          Malawi &           166.239 &             38.545 &              2181 &     (78.392, 92.302) &    (11.759, 13.845) &       (1029, 1211) \\
                            Mali &           106.669 &             24.532 &              1400 &     (42.962, 48.675) &      (6.444, 7.301) &         (564, 639) \\
                      Mauritania &            29.569 &              6.893 &               388 &      (13.923, 15.83) &      (2.088, 2.375) &         (183, 208) \\
                       Mauritius &             2.706 &              0.676 &                35 &        (1.395, 1.62) &      (0.209, 0.243) &           (18, 21) \\
                         Mayotte &             2.152 &              0.506 &                28 &       (0.908, 1.046) &      (0.136, 0.157) &           (12, 13) \\
                      Mozambique &           300.967 &             69.591 &              3949 &    (130.585, 151.98) &    (19.588, 22.797) &       (1713, 1994) \\
                         Namibia &            17.691 &              4.287 &               232 &       (8.581, 9.816) &      (1.287, 1.472) &         (113, 129) \\
                           Niger &           109.446 &             24.464 &              1436 &     (49.729, 57.335) &        (7.459, 8.6) &         (652, 752) \\
                         Nigeria &          1159.397 &            269.397 &             15215 &     (400.67, 447.48) &    (60.101, 67.122) &       (5258, 5873) \\
                          Rwanda &            74.075 &             17.711 &               972 &     (35.425, 42.297) &      (5.314, 6.345) &         (464, 555) \\
                         R\'eunion &             2.570 &              0.633 &                33 &        (1.25, 1.419) &      (0.188, 0.213) &           (16, 18) \\
           Sao Tome and Principe &             1.206 &              0.281 &                16 &       (0.626, 0.727) &      (0.094, 0.109) &            (9, 10) \\
                         Senegal &            74.910 &             17.475 &               983 &     (39.103, 45.341) &      (5.865, 6.801) &         (513, 595) \\
                      Seychelles &             0.272 &              0.069 &                 4 &       (0.146, 0.171) &      (0.022, 0.026) &             (2, 2) \\
                    Sierra Leone &            40.570 &              9.763 &               532 &     (18.513, 21.626) &      (2.777, 3.244) &         (243, 284) \\
                         Somalia &            68.344 &             15.595 &               897 &     (19.689, 21.934) &       (2.953, 3.29) &         (258, 287) \\
                    South Africa &           338.967 &             83.555 &              4448 &   (149.607, 168.932) &     (22.441, 25.34) &       (1963, 2217) \\
                     South Sudan &            63.586 &             14.992 &               835 &      (17.184, 19.16) &      (2.578, 2.874) &         (226, 252) \\
                     Tanzania (United Republic of) &           390.524 &             89.390 &              5124 &   (186.192, 212.806) &    (27.929, 31.921) &       (2443, 2792) \\
                            Togo &            48.580 &             11.313 &               638 &     (21.538, 24.703) &      (3.231, 3.705) &         (283, 324) \\
                          Uganda &           330.125 &             78.227 &              4332 &    (162.81, 193.292) &    (24.421, 28.994) &       (2137, 2537) \\
                          Zambia &           110.407 &             25.320 &              1449 &      (49.302, 57.45) &      (7.395, 8.618) &         (647, 754) \\
                        Zimbabwe &           117.331 &             28.117 &              1540 &     (60.613, 69.218) &     (9.092, 10.383) &         (796, 908) \\
\bottomrule
\label{dalys}
\end{longtable}
}

\section{Methods}\label{sec11}

\subsection{Mathematical model}\label{model}

We developed an age and comorbidity structured model for the transmission of iNTS in sSA.% (\href{run:./iNTS_SI.pdf}{\emph{SI Appendix}, Fig. S1}). 
Our model uses a susceptible-infected-recovered framework \cite{KeelingRohani2011}, where the population is divided in compartments depending on their age and health status, and transitions between compartments over time (Fig. \ref{fig:model}). As iNTS transmission is age-dependent \cite{Stanaway2019, Feasey2010, Ao2015, Balasubramanian2019, Keddy2017, Bornstein2017}, we divide the population in $4$ age groups: 0-6 mos, 7-9 mos, 10-59 mos, $>59$ mos. Individuals in the first age group are assumed to be maternal immune \cite{MacLennan2008, Feasey2012, Lepage1984}. Consistent with the evidence that iNTS transmission depends on the presence of comorbidities \cite{Marchello22, Feasey2012, Park2016, Muthumbi2015, Scott2011, Gordon2011, Greenwood1972, MacLennan2010}  each age group is stratified according to their comorbidity status. For children up to 59 mos we consider both Malaria and HIV, while after 59 mos we only consider immunocompromised as at high risk of developing iNTS disease \cite{Feasey2012, Crump2015CMR}. Carriers have an important role in the transmission of iNTS as they contribute to spread the bacteria despite not being affected by the invasive disease \cite{Post2019, Kariuki2006, Koolman2022, Parsons2013}.  We assume that only healthy individuals above 59 mos can become carriers, while children 7-59 mos (irrespective of their comorbidity status), and those older than 59 mos with HIV can get the invasive disease upon exposure \cite{Muthumbi2015, Levine1991, Gordon2001, Gordon2002}. Model equations and parameters can be found in Supplementary Methods 1 and Supplementary Table 2. %\href{run:./iNTS_SI.pdf}

\renewcommand{\thefigure}{\textbf{Fig. 3 Diagram of the compartmental model.}}
\begin{figure}%[tbhp]
\centering
\includegraphics[width=.9\linewidth]{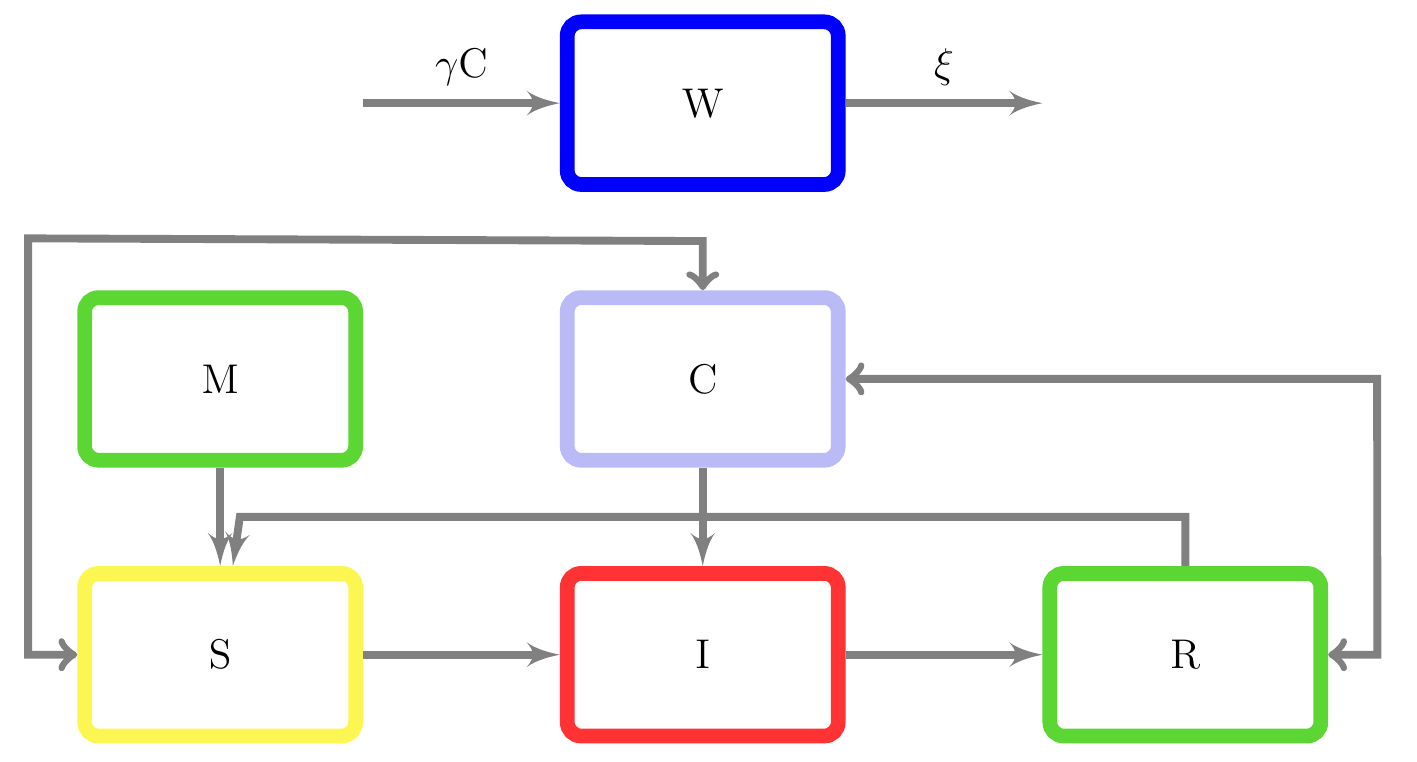}
\caption{Age and comorbidity groups not shown. Individuals are born into the maternal immunity compartment $M$, then transition to the susceptible compartment $S$ where they can become infected $I$ if they are below 5 y or above 5 and with HIV. After the infection they will transition to the recovered compartment $R$, where they are immune, and upon waning of immunity they transition back to $S$. Healthy adult can become carriers $C$, who do not acquire immunity upon recovery and transition back to $S$. As immunity from the invasive disease does not confer immunity from becoming a carrier, recovered adults can transition to $C$. Carriers spread bacteria in the water source $W$ at a rate $\gamma$ and they decay at rate $\xi$.}
\label{fig:model}
\end{figure}

\subsection{Data and calibration}\label{data}
Population pyramids and population growth rates are taken from the United Nations Population Division \cite{Data_UN}. We consider the medium fertility variant (median prediction interval) projections for the population growth. As population pyramids are structured in one year intervals, we use Sprague multipliers to estimate population pyramids structured in four age groups. Malaria reported confirmed cases are obtained from WHO Global Health Observatory Data Repository \cite{Data_WHO}, and the number of children 0-14 y.o. and adults 15-49 y.o. living with HIV is based on UNAIDS estimates \cite{Data_UNAIDS}. Comorbidity data refer to 2016, except a few cases where 2016 data were not available, for which we took the closest year available. For countries where data was not available we assumed they had the same comorbidity level of their most similar neighbour.

UN estimates reports the number of children 0-14 y.o. living with HIV, to compute the percentages of children living with HIV in age group 0-6 months, 7-9 months and 10-59 months, we assumed that the number of children living with HIV is uniformly distributed over age groups in the interval 0-14 years of age. Similarly we assume that Malaria cases are uniformly distributed among different age groups.
For the routine vaccination scenario we use the WHO-UNICEF estimation of DTP1 vaccine coverage for each country, choosing as reference year 2019 \cite{Data_WHOENIC}. For the catch-up vaccination scenario we use the same DTP1 coverage if this is greater than $90\%$, otherwise we assume a $90\%$ coverage for the one-year campaign. Coverage data is summarised in Supplementary Table 1.

We calibrate the age and comorbidity specific transmission parameters to reach a yearly prevalence level of 0.3$\%$ among children age 0-5 years for the aggregate sub-Saharan Africa, using maximum likelihood estimation with normally distributed error. We then simulate the transmission model for each country, using country specific population pyramid, comorbidities and vaccine coverage. We assume that transmission rate in presence of comorbidities is 3.5 times the transmission rate without comorbidities for that age group. We estimate the rate of progression among age groups using mortality rates and annual population growth rate.

\subsection{Intervention simulations}\label{data}
We simulate a one-year catch-up vaccination campaign among children between 9 months old and 5 years old, followed by a routine vaccination campaign at 9 months of age. We simulate vaccine efficacy between 85$\%$ and 95$\%$. We assume no waning of vaccine induced immunity as well as no protection against carriage, as currently there is no evidence regarding these two aspects. Furthermore, we assume that vaccine efficacy is not affected by Malaria or HIV status of the recipient.

\renewcommand{\thefigure}{\textbf{Fig. 4 Cumulative cases prevented, cumulative cases and yearly cases for sSA.}}
\begin{figure}[t]
\centering
\includegraphics[width=1\textwidth]{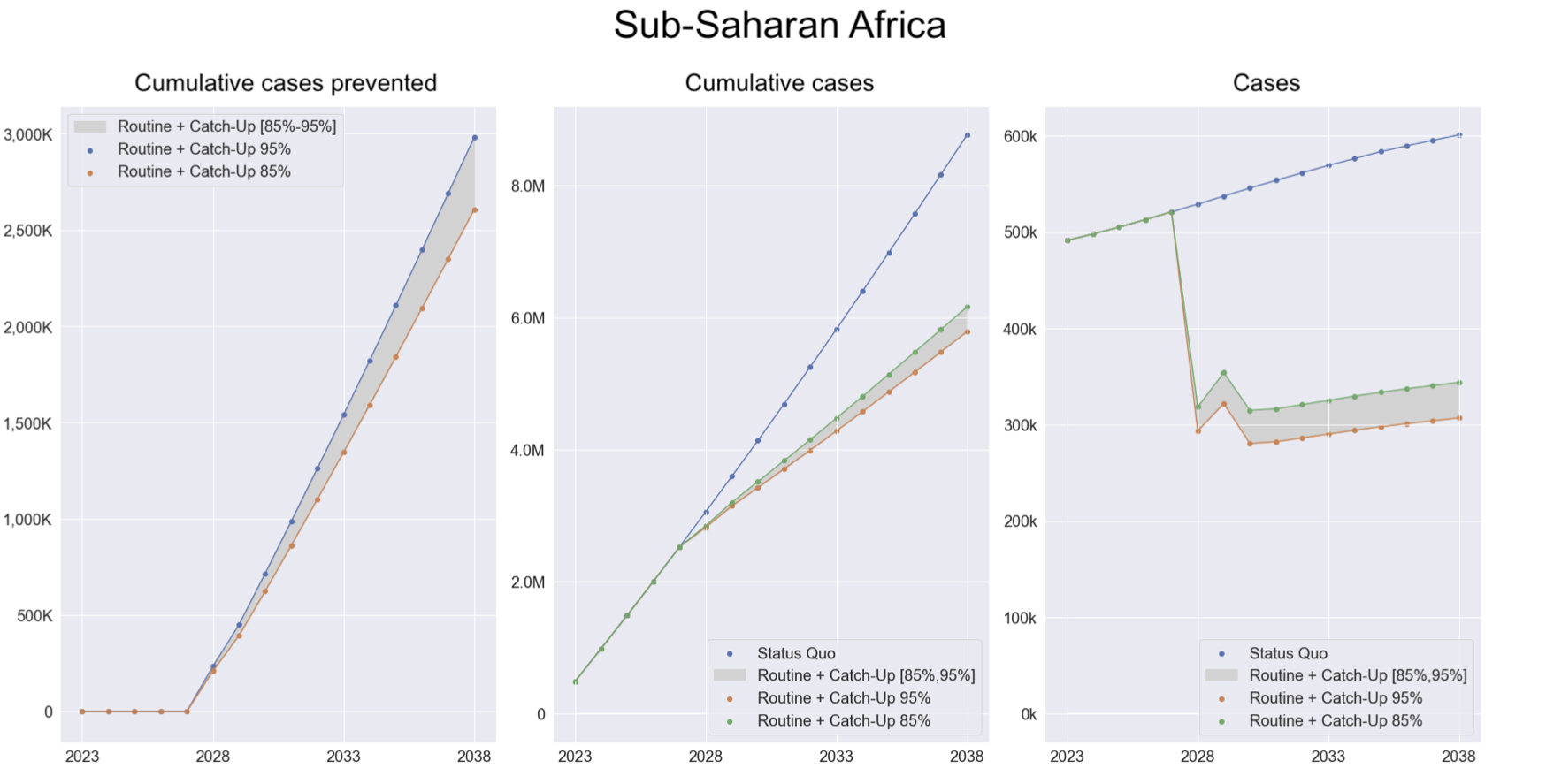}
\caption{Cumulative number of cases  prevented with a vaccination campaign over 10 years, starting with a one year catch-up followed by a 9-years routine vaccination over the period 2028-2038 (left). Cumulative number of cases  in the status quo scenario, without vaccination and in the vaccination scenario (centre). Yearly number of cases for the status quo and vaccination scenario (right). Vaccine efficacy is between $85\%$ and  $95\%$. Cases aggregated across countries in sSA.}\label{fig:perc}
\end{figure}

\section{Discussion}\label{sec12}

Without any health intervention in sSA our model calculated that the annual number of cases in children below 5 years of age will grow, resulting in an estimate cumulative number of cases in children below 5 years of age of 9.7 million by 2038. This huge number is partly due to the expected growth of the population below 5 years in sSA, this age group being the most affected by the disease. Vaccines are one of the most successful public health initiatives in eliminating or reducing the impact of infectious diseases \cite{Timmis2017}, having a great impact on human health and contributing to increase life expectancy and quality \cite{Who2013,Black2013}. Our model computed the impact of different immunization strategies considering different levels of vaccine efficacy. We also show that in evaluating the impact of a large scale vaccination campaign against iNTS is of key importance to focus on a country-level analysis (Supplementary Figures 10-58). %(\href{run:./iNTS_SI.pdf}{\emph{SI Appendix}, Figs. S10-S58}). 
Our analyses indicate that vaccination of children below 5 years of age could effectively and efficiently reduce iNTS burden in sSA. Different levels of vaccine coverage are the main factor behind differences in the reduction of cases following vaccination. For example South Sudan and Somalia, which have the lowest coverage rate for routine vaccination (51$\%$ and 52$\%$ respectively), show the lowest reduction in cumulate infections  (27$\%$-30$\%$ and 29$\%$-32$\%$ depending on vaccine efficacy). Increasing EPI coverage at 90$\%$ for both countries would reduce iNTS cases by (45$\%$-51$\%$) and (47$\%$-53$\%$) respectively. Differences in coverage are not the only factor determining reduction in cases following vaccination: consider Burundi and Senegal, they both have a coverage level of 97$\%$ (both for routine and catch-up) but the reduction in cumulate cases significantly less for Burundi [43$\%$ - 52$\%$] than for Senegal [52$\%$ - 60$\%$], difference that can be explained by the much higher number of Malaria cases among children in Burundi. This stresses the role of comorbidities in the diffusion of iNTS and consequently on the impact of vaccination (Supplemetary Table 3).% (\href{run:./iNTS_SI.pdf}{\emph{SI Appendix}}). 
A caveat must be raised when we compare different countries, given the uncertainty on country-specific incidence and given that our model does not take into account differences in water sanitation levels which can have a significant impact on the diffusion of the disease. Improving water sanitation would curb the diffusion of iNTS by reducing bacteria circulation, but implementing water sanitation improvement on a large scale is a more challenging goal, especially in the short-medium term, while a massive vaccination program could be more easily and quickly implemented and yet provide immediate and consistent benefits, as stressed by the reduction in DALYs. Furthermore the model does not take into account the effects that the COVID-19 pandemic might have had on the diffusion of iNTS in Africa, reducing personal contacts and improving life style, quality of sanitation and hygienic conditions. Although multidrug resistance is becoming a major concern for the epidemiology and treatment of \emph{Salmonella}, possible modification in the iNTS disease transmission due to the \emph{Salmonella} serovar antibiotic resistant were not taken in consideration in the model, most likely leading to a smaller estimate of the number of cases and deaths related to iNTS. Similarly, we also assumed no vaccine waning and no protection against carriage: currently information on the human responses to the vaccination are limited by availability, since the vaccine under study has been just launched in the clinical phases and study are currently in progress. 
In conclusion we evaluate the iNTS diffusion in sSA, by country and age class, considering endemic comorbidities,  and highlight  that the transmission of the iNTS disease among the most fragile age classes below 5 years of age in sSA will be increasing over the next 20 years without the introduction of a iNTS vaccine. By the simulation of different vaccination scenarios we identified the combination of immunization strategies needed to reduce as early as possible the burden of the iNTS disease reducing permanently its incidence. Strategies to address the iNTS burden in sSA include infection prevention, optimised use of antibiotics, improved capacity for microbiological analysis, and vaccine development. Our estimates can be used to help set priorities for vaccine need, demand and development. Overall, our findings bring to the conclusion that until safer sources of water and sanitization will not be widespread distributed and available for all population of sSA, number of cases and deaths will be increasing without the introduction of a vaccine. Vaccination against iNTS, being the faster prevention method and cost-effective medical measure, will be highly beneficial and should be prioritized as primary health intervention. Moreover, from a broader perspective our model offers a very useful tool to monitor the development of iNTS disease, to select the immunization strategies and to identify priority countries for vaccine introduction to adequately address iNTS disease. It also provides estimates on the impact of vaccine introduction at the country level and help us prioritize countries with high expected return on introduction of vaccines given the present prevalence of the disease and current vaccination coverage.

%\newpage
%\printbibliography
\bibliographystyle{acm}
\bibliography{pnas-sample}

\appendix
%dummy comment inserted by tex2lyx to ensure that this paragraph is not empty
\global\long\def\thesection{\Alph{section}}
 \global\long\def\thesubsection{\Alph{section}.\arabic{subsection}}
 \setcounter{figure}{0} \global\long\def\thefigure{{A}.\arabic{figure}}
  \setcounter{table}{0} \global\long\def\thetablee{{A}.\arabic{table}}
% \global\long\def\theexample{\Alph{example}}
% \setcounter{example}{0}

\section*{Supplementary Methods 1: Model specification}

The model stratifies the population of each country in  4 age groups: (1) 0-6 months, (2) 7-9 months, (3) 10-59 months, (4) $>59$ months and 5 compartments according to the disease status: maternal immune ($M$), susceptible ($S$), infected by the invasive disease ($I$), carrier of the disease ($C$) and recovered ($R$). All compartments except carriers are further stratified according to comorbidity status. Carriers are individuals who can get infected upon exposure and contribute to spread the bacteria but who will not get the invasive disease \cite{Gilchrist2019, Hansen2002, Okoro2012}.  We assume that only individuals older than 59 months without comorbidities can be carriers, while individuals in the first three age groups, both with and without comorbidities, and individuals older than 59 months with HIV can get the invasive disease upon exposure. This is consistent with evidence that children are the most at risk to get iNTS. 

We do not model the transmission dynamics of comorbidities (Malaria and HIV), instead we estimate the fraction of people with comorbidities in each age group and we assume it constant. Individuals are born maternally immune, so they enter the model in compartment M where they stay for their first 6 months of life, then transition into the susceptible compartment. The susceptible compartment  $S_{a,m}$ is stratified in age groups $a = \{2, 3, 4\}$ (first age group not included because of maternal immunity) and in comorbidity status $m = \{n,c\}$, where $n$ stands for \emph{no comorbidity} and $c$ for \emph{with comorbidity}. Susceptible individuals below 5 years old (both with or without comorbidities) and above 5 years old with comorbidities can become infected with force of infection $\lambda_{a,m}$ depending on their age group and comorbidity status.  Once infected, individuals remain in the corresponding infected compartment $I_{a,m}$ for 14 days, then transition into the recovered compartment $R_{a,m}$. Following natural immunity waning, individuals will transition from $R_{a,m}$ to the susceptible compartment $S_{a,m}$ at a rate $\phi$. Susceptible individuals older than 5 years of age and without comorbidities will transition to the carrier compartment $C_{4,n}$ at a rate $\lambda_{4,n}$. We assume that infected individuals do not transmit the infection while carrier individuals both transmit the infection and shed the bacteria at a rate $\gamma$ in the water supply.
We allow both recovered and vaccinated individuals, with and without comorbidities, to become carriers upon exposure, which corresponds to assuming that both natural immunity and vaccine induced immunity protects from the invasive disease, but do not prevent becoming carrier. The demographic of the population is modelled letting individual transition within their compartment from age group $i$ to age group $i+1$ at rate $K_{i,m}$ and die at rate $d_{i,m}$ where $m$ is comorbidity status, such that total population, age structure and comorbidity status remain constant \cite{Karachaliou2015}.
The transition between compartments is captured by a system of differential equations, that in the case without vaccination read as follows (where $\dot{x} = \frac{d x}{d t}$):

\begin{align}
\dot{M}_{1,m}& = q_m  - (d_{1,m} + K_{1,m})M_{1,m} 
%\nonumber \\
\nonumber \\ 
\dot{S}_{2,m} & =  K_{1,m} M_{1,m}  + \phi R_{2,m} - (\lambda_{2,m} + d_{2,m} + K_{2,m})S_{2,m}  
% \nonumber \\
\nonumber \\
\dot{S}_{3,m} & =  K_{2,m} S_{2,m}  + \phi R_{3,m}  - (\lambda_{3,m} + d_{3,m} + K_{3,m})S_{3,m}
%\nonumber \\ 
\nonumber \\
\dot{S}_{4,n} & =  K_{3,n} S_{3,n}  + \phi R_{4,n} + \psi CS_{4,n} - (\lambda_{4,n} + d_{4,n} )S_{4,n}
%\nonumber \\ 
\nonumber \\
\dot{S}_{4,c} & =  K_{3,c} S_{3,c}  + \phi R_{4,c} - (\lambda_{4,c} + d_{4,c} )S_{4,c}
%\nonumber \\ 
\nonumber \\
\dot{I}_{2,m} & =  \lambda_{2,m} S_{2,m} - (\rho + d_{2,m} +K_{2,m})  I_{2,m}  \ \  
%\nonumber  \\
\nonumber \\
\dot{I}_{3,m} & =  K_{2,m} I_{2,m} + \lambda_{3,m} S_{3,m}  - (\rho + d_{3,m} + K_{3,m})  I_{3,m} \ \  
% \nonumber  \\
\nonumber \\
\dot{I}_{4,n} & =  K_{3,n} I_{3,n}  - (\rho + d_{4,n})  I_{4,n} \ \   
%\nonumber  \\
\nonumber \\
\dot{I}_{4,c} & =  K_{3,c} I_{3,c} + \lambda_{4,c} S_{4,c} -  (\rho + d_{4,c})  I_{4,c} \ \   
%\nonumber \\
\nonumber  \\
\dot{R}_{2,m}& =  \rho I_{2,m}  -  (\phi + d_{2,m} + K_{2,m} )R_{2,m}  \ \ 
%\nonumber  \\
\nonumber \\
\dot{R}_{3,m}& =  K_{2,m} R_{2,m} + \rho I_{2,m}  -  (\phi + d_{2,m} + K_{2,m})R_{2,m}  \ \   
%\nonumber  \\
\nonumber \\
\dot{R}_{4,m} & =  K_{3,m} R_{3,m} + \rho I_{4,m} + \psi CR_{4,m} -  (\phi + d_{4,m}  + \lambda_{4,m})R_{4,m} 
%  \nonumber \\
\nonumber \\
%\frac{d R_{a,c}}{dt} & =   K_{1,-c} R_{1,-c} + \rho I_{2,-c} + \alpha C_{2,-c} - (\phi  + d_{2 ,-c}+ K_{2,-c}-\psi)R_{2,-c}+\zeta   R_{a,c}\\ 
\dot{CS}_{4,n}& =  \lambda_{4,n} S_{4,n}  - ( d_{4,n} + \psi)CS_{4,n}   
%\nonumber \\
\nonumber \\
\dot{CR}_{4,m}& =  \lambda_{4,m}  R_{4,m}   - (a_{4,m} +  d_{4,m} + \psi)CR_{4,m}  
%\nonumber \\ 
%\dot{d VI_{a,m}} & = v_a I_{a,m} + K_{a-1,m} VI_{a-1,m} - (d_{a,m} + K_{a,m} + h_a) VI_{a,m} \\
 \nonumber \\
 \dot{W} & = \gamma (CS_{4,n} + \sum_m CR_{4,m}) - \xi W  
 \end{align}

$W$ represents the contamination of the water supply due to the shedding of carriers, $q_m$ is birth rate for comorbidity status (we are assuming that individuals are born with or without comorbidities, as we do not explicitly model the dynamics of Malaria and HIV), $d_{a,m}$ are age and comorbidity specific death rates, $K_{a,m}$ transition rates between age classes, $\rho_{a,m}$ the natural recovery rate, $\phi$ rate of natural immunity loss. Finally $\xi$ is the rate at which bacteria are eliminated from the water supply. 

The force of infection for iNTS depends on seasonality, age and comorbidity status according to the equation:

\begin{equation}
\lambda_{a,m}  =  \beta_{a,m} \Bigl [CS_{4,n} + \sum_m CR_{4,m} + \Bigl (1 + f \cos \Bigl (\frac{g\pi}{365}(t - p) \Bigr) \Bigr )W \Bigr ]
\end{equation}

Where $CS$ and $CR$ are adult susceptible and recovered  individuals respectively who became carriers. The transmission rates $\beta_{a,m}$ capture that susceptibility differs because of age and comorbidities, and  the term in brackets before $W$ captures seasonal variation: $g$ is a cosine scaling, $p$ is a seasonal offset and $f$ the seasonal amplitude. Note that we assume homogeneous mixing between age groups.

% death rates
%
%\begin{equation}
%\begin{split}
%d_{1,m} & = \frac{q_m -  K_{1,m} age_{1,n}}{age_{1,n} } \\
%d_{2,m} & = \frac{K_{a-1,m} age_{a-1,m} - K_{a,m}age_{a,m}}{age_{a,m}} \quad a \in [2,3] \\
%d_{4,m} & = \frac{ K_{3,m}age_{3,m}}{age_{4,m}}
%\end{split}
%\end{equation}

%\begin{equation} 
%\omega(t)  = 1 + f \cos \Bigl(\frac{g\pi}{365}(t - p) \Bigr)
%\end{equation}

\noindent
In the model with vaccination the population is further stratified to include vaccine related compartments. In our simulation scenario we vaccinate either at 9 months (in the routine scenario), or between 9 months and 5 years (in the catch-up scenario), hence both the maternally immune $M$ and carrier  $C$ do not have their corresponding vaccinated compartment, even if we do keep track of vaccinated susceptibles and vaccinated recovered that become carriers. \ref{disegnino} shows the compartments and the transition between them.

\renewcommand{\thefigure}{\textbf{Supplementary Fig. 1 Model Compartments}}
\bigskip  
\begin{figure}[h!]
\begin{tikzpicture}[node distance = 3cm, auto]
\centering
    % Place nodes
    \node [input, name=input] {};
    \node [block, draw=green!80!red!80!, line width=1.5mm] (Mat) {M};
    \node [block, draw=yellow!90!brown!70!, line width=1.5mm, below = 1cm of Mat] (Sus) {S};
    \node [block, draw=green!80!red!80!, line width=1.5mm, below=1cm of Sus] (VS) {VS};
     \node [block,draw=red!80!, line width=1.5mm,  right =2cm of Sus] (Inf) {I};
      \node [block, draw=green!80!red!80!, line width=1.5mm, below=1cm of Inf] (VI) {VI};
     \node [block,draw=green!80!red!80!, line width=1.5mm, right =2cm of Inf] (Rec) {R};
      \node [block, draw=green!80!red!80!, line width=1.5mm, below=1cm of Rec] (VR) {VR};
     \node [block,draw=blue!80!gray!30!, line width=1.5mm, right=2cm of Mat] (Car) {C};
   %  \node [block,draw=green!80!red!80!, line width=1.5mm, above=1cm of Mat] (VM) {VM};
%      \node [block,draw=green!80!red!80!, line width=1.5mm, above=1cm of Car] (VC) {VC};
     \node [block,draw=blue!, line width=1.5mm, above=1.2cm of Car] (W) {W};

     %Place arrows
     
      % M --> S
      \path [line,draw=gray!, line width=0.8mm] (Mat)  ([shift={(0mm,0mm)}]Mat.south)-- ([shift={(0mm,0mm)}]Sus.north)(Sus);
     
     % M <--> VM
    %  \path [line,draw=gray!, line width=0.8mm][<->] (Mat)  ([shift={(0mm,0mm)}]Mat.north)-- ([shift={(0mm,0mm)}]VM.south)(VM);

     % S <--> I
      \path [line,draw=gray!, line width=0.8mm] (Sus)  ([shift={(0mm,0mm)}]Sus.east)-- ([shift={(-5mm,0mm)}]Inf.west)|- (Inf);
      
      % I --> R
       \path [line,draw=gray!, line width=0.8mm] (Inf)  ([shift={(0mm,0mm)}]Inf.east)-- ([shift={(0mm,0mm)}]Rec.west) (Rec);
       
       % S <--> C
\path [line,draw=gray!, line width=0.8mm][<->] (Sus)  -|([shift={(-5mm, 42.3mm)}]Sus.west)-- ([shift={(0mm,5mm)}]Car.north)-| (Car);
 
    % S <--> VS
\path [line,draw=gray!, line width=0.8mm][<->] (Sus)  ([shift={(0mm,0mm)}]Sus.south)-- ([shift={(0mm,0mm)}]VS.north) (VS);

     % C <--> I
 \path [line,draw=gray!, line width=0.8mm] (Car)  |-([shift={(0mm,0mm)}]Car.south)-- ([shift={(0mm,0mm)}]Inf.north) (Inf);
 
     % R <--> S
 \path [line,draw=gray!!, line width=0.8mm] (Rec)  -l([shift={(0mm,5mm)}]Rec.north)-- ([shift={(2mm,5mm)}]Sus.north)-l (Sus);
 
    % R <--> VR
  \path [line,draw=gray!, line width=0.8mm][<->] (Rec)  ([shift={(0mm,0mm)}]Rec.south)-- ([shift={(0mm,0mm)}]VR.north) (VR);
  
  % I <--> VI
  \path [line,draw=gray!, line width=0.8mm][<->] (Inf)  ([shift={(0mm,0mm)}]Inf.south)-- ([shift={(0mm,0mm)}]VI.north) (VI);
  
  % R <--> C
  \path [line,draw=gray!, line width=0.8mm, dashed][<->]  (Rec)  -l([shift={(5mm,0mm)}]Rec.east)-- ([shift={(54mm,0mm)}]Car.east)-l (Car);
  
  % VR <--> C
  \path [line,draw=gray!, line width=0.8mm, dashed][<->]  (VR)  -l([shift={(7mm,0mm)}]VR.east)-- ([shift={(56mm,0mm)}]Car.east)-l (Car);
  
  % VI <--> C
  \path [line,draw=gray!, line width=0.8mm, dashed][<->]  (VI)  -l([shift={(5mm,0mm)}]VI.east)-- ([shift={(5mm,0mm)}]Car.east)-l (Car);
  
  % VS <--> C
  \path [line,draw=gray!, line width=0.8mm, dashed][<->]  (VS)  -l([shift={(10mm,0mm)}]VS.east)-- ([shift={(-10mm,0mm)}]Car.west)-l (Car); 
  
  % C <-> VC
%  \path [line,draw=gray!, line width=0.8mm][<->]  (VC)  ([shift={(00mm,0mm)}]VC.south)-- ([shift={(0mm,0mm)}]Car.north)(Car); 
  
  % bacteria to -> W
  \coordinate[left=2cm of W] (bacteria);
  \path [line,draw=gray!, line width=0.8mm]  (bacteria) --node[midway,above]{$\gamma$C} (W);
  
  % W -> out
 \coordinate[right=2cm of W] (out);
  \draw[line,draw=gray!, line width=0.8mm] (W) --node[midway,above,sloped]{$\xi$} (out);
  
  % Births -> M
  \coordinate[left=2cm of Mat] (birth);
  \path[line,draw=gray!, line width=0.8mm] (birth) --node[midway,above,sloped]{$\mu$} (Mat);
  
  %Deaths
  
  %DVM
  %\coordinate[below left=0.6cm of VM] (dvm);
 % \path[line,draw=gray!, line width=0.5mm] (VM) --node[midway, above, sloped]{$\delta$} (dvm);
  
  %DM
   \coordinate[below left=0.6cm of Mat] (dm);
  \path[line,draw=gray!, line width=0.5mm] (Mat) --node[midway, above, sloped]{$\delta$} (dm);
  
   %DC
  \coordinate[below left=0.6cm of Car] (dc);
  \path[line,draw=gray!, line width=0.5mm] (Car) --node[midway, above, sloped]{$\delta$} (dc);
  
  %DS
  \coordinate[below left=0.6cm of Sus] (ds);
  \path[line,draw=gray!, line width=0.5mm] (Sus) --node[midway, above, sloped]{$\delta$} (ds);
  
  %DI
  \coordinate[below left=0.6cm of Inf] (di);
  \path[line,draw=gray!, line width=0.5mm] (Inf) --node[midway, above, sloped]{$\delta$} (di);
  
  %DR
  \coordinate[below left=0.6cm of Rec] (dr);
  \path[line,draw=gray!, line width=0.5mm] (Rec) --node[midway, above, sloped]{$\delta$} (dr);
  
  %DVS
  \coordinate[below left=0.6cm of VS] (dvs);
  \path[line,draw=gray!, line width=0.5mm] (VS) --node[midway, above, sloped]{$\delta$} (dvs);
  
  %DVI
  \coordinate[below left=0.6cm of VI] (dvi);
  \path[line,draw=gray!, line width=0.5mm] (VI) --node[midway, above, sloped]{$\delta$} (dvi);
  
  %DVR
  \coordinate[below left=0.6cm of VR] (dvr);
  \path[line,draw=gray!, line width=0.5mm] (VR) --node[midway, above, sloped]{$\delta$} (dvr);
  
  %DVC
%  \coordinate[below left=0.6cm of VC] (dvc);
%  \path[line,draw=gray!, line width=0.5mm] (VC) --node[midway, above, sloped]{$\delta$} (dvc);
 
\end{tikzpicture}
\caption{Scheme of the compartmental model (age structure and comorbidities not shown) with vaccination. The green boxes (Maternal Immune, Recovered and Vaccinated) identify groups that are immune to the invasive disease. Notice that we assume that neither the vaccine nor immunity after recovering prevent from becoming carrier if exposed to the bacteria. (The lines are dashed as only adults can become carriers).}\label{disegnino}
\end{figure}
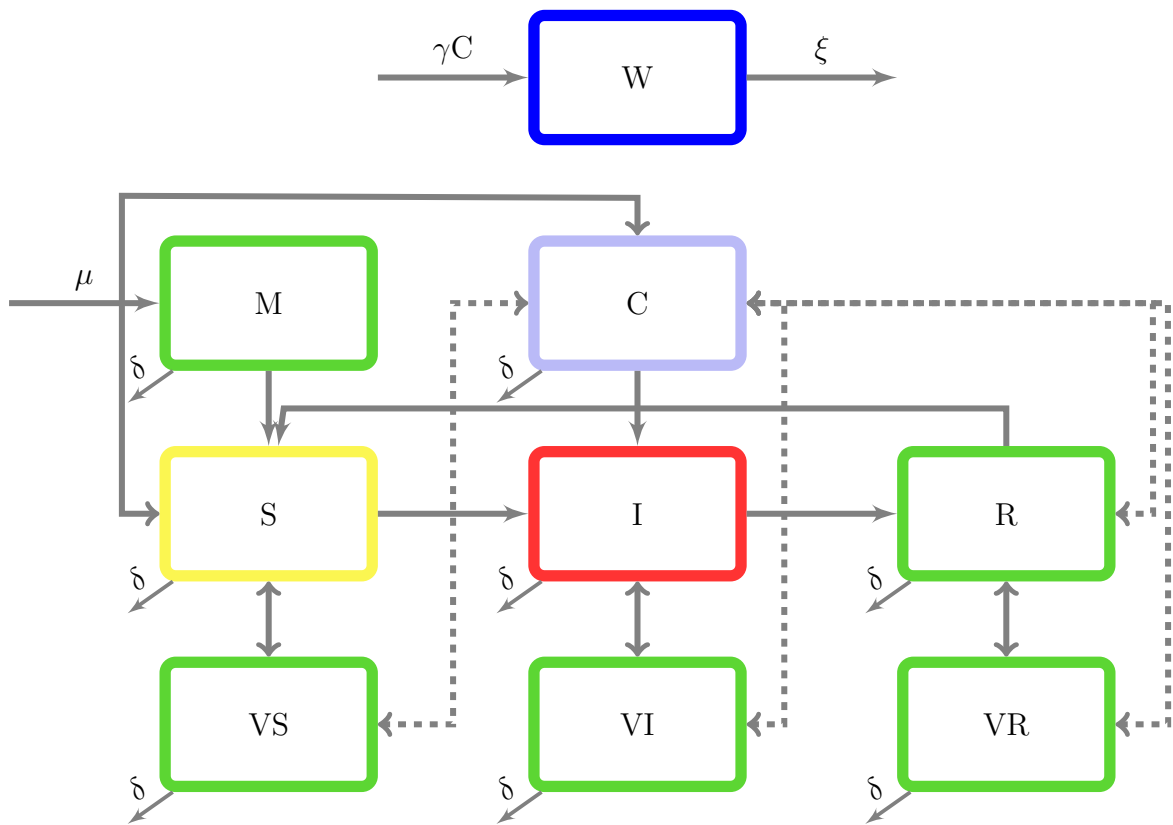

\newpage
When we introduce vaccination, the transmission model is governed by the following set of differential equations: 

\begin{align}
\dot{M}_{1,m} & = q_m  - (d_{1,m} + K_{1,m})M_{1,m} 
%\nonumber \\
\nonumber \\ 
\dot{S}_{2,m} & =  K_{1,m} M_{1,m}  + \phi R_{2,m} - (\lambda_{2,m} + d_{2,m} + K_{2,m} )S_{2,m}  
%\nonumber \\
\nonumber \\
\dot{S}_{3,m} & =  K_{2,m} S_{2,m}  + \phi R_{3,m}  - (\lambda_{3,m} + d_{3,m} + K_{3,m} - v)S_{3,m}
%\nonumber \\ 
\nonumber \\
\dot{S}_{4,n} & =  K_{3,n} S_{3,n}  + \phi R_{4,n} + \psi CS_{4,n} - (\lambda_{4,n} + d_{4,n} )S_{4,n}
%\nonumber \\ 
\nonumber \\
\dot{S}_{4,c} & =  K_{3,c} S_{3,c}  + \phi R_{4,c} - (\lambda_{4,c} + d_{4,c} )S_{4,c}
%\nonumber \\ 
\nonumber \\
\dot{I}_{2,m} & =  \lambda_{2,m} S_{2,m} - (\rho + d_{2,m} +K_{2,m} )  I_{2,m}  \ \  
%\nonumber  \\
\nonumber \\
\dot{I}_{3,m} & =  K_{2,m} I_{2,m} + \lambda_{3,m} S_{3,m}  - (\rho + d_{3,m} + K_{3,m} -v)  I_{3,m} \ \  
%\nonumber  \\
\nonumber \\
\dot{I}_{4,n} & =  K_{3,n} I_{3,n}  - (\rho + d_{4,n})  I_{4,n} \ \   
%\nonumber  \\
\nonumber \\
\dot{I}_{4,c} & =  K_{3,c} I_{3,c} + \lambda_{4,c} S_{4,c} -  (\rho + d_{4,c})  I_{4,c} \ \   
%\nonumber \\
\nonumber  \\
\dot{R}_{2,m} & =  \rho I_{2,m}  -  (\phi + d_{2,m} + K_{2,m})R_{2,m}  \ \ 
%\nonumber  \\
\nonumber \\
\dot{R}_{3,m} & =  K_{2,m} R_{2,m} + \rho I_{2,m}  -  (\phi + d_{2,m} + K_{2,m} - v)R_{2,m}  \ \   
%\nonumber  \\
\nonumber \\
\dot{R}_{4,m} & =  K_{3,m} R_{3,m} + \rho I_{4,m} + \psi CR_{4,m} -  (\phi + d_{4,m}  + \lambda_{4,m})R_{4,m}   
%\nonumber \\
\nonumber \\
%\frac{d R_{a,c}}{dt} & =   K_{1,-c} R_{1,-c} + \rho I_{2,-c} + \alpha C_{2,-c} - (\phi  + d_{2 ,-c}+ K_{2,-c}-\psi)R_{2,-c}+\zeta   R_{a,c}\\ 
\dot{CS}_{4,n}& =  \lambda_{4,n} S_{4,n}  - ( d_{4,n} + \psi)CS_{4,n}   
%\nonumber \\
\nonumber \\
\dot{CR}_{4,m}& =  \lambda_{4,m}  R_{4,m}   - (a_{4,m} +  d_{4,m} + \psi)CR_{4,m} 
% \end{align}
% 
%\begin{align}
\nonumber \\
\dot{CVS}_{4,m}& =  \lambda_{4,n} VS_{4,m}  - ( d_{4,m} + \psi)CVS_{4,m}  
% \nonumber \\
\nonumber \\
\dot{CVR}_{4,m}& =  \lambda_{4,m}  VR_{4,m}   - (a_{4,m} +  d_{4,m} + \psi)CVR_{4,m} 
% \nonumber \\ 
%\dot{d VI_{a,m}} & = v_a I_{a,m} + K_{a-1,m} VI_{a-1,m} - (d_{a,m} + K_{a,m} + h_a) VI_{a,m} \\
 \nonumber \\
 \dot{VS}_{3, m}& =  v S_{3,m}  - (d_{3, m}  +\lambda_{3,m})  VS_{3, m} 
  \nonumber \\
 \dot{VS}_{4, m}& =  K_{3,m} VS_{3,m}  - (d_{4, m}  + \lambda_{4,m})  VS_{4, m} 
 %\nonumber \\ 
% \nonumber \\
% \dot{VI}_{3, m}& =  v I_{3,m} - (d_{3, m} + \lambda_{3,m})  VI_{3, m}
 % \nonumber \\ 
 \nonumber \\
 \dot{VR}_{3, m}& =  v R_{3,m}  - (d_{3, m} + \lambda_{3,m})  VR_{3, m} 
 \nonumber \\
 \dot{VR}_{4, m}& =  K_{3,m} VR_{3,m}  - (d_{4, m} + \lambda_{4,m})  VR_{4, m} 
 %\nonumber \\ 
  \nonumber \\
 \dot{W} & = \gamma (CS_{4,n} + \sum_m CR_{4,m} + \sum_m CVS_{4,m} + \sum_m CVR_{4,m} ) - \xi W 
 \end{align}

where $v$ captures country-specific coverage rates. $CVS$ and $CVR$ are adult vaccinated susceptible (recovered) individuals who become carriers: they are immune to the invasive disease but not to carriage. $VS$ and $VR$ are vaccinated susceptible and recovered individuals respectively. We model two vaccination scenarios, routine and catch-up. In the routine scenario we assume that iNTS vaccination follows the EPI schedule set by WHO, with the first vaccination either at week 6, 10 or 14 and the second vaccination at 9 months. For simplicity we assume that the first vaccination does not induce any immunity, so it is only the second dose at 9 months gives coverage: as soon as individuals enter age group 3 they are vaccinated, and so removed from the susceptible compartment. For the catch-up scenario, we assumed one dose only, vaccinating children below 5 years of age.

\renewcommand{\thetable}{\textbf{Supplementary Table 1 Vaccine Coverage (DTP1)}}
\newpage
\begin{table}[h!]
\centering
\caption{Country specific vaccine coverage for routine (DTP1) and catch-up.}
\medskip
%\begin{longtable}{lrr}
\begin{adjustbox}{width=1\textwidth}
\begin{tabular}[t]{lrr}
\toprule
                          Country &  Coverage DTP1  &  Coverage catch-up \\
\midrule
                           Angola &               67 &                 90 \\ 
                          Benin &               84 &                 90 \\
                          Botswana &               98 &                 98 \\
                          Burkina Faso &               95 &                 95\\
                          Burundi &               97 &                 97 \\
                          Cabo Verde &               96 &                 96\\
                          Cameroon &               75 &                 90 \\
                          CAR &               69 &           90 \\
                          Chad &               65 &                 90\\
                           Comoros &               96 &                 96\\
                           Congo &               82 &                 90\\
                          C\^ote d'Ivoire &               98 &                 98 \\
                           DRC &               66 &                 90 \\
                           Djibouti &               90 &                 90 \\
                           Equatorial Guinea &               77 &                 90 \\
                           Eritrea &               97 &                 97 \\
                           Eswatini &               96 &                 96 \\
                            Ethiopia &               80 &                 90 \\
                            Gabon &               77 &                 90 \\
                            Gambia &               93 &                 93 \\
                             Ghana &               97 &                 97 \\
                            Guinea &               62 &                 90 \\
                            Guinea-Bissau &               85 &                 90 \\
                            Kenya &               97 &                 97 \\
                            Lesotho &               92 &                 92 \\
                             Liberia &               94 &                 94 \\
                             \bottomrule
\end{tabular}
\begin{tabular}[t]{lrr}
\toprule
                          Country &  Coverage DTP1  &  Coverage catch-up \\
\midrule

                              Madagascar &               85 &                 90 \\
                              Malawi &               97 &                 97 \\
                              Mali &               82 &                 90 \\
                              Mauritania &               89 &                 90\\
                              Mauritius &               97 &                 97 \\
                              Mayotte &               90 &                 90 \\
                               Mozambique &               93 &                 93  \\
                                Namibia &               92 &                 92 \\ 
                                Niger &               92 &                 92 \\
                                R\'eunion &               90 &                 90 \\
                                 Nigeria &               65 &                 90 \\
                                  Rwanda &               99 &                 99 \\
                                  Sao Tome &               97 &                 97 \\
                                  Senegal &               97 &                 97 \\
                                   Seychelles &               99 &                 99 \\
                                   Sierra Leone &               95 &                 95 \\
                                   Somalia &               52 &                 90 \\
                                   South Africa &               84 &                 90 \\
                                   South Sudan &               51 &                 90 \\
                              Togo &               90 &                 90 \\
                              Uganda &               99 &                 99 \\
                              UR of Tanzania &               91 &                 91\\ 
                              Zambia &               94 &                 94 \\
                              Zimbabwe &               94 &                 94 \\
                              & & \\
               SUB-SAHARAN AFRICA &          92 &                 92 \\
\bottomrule
\end{tabular}
\end{adjustbox} 
\label{tab: coverage}
\end{table}

%\section*{Supplementary Methods 2: Parameters choice}

%We calibrate the age and comorbidity specific transmission parameters to reach a yearly prevalence level of 0.3$\%$ among children age 0-5 years for the aggregate sub-Saharan Africa, using maximum likelihood estimation with normally distributed error. We then simulate the transmission model for each country, using country specific population pyramid, comorbidities and vaccine coverage. The following table summarises all model parameters. Note that we assume that transmission rate in presence of comorbidities is 3.5 times the transmission rate without comorbidities for that age group. We estimate the rate of progression among age groups using mortality rates and annual population growth rate.

\renewcommand{\thetable}{\textbf{Supplementary Table 2 Model Parameters}}
\begin{table}[h!]
\caption{Parameters choice and their justification.}
\medskip
      \begin{tabular}{lcc}
        \hline
            \textbf{Description}  & \textbf{Value}  & \textbf{Justification} \\ 
            \hline
        Transmission rate $0-6$  months, no comorbidity &$\beta_{1,n} = 0 $ & Assumption\\
       Transmission rate $0-6$  months, comorbidity &$\beta_{1,c} =  0$  & Assumption\\
        Transmission rate $7-9$  months, no comorbidity &$\beta_{2,n} = 0.00022 $  & Calibrated \\
       Transmission rate $7-9$  months,  comorbidity& $\beta_{2,c} = 0.00077$ & Assumption\\
       Transmission rate $10-59$  months, no comorbidity &$\beta_{3,n} = 0.00022 $ & Calibrated\\
       Transmission rate $10-59$  months, comorbidity &$\beta_{3,c} =  0.00077  $  & Assumption\\
       Transmission rate $>  59$  months, no comorbidity &$\beta_{4,n} = 0.0082$  & Calibrated \\
        Transmission rate $>  59$  months,  comorbidity& $\beta_{4,c} = $ 0.0290 & Assumption\\
         \hline
         Duration of infection for Carrier &$\psi = 20$ days & Assumption\\
        % Duration of invasive infection, all ages, no comorbidity &1/$\rho_{n} = 14$ days & Assumption\\
      Duration of invasive infection, all ages &$1/\rho = 14$ days & Assumption\\
      \hline
      Duration of natural immunity &$1/\phi = 7$ years & Assumption\\
      Waning rate of vaccine induced immunity  & No waning & Assumption\\
      \hline
      Survival time of bacteria in water &$1/\xi = 50$ days & \cite{Moore}\\
      Rate of shedding of bacteria into water supply &$\gamma = 1$ infectious unit p.w. & Assumption\\
        \hline
        Birth rate &$q = 0.0363$ yearly & Data \cite{Data_UN}\\
        \hline
        Peak of seasonal forcing &$1/p = 34$ weeks & Assumption\\
        Seasonal forcing &$f = 0.1$ & Assumption\\
         Cos scaling &$g = 2.1$ & Assumption\\
         Rate of progression between age groups & Age-specific & Calibrated \\ 
         
	\label{par}
      \end{tabular}
\end{table}

\clearpage
\section*{Supplementary Methods 2: Sensitivity Analysis}

To assess the impact of the main parameters on the model output (the yearly number of infected children below 5 y.o.)  we conducted global sensitivity analysis using two methods: the extended Fourier amplitude sensitivity test (eFAST) \cite{Saltelli2007}, which decomposes the variance of the output into the contribution of each parameter, and the partial rank correlation coefficient (PRCC) \cite{Marino2008}, a sampling method that measures the independent effect of each parameter on the number of cases. We implement eFAST sensitivity analysis using SALib library \cite{Herman2017}, generating parameters' samples using a revised  Saltelli sampling \cite{Saltelli2007}.  We run 16.366 experiments for the aggregate sub-Saharan Africa and compute the first order index (S1), which captures the contribution to the variance of the output of each parameter averaged over variation of the other parameters and the total order index (ST), which captures the contribution of a parameter and of its interactions with other parameters to the variance of the output. We choose to perform sensitivity analysis on 7 parameters: the transmission rates of children and adult without comorbidities, seasonal forcing, the carrier and infected recovery rates,  the waning rate of bacteria from water and the waning rate of natural immunity. Samples are drawn from the uniform distributions over (0.00020,0.00024) for $\beta_{2,n}$, (0.0080,0.00084) for $\beta_{4,n}$, (0.05,0.1) for $\rho$, (0.0485, 0.055) for $\psi$, (0.018,0.022) for $\xi$, (0.00027, 0.00054)  for $\phi$ and (0.088,0.12) for $f$.

\renewcommand{\thefigure}{\textbf{Supplementary Fig. 2 eFAST sensitivity analysis}}
\begin{figure}
\centering
\includegraphics[width = 0.8\textwidth]{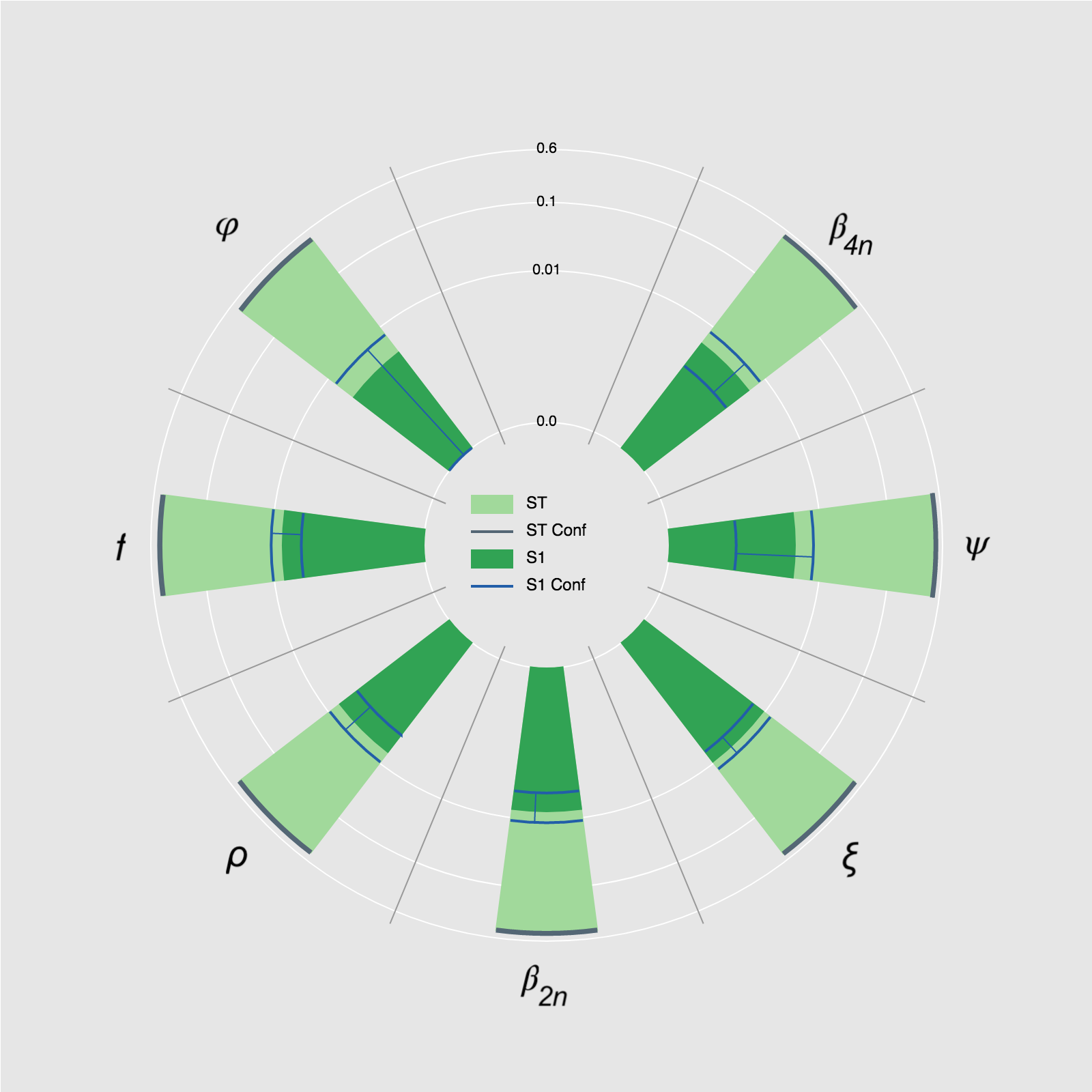}
\caption{eFAST sensitivity analysis: S1 captures the contribution of each parameter alone averaged over variation of other input parameters, ST captures the contribution of each parameter including the effect that its interaction has with all the other parameters. Confidence intervals at 95$\%$}
\label{eFAST}
\end{figure}

The parameters showing highest first order index are the time it takes for bacteria to decay in the water supply $\xi$ and the below 5 y.o. transmission parameter $\beta_{2,n}$, followed by forcing $f$ and over 5 y.o. transmission parameter $\beta_{4,n}$, with the latter showing slightly larger confidence bounds. Each parameter, when considered together with all others, contributes significantly to the output variance as can be seen by the high total order indices and the narrow confidence bounds.

\renewcommand{\thefigure}{\textbf{Supplementary Fig. 3 PRCC sensitivity analysis}}
\begin{figure}\label{prcc}
\centering
\includegraphics[width = 0.8\textwidth]{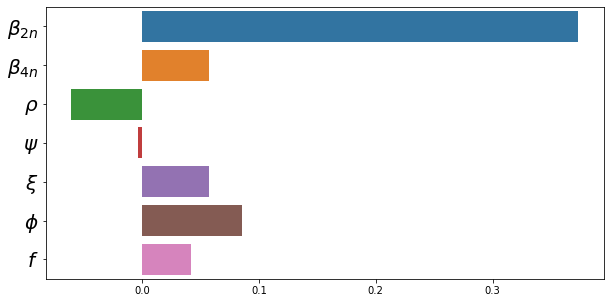}
\caption{Sensitivity of iNTS to model parameters, measured by partial rank correlation coefficients.}
\end{figure}

PRCC helps giving a clearer picture of the role of parameters in the model. The graph in \ref{eFAST} shows that the most important parameter in determining the number of infected is $\beta_{2,n}$, while forcing is less important than the eFAST analysis would suggest: this tells us that the high impact of $f$ on the output is explained mostly through its role in directly increasing children infections, as it acts as a rescaling factor for $\beta_{2,n}$, more than through its impact on the number of carriers. This is confirmed by a high interaction between $f$ and $\beta_{2,n}$,  and a weak interaction between $f$ and $\beta_{4,n}$ as captured by second order Sobol indices (not reported here). The parameters that follows in order of importance are $\phi$ and $\rho$: a shorter duration of natural immunity (higher $\phi$)  increases the number of paediatric infections and a shorter duration of the infection period (higher $\rho$) decreases them. Finally  $\beta_{4,n}$ and  $\xi$ show roughly the same effect on the output.

%In addition we also run 1,000 experiments for each country separately, letting only the transmission rates $\beta_{2,n}$ and  $\beta_{4,n}$ vary over the above specified intervals, and we report the $25^{th}$ and $75^{th}$ percentile of these in the following section.

\newpage
\section*{Supplementary Results 1 Different vaccination scenarios and factors driving cases reduction}

\ref{cp} and \ref{da}  illustrate the reduction in cases and deaths respectively for each country, under the scenario of a routine and catch-up campaign with $95\%$ vaccine efficacy. There are considerable differences across countries, mostly attributable to their different EPI vaccine coverage levels. This is better captured in \ref{perc}, which reports the reduction in cumulative iNTS cases in the period 2028-2038 in percentage terms, to make it independent of the population size. The colour of each bar corresponds to a coverage level in that country, as captured by the colormap on the righthand side of  the graph. There is a clear correspondence between low EPI coverage (below 60$\%$) and lower reduction in cases prevented. To clarify this point even further, we simulate what would happen if the routine coverage level would be at least $90\%$ in any country. The bottom graph in \ref{perc} shows how in almost all countries there is a reduction in cases of at least 50\%. Differences among countries in this scenario are then due to differences in population pyramids and comorbidities, and marginally to residual difference in catch-up coverage levels. \ref{causes} reports the dependence (simple OLS regression) of cumulate percentual reduction in cases (status quo versus routine + catch up vaccination with $95\%$ efficacy) over malaria incidence, percentage of children below 5 years old, percentages of children with HIV and routine coverage: once controlling for coverage, a younger population and higher levels of comorbidities explain lower percentual reduction in cases post vaccination.

\renewcommand{\thetable}{\textbf{Supplementary Table 3 Vaccine efficacy causes}}
\begin{table}[htb]
    \centering
\sisetup{table-number-alignment = center, % <-- added/changed
         table-space-text-pre ={(},
         table-space-text-post={\textsuperscript{***}},
         input-open-uncertainty={[},
         input-close-uncertainty={]},
         table-align-text-pre=false,
         table-align-text-post=false}
\begin{threeparttable}
    \caption{Vaccine efficacy causes}
    \label{causes}
\begin{tabular}{r 
                S[table-format=-2.3] % <-- adopted to number of digits in numbers in cells
                S[table-format=-1.4] % <-- adopted ...
                S[table-format=-1.5] % <-- adopted ...
                 }
\toprule
& \multicolumn{3}{c}{Percentage reduction in cases}    \\
\midrule
Intercept        &  0.0048   & (0.018)     \\

\midrule 
Malaria  &   -0.1935***\tnote{}            &  (0.017)  \\
Children below 5  y.o.   &          -0.1348*       &    (0.067)          \\
Children with HIV  &        -0.3854**         &  (0.155)             \\
Coverage DTP1    &        0.6274***         &  (0.014)            \\
\midrule
Observations        & {49}         &       \\
R-squared        & 0.983           &         \\
Joint significance (p-value F-statistics) & 0.00 &    \\
\bottomrule
\end{tabular}
    \smallskip
    \footnotesize
standard error in parentheses\par
\begin{tablenotes}[para,flushleft]
    \item[*]    $p < 0.10$,
    \item[**]   $p < 0.05$,
    \item[***]  $p < 0.01$
    \end{tablenotes}\par
\end{threeparttable}
\end{table}

%\autoref{yc} and \autoref{yd} report the average yearly number of cases and deaths respectively for all countries in SSA under three scenarios: no intervention (the blue line), only routine vaccination (the orange line) and routine + catch-up vaccination (the green line). Up to 2027 the three lines coincide; in 2028 it can be seen the effect of the catch-up campaign, and how that yields an additional reduction in the number of cases compared with the scenario with routine only. Note that as we take yearly averages, the oscillation due to seasonality cannot be seen in these graphs.

We also simulated a vaccination scenario with only a catch-up campaign, and one with only a routine campaign, and both prove to be  less effective than a catch-up plus routine campaign, as shown in \ref{c3scenarios} and \ref{allsSA}, which report the cumulative cases prevented for the entire sSA under each scenarios, for vaccine efficacy between 85\% and 95\%. A catch-up only campaign is clearly the least effective, confirming that in absence of other interventions, like water sanitation, the only way to keep paediatric cases and deaths under control is through sustained routine vaccination. %Finally we report the cumulated cases, deaths and DALYs for each country over the period 2028-2038, as well as the cumulated cases prevented, deaths and DALYs averted under a routine + catch-up campaign with vaccine efficacy between $85\%$ and $95\%$ \ref{dalys}.

%\autoref{cprc} and \autoref{dprc} report the yearly average number of cases prevented and deaths averted under the scenario routine + catch-up. It is evident that both cases prevented and deaths averted increase as the population grows. Comparing the two figures the effect of an initial catch-up campaign in 2028 can be clearly seen.
%Under the model assumption the prevalence level pre-vaccination is going to be different because of differences in population pyramids and comorbidities. Vaccination is also going to have a different impact due to these differences and to the different coverage levels. \autoref{ranking} clarifies these points showing the ranking of SSA countries per infection levels pre-vaccination and post vaccination.

\renewcommand{\thefigure}{\textbf{Supplementary Fig. 4 Cases in 2021}}
\begin{figure}
\centering
\includegraphics[width = 0.8\textwidth]{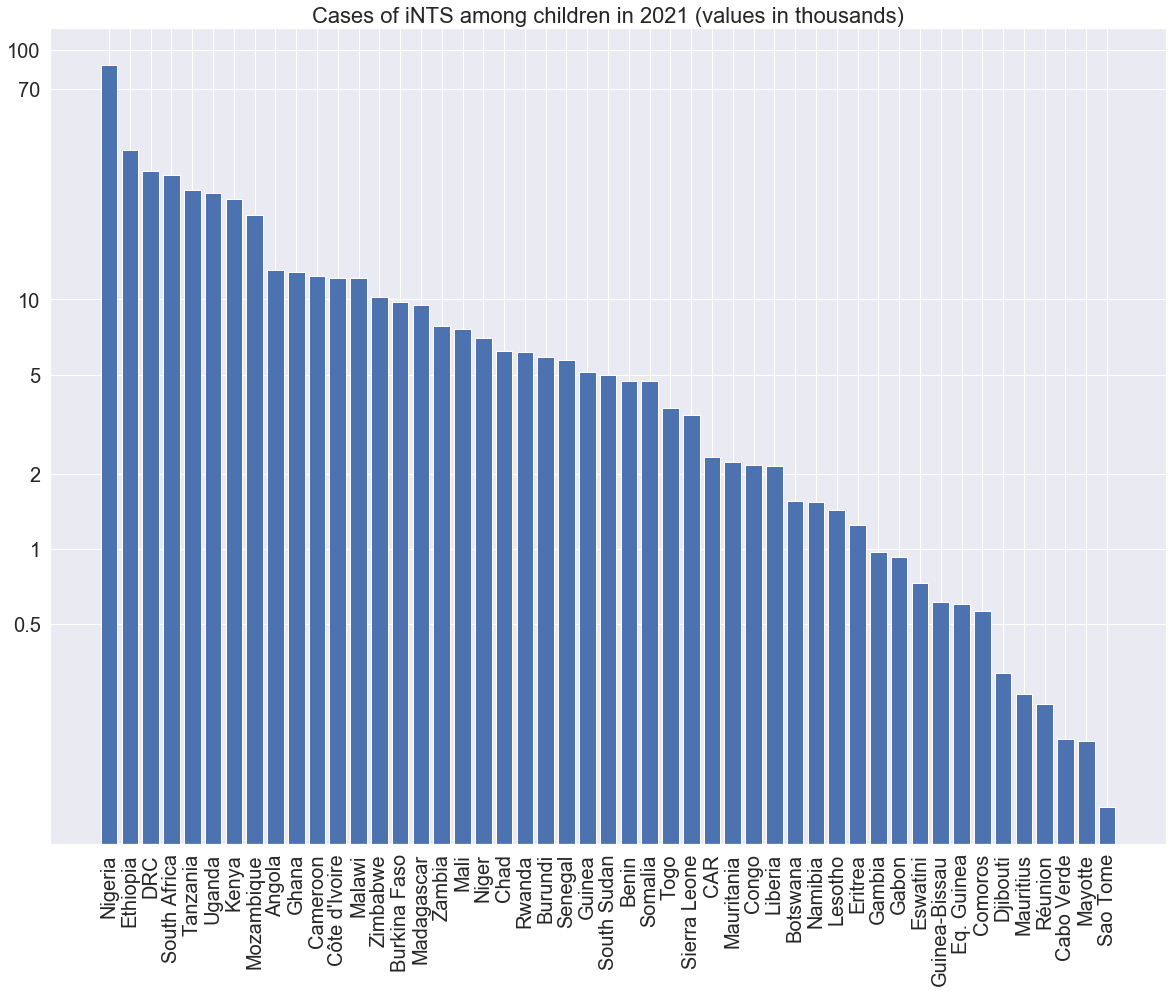}
\caption{Yearly number of cases in 2021 for each SSA Country, where the $y$-axis is in logarithmic scale.}
\label{sqln}
\end{figure}
\renewcommand{\thefigure}{\textbf{Supplementary Fig. 5 Cumulative cases up to 2038}}
\begin{figure}
\centering
\includegraphics[width = 0.8\textwidth]{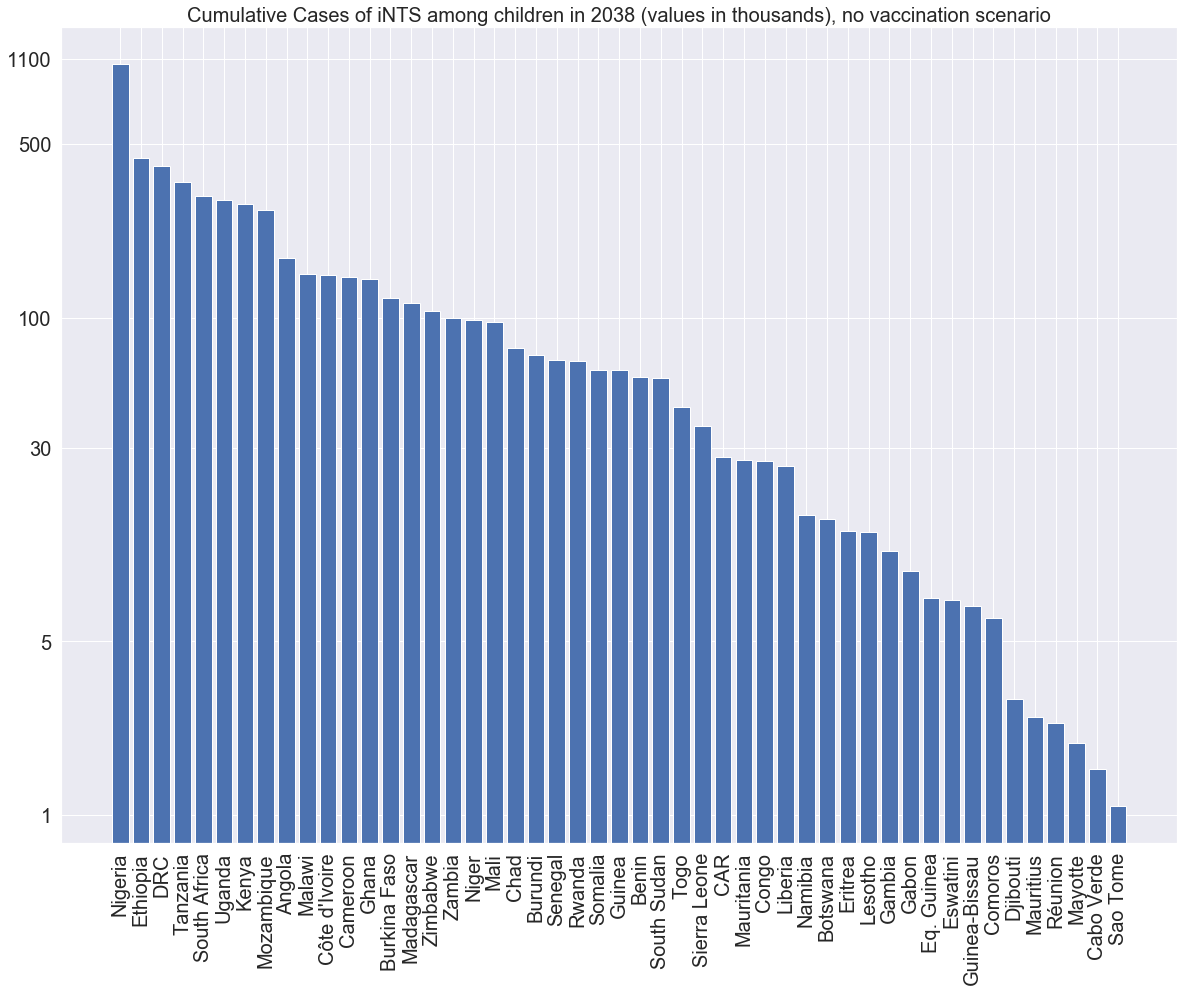}
\caption{Cumulative iNTS cases among children below 5 years old up to 2038 for each country, where the $y$-axis is in logarithmic scale.}
\label{cnvl}
\end{figure}
\renewcommand{\thefigure}{\textbf{Supplementary Fig. 6 Cumulative cases prevented 2028-2038}}
\begin{figure}
\centering
\includegraphics[width = 0.8\textwidth]{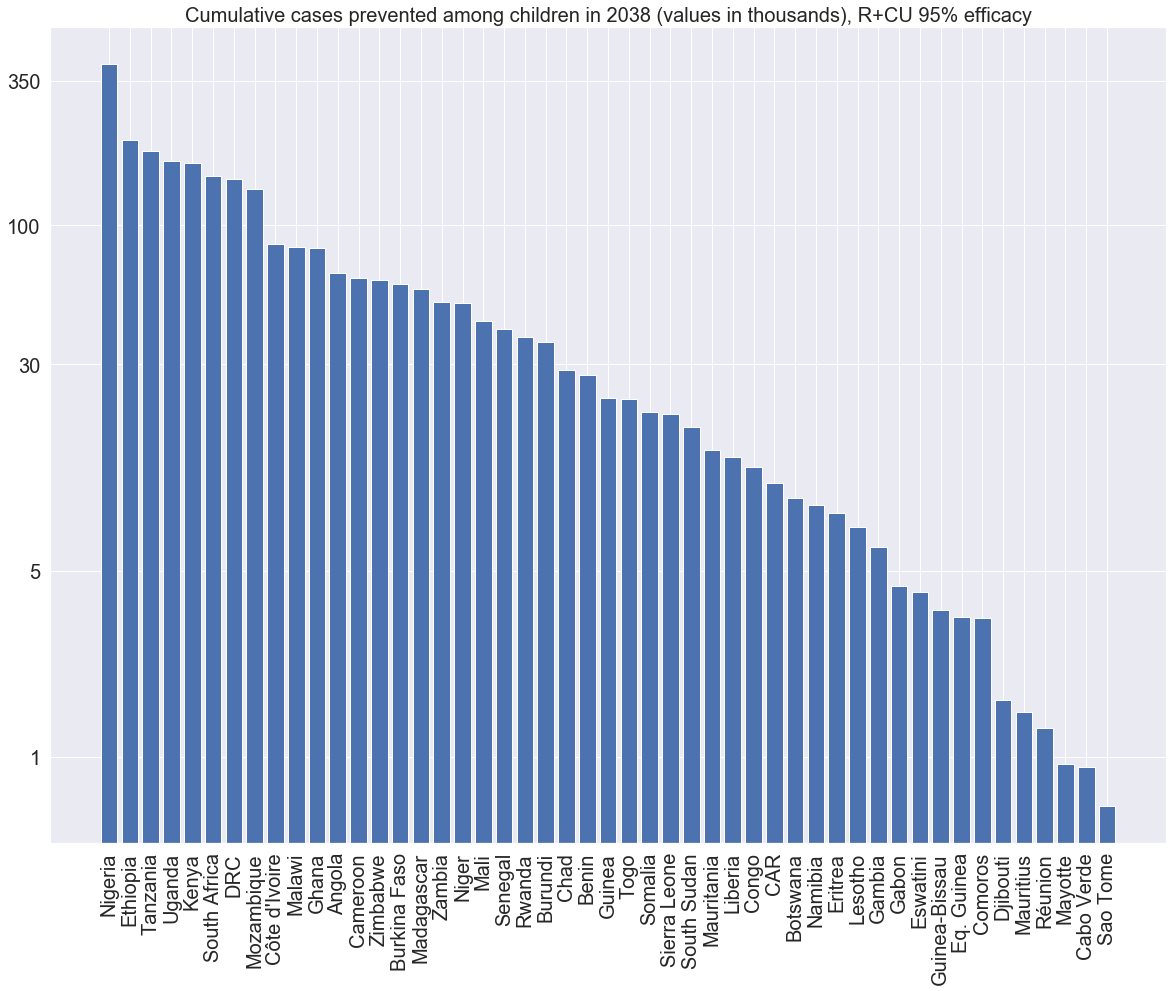}
\caption{Cumulative cases prevented for each country during the period 2028-2038. Routine + catch-up campaign with $95\%$ vaccine efficacy. $y$-axis is on a logarithmic scale.}
\label{cp}
\end{figure}
\renewcommand{\thefigure}{\textbf{Supplementary Fig. 7 Cumulative deaths averted 2028-2038}}
\begin{figure}
\centering
\includegraphics[width = 0.8\textwidth]{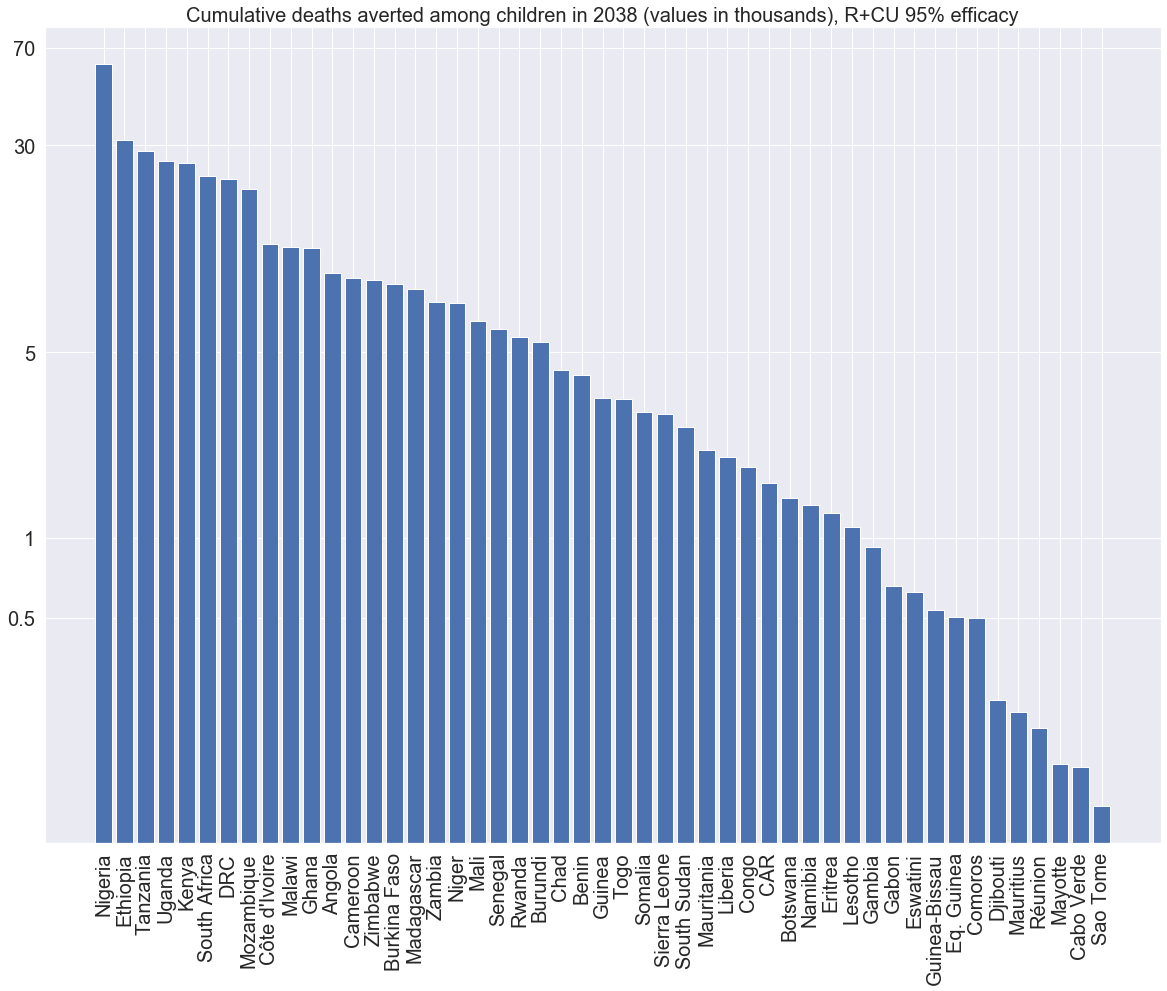}
\caption{Cumulative deaths averted for each country during the period 2028-2038. Routine + catch-up campaign with $95\%$ vaccine efficacy and $20\%$ mortality. $y$-axis is on a logarithmic scale.}
\label{da}
\end{figure}

%\begin{figure}
%\centering
%\includegraphics[width = 0.8\textwidth]{Images/Hiv_/yearlycases}
%\caption{Average number of iNTS cases among children below 5 years old per year for three simulated scenarios: no vaccination, only Routine, Routine + Catch-up, all SSA. Figures in thousands.}
%\label{yc}
%\end{figure}
%
%\begin{figure}
%\centering
%\includegraphics[width = 0.8\textwidth]{Images/Hiv_/yearlydeaths}
%\caption{Average number of iNTS deaths among children below 5 years old per year for three simulated scenarios: no vaccination, only Routine, Routine + Catch-up, all SSA. Figures in thousands.}
%\label{yd}
%\end{figure}

\newpage
\renewcommand{\thefigure}{\textbf{Supplementary Fig. 8 Importance of vaccine coverage}}
\begin{figure}
\centering
\includegraphics[width = 1\textwidth]{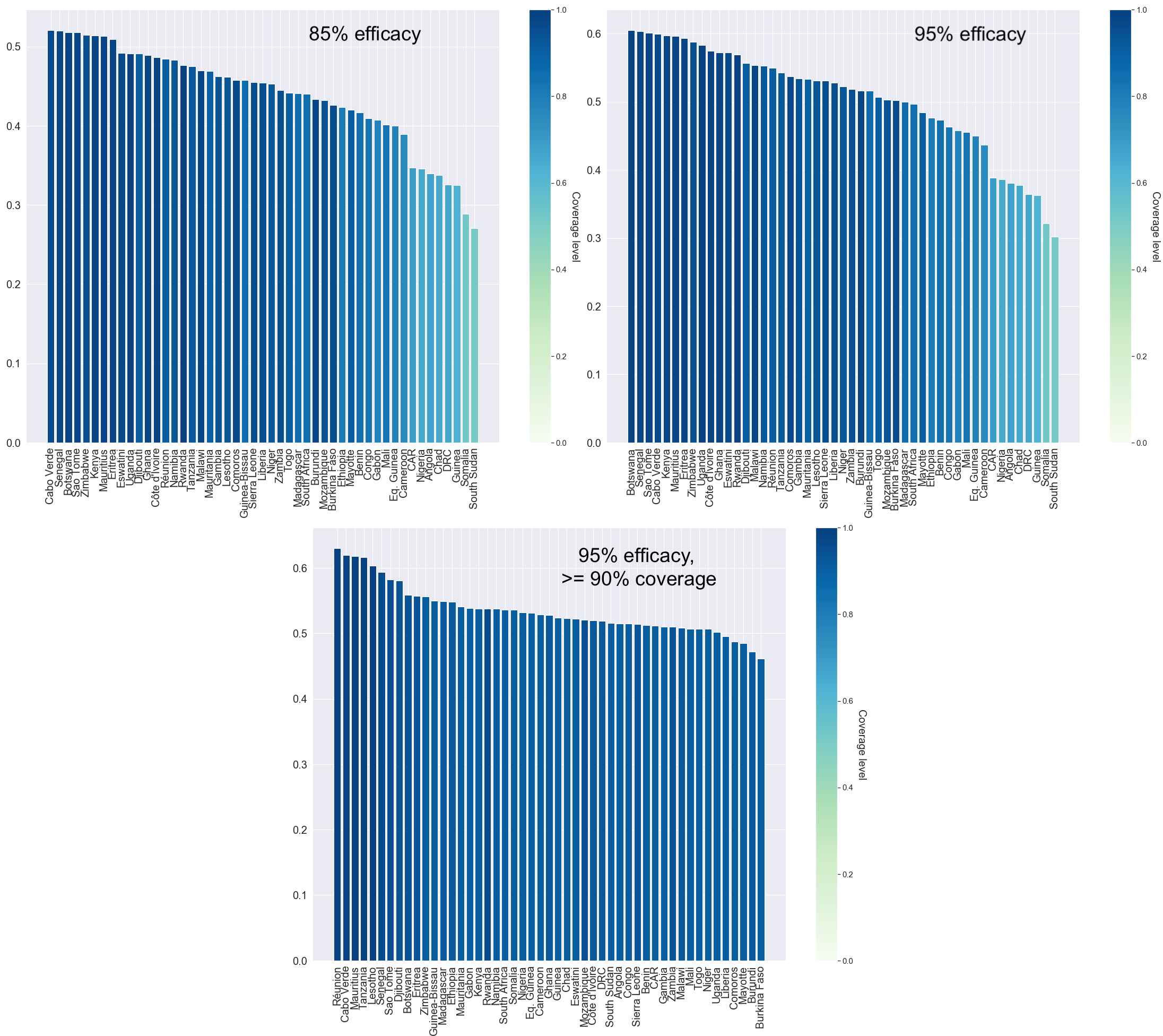}
\caption{Percentage reduction of iNTS cases compared to status quo following  a routine + catch-up campaign between 2028 and 2038 at actual coverage levels (top) with $85\%$ (left), $95\%$ (right) vaccine efficacy and at coverage levels at or above $90\%$ (bottom) with $95\%$ vaccine efficacy. Bar colors identify coverage levels.}
\label{perc}
\end{figure}

\renewcommand{\thefigure}{\textbf{Supplementary Fig. 9 Cumulative cases prevented in the three vaccination scenarios}}
\begin{figure}
\centering
\includegraphics[width = 0.8\textwidth]{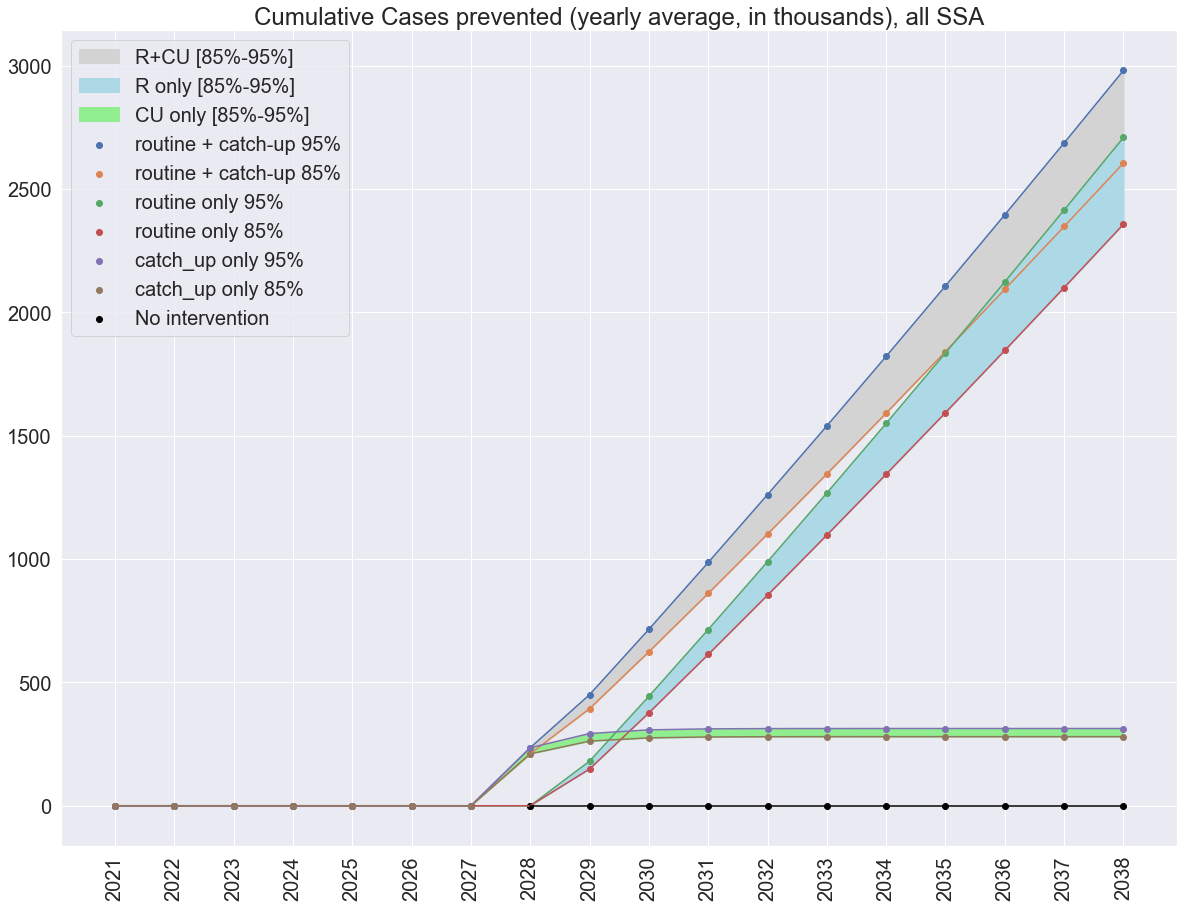}
\caption{The three vaccination scenarios are routine only, routine + catch-up and catch-up only. Vaccine efficacy between $85\%$ and $95\%$. Values in thousands}
\label{c3scenarios}
\end{figure}

\renewcommand\thetable{\textbf{Supplementary Table 4 Yearly cases for all sSA}}
\begin{table}
\centering
\caption{Cases (thousands) per year among children below 5 y.o., for the aggregate sSA, all scenarios.}
\begin{tabular}{rrlll}
\toprule
  Year &  No vac &      CU + R &      R only &     CU only \\
\midrule
      2021 & 478.152 &  - & - &  - \\
      2022 & 484.596 &  - & - &  - \\
      2023 & 491.225 &  - & - &  - \\
      2024 & 498.071 &  - & - &  - \\
      2025 & 505.150 &   - & - &  - \\
      2026 & 512.928 &  - & - &  - \\
      2027 & 520.900 &     - & - &  - \\
      2028 & 528.813 & [318.455, 293.699] & [528.813, 528.813] & [318.455, 293.699] \\
      2029 & 537.225 & [354.151, 322.101] & [388.426, 356.546] & [485.705, 479.643] \\
      2030 & 545.568 & [315.018, 280.579] &  [317.84, 282.759] & [532.167, 530.589] \\
      2031 & 553.562 & [316.668, 282.525] & [316.918, 282.687] &  [549.99, 549.568] \\
      2032 & 561.419 & [320.999, 286.475] & [321.008, 286.476] & [560.462, 560.353] \\
      2033 & 569.080 & [325.438, 290.464] & [325.424, 290.456] & [568.833, 568.802] \\
      2034 & 576.480 & [329.737, 294.325] & [329.726, 294.317] & [576.419, 576.413] \\
      2035 & 583.579 & [333.863, 298.034] & [333.855, 298.026] & [583.572, 583.569] \\
      2036 & 589.632 & [337.409, 301.218] &   [337.4, 301.207] & [589.638, 589.637] \\
      2037 & 595.429 &  [340.79, 304.263] & [340.783, 304.255] & [595.436, 595.438] \\
      2038 & 601.016 & [344.061, 307.202] & [344.057, 307.201] & [601.024, 601.025] \\
      \midrule
     Total 2021-2038 & 9732.825 &  &  &  \\
     Total 2028-2038 & 6241.803 & [3636.589, 3260.885] & [3636.589, 3532.743] & [5928.736, 5961.701] \\
\bottomrule
\label{allsSA}
\end{tabular} 
\end{table}

\clearpage
\section*{Supplementary Result 2: Per-country vaccine effectiveness}

\begin{figure}[htbp]
\renewcommand{\thefigure}{\textbf{Supplementary Fig. 10 Angola cumulative and yearly iNTS cases}}
\begin{subfigure}[b]{\textwidth}
\centering
\includegraphics[width=0.7\linewidth]{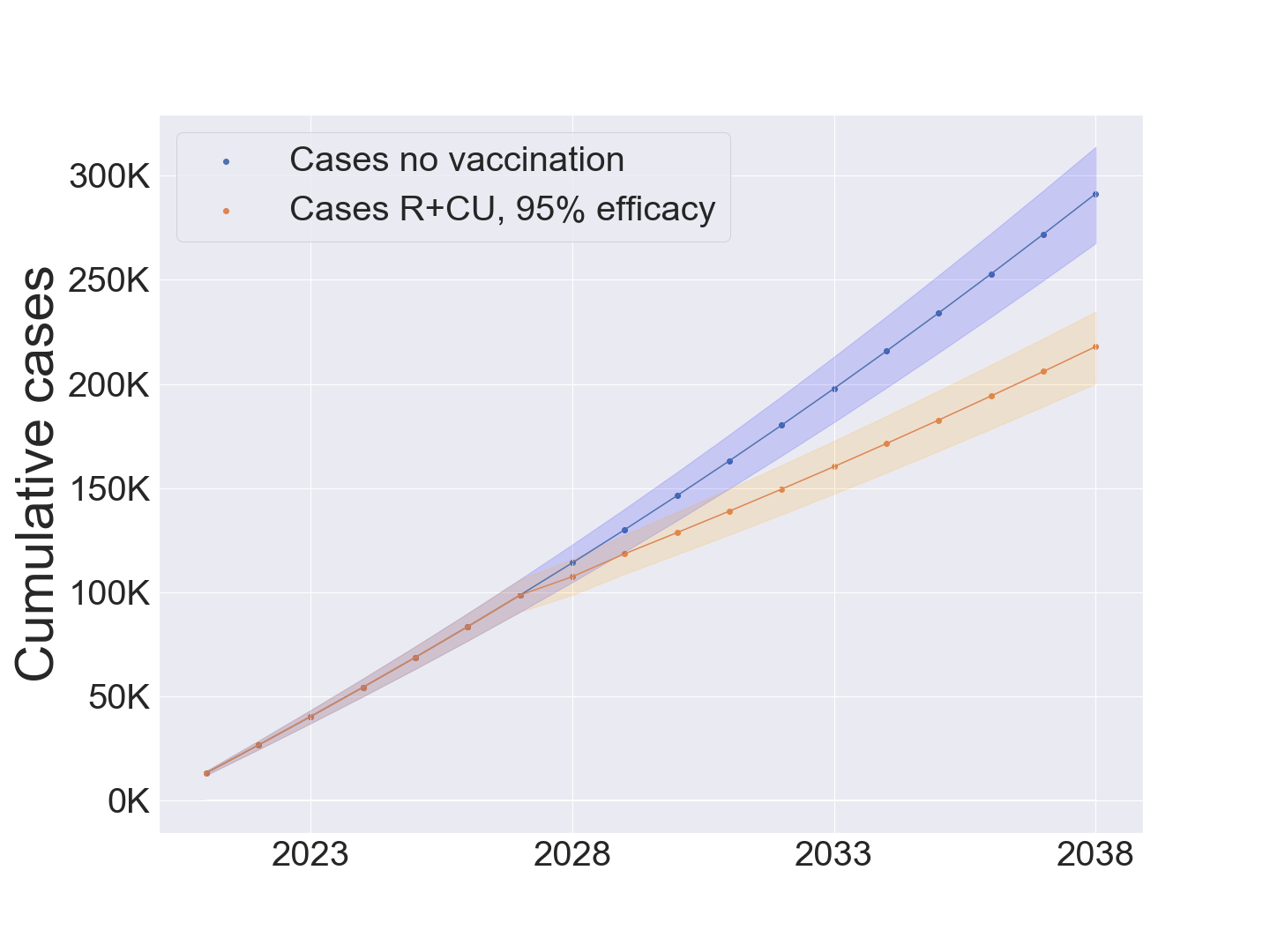}
\includegraphics[width=0.7\linewidth]{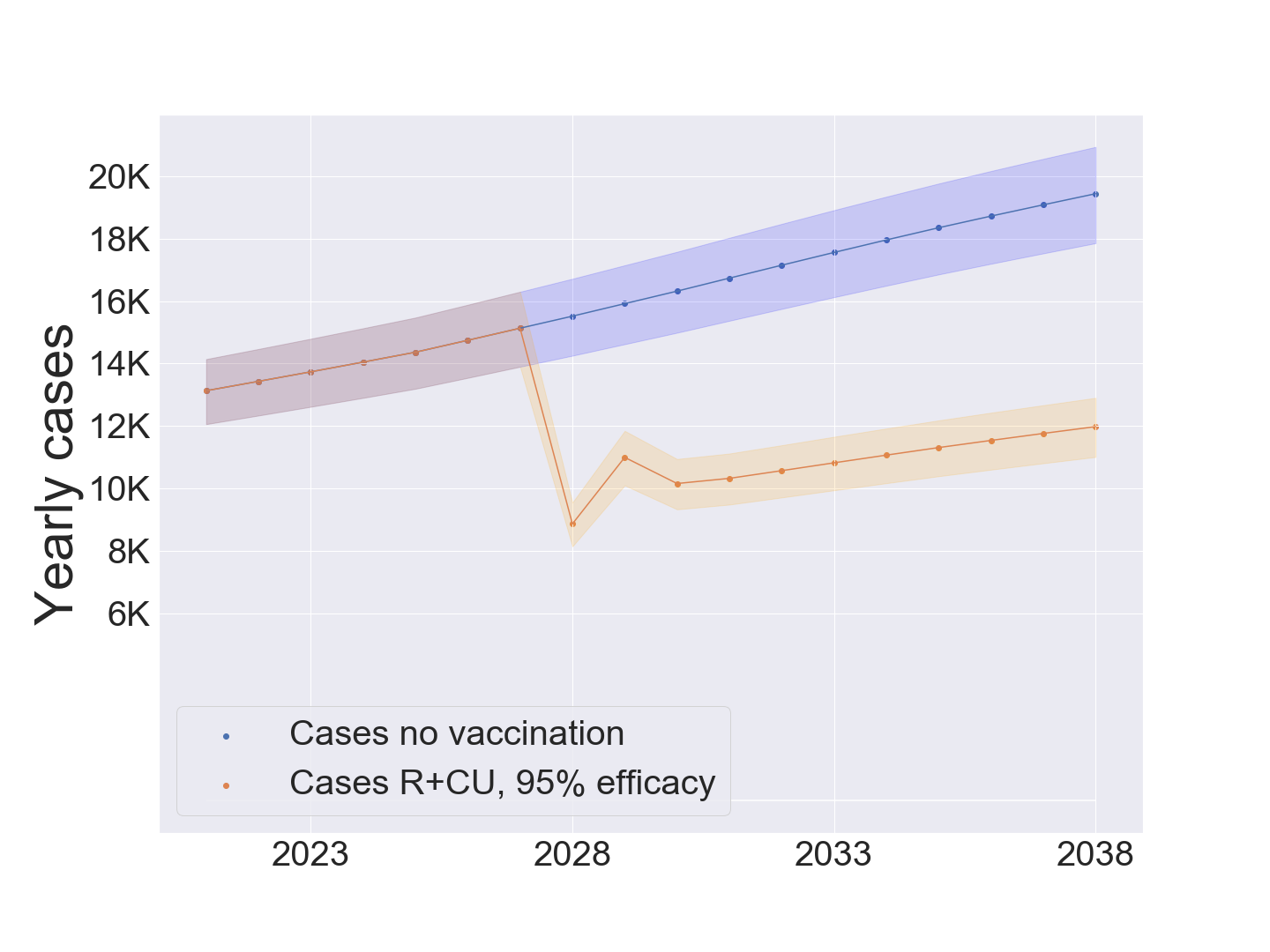}
\end{subfigure}
\caption{Angola cumulative (top)  and yearly (bottom) iNTS cases under the status quo and routine + catch-up vaccination ($95\%$ efficacy) scenarios. Shaded areas show the 25th and 75th percentiles, line shows the median over 1000 experiments, samples drawn from uniform distributions over (0.00020,0.00024) for $\beta_{2,n}$ and (0.0080,0.0084) for $\beta_{4,n}$. }\label{fig:Angola}
\end{figure}

\begin{figure}[htbp]
\renewcommand{\thefigure}{\textbf{Supplementary Fig. 11 Benin cumulative and yearly iNTS cases}}
\begin{subfigure}[b]{\textwidth}
\centering
\includegraphics[width=0.7\linewidth]{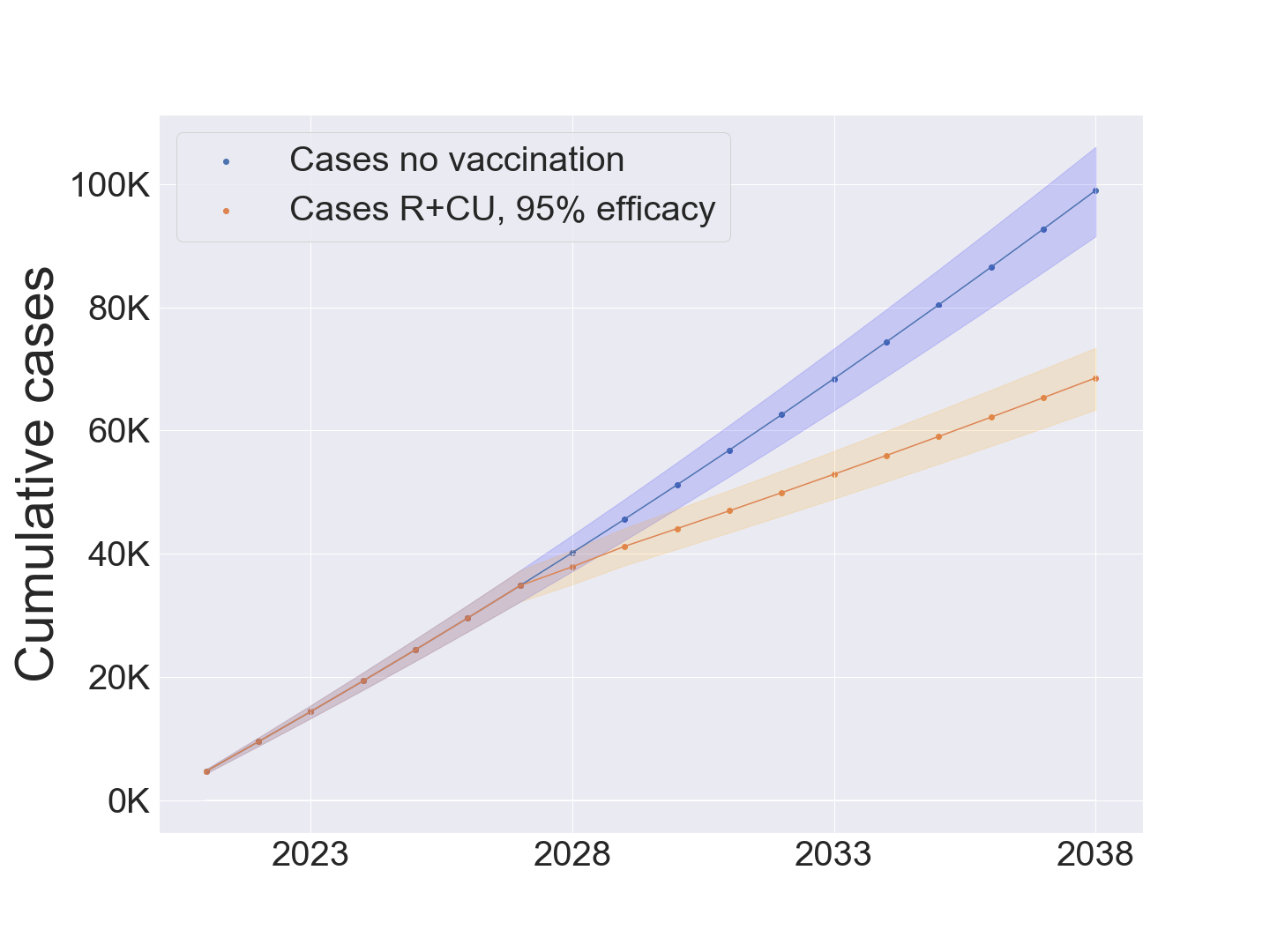}
\includegraphics[width=0.7\linewidth]{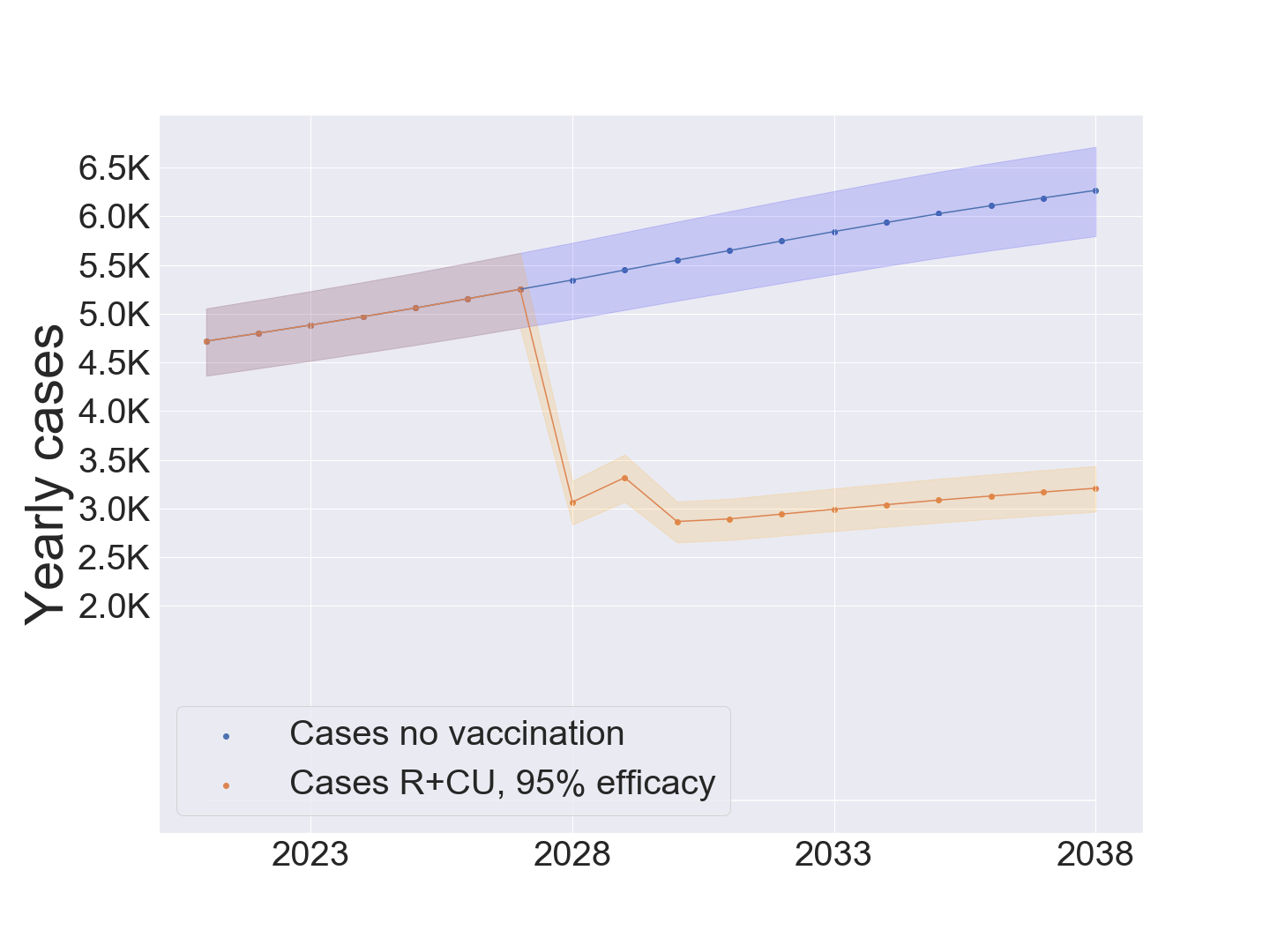}
\end{subfigure}
\caption{Benin cumulative (top)  and yearly (bottom) iNTS cases under the status quo and routine + catch-up vaccination ($95\%$ efficacy) scenarios. Shaded areas show the 25th and 75th percentiles, line shows the median over 1000 experiments, samples drawn from uniform distributions over (0.00020,0.00024) for $\beta_{2,n}$ and (0.0080,0.0084) for $\beta_{4,n}$. }\label{fig:Benin}
\end{figure}

\begin{figure}[htbp]
\renewcommand{\thefigure}{\textbf{Supplementary Fig. 12 Botswana cumulative and yearly iNTS cases}}
\begin{subfigure}[b]{\textwidth}
\centering
\includegraphics[width=0.7\linewidth]{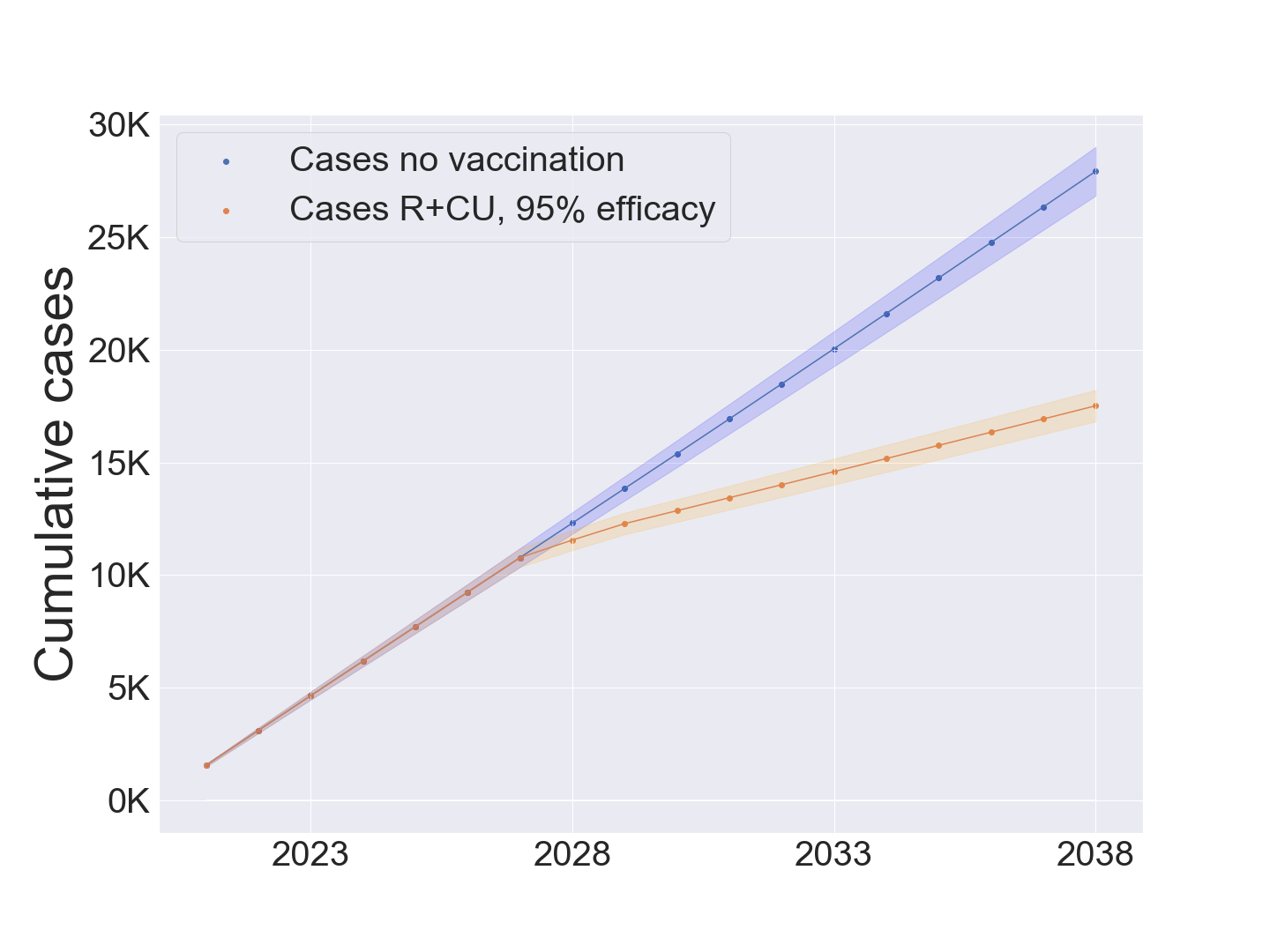}
\includegraphics[width=0.7\linewidth]{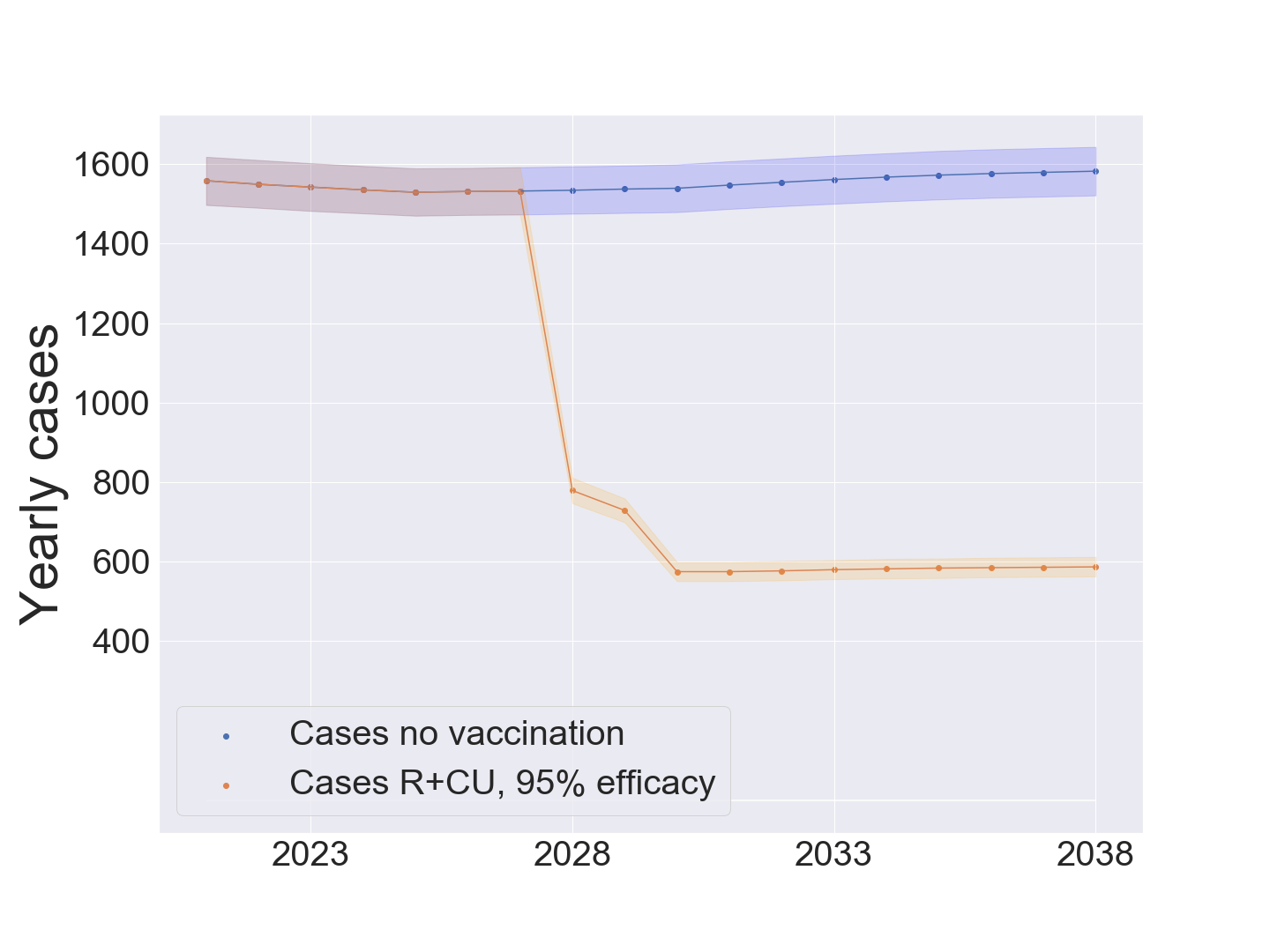}
\end{subfigure}
\caption{Botswana cumulative (top)  and yearly (bottom) iNTS cases under the status quo and routine + catch-up vaccination ($95\%$ efficacy) scenarios. Shaded areas show the 25th and 75th percentiles, line shows the median over 1000 experiments, samples drawn from uniform distributions over (0.00020,0.00024) for $\beta_{2,n}$ and (0.0080,0.0084) for $\beta_{4,n}$. }\label{fig:Botswana}
\end{figure}

\begin{figure}[htbp]
\renewcommand{\thefigure}{\textbf{Supplementary Fig. 13 Burkina Faso cumulative and yearly iNTS cases}}
\begin{subfigure}[b]{\textwidth}
\centering
\includegraphics[width=0.7\linewidth]{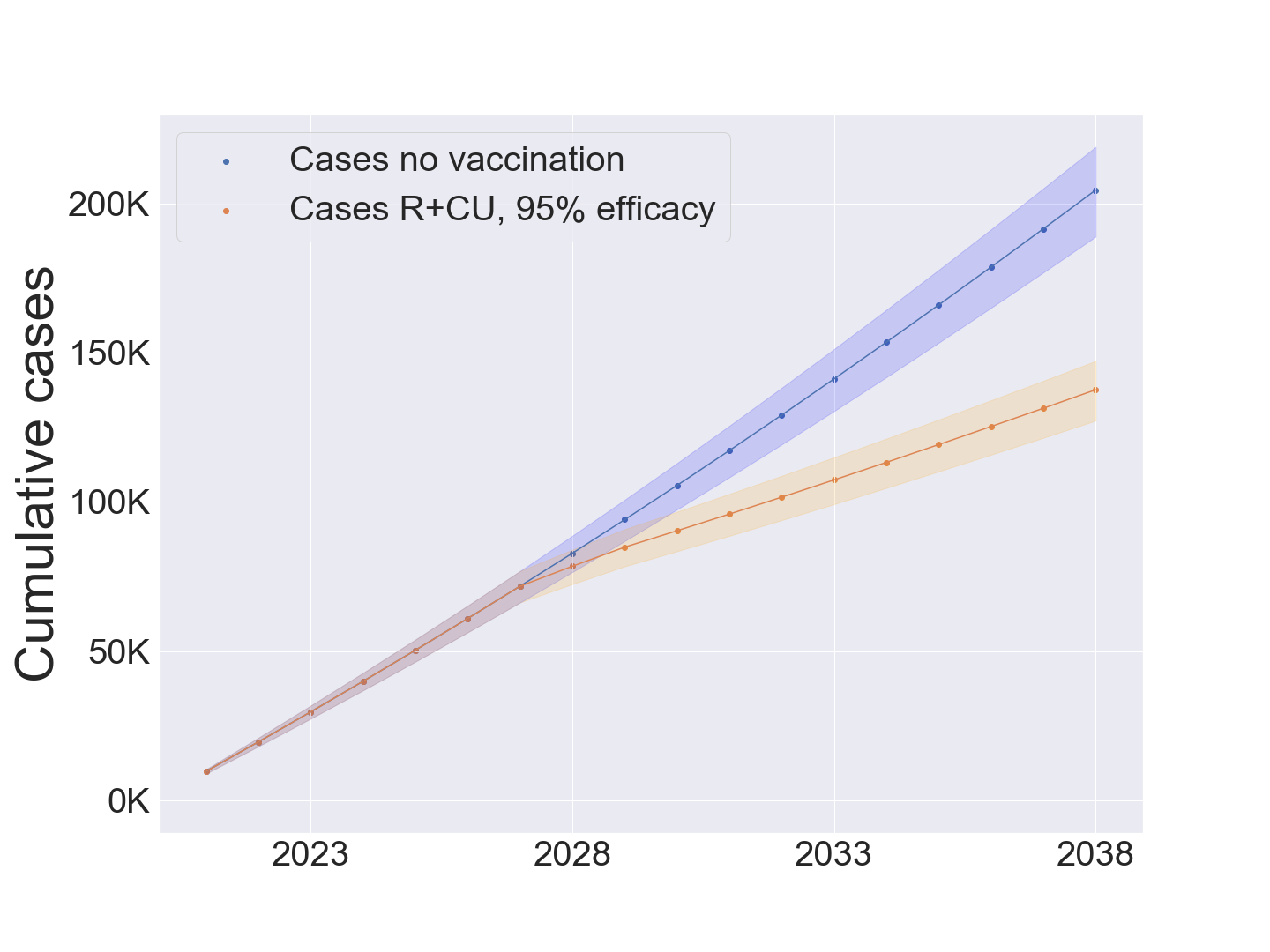}
\includegraphics[width=0.7\linewidth]{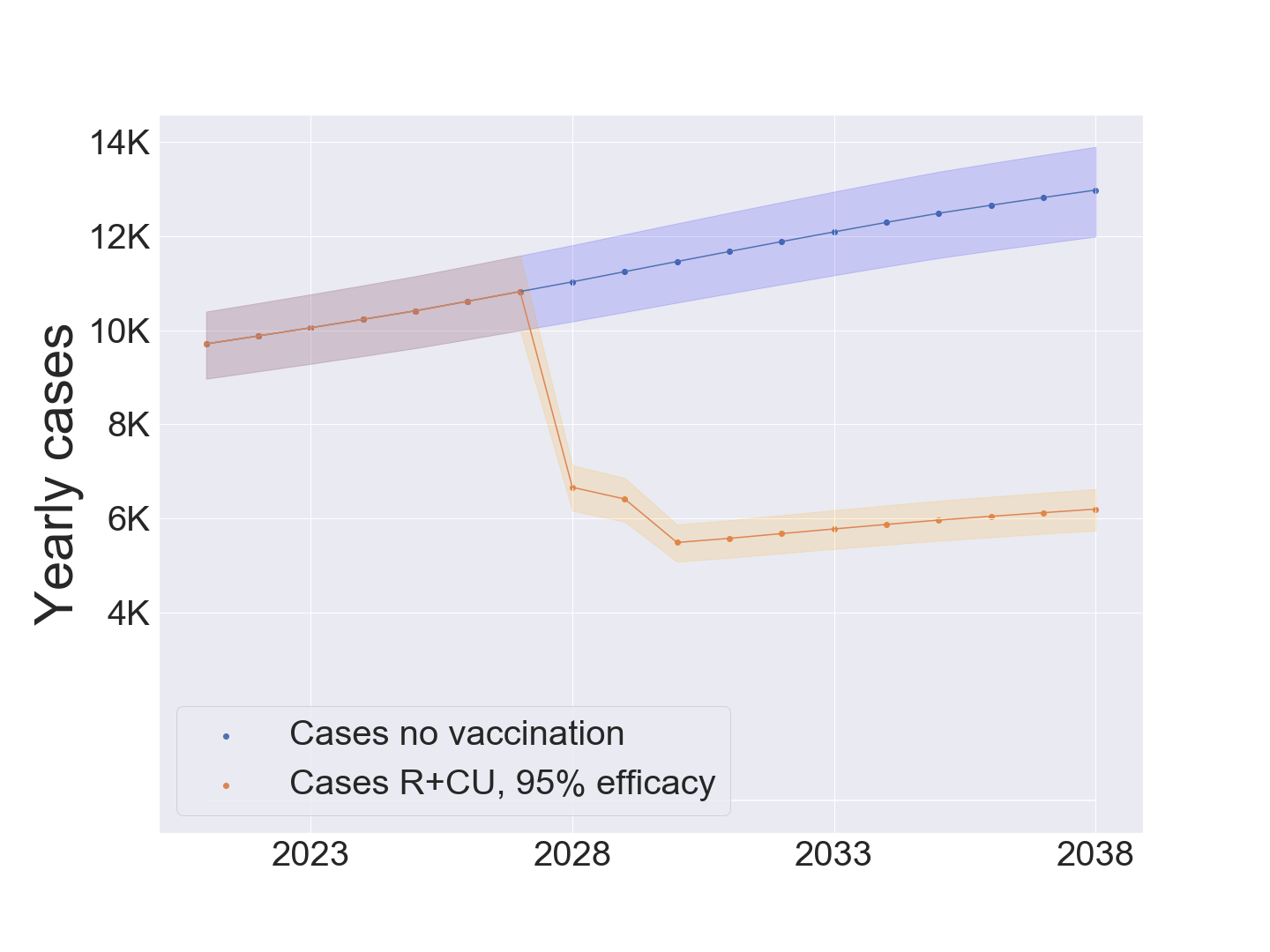}
\end{subfigure}
\caption{Burkina Faso cumulative (top)  and yearly (bottom) iNTS cases under the status quo and routine + catch-up vaccination ($95\%$ efficacy) scenarios. Shaded areas show the 25th and 75th percentiles, line shows the median over 1000 experiments, samples drawn from uniform distributions over (0.00020,0.00024) for $\beta_{2,n}$ and (0.0080,0.0084) for $\beta_{4,n}$. }\label{fig:BurkinaFaso}
\end{figure}

\begin{figure}[htbp]
\renewcommand{\thefigure}{\textbf{Supplementary Fig. 14 Burundi cumulative and yearly iNTS cases}}
\begin{subfigure}[b]{\textwidth}
\centering
\includegraphics[width=0.7\linewidth]{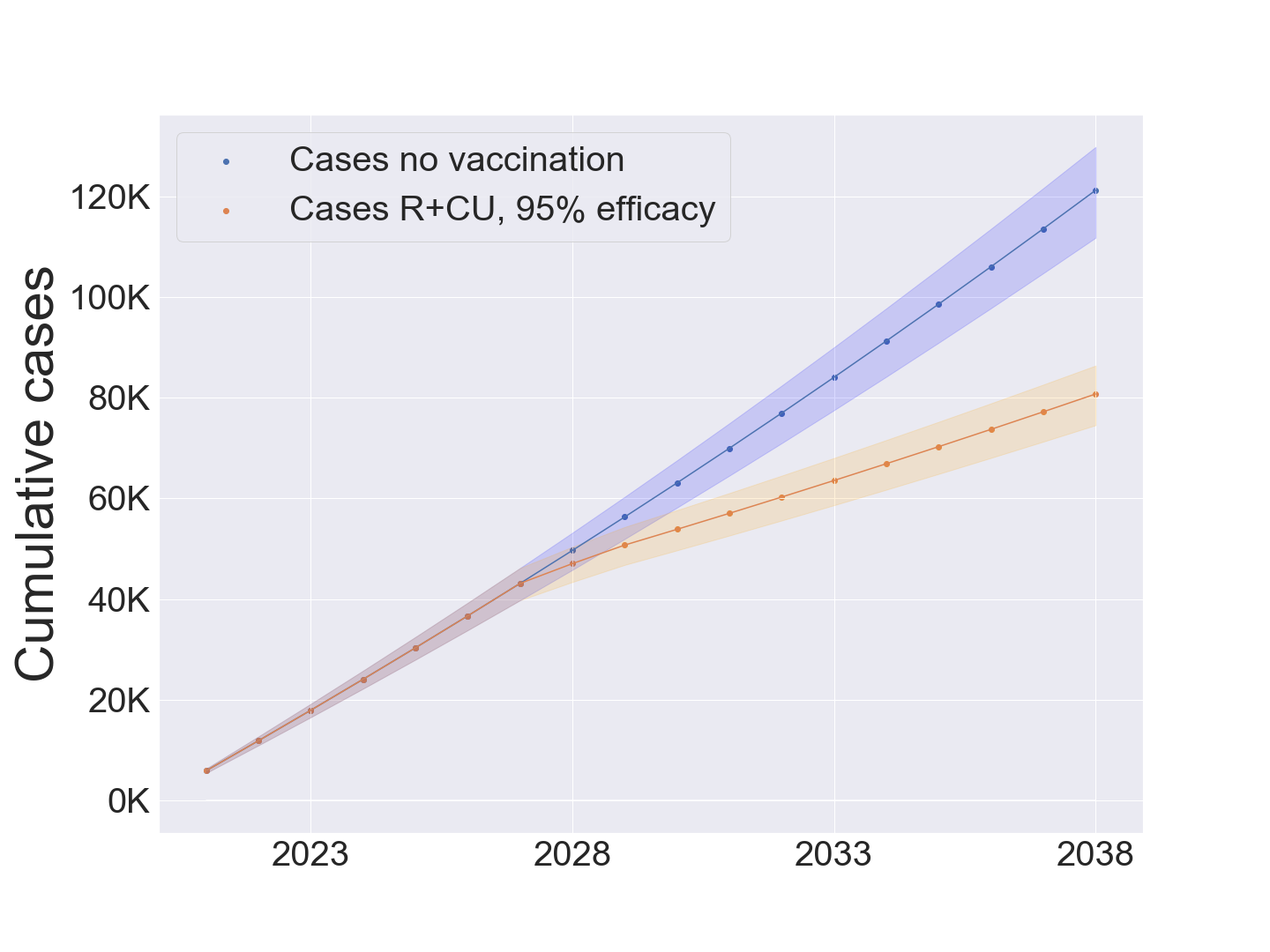}
\includegraphics[width=0.7\linewidth]{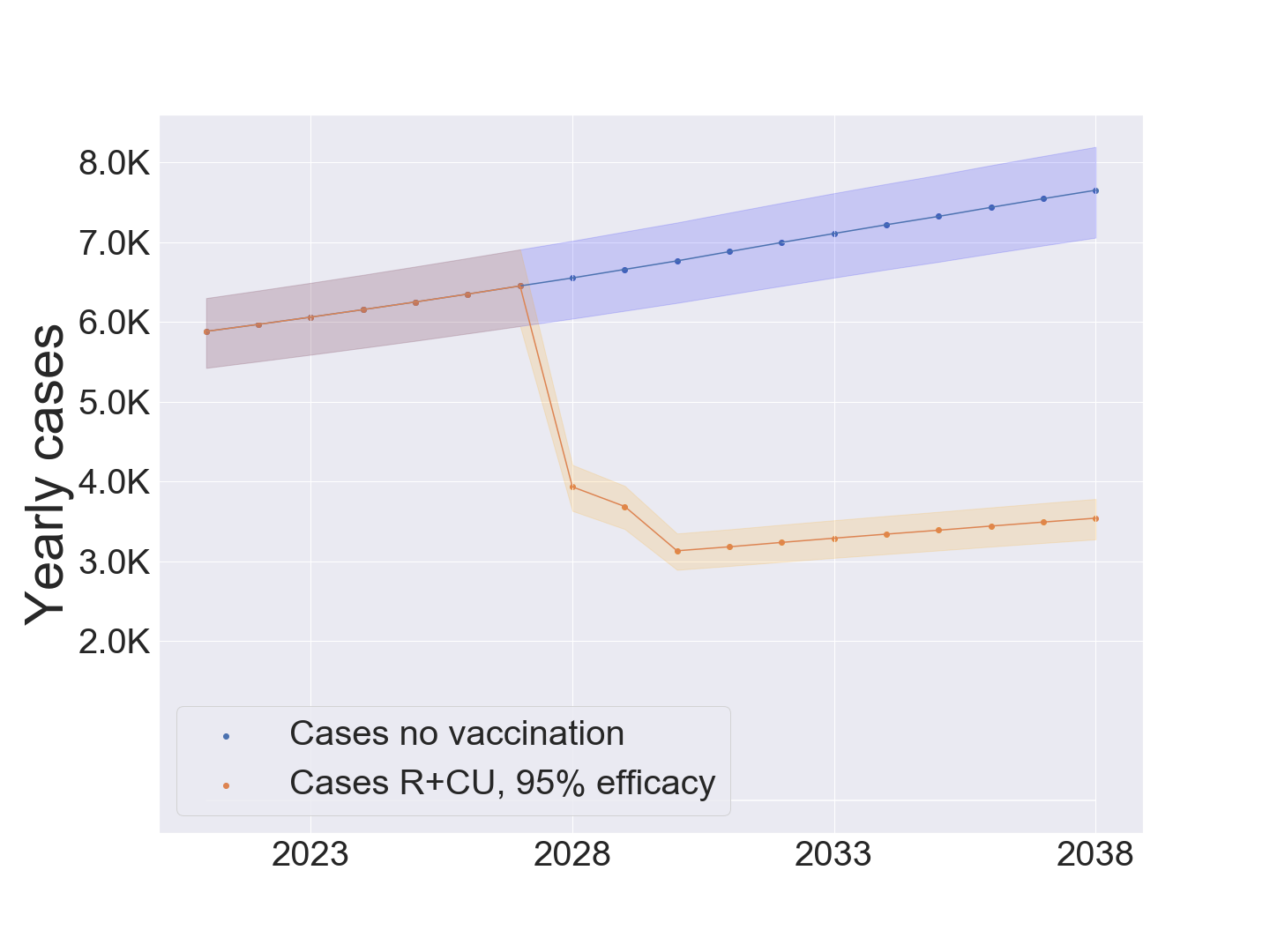}
\end{subfigure}
\caption{Burundi cumulative (top)  and yearly (bottom) iNTS cases under the status quo and routine + catch-up vaccination ($95\%$ efficacy) scenarios. Shaded areas show the 25th and 75th percentiles, line shows the median over 1000 experiments, samples drawn from uniform distributions over (0.00020,0.00024) for $\beta_{2,n}$ and (0.0080,0.0084) for $\beta_{4,n}$. }\label{fig:Burundi}
\end{figure}

\begin{figure}[htbp]
\renewcommand{\thefigure}{\textbf{Supplementary Fig. 15 Central African Republic cumulative and yearly iNTS cases}}
\begin{subfigure}[b]{\textwidth}
\centering
\includegraphics[width=0.7\linewidth]{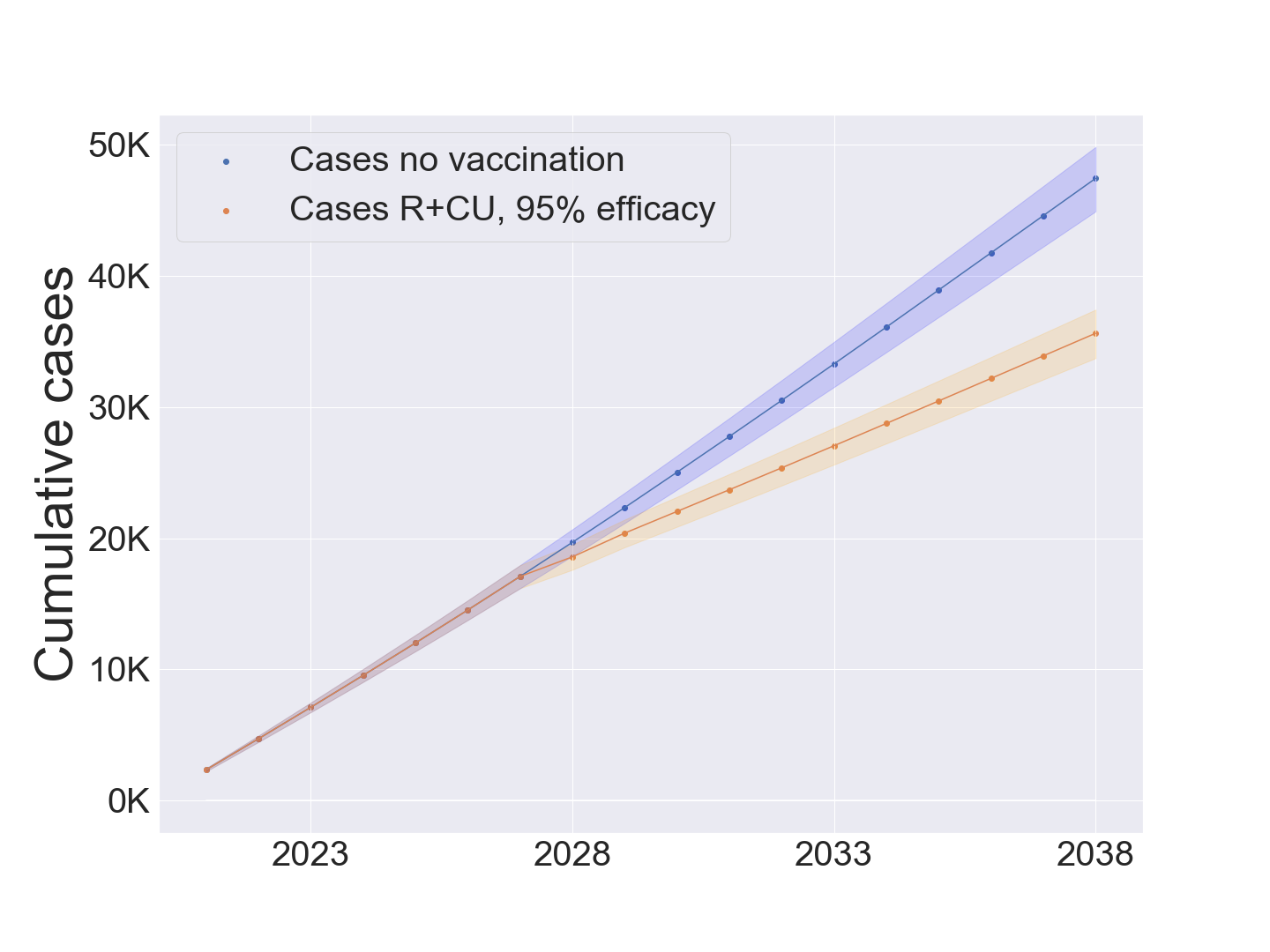}
\includegraphics[width=0.7\linewidth]{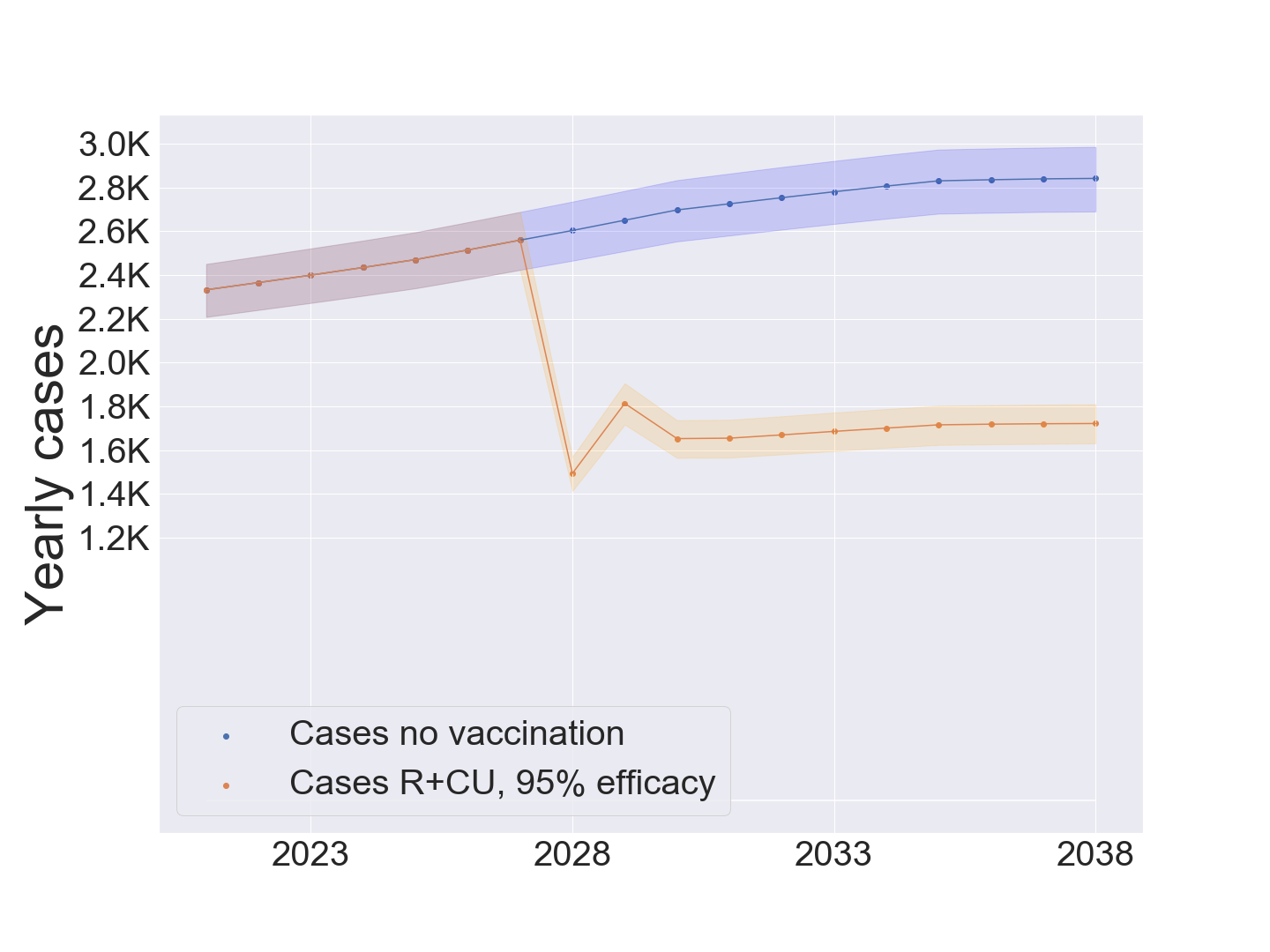}
\end{subfigure}
\caption{Central African Republic cumulative (top)  and yearly (bottom) iNTS cases under the status quo and routine + catch-up vaccination ($95\%$ efficacy) scenarios. Shaded areas show the 25th and 75th percentiles, line shows the median over 1000 experiments, samples drawn from uniform distributions over (0.00020,0.00024) for $\beta_{2,n}$ and (0.0080,0.0084) for $\beta_{4,n}$. }\label{fig:CAR}
\end{figure}

\begin{figure}[htbp]
\renewcommand{\thefigure}{\textbf{Supplementary Fig. 16 Cabo Verde cumulative and yearly iNTS cases}}
\begin{subfigure}[b]{\textwidth}
\centering
\includegraphics[width=0.7\linewidth]{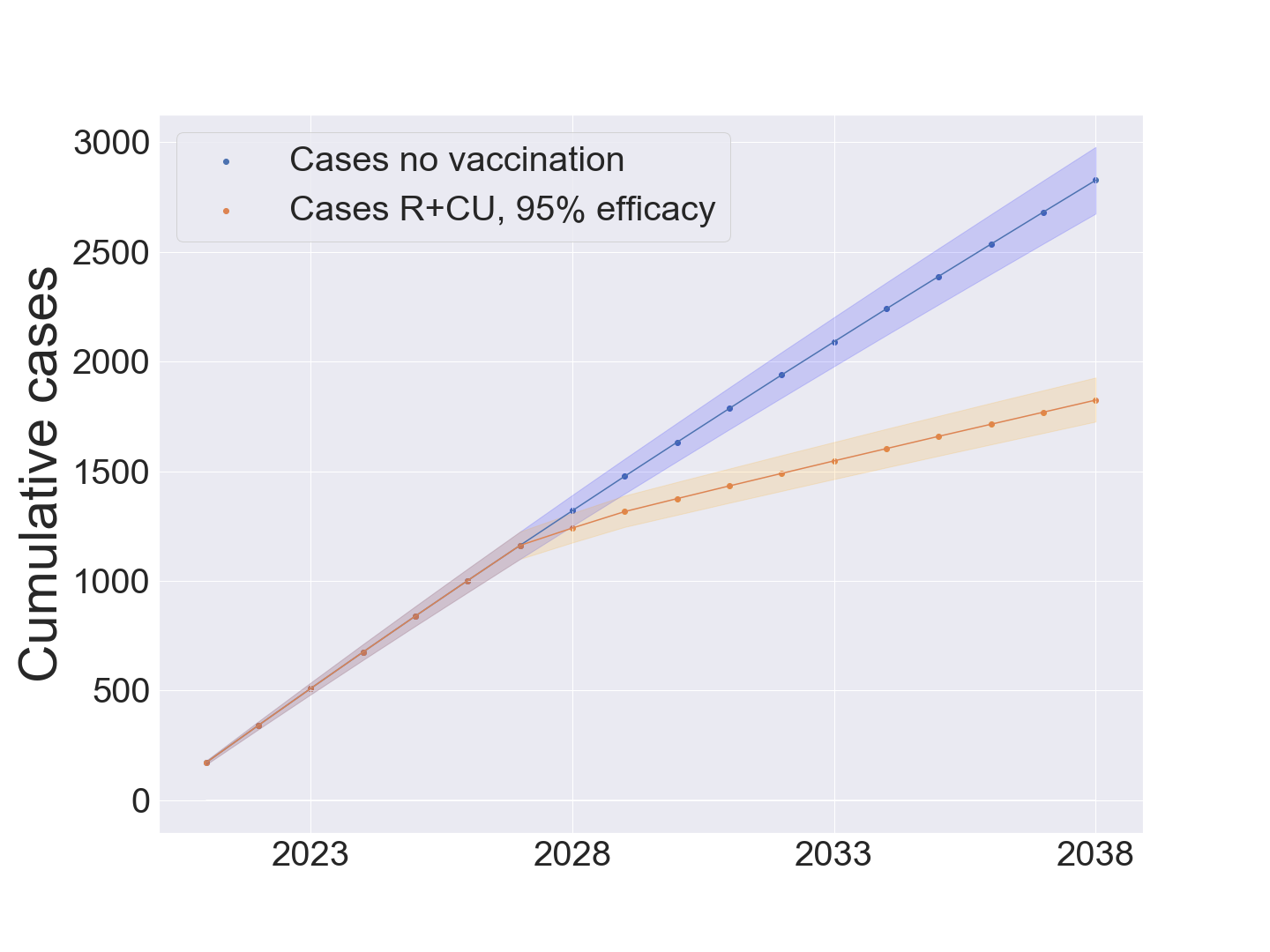}
\includegraphics[width=0.7\linewidth]{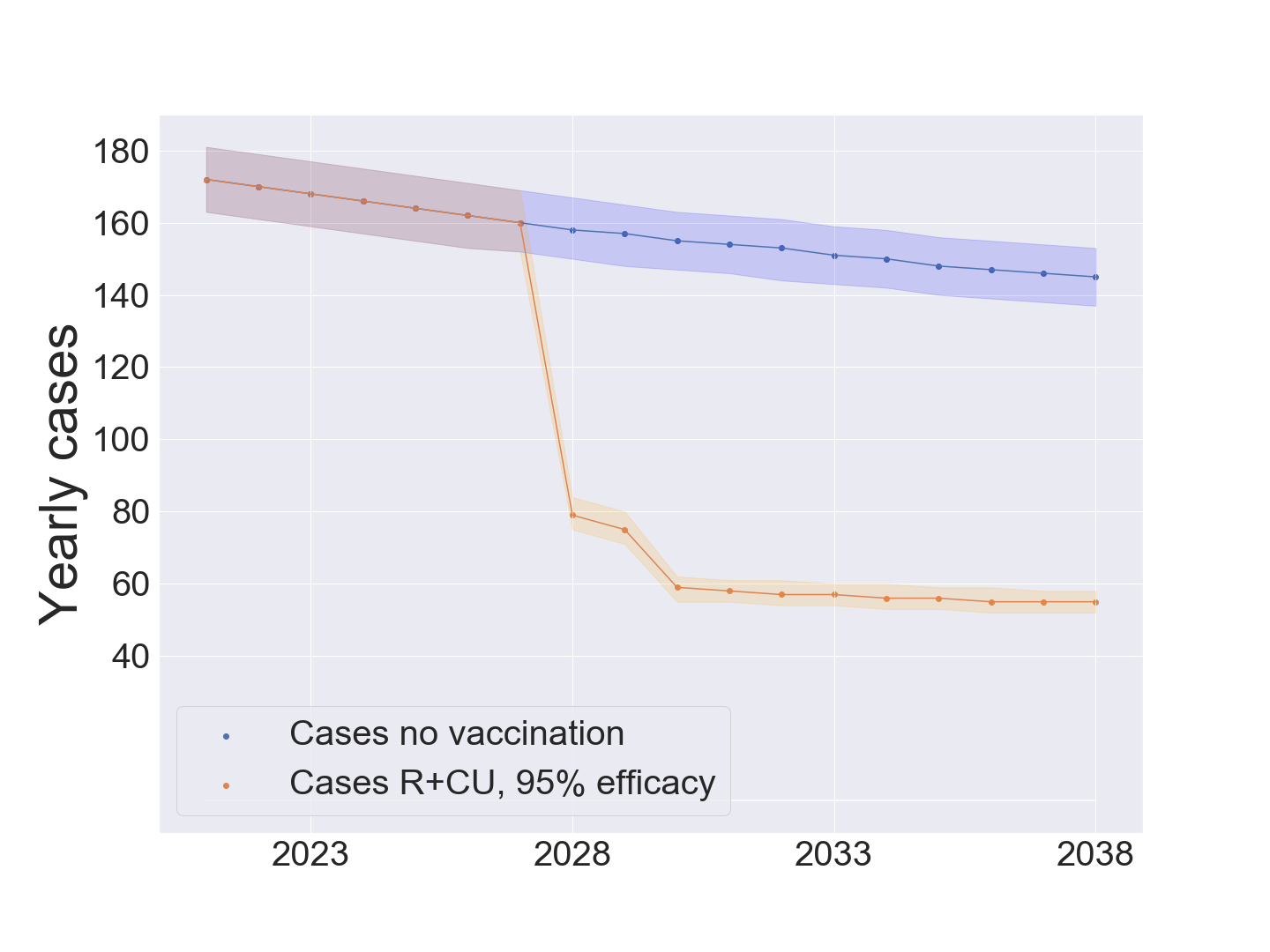}
\end{subfigure}
\caption{Cabo Verde cumulative (top)  and yearly (bottom) iNTS cases under the status quo and routine + catch-up vaccination ($95\%$ efficacy) scenarios. Shaded areas show the 25th and 75th percentiles, line shows the median over 1000 experiments, samples drawn from uniform distributions over (0.00020,0.00024) for $\beta_{2,n}$ and (0.0080,0.0084) for $\beta_{4,n}$. }\label{fig:CaboVerde}
\end{figure}

\begin{figure}[htbp]
\renewcommand{\thefigure}{\textbf{Supplementary Fig. 17 Cameroon cumulative and yearly iNTS cases}}
\begin{subfigure}[b]{\textwidth}
\centering
\includegraphics[width=0.7\linewidth]{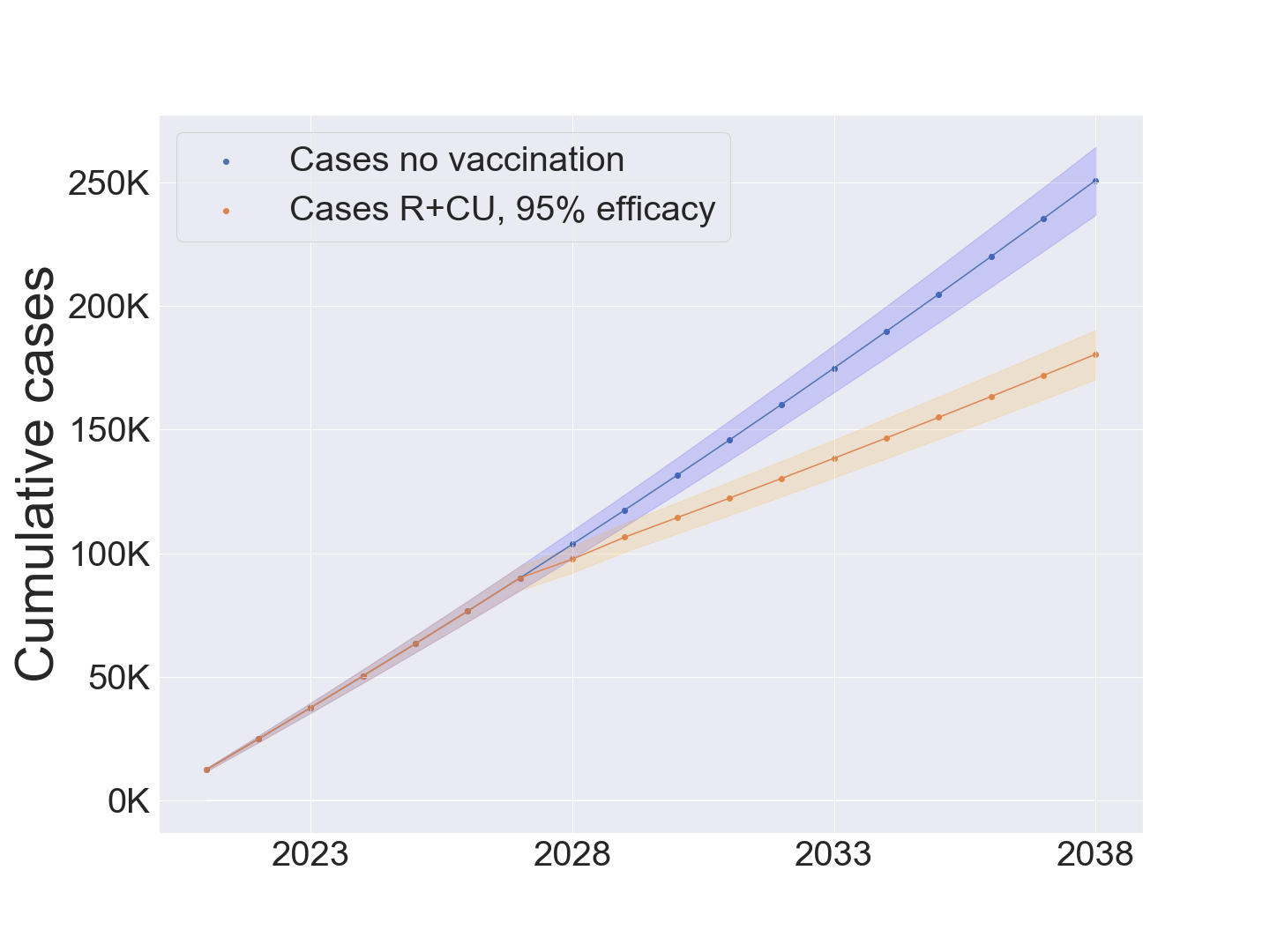}
\includegraphics[width=0.7\linewidth]{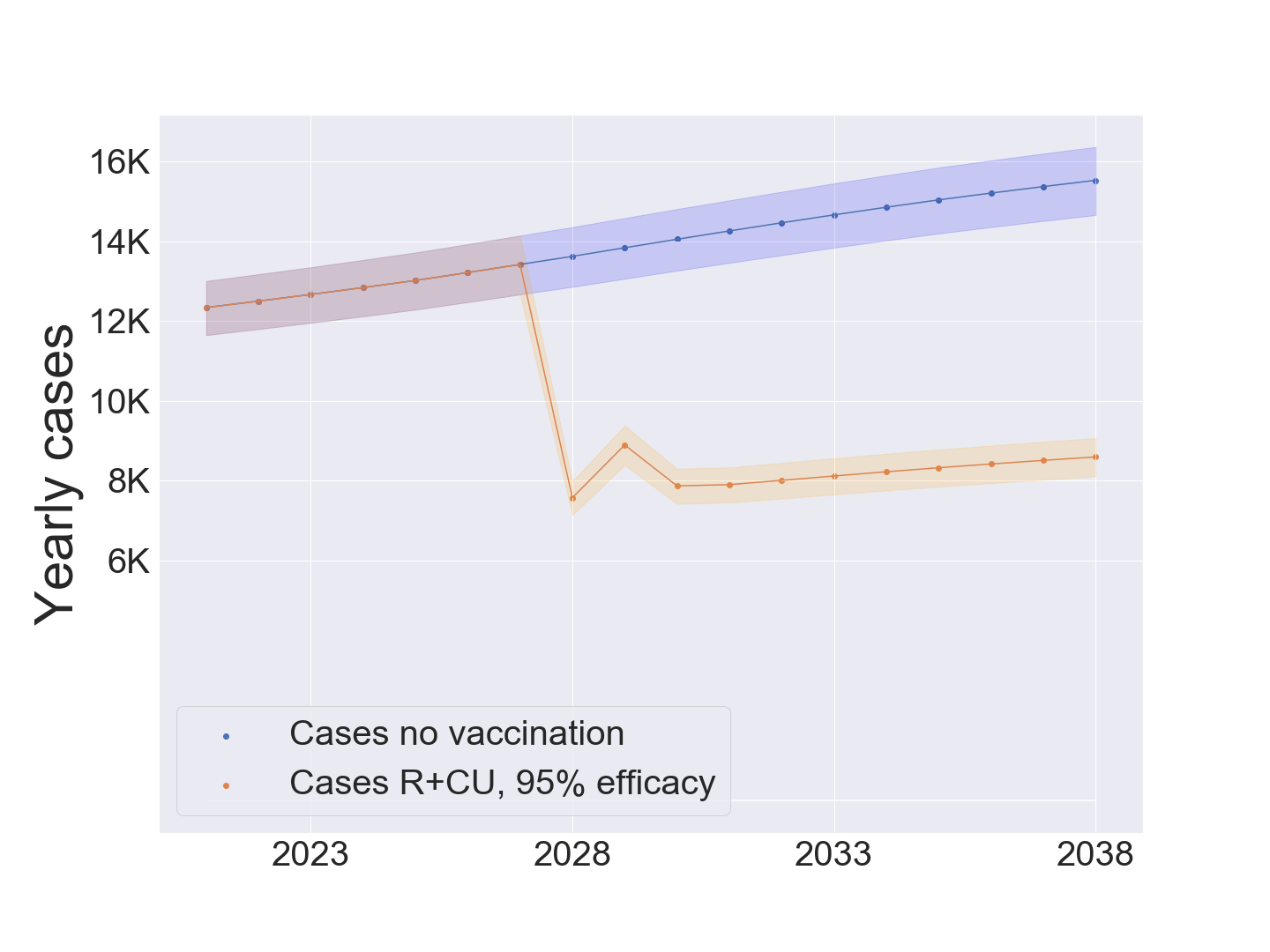}
\end{subfigure}
\caption{Cameroon cumulative (top)  and yearly (bottom) iNTS cases under the status quo and routine + catch-up vaccination ($95\%$ efficacy) scenarios. Shaded areas show the 25th and 75th percentiles, line shows the median over 1000 experiments, samples drawn from uniform distributions over (0.00020,0.00024) for $\beta_{2,n}$ and (0.0080,0.0084) for $\beta_{4,n}$. }\label{fig:Cameroon}
\end{figure}

\begin{figure}[htbp]
\renewcommand{\thefigure}{\textbf{Supplementary Fig. 18 C\^ote d'Ivoire cumulative and yearly iNTS cases}}
\begin{subfigure}[b]{\textwidth}
\centering
\includegraphics[width=0.7\linewidth]{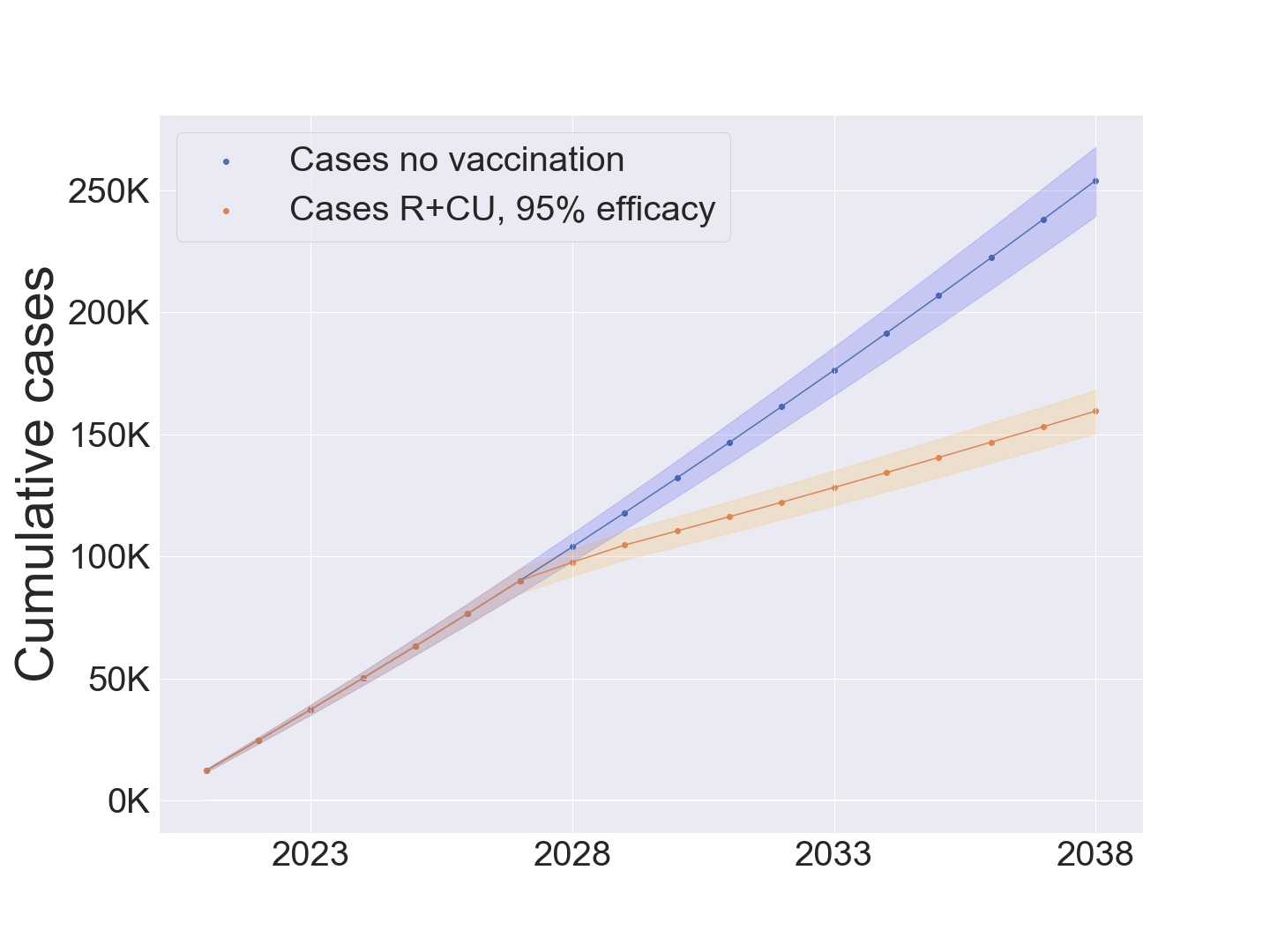}
\includegraphics[width=0.7\linewidth]{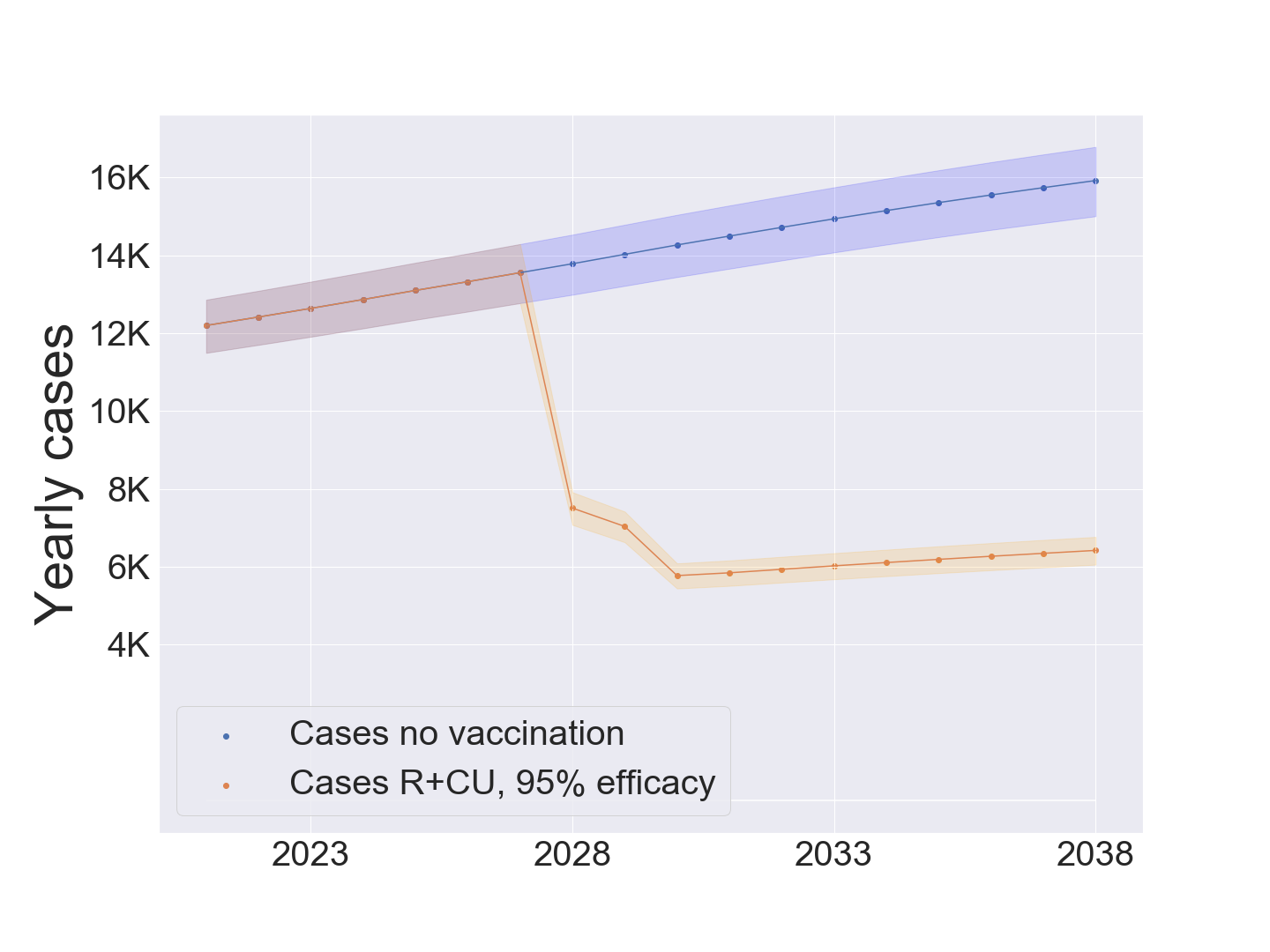}
\end{subfigure}
\caption{C\^ote d'Ivoire cumulative (top)  and yearly (bottom) iNTS cases under the status quo and routine + catch-up vaccination ($95\%$ efficacy) scenarios. Shaded areas show the 25th and 75th percentiles, line shows the median over 1000 experiments, samples drawn from uniform distributions over (0.00020,0.00024) for $\beta_{2,n}$ and (0.0080,0.0084) for $\beta_{4,n}$. }\label{fig:CdIvoire}
\end{figure}

\begin{figure}[htbp]
\renewcommand{\thefigure}{\textbf{Supplementary Fig. 19 Chad cumulative and yearly iNTS cases}}
\begin{subfigure}[b]{\textwidth}
\centering
\includegraphics[width=0.7\linewidth]{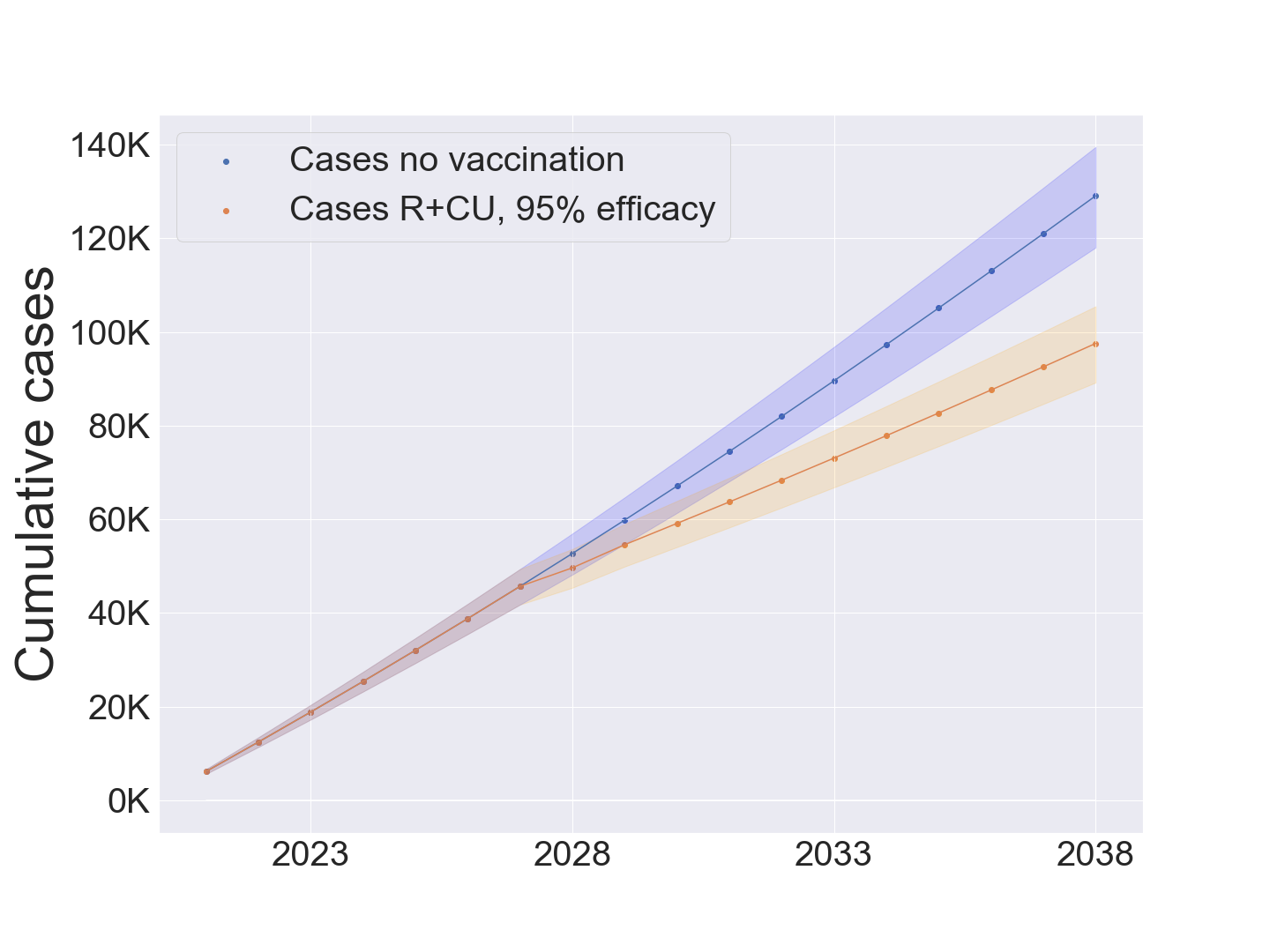}
\includegraphics[width=0.7\linewidth]{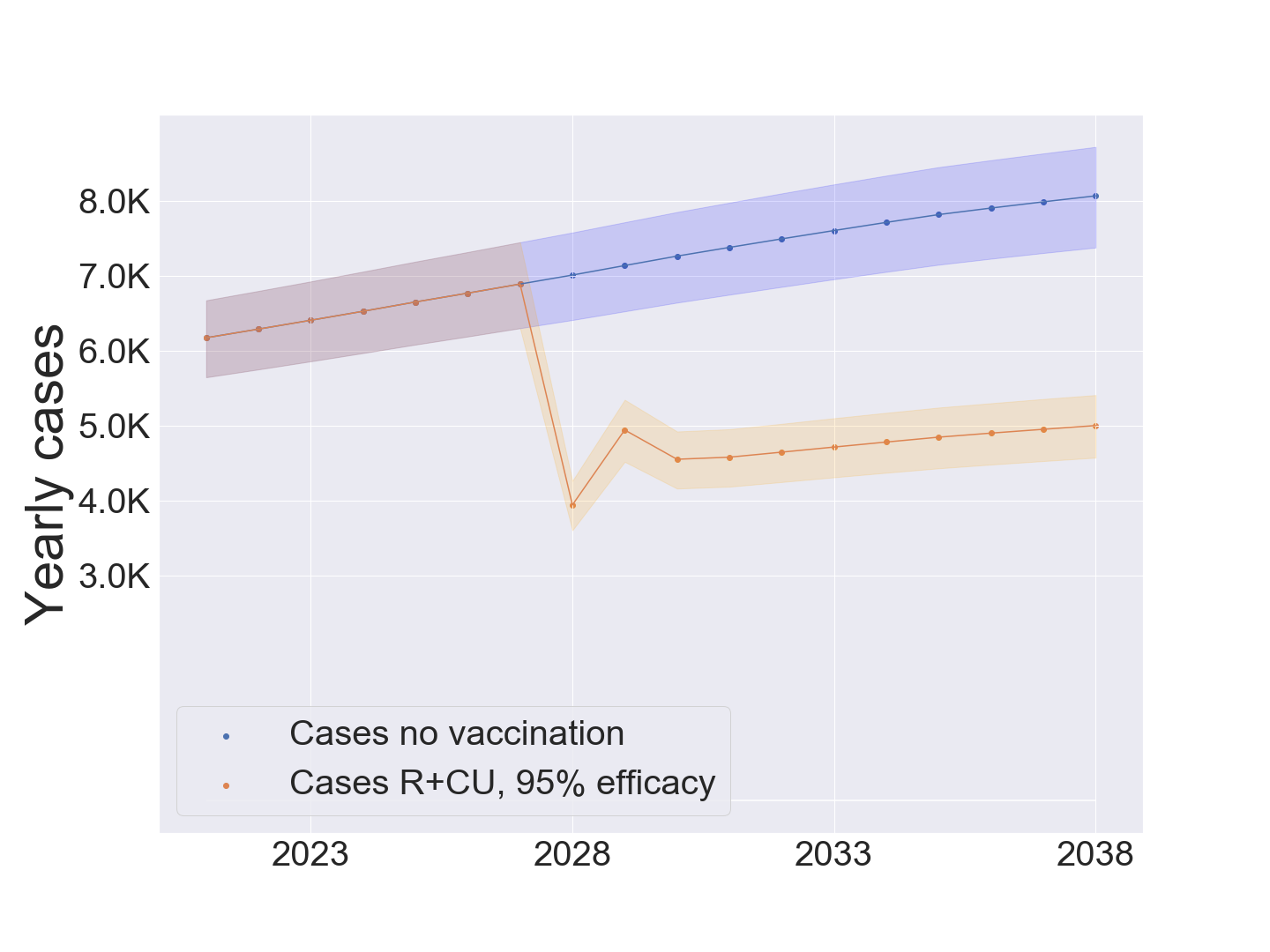}
\end{subfigure}
\caption{Chad cumulative (top)  and yearly (bottom) iNTS cases under the status quo and routine + catch-up vaccination ($95\%$ efficacy) scenarios. Shaded areas show the 25th and 75th percentiles, line shows the median over 1000 experiments, samples drawn from uniform distributions over (0.00020,0.00024) for $\beta_{2,n}$ and (0.0080,0.0084) for $\beta_{4,n}$. }\label{fig:Chad}
\end{figure}

\begin{figure}[htbp]
\renewcommand{\thefigure}{\textbf{Supplementary Fig. 20 Comoros cumulative and yearly iNTS cases}}
\begin{subfigure}[b]{\textwidth}
\centering
\includegraphics[width=0.7\linewidth]{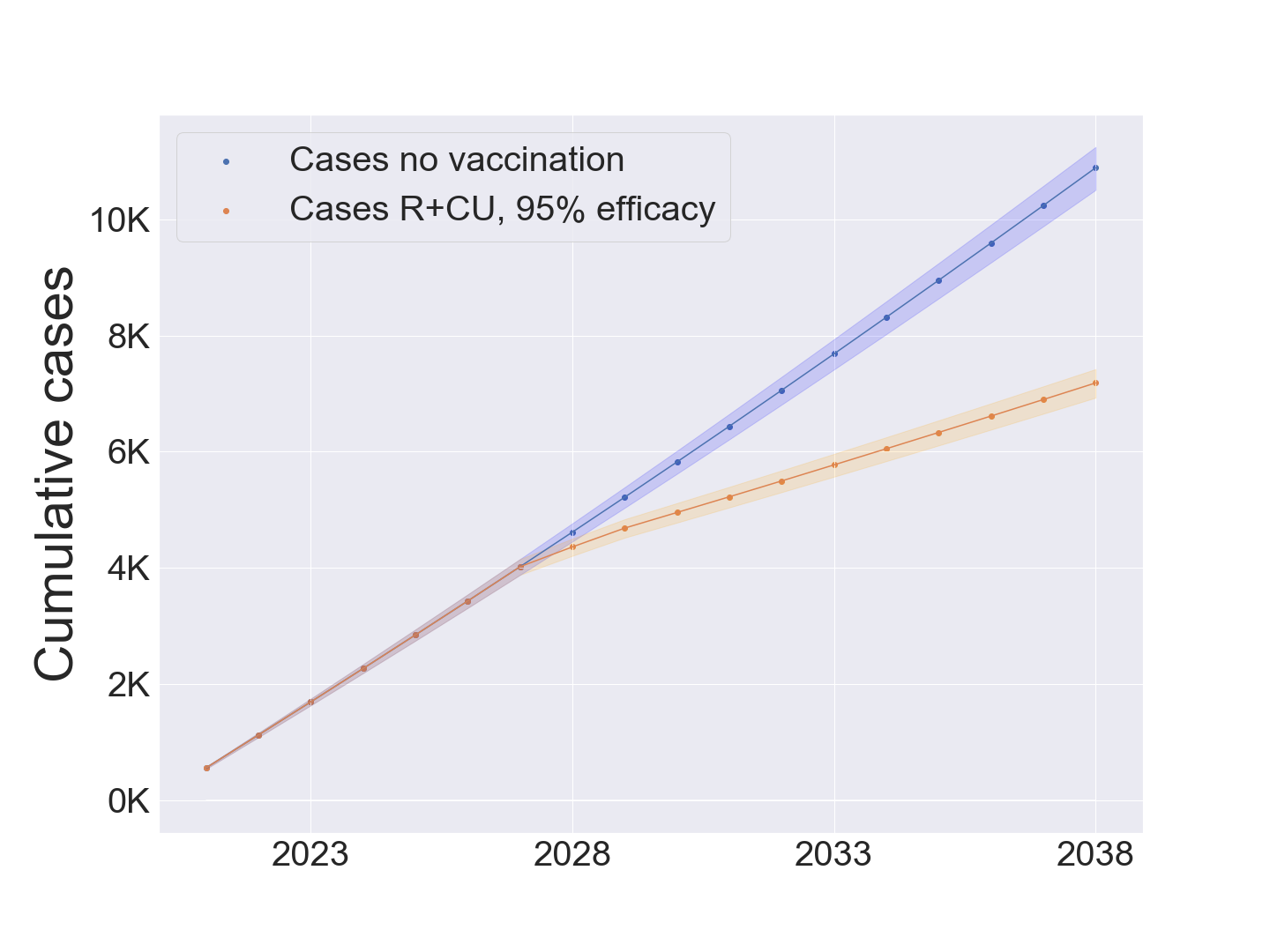}
\includegraphics[width=0.7\linewidth]{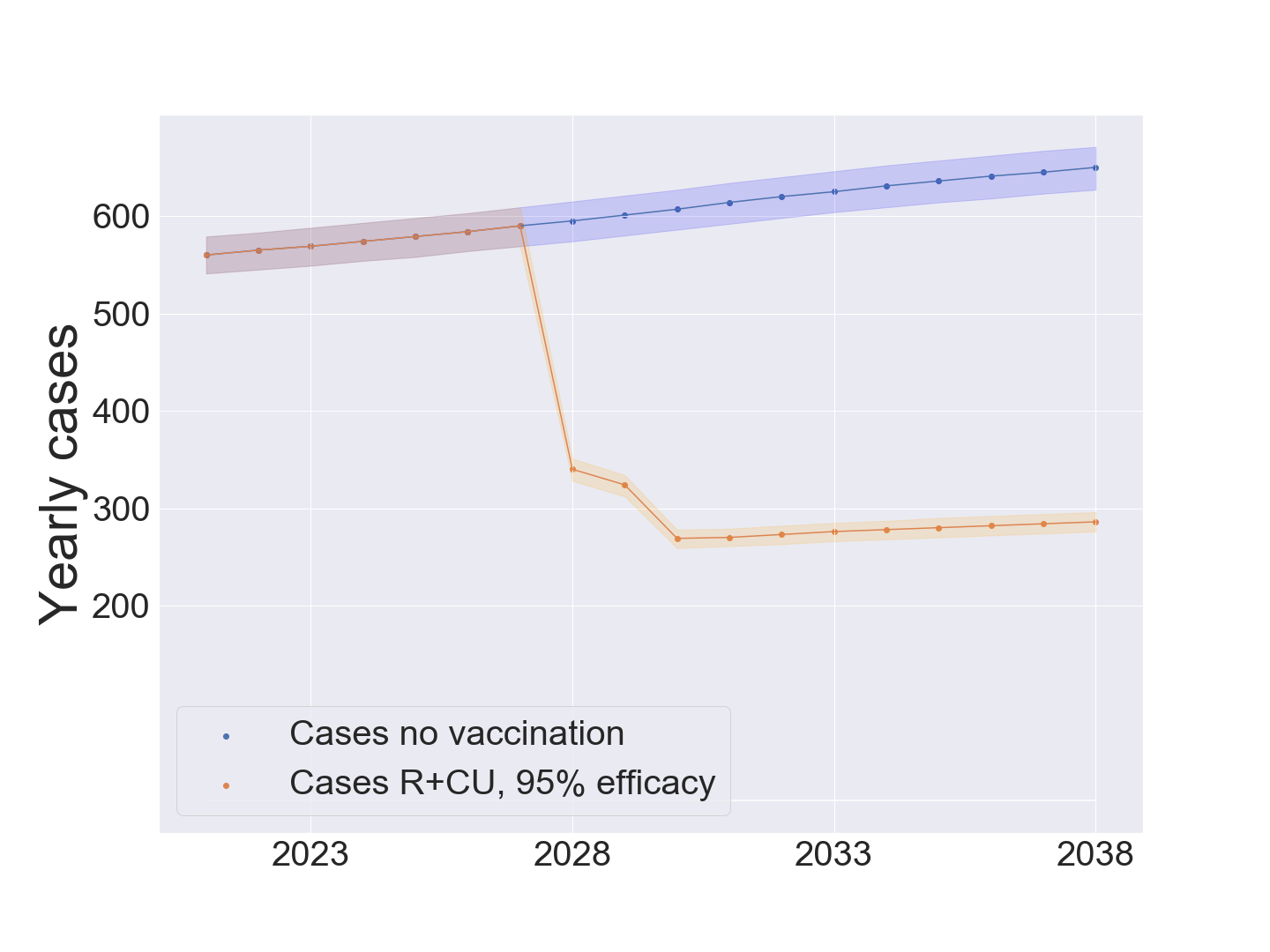}
\end{subfigure}
\caption{Comoros cumulative (top)  and yearly (bottom) iNTS cases under the status quo and routine + catch-up vaccination ($95\%$ efficacy) scenarios. Shaded areas show the 25th and 75th percentiles, line shows the median over 1000 experiments, samples drawn from uniform distributions over (0.00020,0.00024) for $\beta_{2,n}$ and (0.0080,0.0084) for $\beta_{4,n}$. }\label{fig:Comoros}
\end{figure}

\begin{figure}[htbp]
\renewcommand{\thefigure}{\textbf{Supplementary Fig. 21 Congo cumulative and yearly iNTS cases}}
\begin{subfigure}[b]{\textwidth}
\centering
\includegraphics[width=0.7\linewidth]{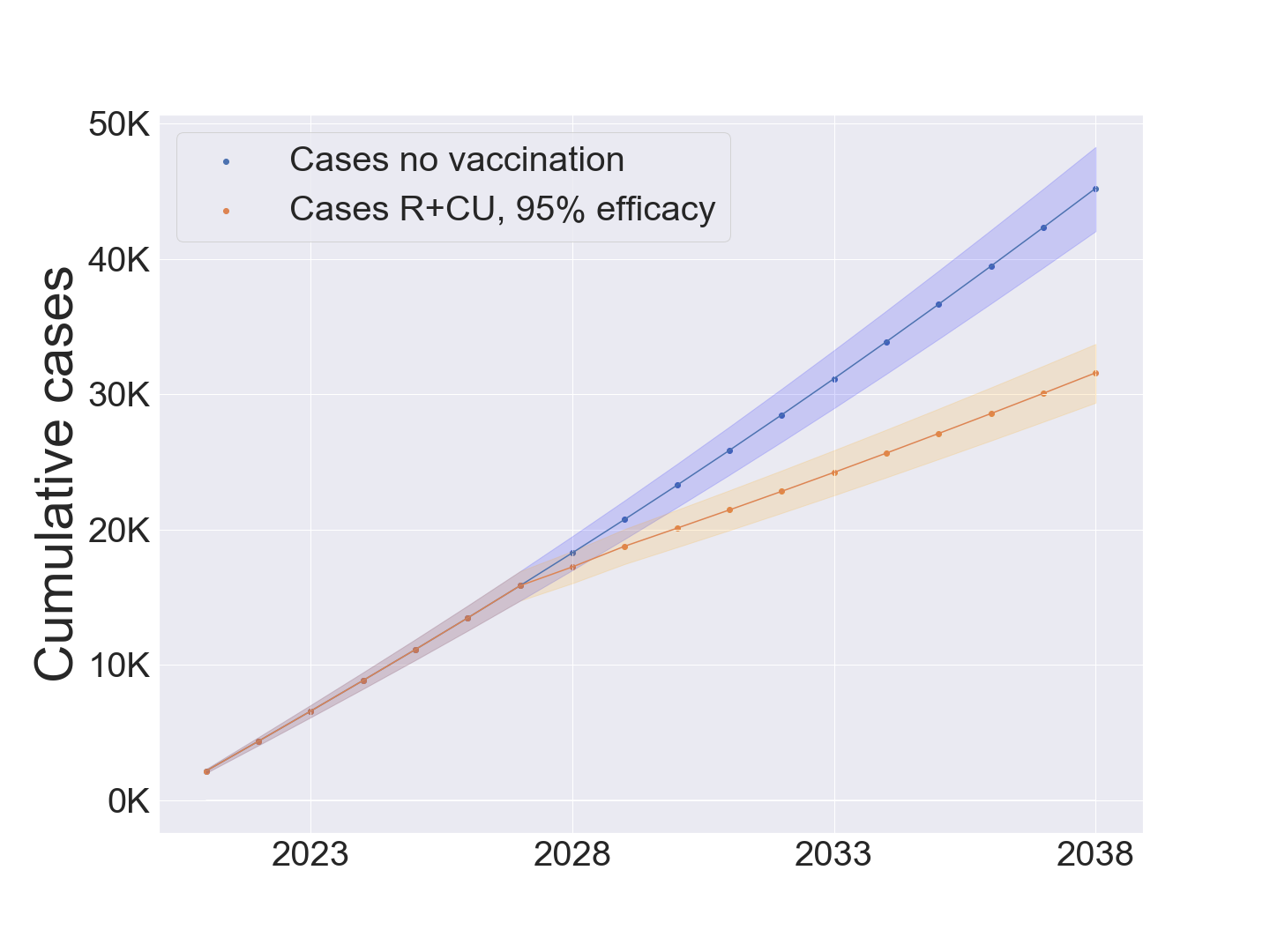}
\includegraphics[width=0.7\linewidth]{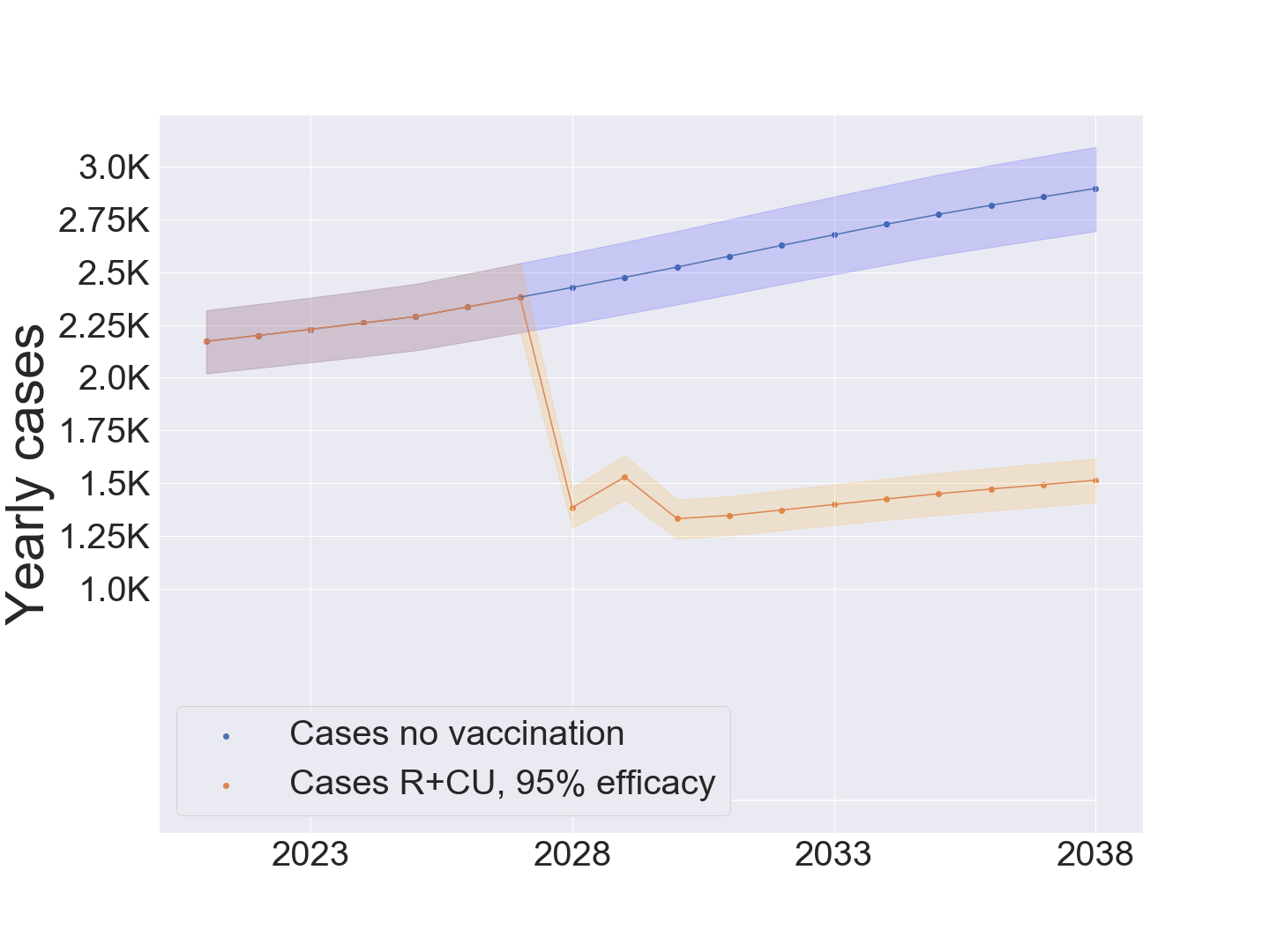}
\end{subfigure}
\caption{Congo cumulative (top)  and yearly (bottom) iNTS cases under the status quo and routine + catch-up vaccination ($95\%$ efficacy) scenarios. Shaded areas show the 25th and 75th percentiles, line shows the median over 1000 experiments, samples drawn from uniform distributions over (0.00020,0.00024) for $\beta_{2,n}$ and (0.0080,0.0084) for $\beta_{4,n}$. }\label{fig:Congo}
\end{figure}

\begin{figure}[htbp]
\renewcommand{\thefigure}{\textbf{Supplementary Fig. 22 Democratic Republic of the Congo cumulative and yearly iNTS cases}}
\begin{subfigure}[b]{\textwidth}
\centering
\includegraphics[width=0.7\linewidth]{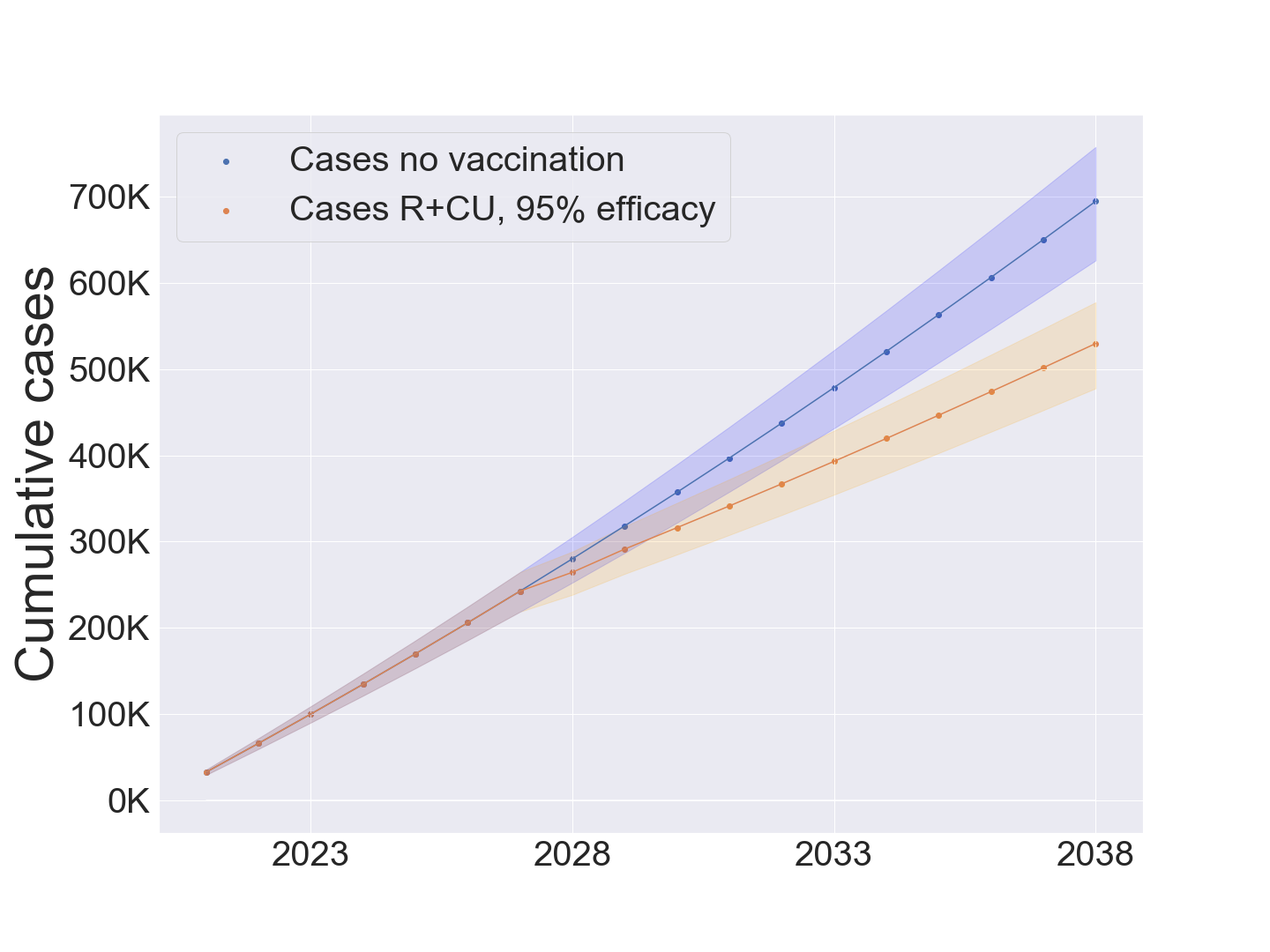}
\includegraphics[width=0.7\linewidth]{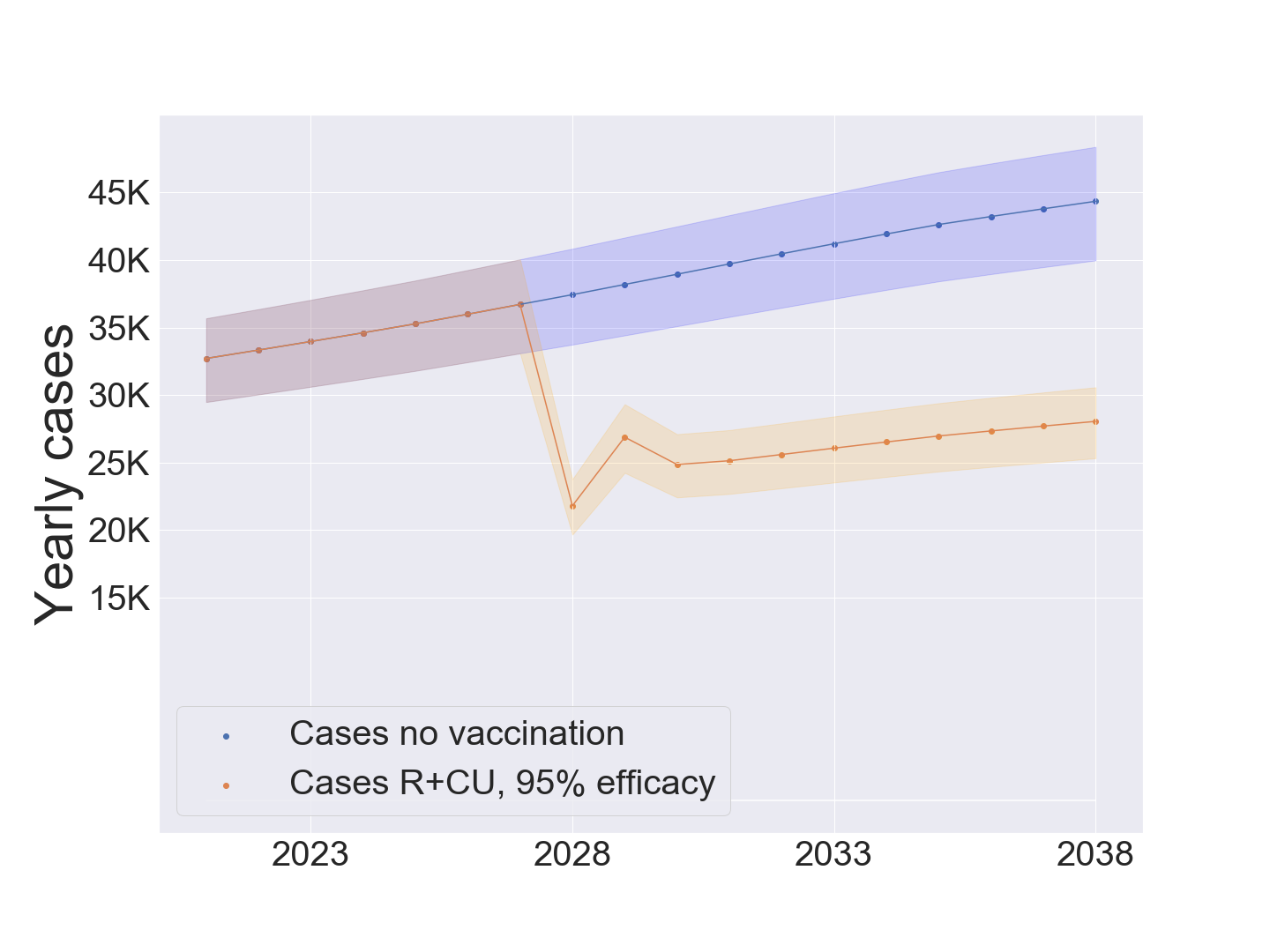}
\end{subfigure}
\caption{Democratic Republic of the Congo cumulative (top)  and yearly (bottom) iNTS cases under the status quo and routine + catch-up vaccination ($95\%$ efficacy) scenarios. Shaded areas show the 25th and 75th percentiles, line shows the median over 1000 experiments, samples drawn from uniform distributions over (0.00020,0.00024) for $\beta_{2,n}$ and (0.0080,0.0084) for $\beta_{4,n}$. }\label{fig:DRC}
\end{figure}

\begin{figure}[htbp]
\renewcommand{\thefigure}{\textbf{Supplementary Fig. 23 Dijbouti cumulative and yearly iNTS cases}}
\begin{subfigure}[b]{\textwidth}
\centering
\includegraphics[width=0.7\linewidth]{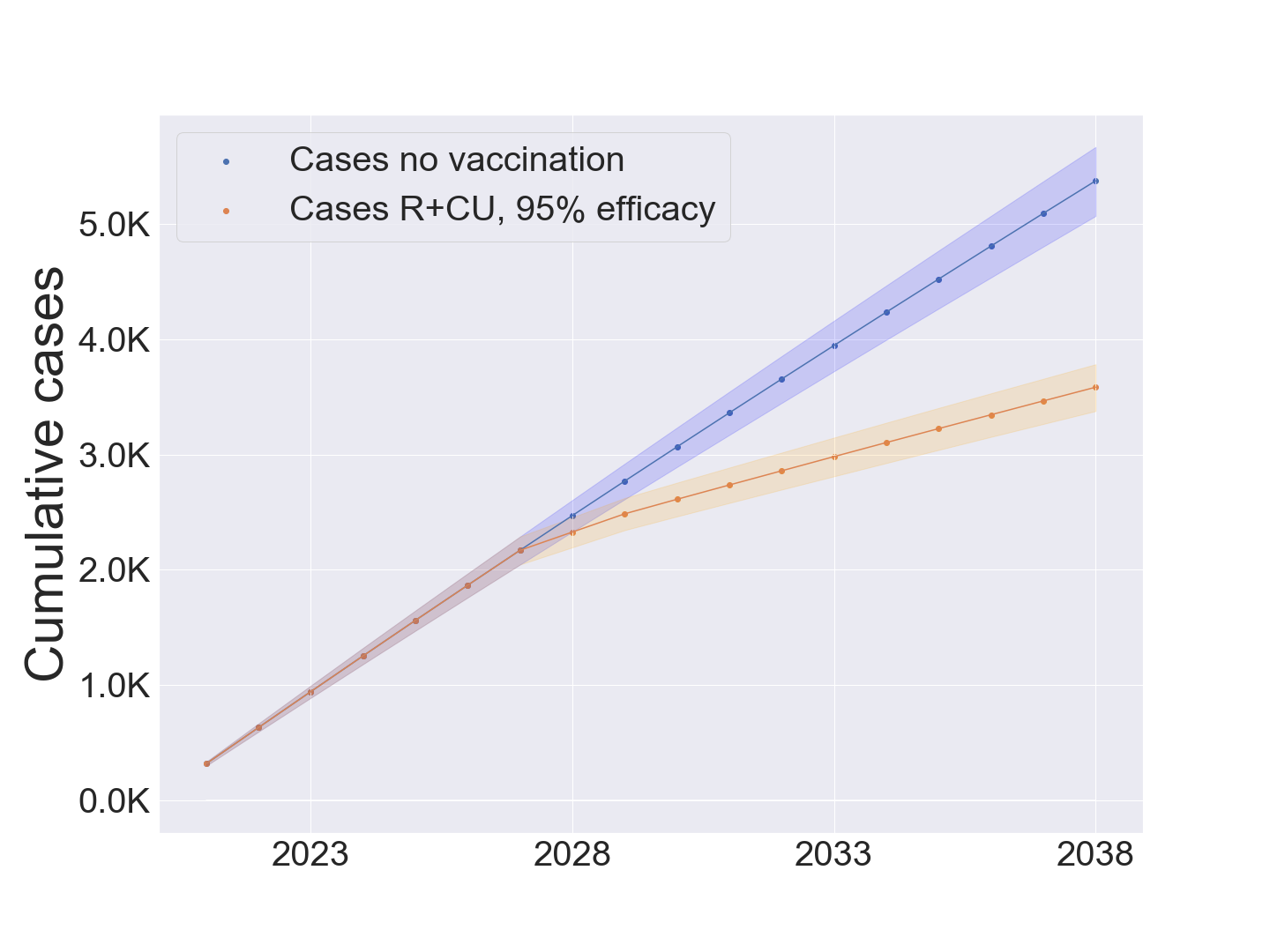}
\includegraphics[width=0.7\linewidth]{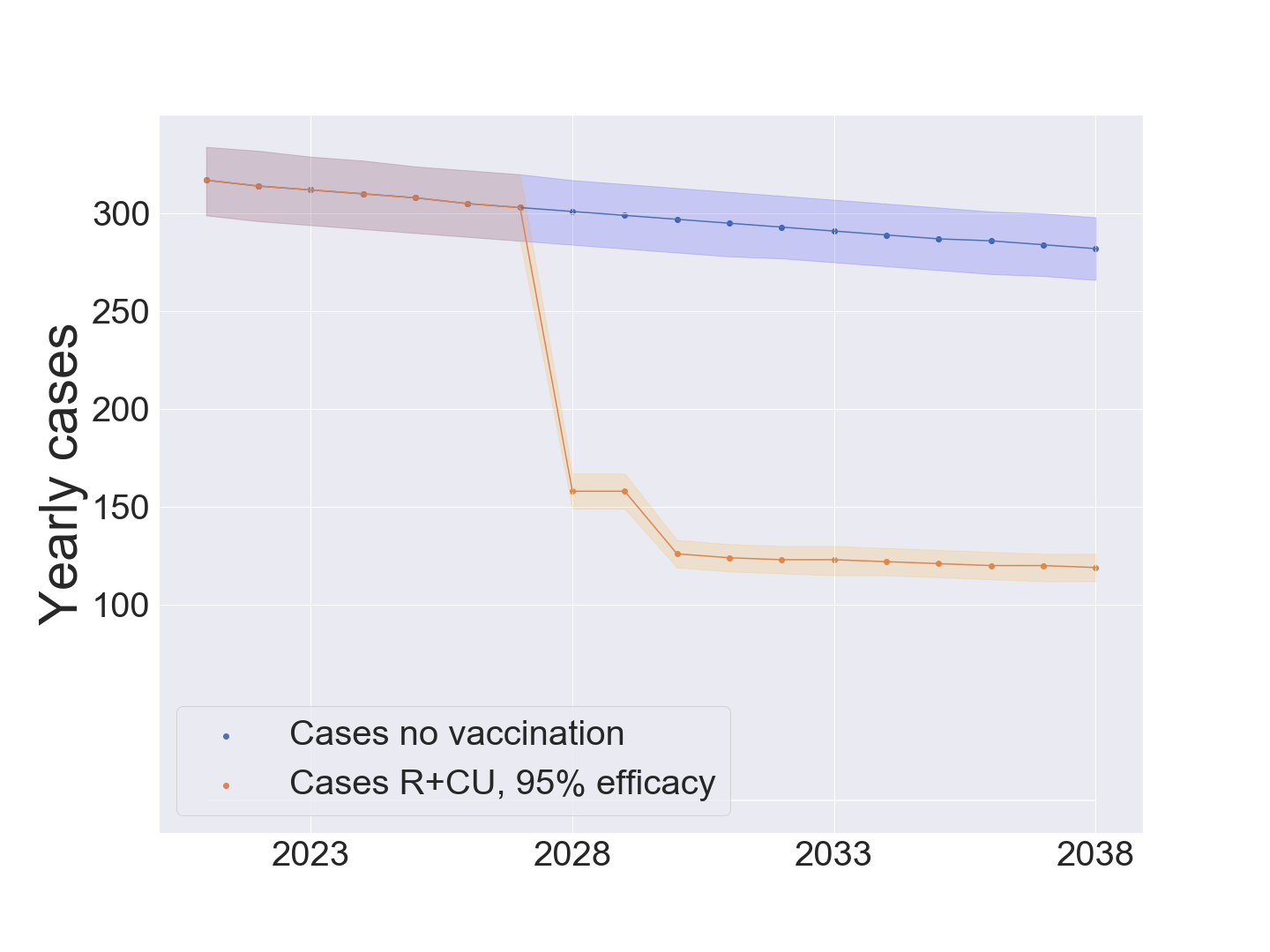}
\end{subfigure}
\caption{Djibouti cumulative (top)  and yearly (bottom) iNTS cases under the status quo and routine + catch-up vaccination ($95\%$ efficacy) scenarios. Shaded areas show the 25th and 75th percentiles, line shows the median over 1000 experiments, samples drawn from uniform distributions over (0.00020,0.00024) for $\beta_{2,n}$ and (0.0080,0.0084) for $\beta_{4,n}$. }\label{fig:Djibouti}
\end{figure}

\begin{figure}[htbp]
\renewcommand{\thefigure}{\textbf{Supplementary Fig. 24 Equatorial Guinea cumulative and yearly iNTS cases}}
\begin{subfigure}[b]{\textwidth}
\centering
\includegraphics[width=0.7\linewidth]{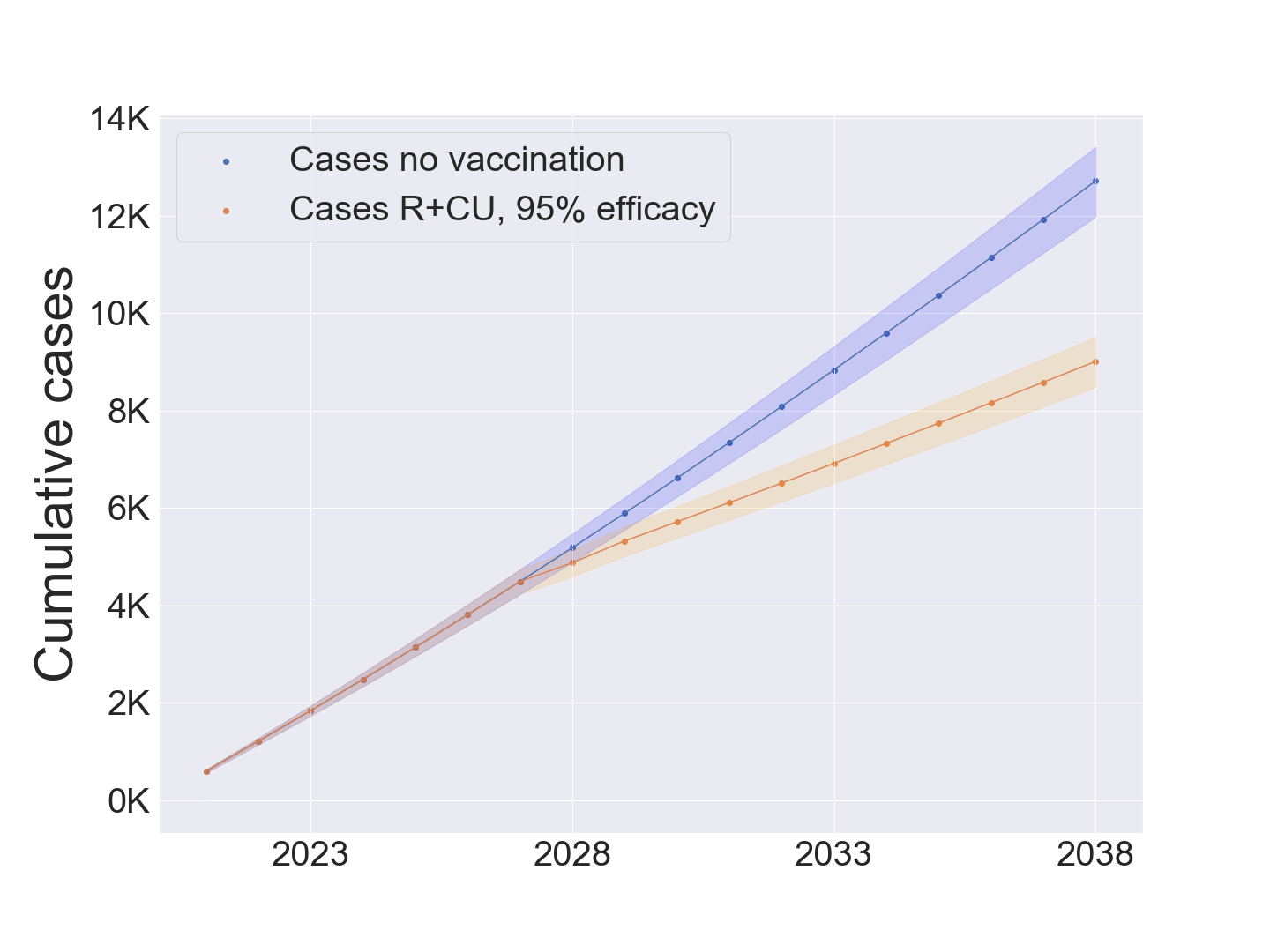}
\includegraphics[width=0.7\linewidth]{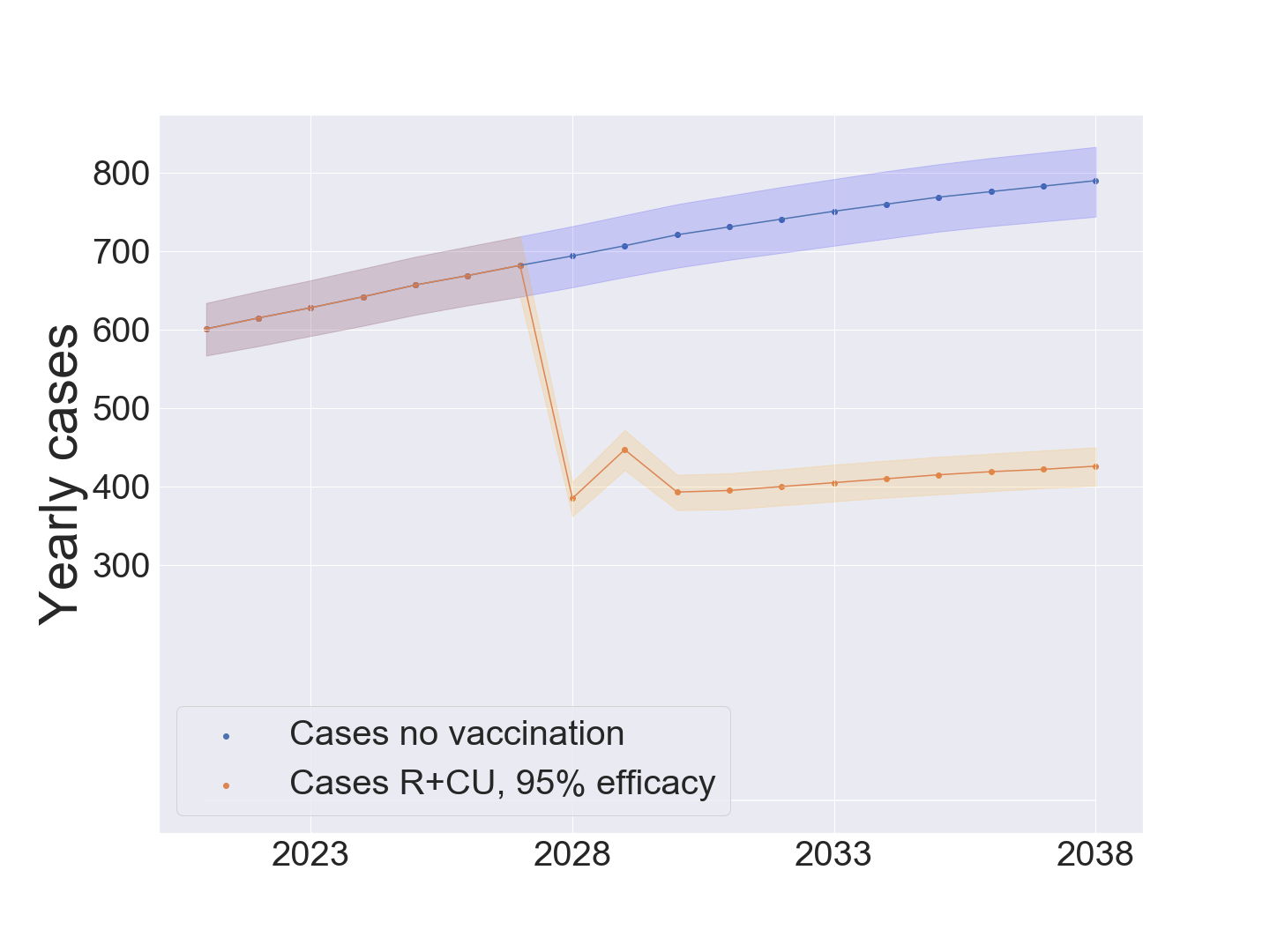}
\end{subfigure}
\caption{Equatorial Guinea cumulative (top)  and yearly (bottom) iNTS cases under the status quo and routine + catch-up vaccination ($95\%$ efficacy) scenarios. Shaded areas show the 25th and 75th percentiles, line shows the median over 1000 experiments, samples drawn from uniform distributions over (0.00020,0.00024) for $\beta_{2,n}$ and (0.0080,0.0084) for $\beta_{4,n}$. }\label{fig:EquatorialGuinea}
\end{figure}

\begin{figure}[htbp]
\renewcommand{\thefigure}{\textbf{Supplementary Fig. 25 Eritrea cumulative and yearly iNTS cases}}
\begin{subfigure}[b]{\textwidth}
\centering
\includegraphics[width=0.7\linewidth]{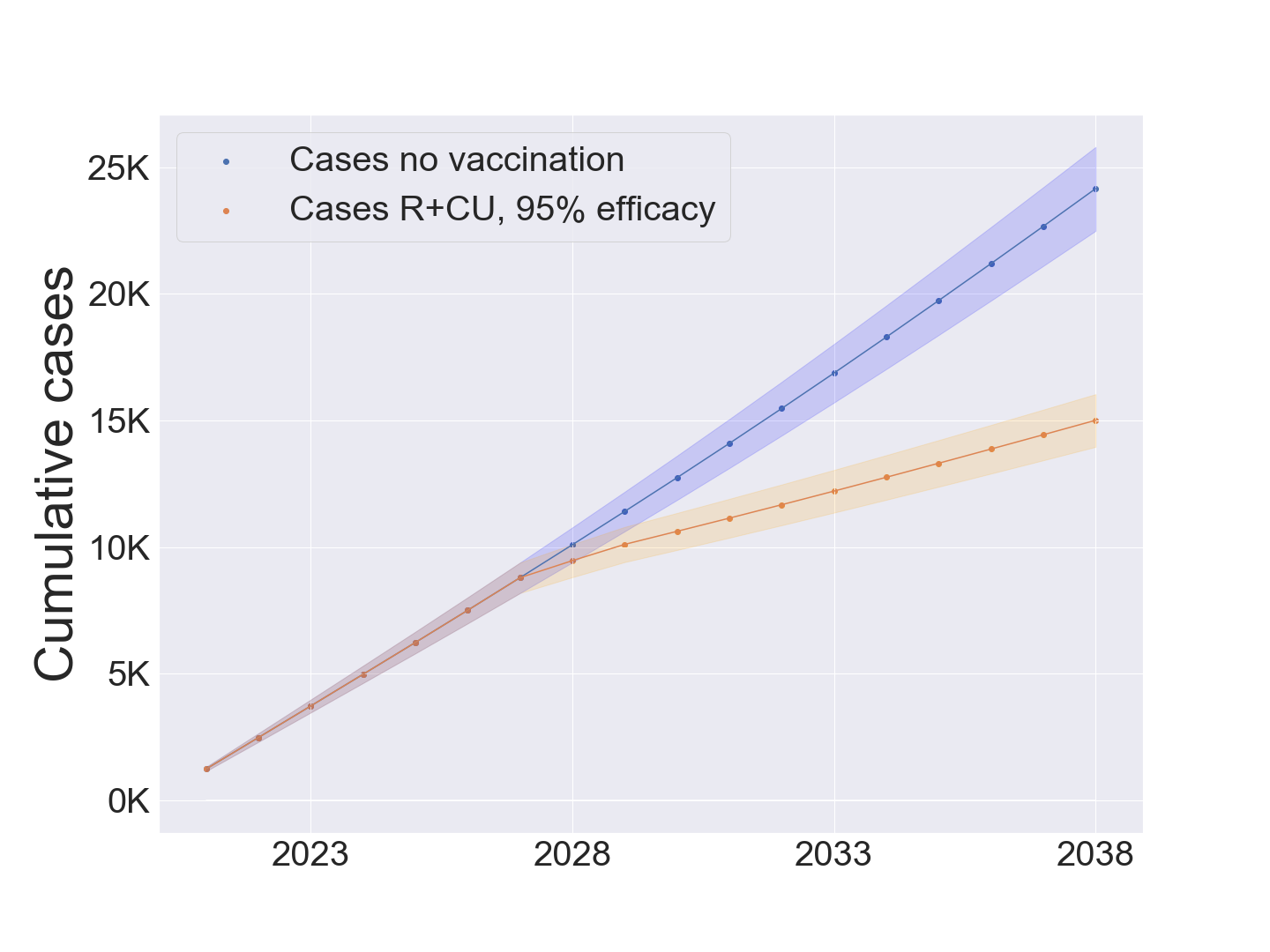}
\includegraphics[width=0.7\linewidth]{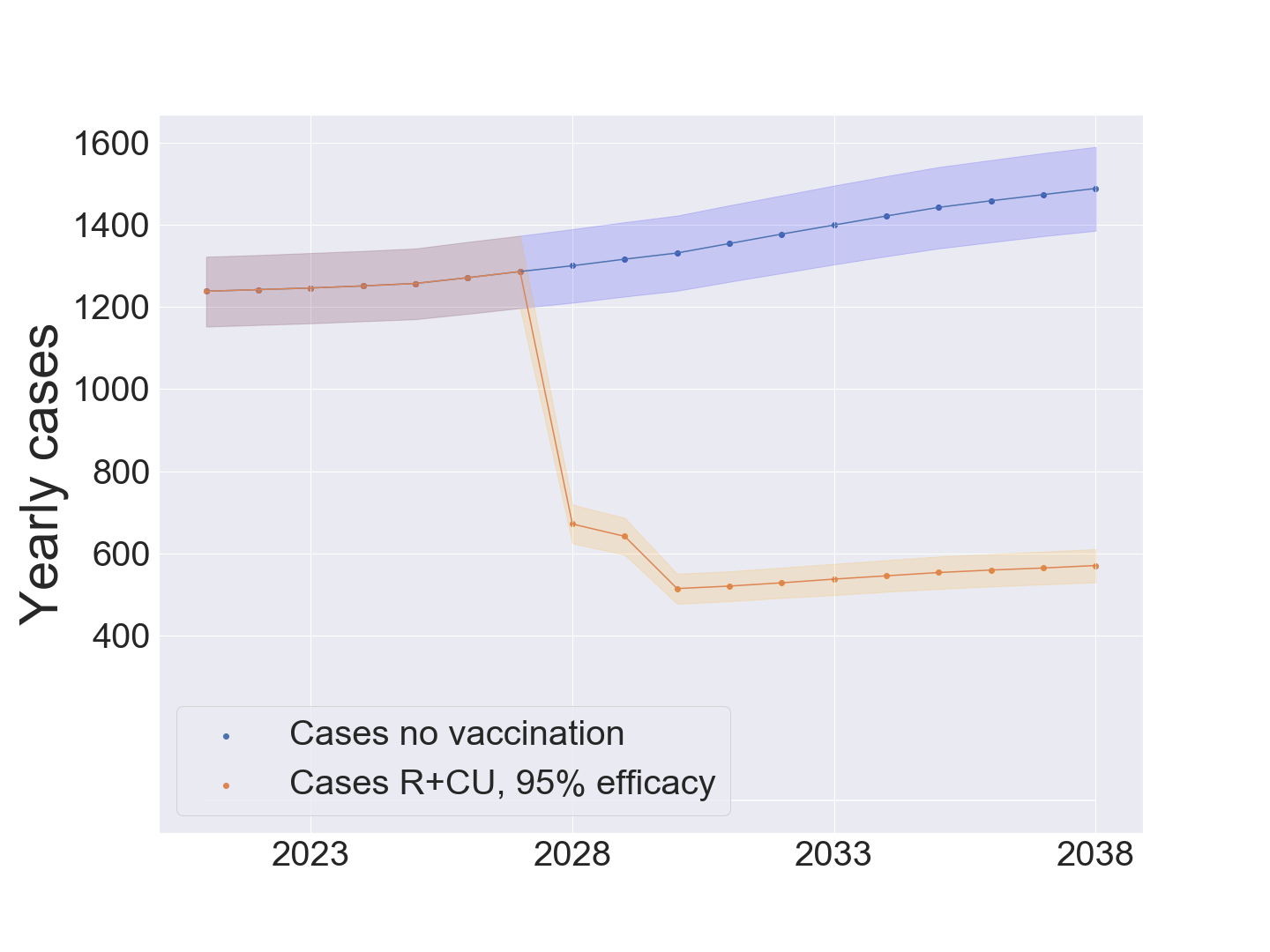}
\end{subfigure}
\caption{Eritrea cumulative (top)  and yearly (bottom) iNTS cases under the status quo and routine + catch-up vaccination ($95\%$ efficacy) scenarios. Shaded areas show the 25th and 75th percentiles, line shows the median over 1000 experiments, samples drawn from uniform distributions over (0.00020,0.00024) for $\beta_{2,n}$ and (0.0080,0.0084) for $\beta_{4,n}$. }\label{fig:Eritrea}
\end{figure}

\begin{figure}[htbp]
\renewcommand{\thefigure}{\textbf{Supplementary Fig. 26 Eswatini cumulative and yearly iNTS cases}}
\begin{subfigure}[b]{\textwidth}
\centering
\includegraphics[width=0.7\linewidth]{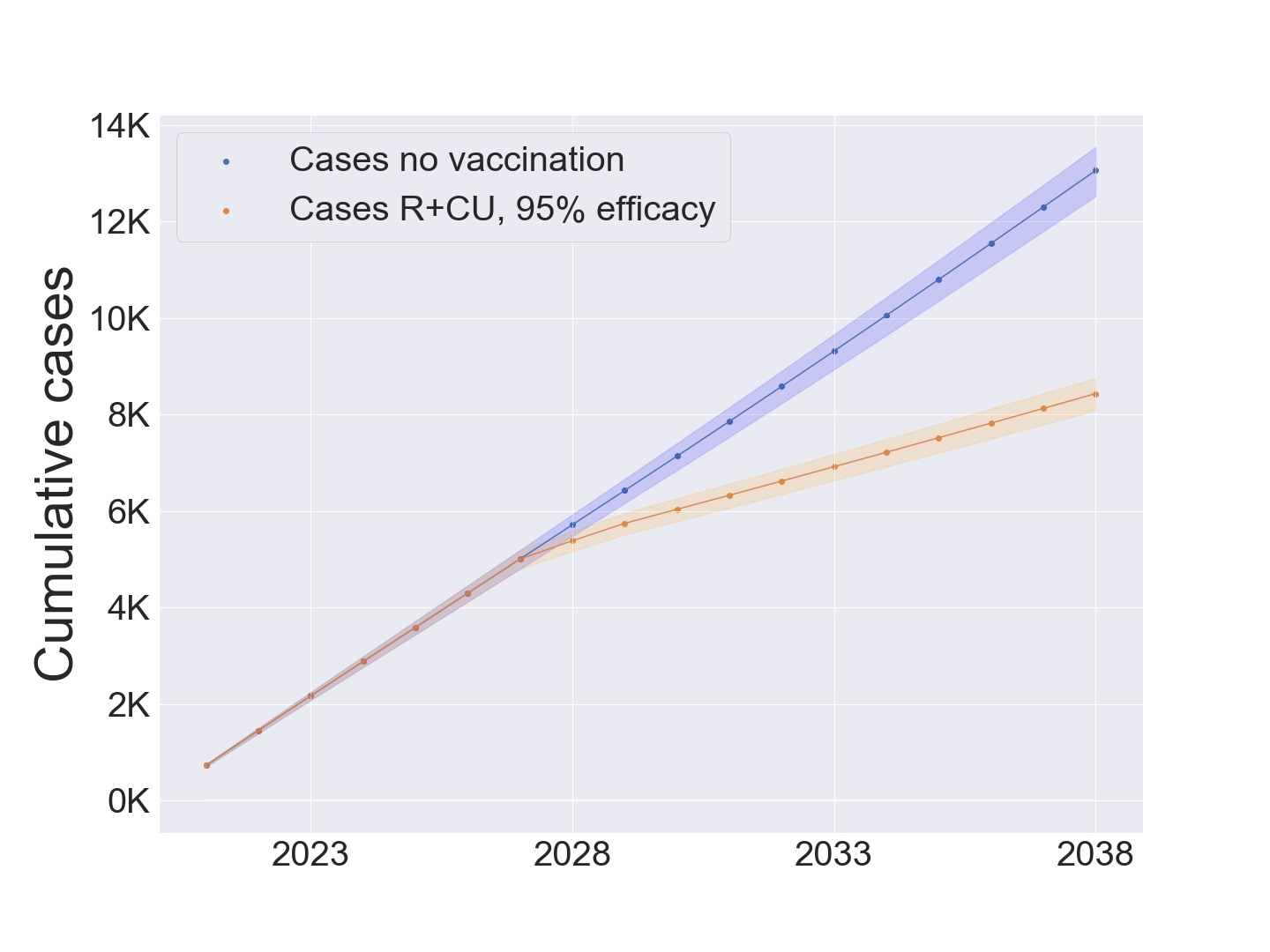}
\includegraphics[width=0.7\linewidth]{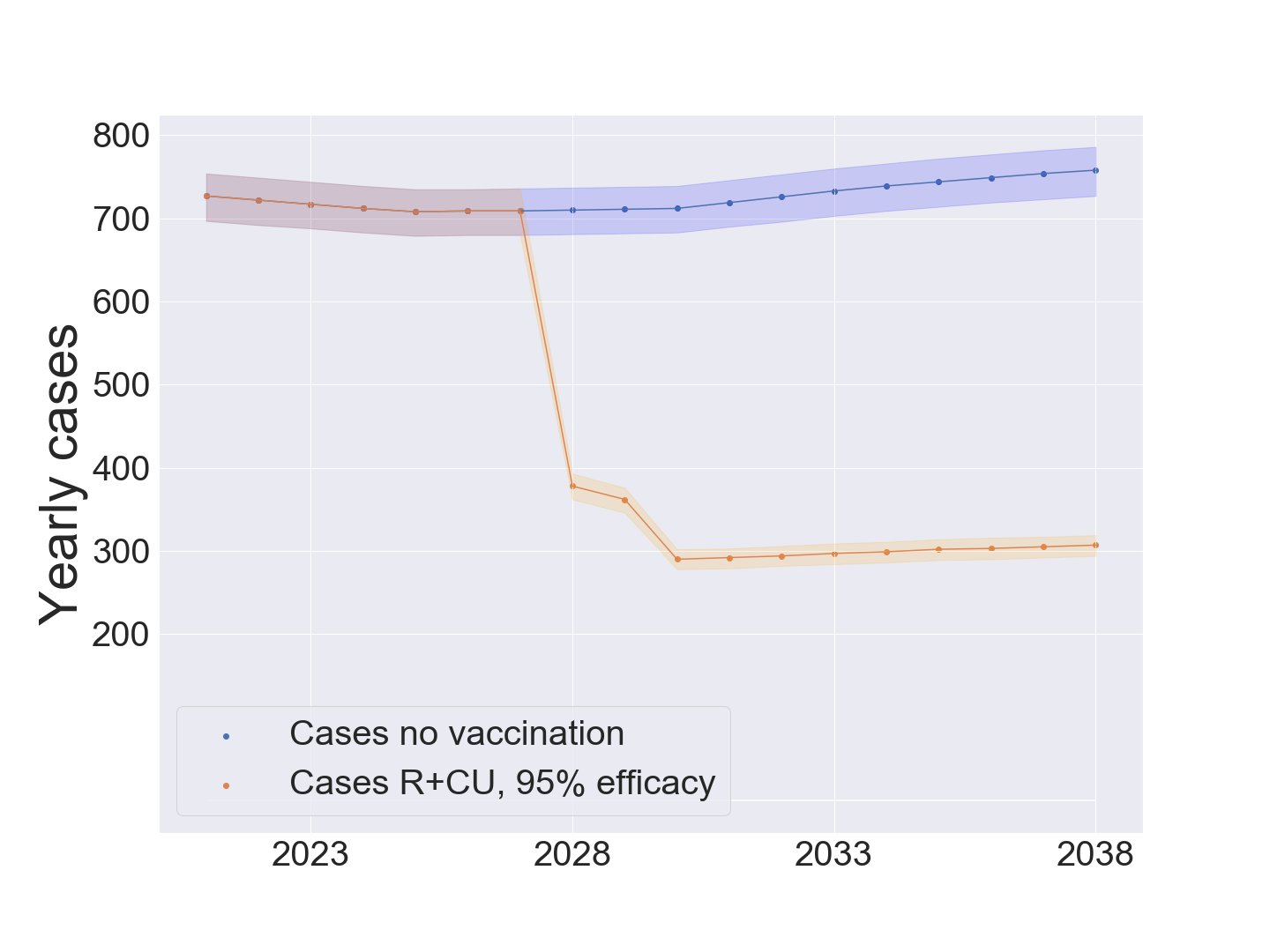}
\end{subfigure}
\caption{Eswatini cumulative (top)  and yearly (bottom) iNTS cases under the status quo and routine + catch-up vaccination ($95\%$ efficacy) scenarios. Shaded areas show the 25th and 75th percentiles, line shows the median over 1000 experiments, samples drawn from uniform distributions over (0.00020,0.00024) for $\beta_{2,n}$ and (0.0080,0.0084) for $\beta_{4,n}$. }\label{fig:Eswatini}
\end{figure}

\begin{figure}[htbp]
\renewcommand{\thefigure}{\textbf{Supplementary Fig. 27 Ethiopia cumulative and yearly iNTS cases}}
\begin{subfigure}[b]{\textwidth}
\centering
\includegraphics[width=0.7\linewidth]{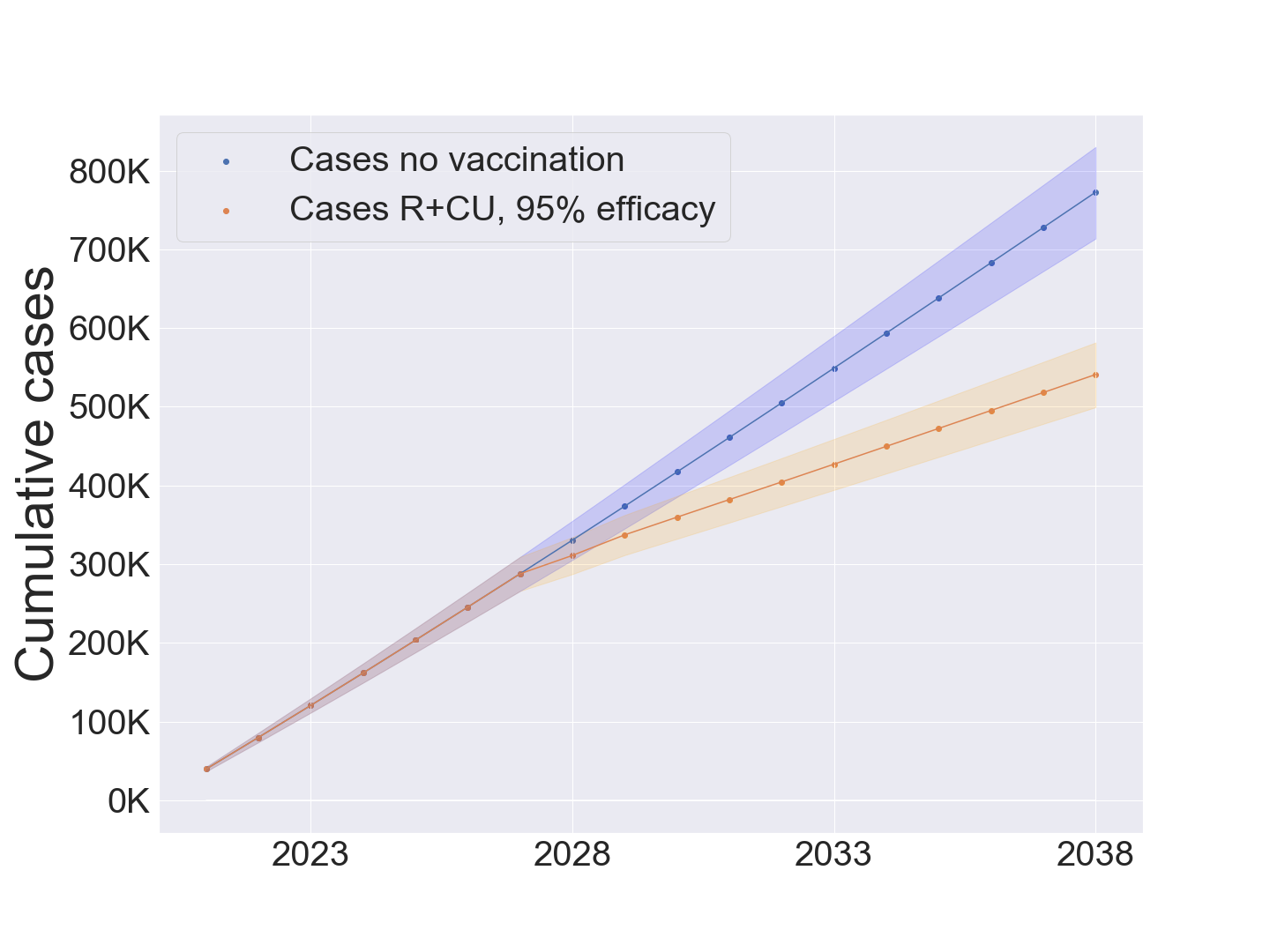}
\includegraphics[width=0.7\linewidth]{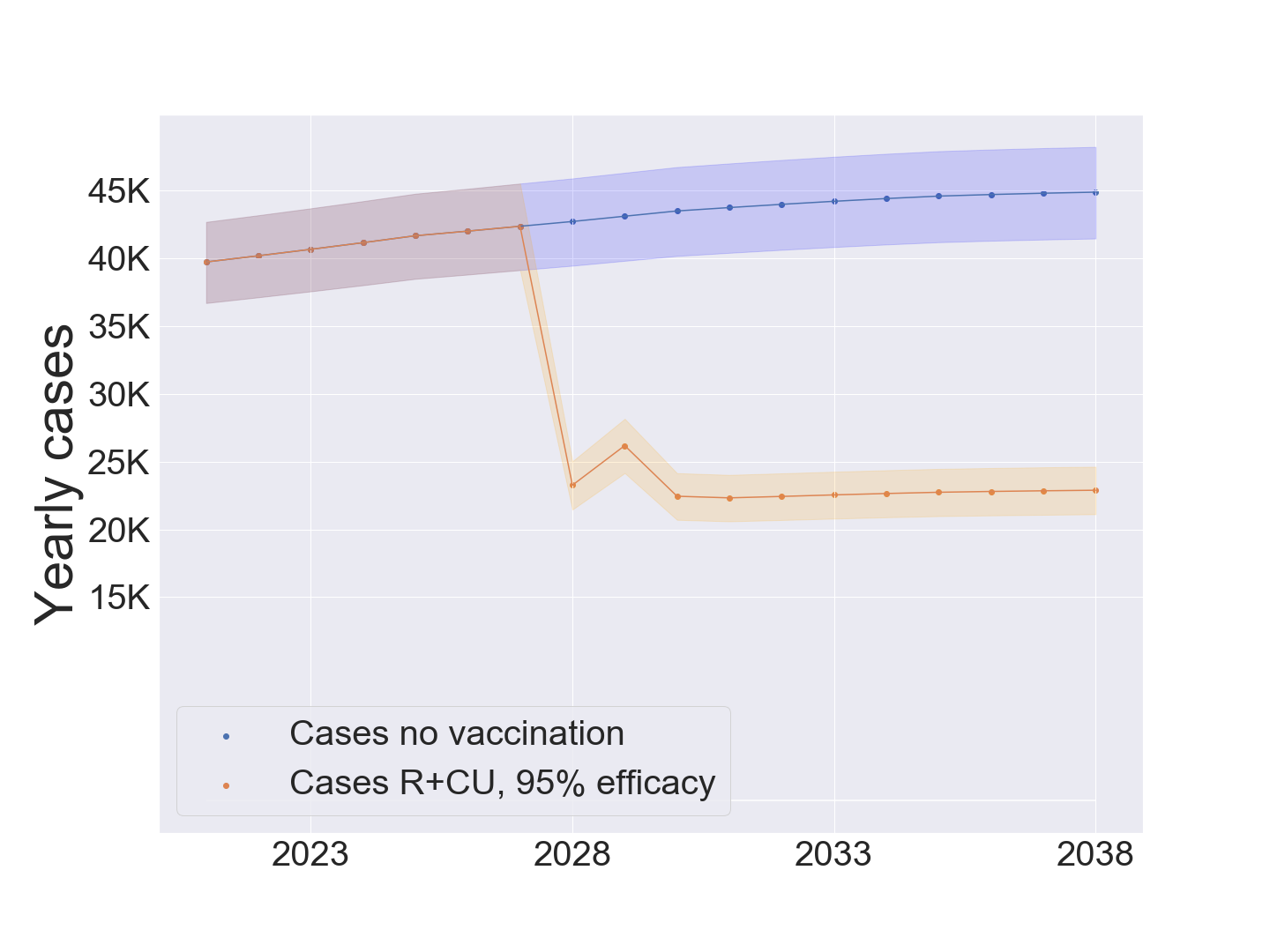}
\end{subfigure}
\caption{Ethiopia cumulative (top)  and yearly (bottom) iNTS cases under the status quo and routine + catch-up vaccination ($95\%$ efficacy) scenarios. Shaded areas show the 25th and 75th percentiles, line shows the median over 1000 experiments, samples drawn from uniform distributions over (0.00020,0.00024) for $\beta_{2,n}$ and (0.0080,0.0084) for $\beta_{4,n}$. }\label{fig:Ethiopia}
\end{figure}

\begin{figure}[htbp]
\renewcommand{\thefigure}{\textbf{Supplementary Fig. 28 Gabon cumulative and yearly iNTS cases}}
\begin{subfigure}[b]{\textwidth}
\centering
\includegraphics[width=0.7\linewidth]{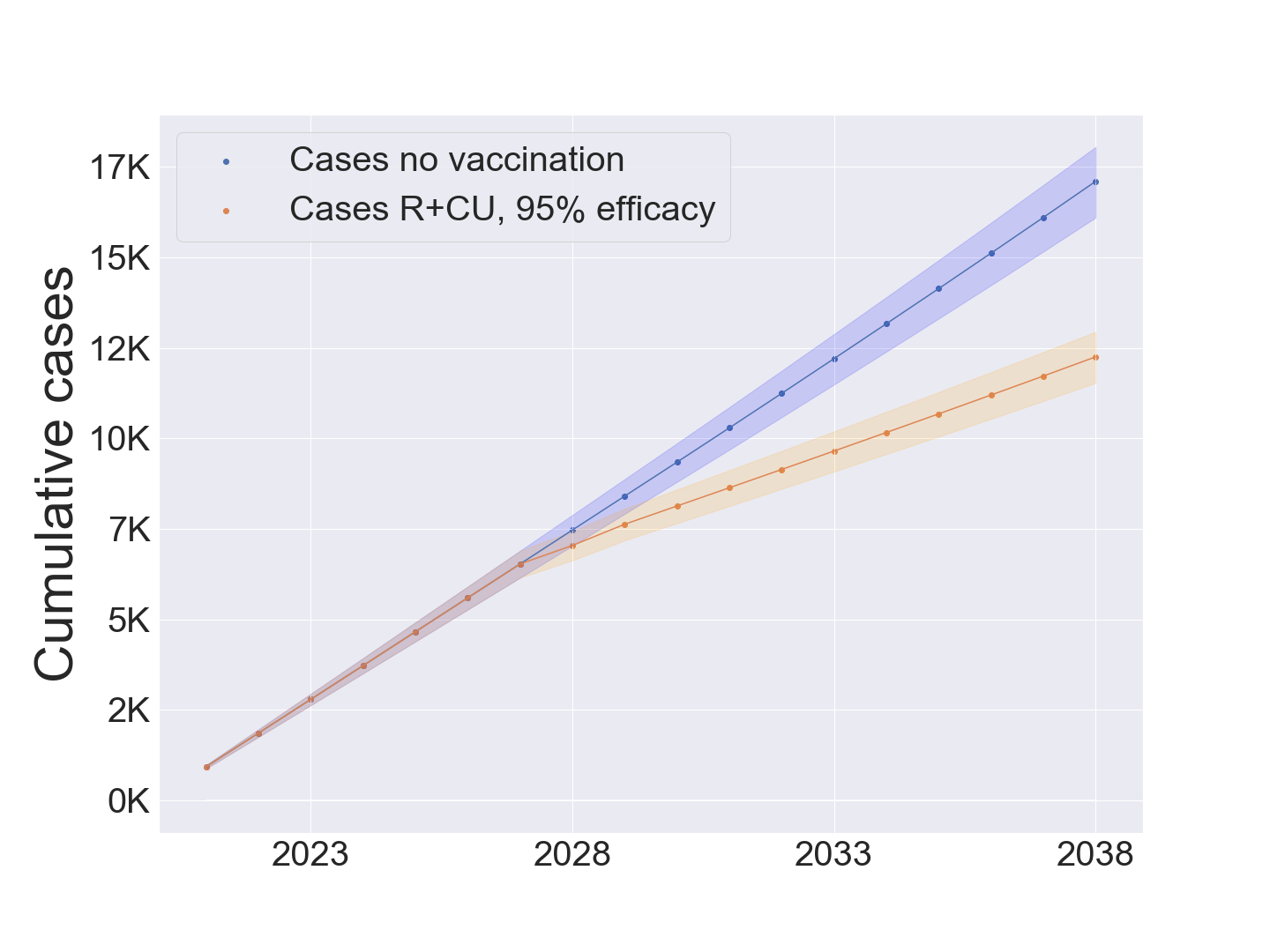}
\includegraphics[width=0.7\linewidth]{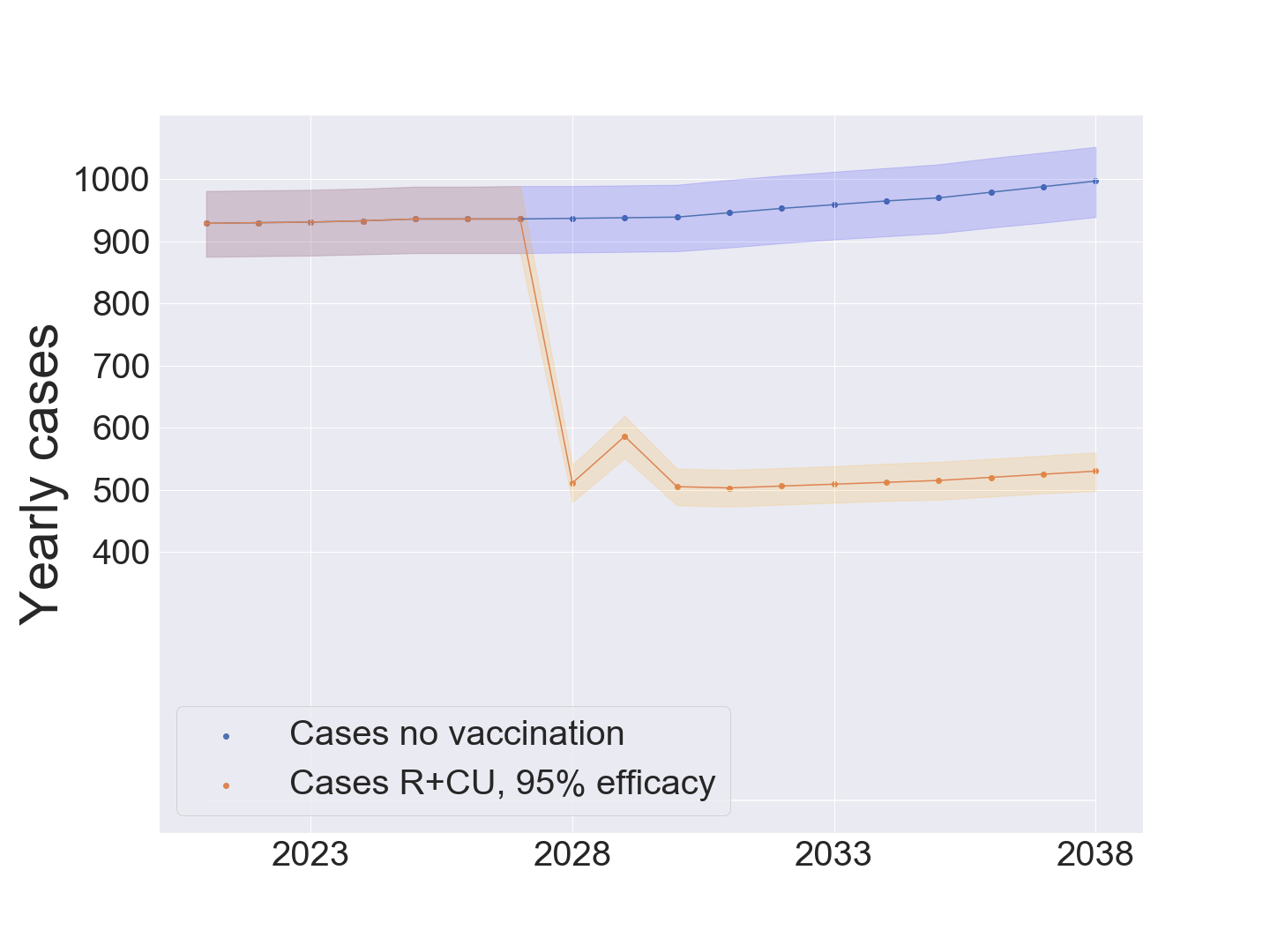}
\end{subfigure}
\caption{Gabon cumulative (top)  and yearly (bottom) iNTS cases under the status quo and routine + catch-up vaccination ($95\%$ efficacy) scenarios. Shaded areas show the 25th and 75th percentiles, line shows the median over 1000 experiments, samples drawn from uniform distributions over (0.00020,0.00024) for $\beta_{2,n}$ and (0.0080,0.0084) for $\beta_{4,n}$. }\label{fig:Gabon}
\end{figure}

\begin{figure}[htbp]
\renewcommand{\thefigure}{\textbf{Supplementary Fig. 29 Gambia cumulative and yearly iNTS cases}}
\begin{subfigure}[b]{\textwidth}
\centering
\includegraphics[width=0.7\linewidth]{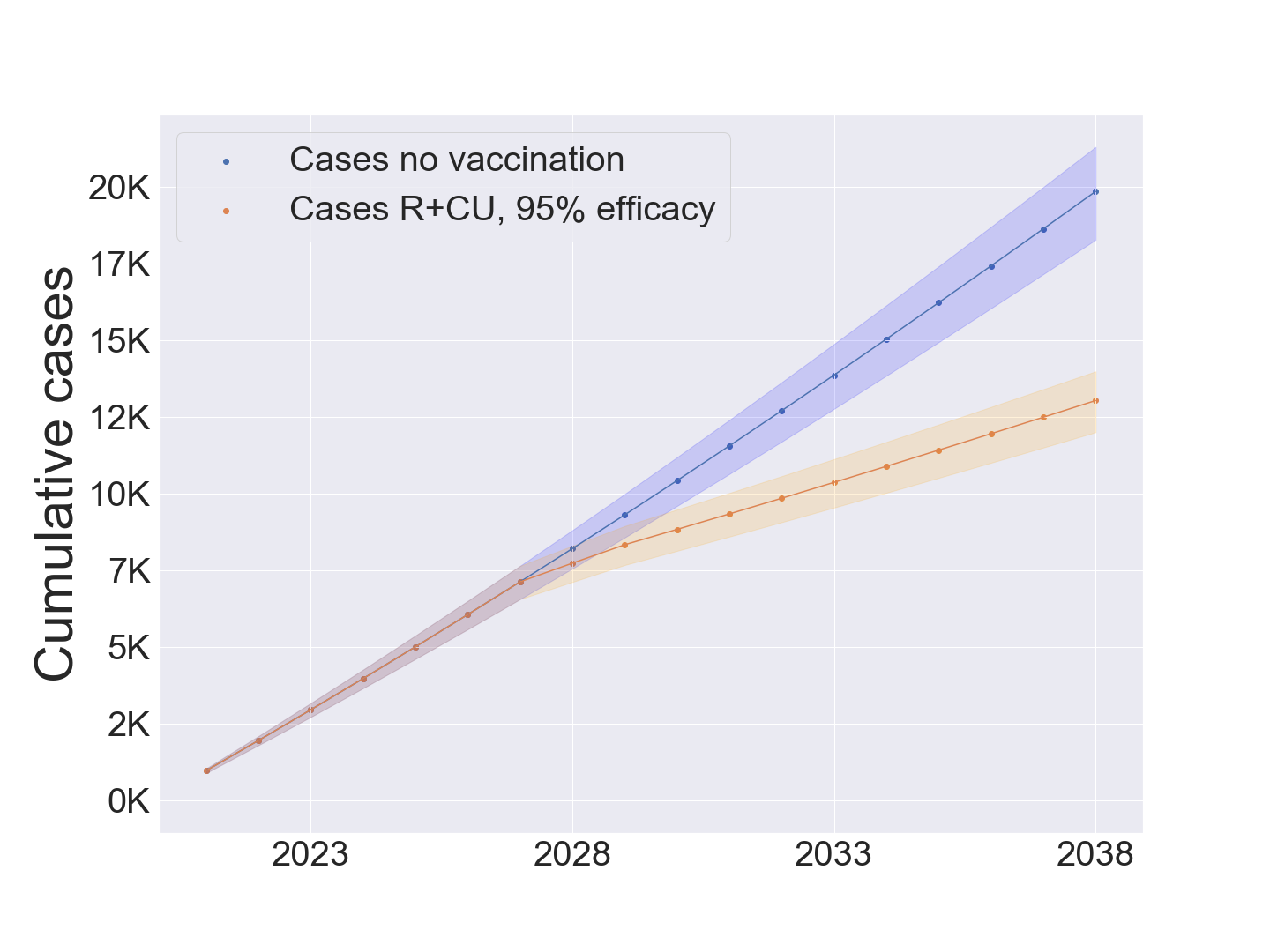}
\includegraphics[width=0.7\linewidth]{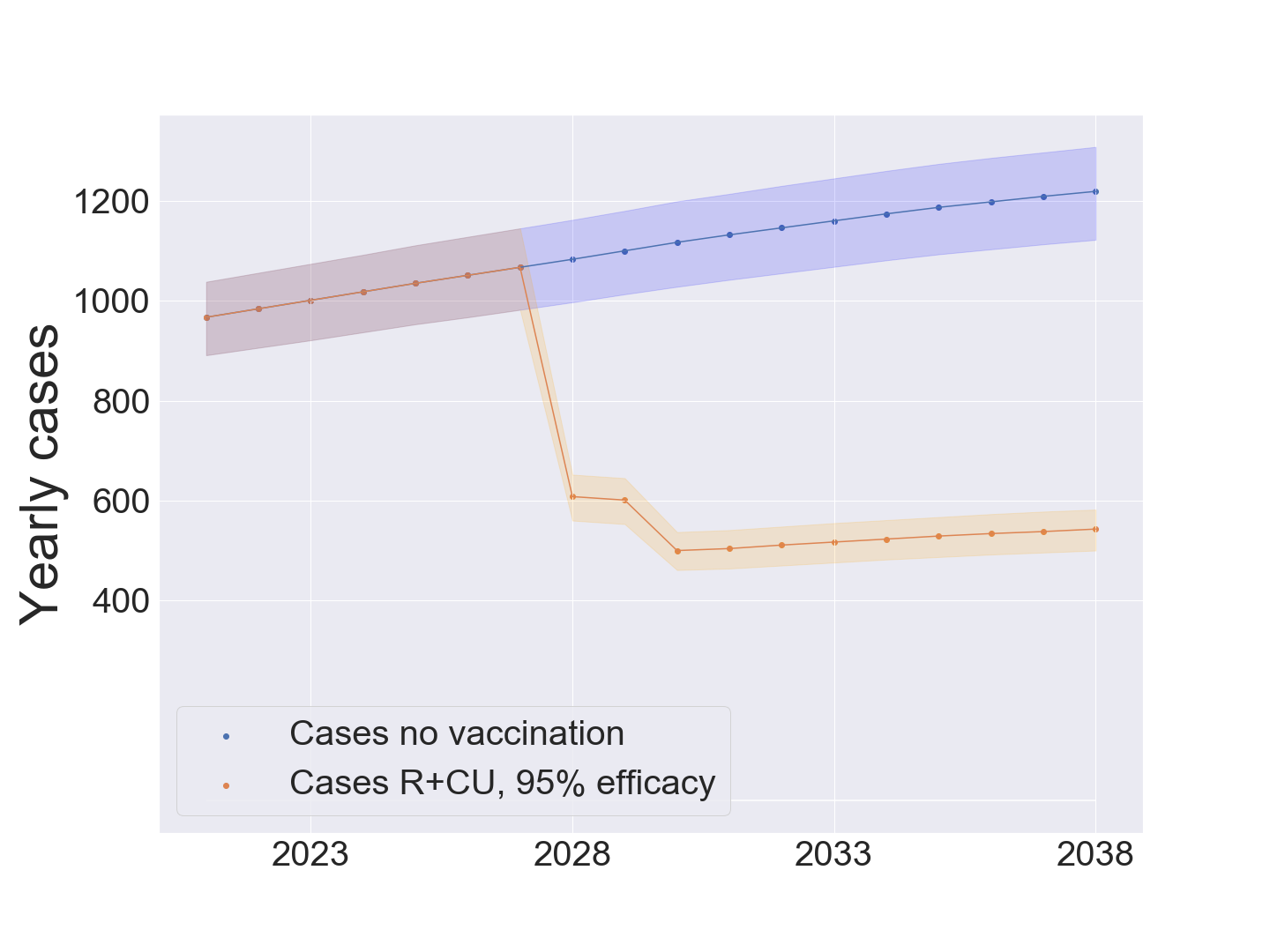}
\end{subfigure}
\caption{Gambia cumulative (top)  and yearly (bottom) iNTS cases under the status quo and routine + catch-up vaccination ($95\%$ efficacy) scenarios. Shaded areas show the 25th and 75th percentiles, line shows the median over 1000 experiments, samples drawn from uniform distributions over (0.00020,0.00024) for $\beta_{2,n}$ and (0.0080,0.0084) for $\beta_{4,n}$. }\label{fig:Gambia}
\end{figure}

\begin{figure}[htbp]
\renewcommand{\thefigure}{\textbf{Supplementary Fig. 30 Ghana cumulative and yearly iNTS cases}}
\begin{subfigure}[b]{\textwidth}
\centering
\includegraphics[width=0.7\linewidth]{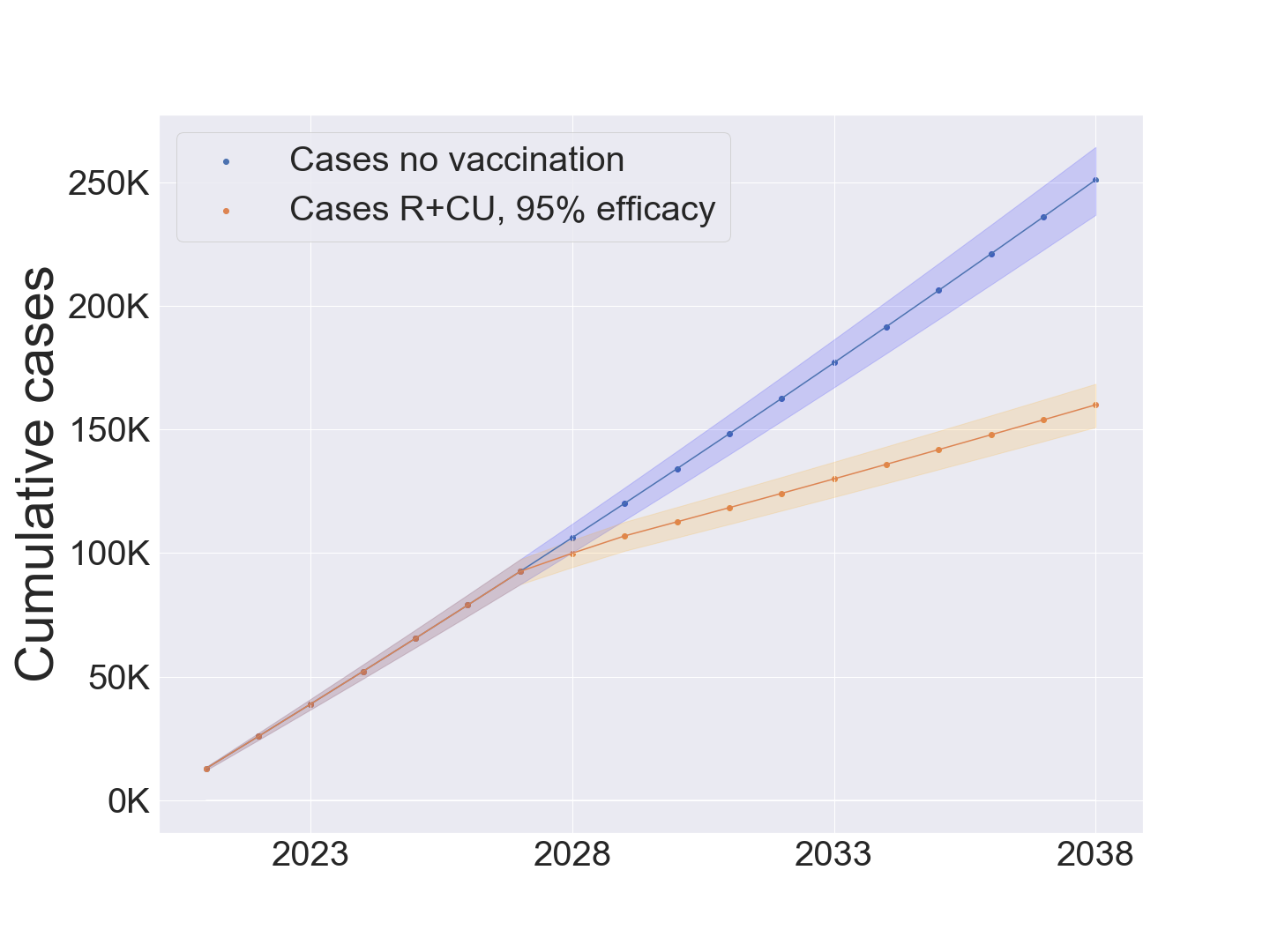}
\includegraphics[width=0.7\linewidth]{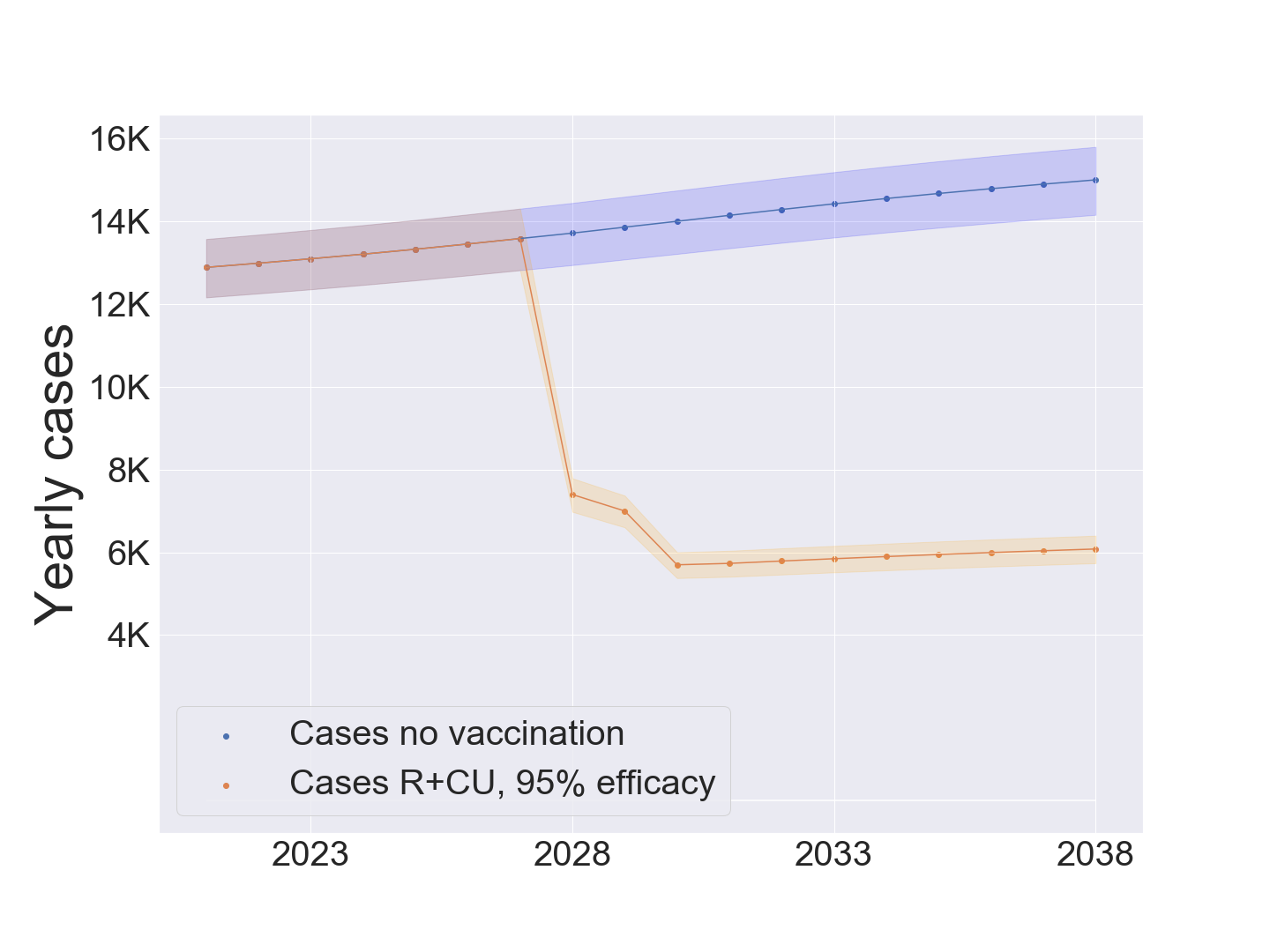}
\end{subfigure}
\caption{Ghana cumulative (top)  and yearly (bottom) iNTS cases under the status quo and routine + catch-up vaccination ($95\%$ efficacy) scenarios. Shaded areas show the 25th and 75th percentiles, line shows the median over 1000 experiments, samples drawn from uniform distributions over (0.00020,0.00024) for $\beta_{2,n}$ and (0.0080,0.0084) for $\beta_{4,n}$. }\label{fig:Ghana}
\end{figure}

\begin{figure}[htbp]
\renewcommand{\thefigure}{\textbf{Supplementary Fig. 31 Guinea cumulative and yearly iNTS cases}}
\begin{subfigure}[b]{\textwidth}
\centering
\includegraphics[width=0.7\linewidth]{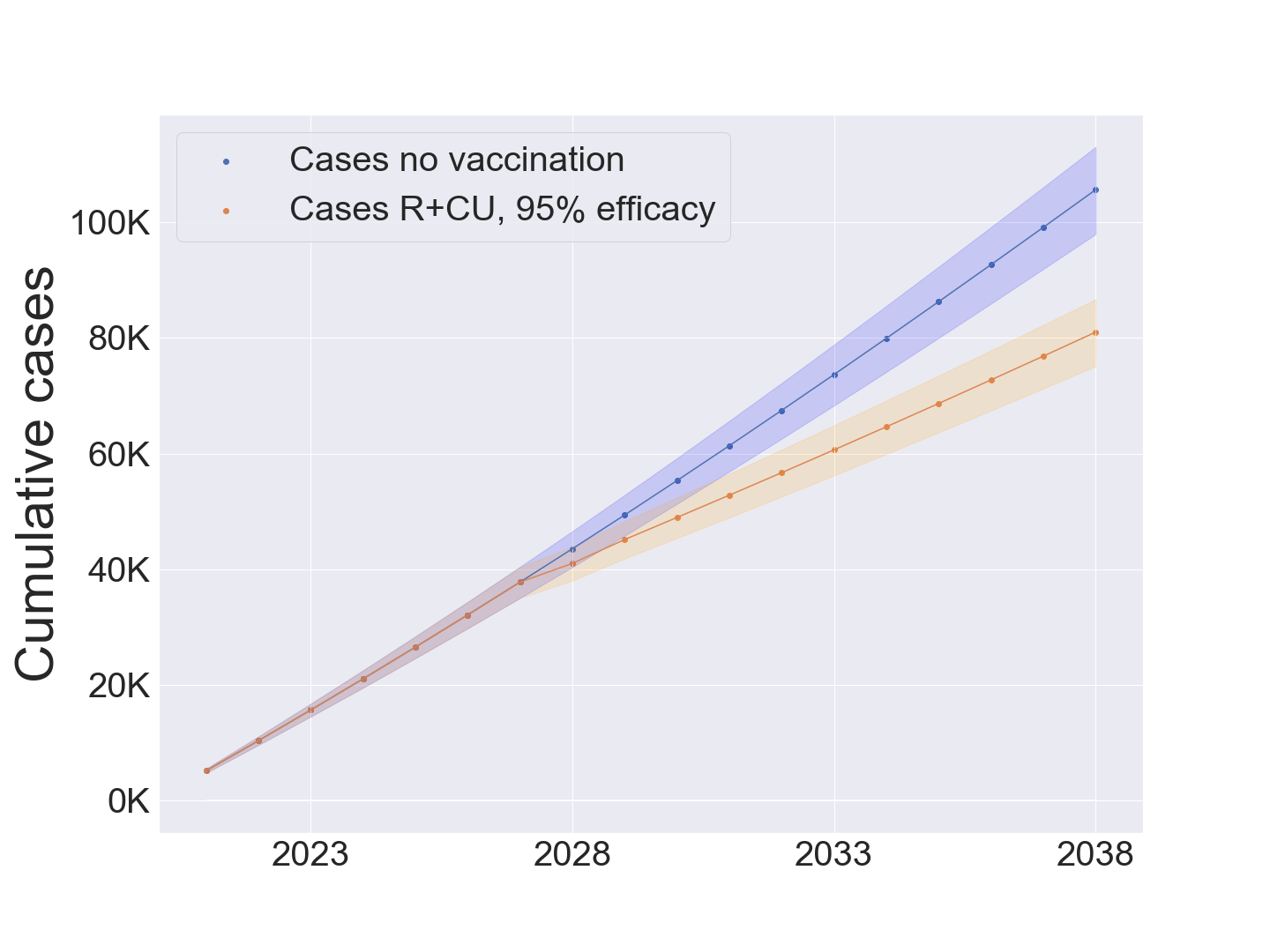}
\includegraphics[width=0.7\linewidth]{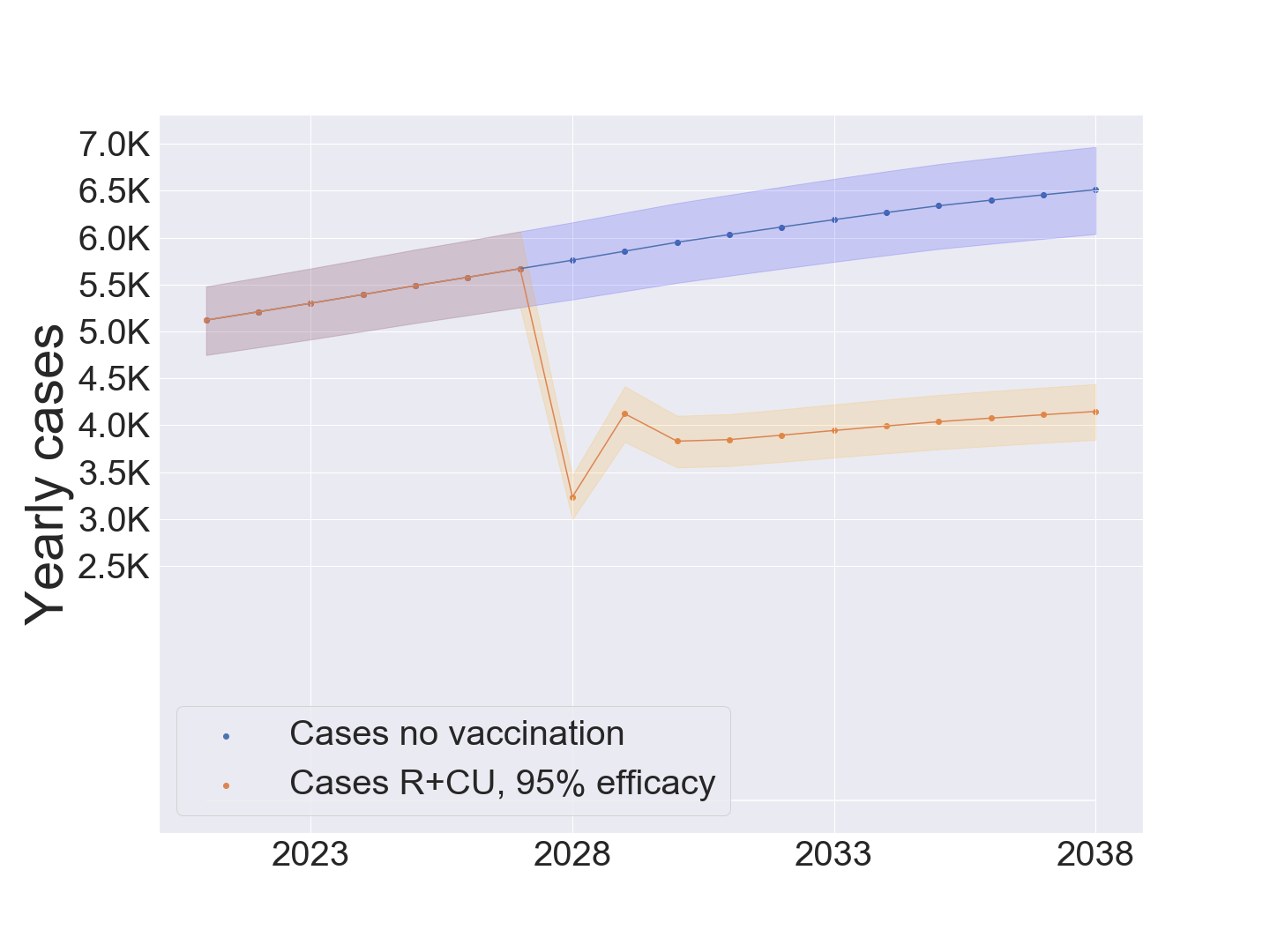}
\end{subfigure}
\caption{Guinea cumulative (top)  and yearly (bottom) iNTS cases under the status quo and routine + catch-up vaccination ($95\%$ efficacy) scenarios. Shaded areas show the 25th and 75th percentiles, line shows the median over 1000 experiments, samples drawn from uniform distributions over (0.00020,0.00024) for $\beta_{2,n}$ and (0.0080,0.0084) for $\beta_{4,n}$. }\label{fig:Guinea}
\end{figure}

\begin{figure}[htbp]
\renewcommand{\thefigure}{\textbf{Supplementary Fig. 32 Guinea-Bissau cumulative and yearly iNTS cases}}
\begin{subfigure}[b]{\textwidth}
\centering
\includegraphics[width=0.7\linewidth]{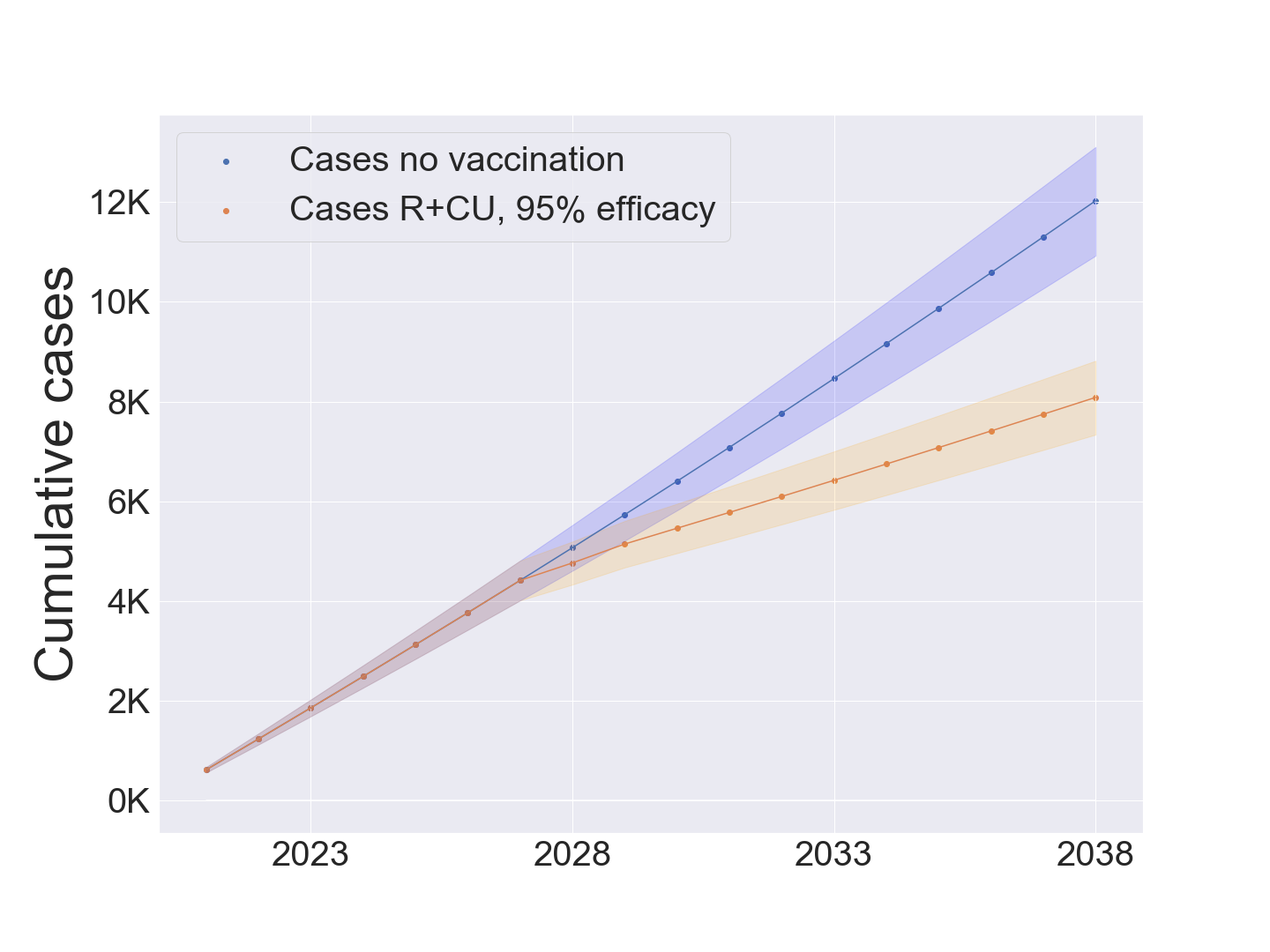}
\includegraphics[width=0.7\linewidth]{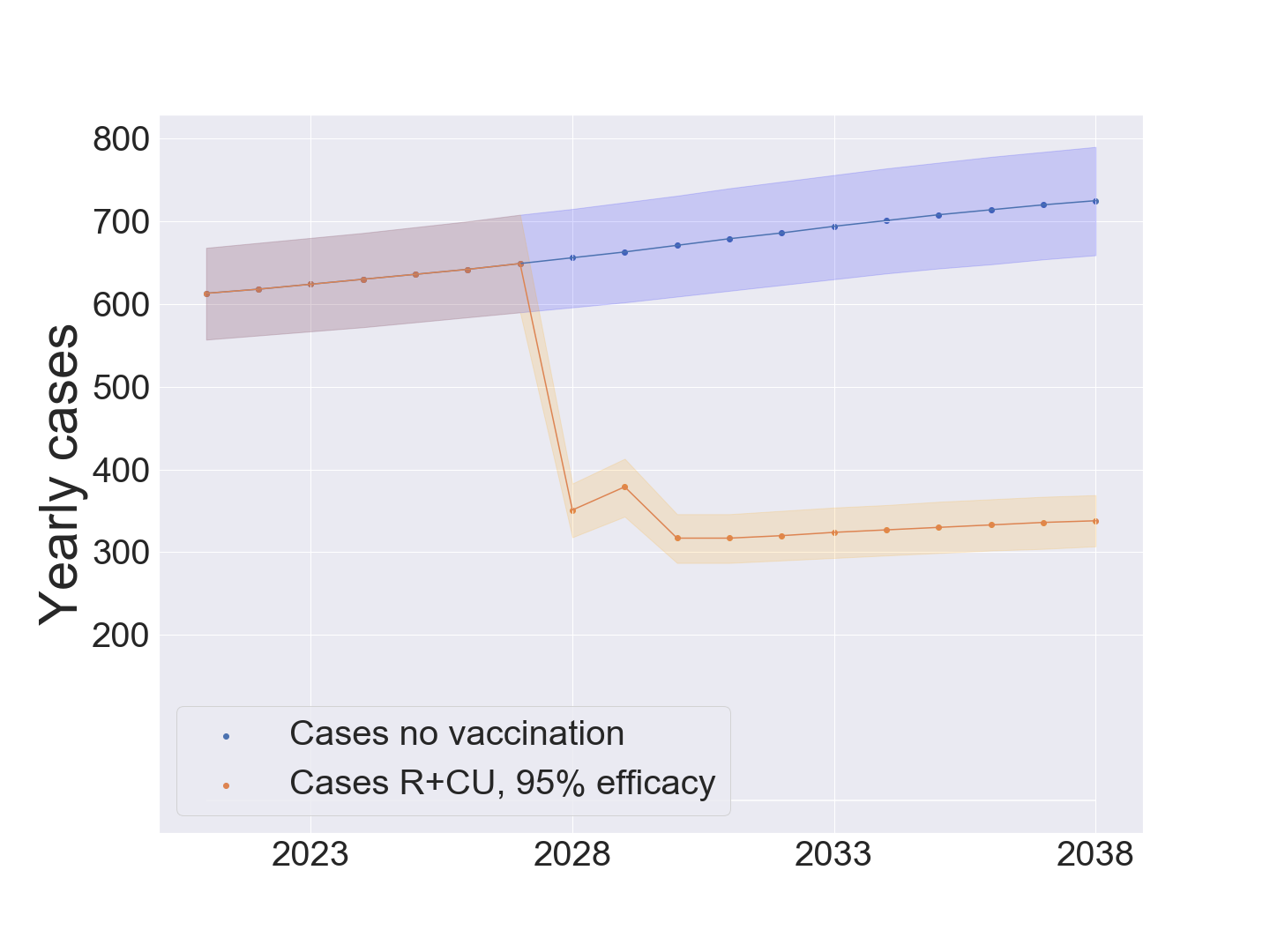}
\end{subfigure}
\caption{Guinea-Bissau cumulative (top)  and yearly (bottom) iNTS cases under the status quo and routine + catch-up vaccination ($95\%$ efficacy) scenarios. Shaded areas show the 25th and 75th percentiles, line shows the median over 1000 experiments, samples drawn from uniform distributions over (0.00020,0.00024) for $\beta_{2,n}$ and (0.0080,0.0084) for $\beta_{4,n}$. }\label{fig:Guinea-Bissau}
\end{figure}

\begin{figure}[htbp]
\renewcommand{\thefigure}{\textbf{Supplementary Fig. 33 Kenya cumulative and yearly iNTS cases}}
\begin{subfigure}[b]{\textwidth}
\centering
\includegraphics[width=0.7\linewidth]{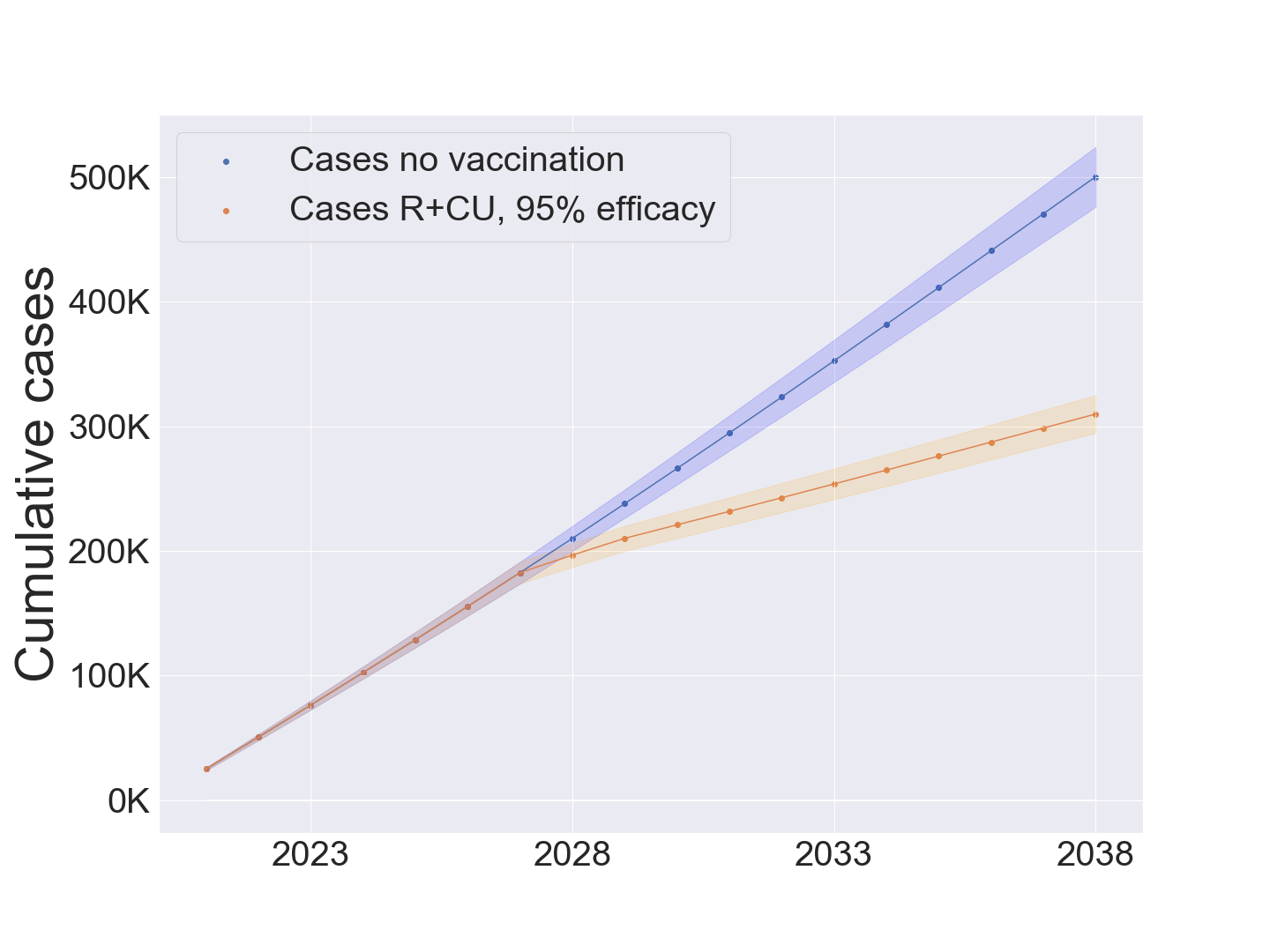}
\includegraphics[width=0.7\linewidth]{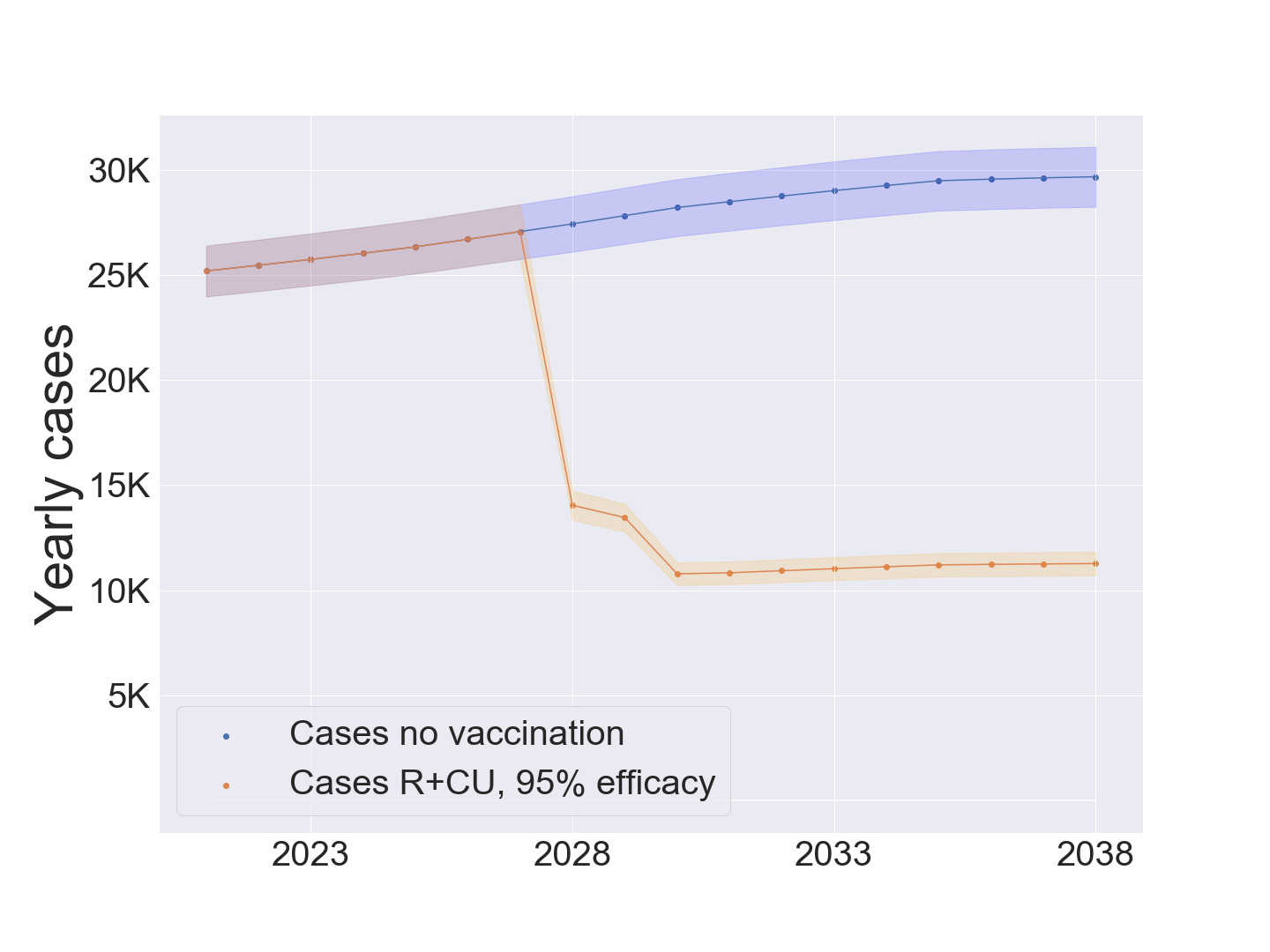}
\end{subfigure}
\caption{Kenya cumulative (top)  and yearly (bottom) iNTS cases under the status quo and routine + catch-up vaccination ($95\%$ efficacy) scenarios. Shaded areas show the 25th and 75th percentiles, line shows the median over 1000 experiments, samples drawn from uniform distributions over (0.00020,0.00024) for $\beta_{2,n}$ and (0.0080,0.0084) for $\beta_{4,n}$. }\label{fig:Kenya}
\end{figure}

\begin{figure}[htbp]
\renewcommand{\thefigure}{\textbf{Supplementary Fig. 34 Lesotho cumulative and yearly iNTS cases}}
\begin{subfigure}[b]{\textwidth}
\centering
\includegraphics[width=0.7\linewidth]{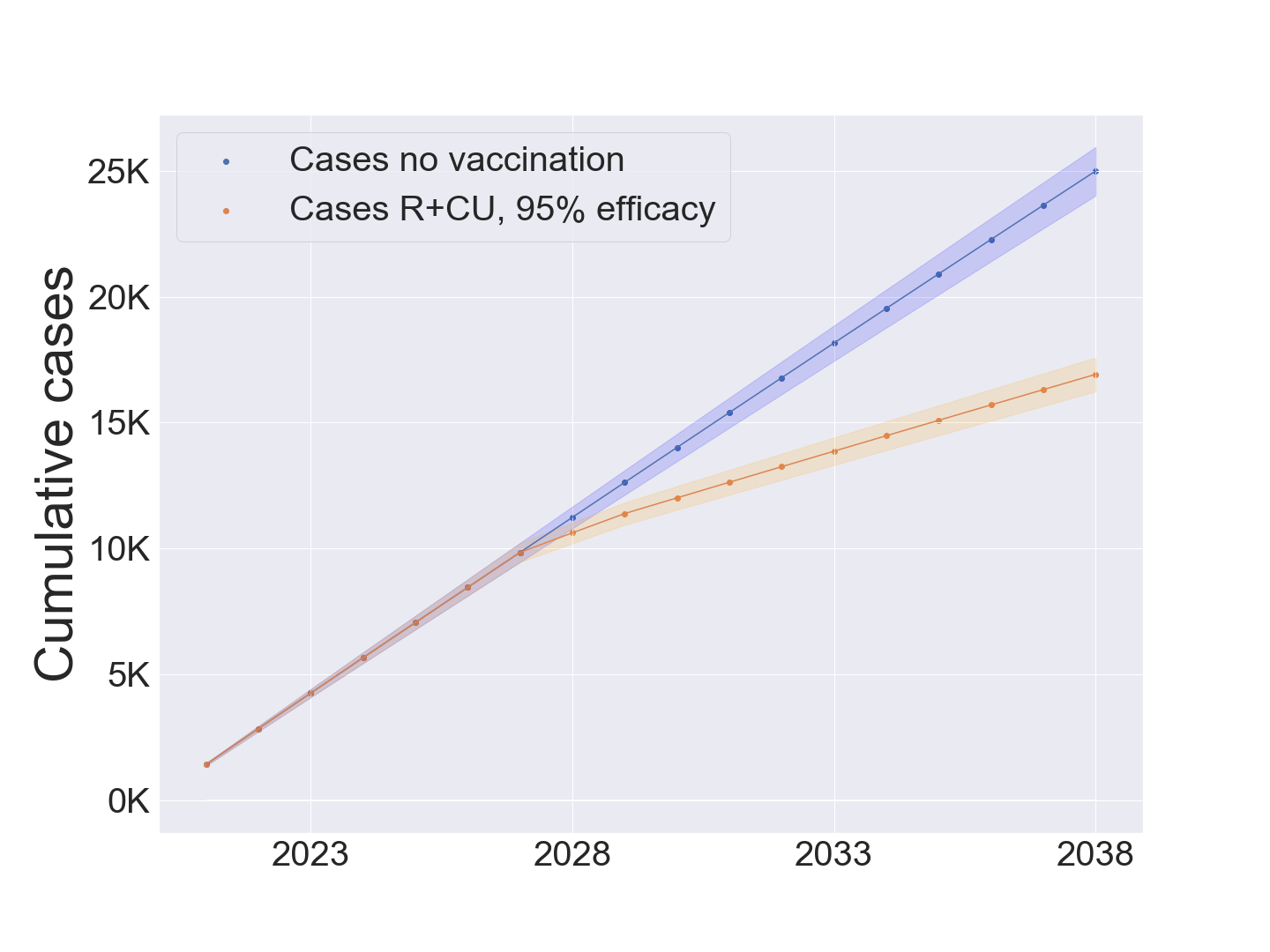}
\includegraphics[width=0.7\linewidth]{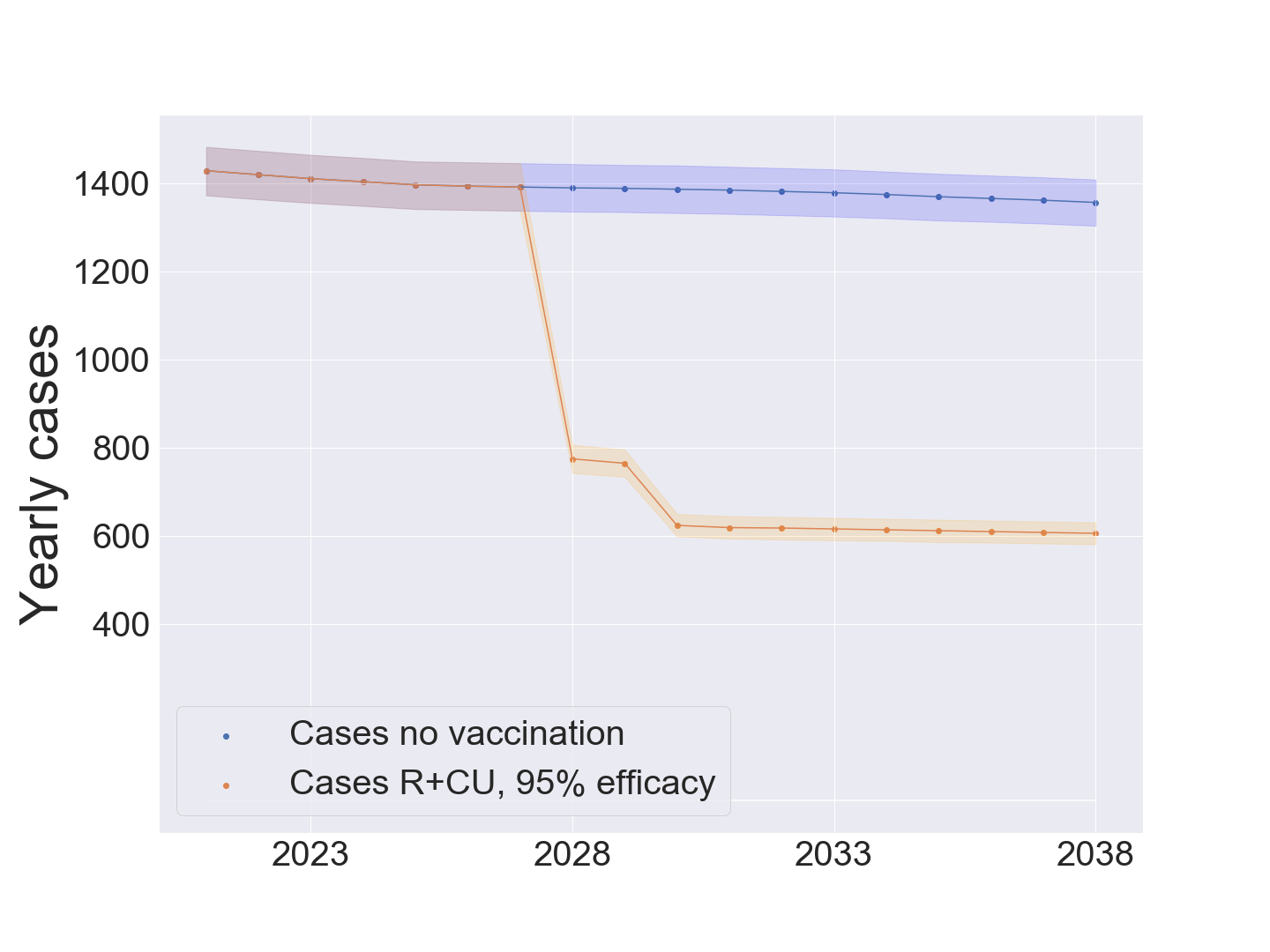}
\end{subfigure}

\caption{Lesotho cumulative (top)  and yearly (bottom) iNTS cases under the status quo and routine + catch-up vaccination ($95\%$ efficacy) scenarios. Shaded areas show the 25th and 75th percentiles, line shows the median over 1000 experiments, samples drawn from uniform distributions over (0.00020,0.00024) for $\beta_{2,n}$ and (0.0080,0.0084) for $\beta_{4,n}$. }\label{fig:Lesotho}
\end{figure}

\begin{figure}[htbp]
\renewcommand{\thefigure}{\textbf{Supplementary Fig. 35 Liberia cumulative and yearly iNTS cases}}
\begin{subfigure}[b]{\textwidth}
\centering
\includegraphics[width=0.7\linewidth]{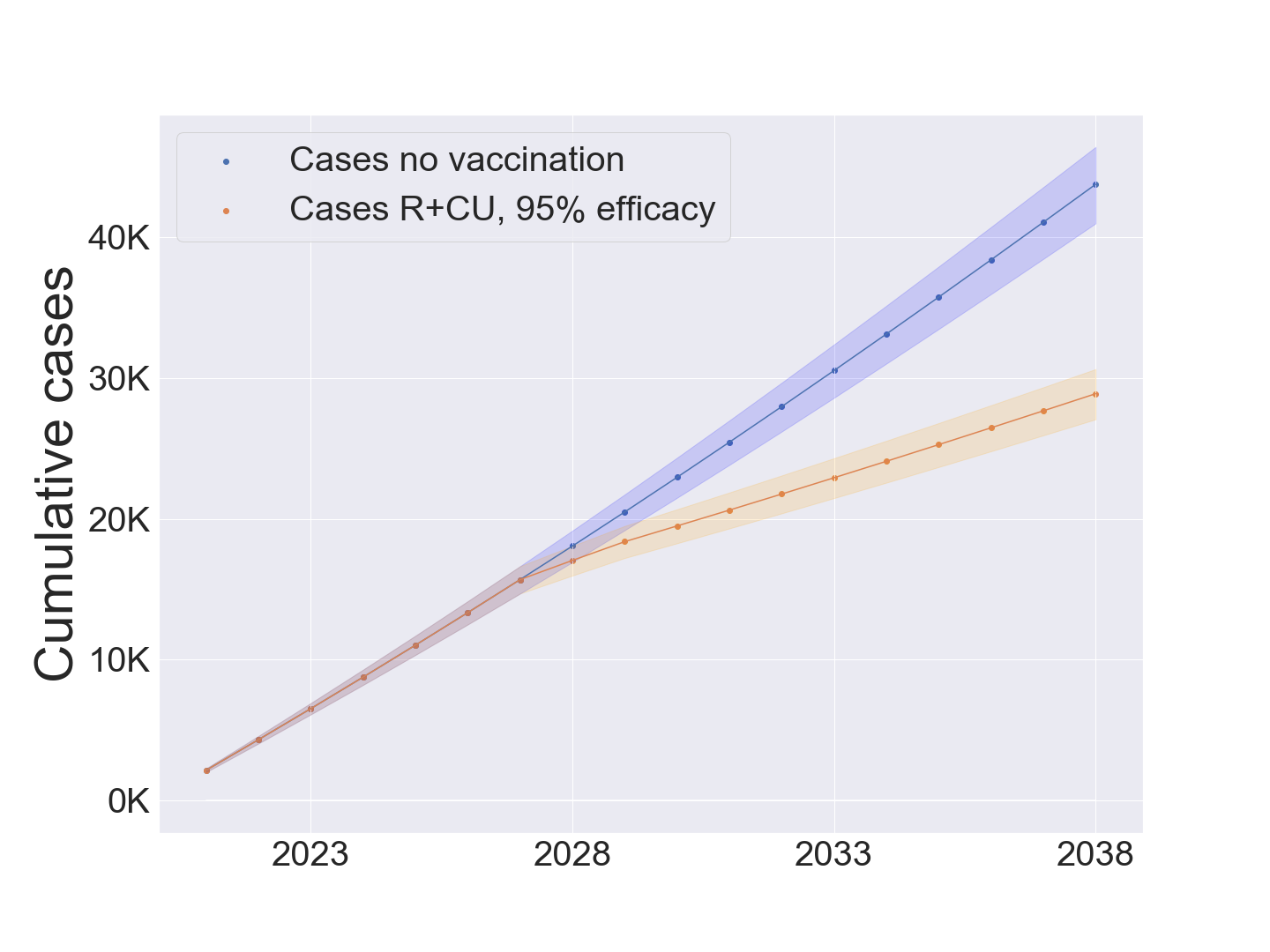}
\includegraphics[width=0.7\linewidth]{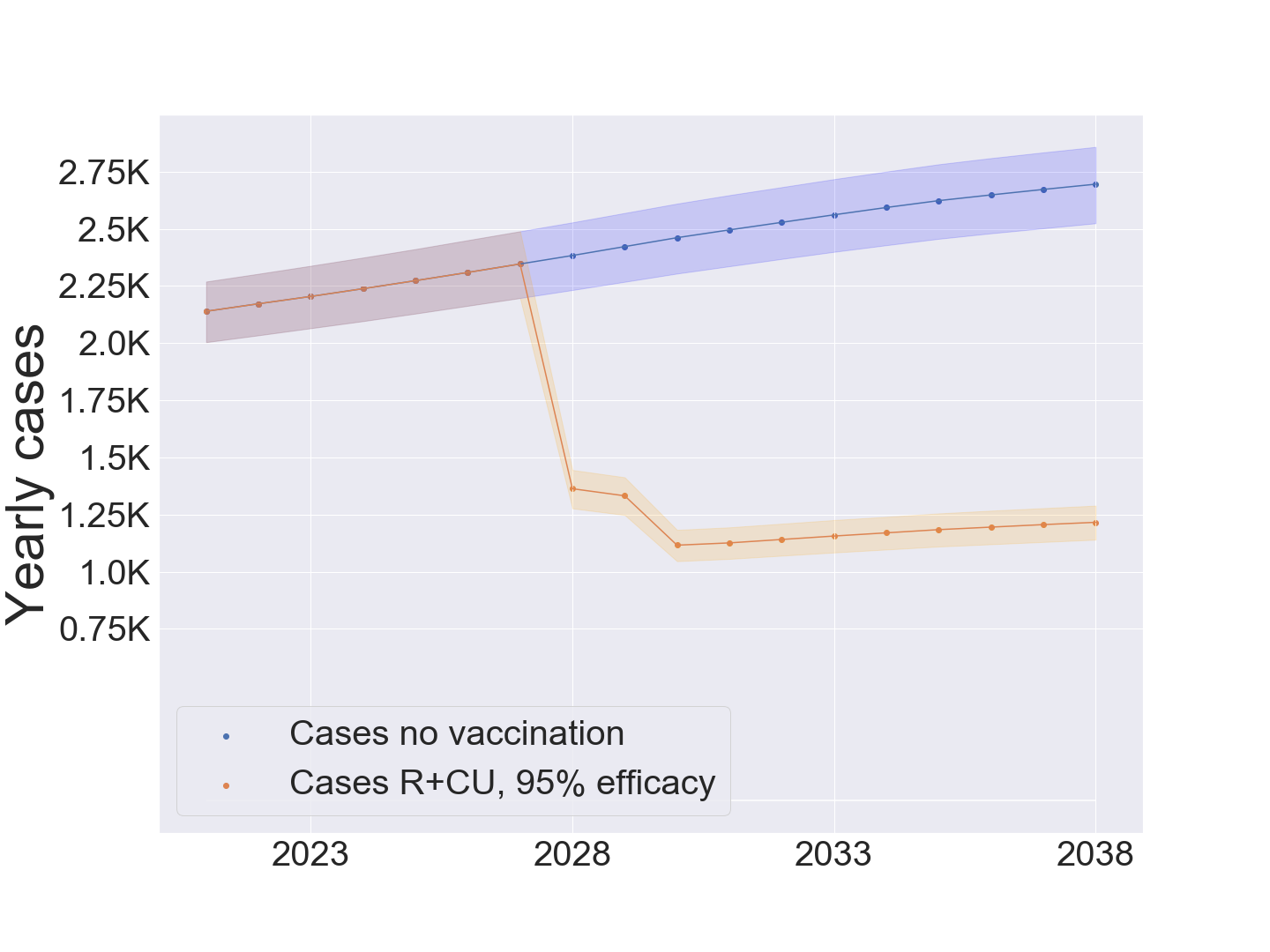}
\end{subfigure}
\caption{Liberia cumulative (top)  and yearly (bottom) iNTS cases under the status quo and routine + catch-up vaccination ($95\%$ efficacy) scenarios. Shaded areas show the 25th and 75th percentiles, line shows the median over 1000 experiments, samples drawn from uniform distributions over (0.00020,0.00024) for $\beta_{2,n}$ and (0.0080,0.0084) for $\beta_{4,n}$. }\label{fig:Liberia}
\end{figure}

\begin{figure}[htbp]
\renewcommand{\thefigure}{\textbf{Supplementary Fig. 36 Madagascar cumulative and yearly iNTS cases}}
\begin{subfigure}[b]{\textwidth}
\centering
\includegraphics[width=0.7\linewidth]{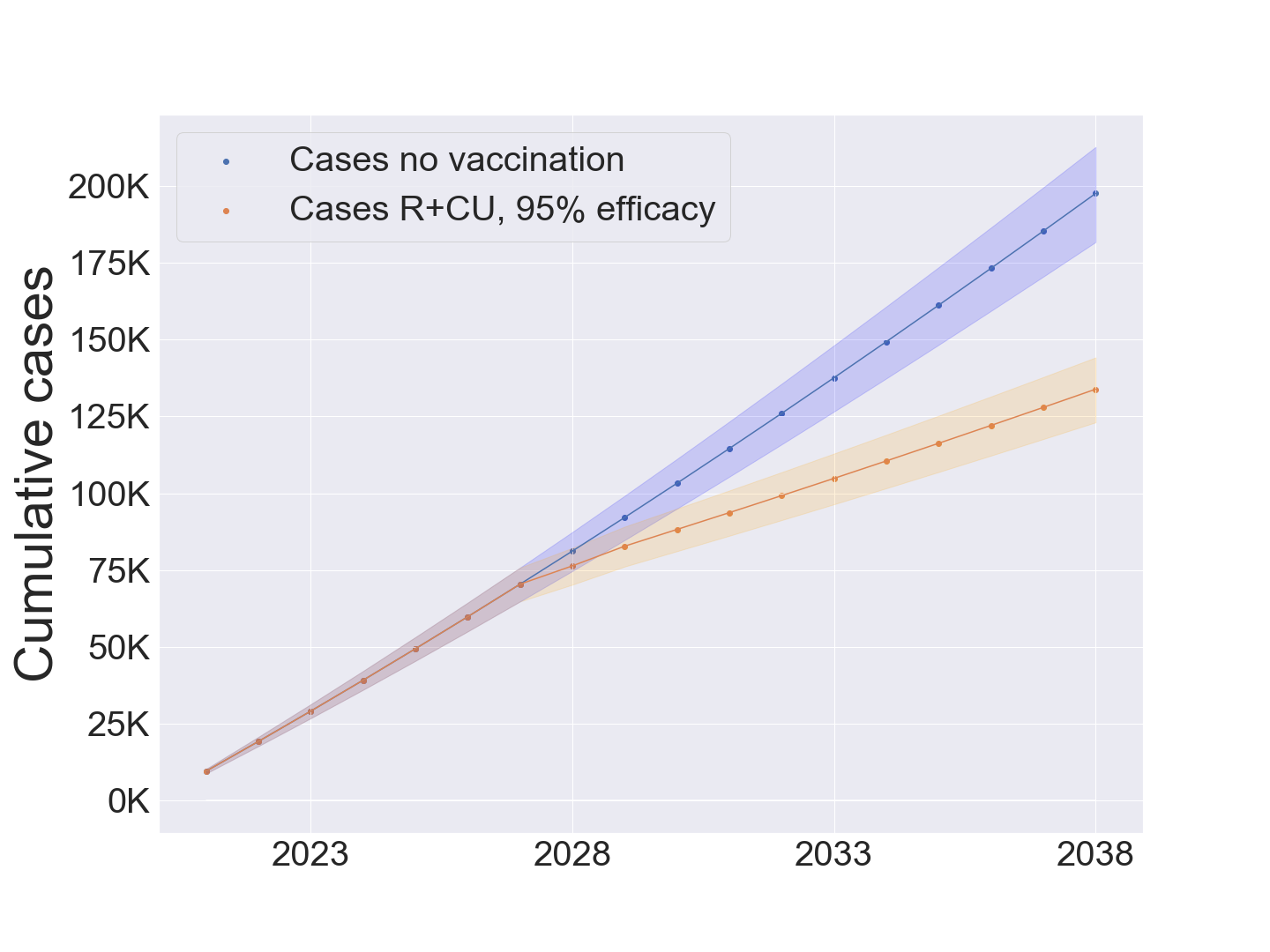}
\includegraphics[width=0.7\linewidth]{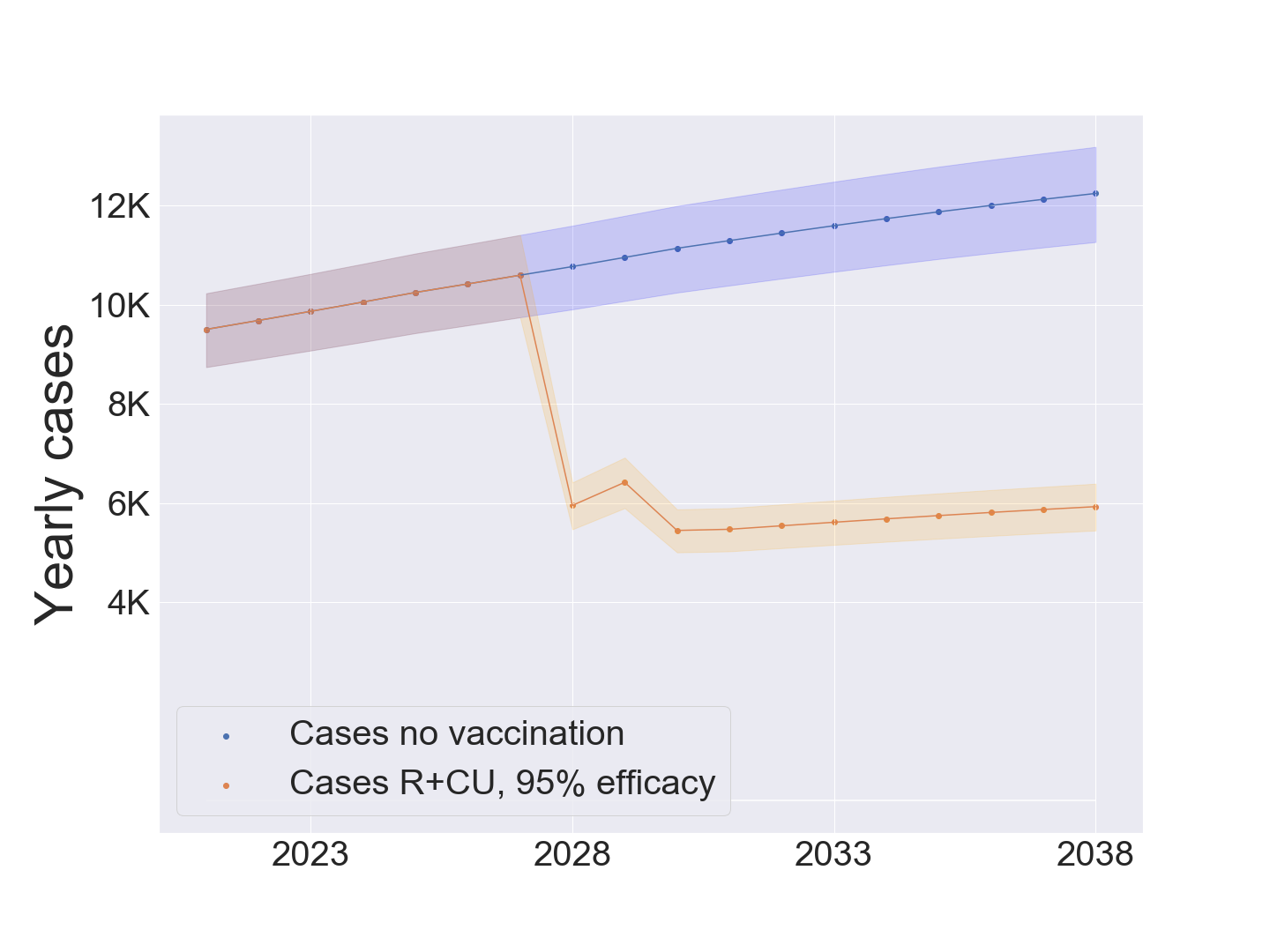}
\end{subfigure}
\caption{Madagascar cumulative (top)  and yearly (bottom) iNTS cases under the status quo and routine + catch-up vaccination ($95\%$ efficacy) scenarios. Shaded areas show the 25th and 75th percentiles, line shows the median over 1000 experiments, samples drawn from uniform distributions over (0.00020,0.00024) for $\beta_{2,n}$ and (0.0080,0.0084) for $\beta_{4,n}$. }\label{fig:Madagascar}
\end{figure}

\begin{figure}[htbp]
\renewcommand{\thefigure}{\textbf{Supplementary Fig. 37 Malawi cumulative and yearly iNTS cases}}
\begin{subfigure}[b]{\textwidth}
\centering
\includegraphics[width=0.7\linewidth]{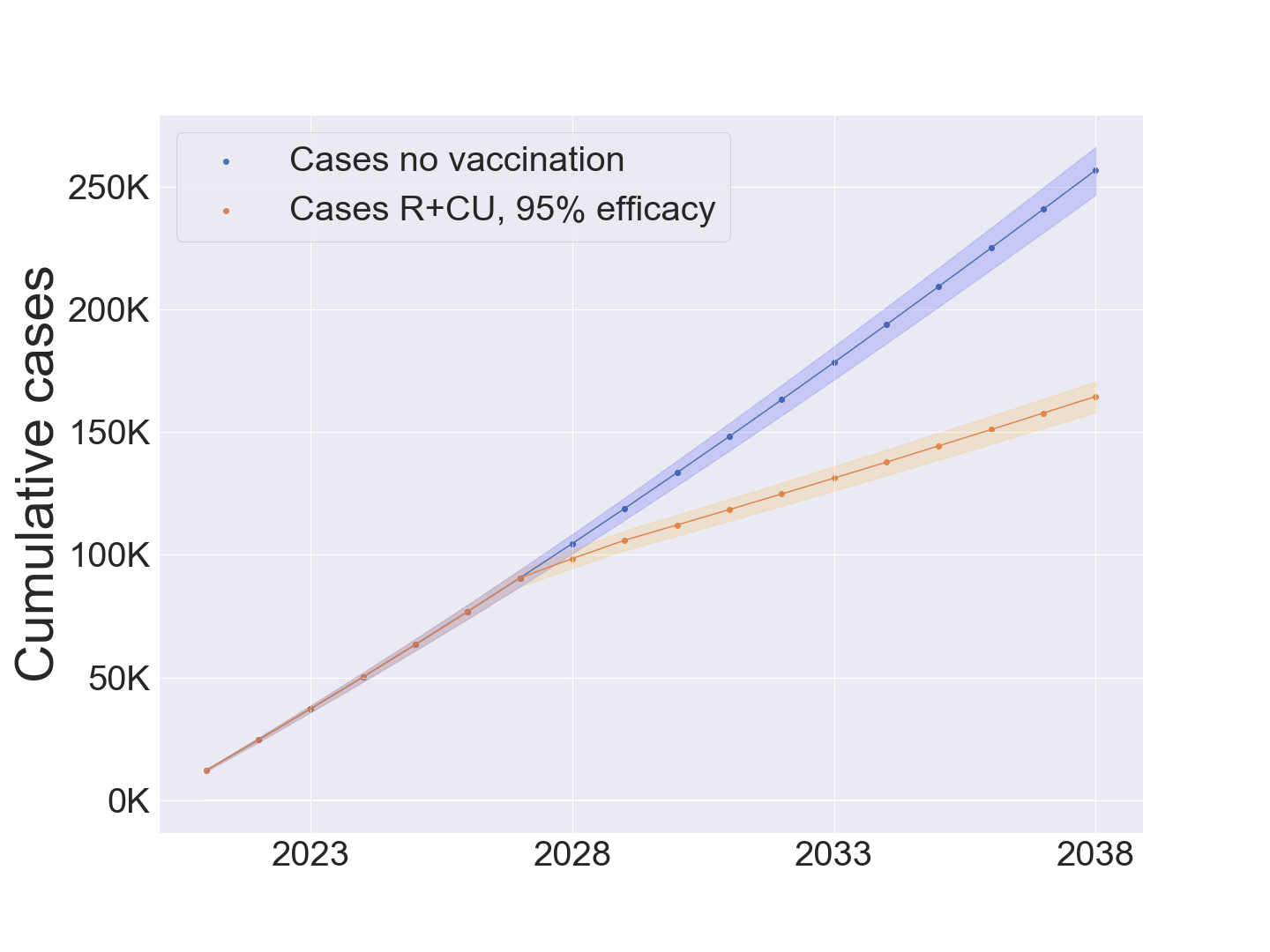}
\includegraphics[width=0.7\linewidth]{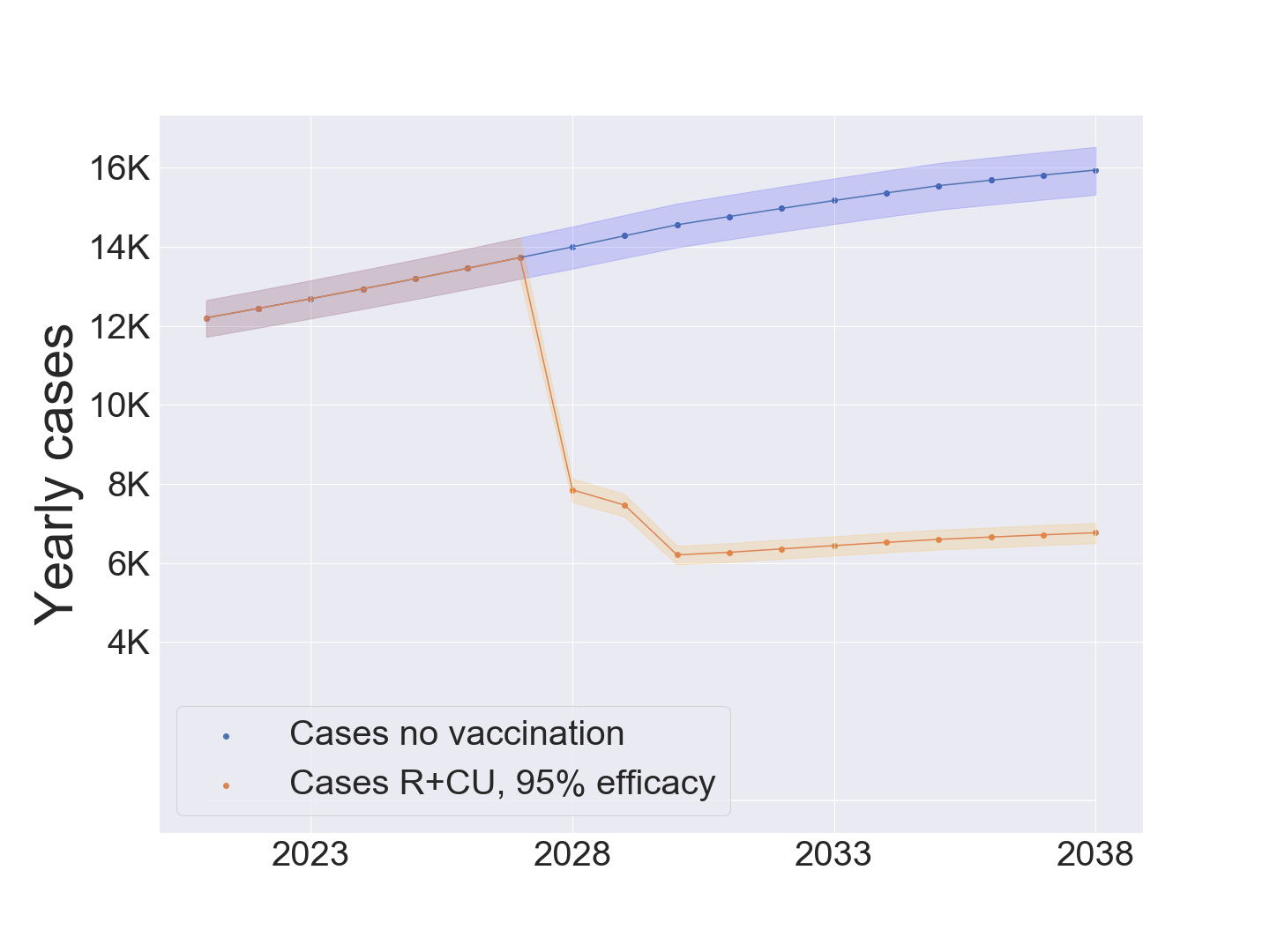}
\end{subfigure}
\caption{Malawi cumulative (top)  and yearly (bottom) iNTS cases under the status quo and routine + catch-up vaccination ($95\%$ efficacy) scenarios. Shaded areas show the 25th and 75th percentiles, line shows the median over 1000 experiments, samples drawn from uniform distributions over (0.00020,0.00024) for $\beta_{2,n}$ and (0.0080,0.0084) for $\beta_{4,n}$. }\label{fig:Malawi}
\end{figure}

\begin{figure}[htbp]
\renewcommand{\thefigure}{\textbf{Supplementary Fig. 38 Mali cumulative and yearly iNTS cases}}
\begin{subfigure}[b]{\textwidth}
\centering
\includegraphics[width=0.7\linewidth]{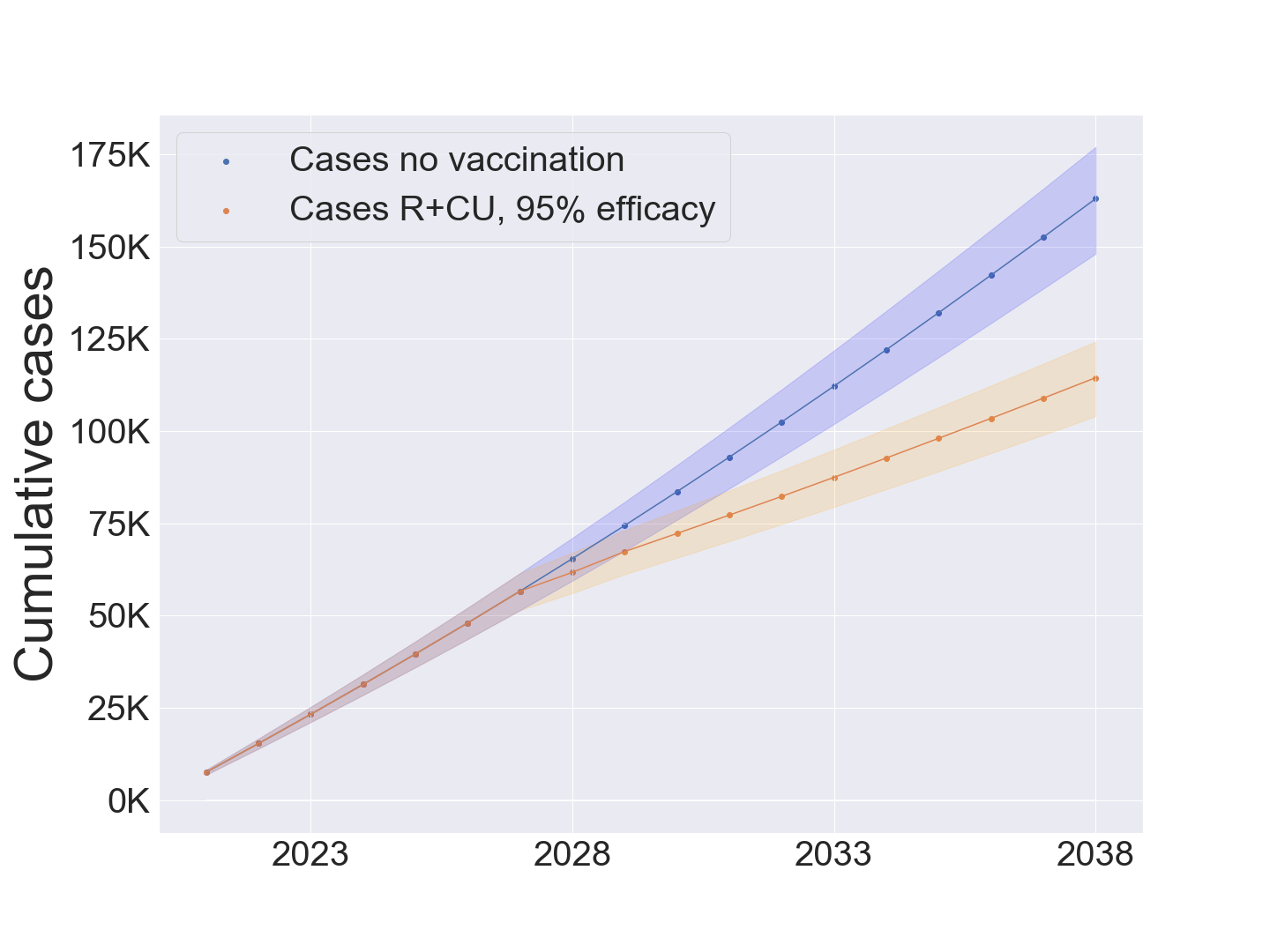}
\includegraphics[width=0.7\linewidth]{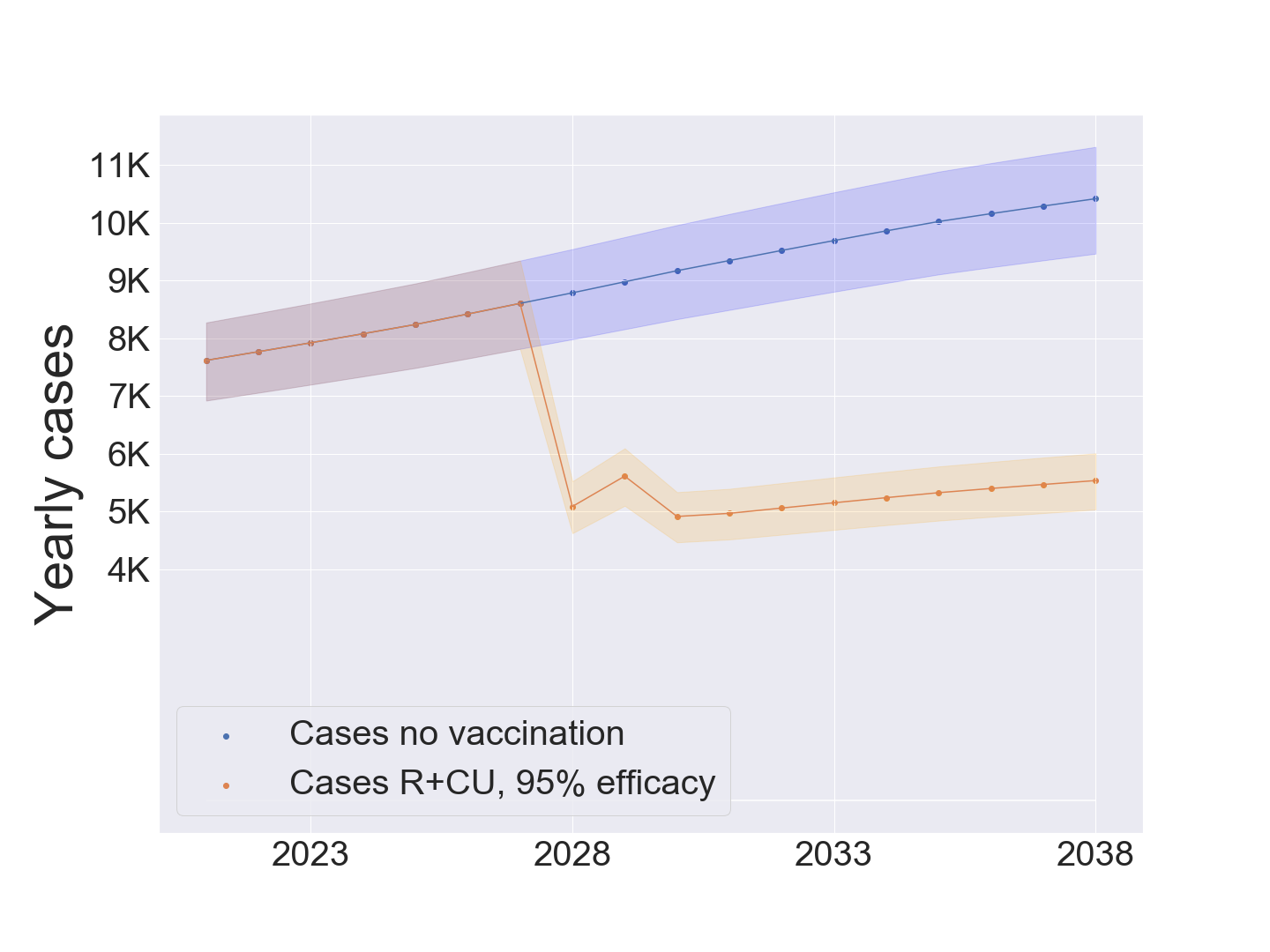}
\end{subfigure}
\caption{Mali cumulative (top)  and yearly (bottom) iNTS cases under the status quo and routine + catch-up vaccination ($95\%$ efficacy) scenarios. Shaded areas show the 25th and 75th percentiles, line shows the median over 1000 experiments, samples drawn from uniform distributions over (0.00020,0.00024) for $\beta_{2,n}$ and (0.0080,0.0084) for $\beta_{4,n}$. }\label{fig:Mali}
\end{figure}

\begin{figure}[htbp]
\renewcommand{\thefigure}{\textbf{Supplementary Fig. 39 Mauritania cumulative and yearly iNTS cases}}
\begin{subfigure}[b]{\textwidth}
\centering
\includegraphics[width=0.7\linewidth]{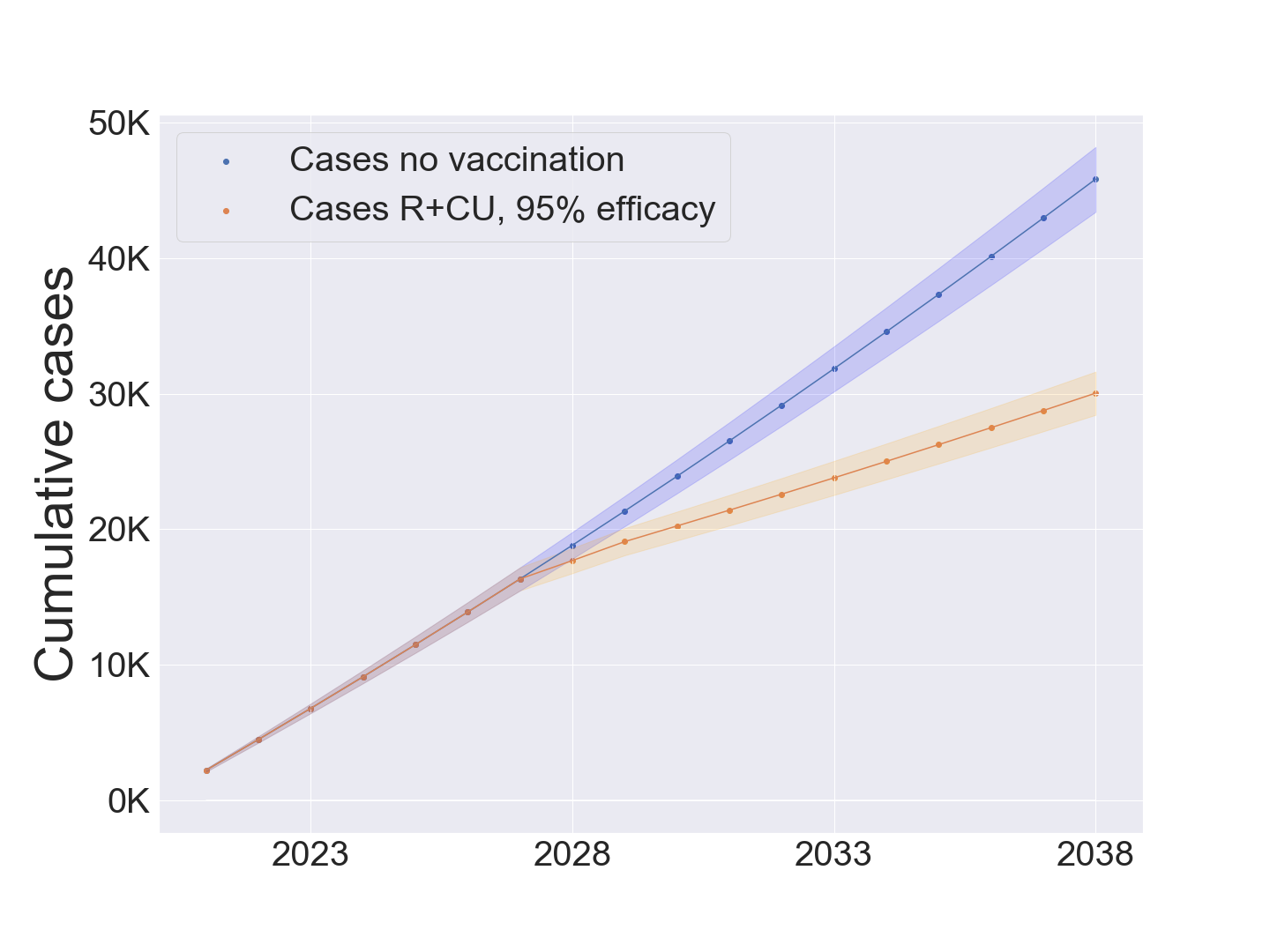}
\includegraphics[width=0.7\linewidth]{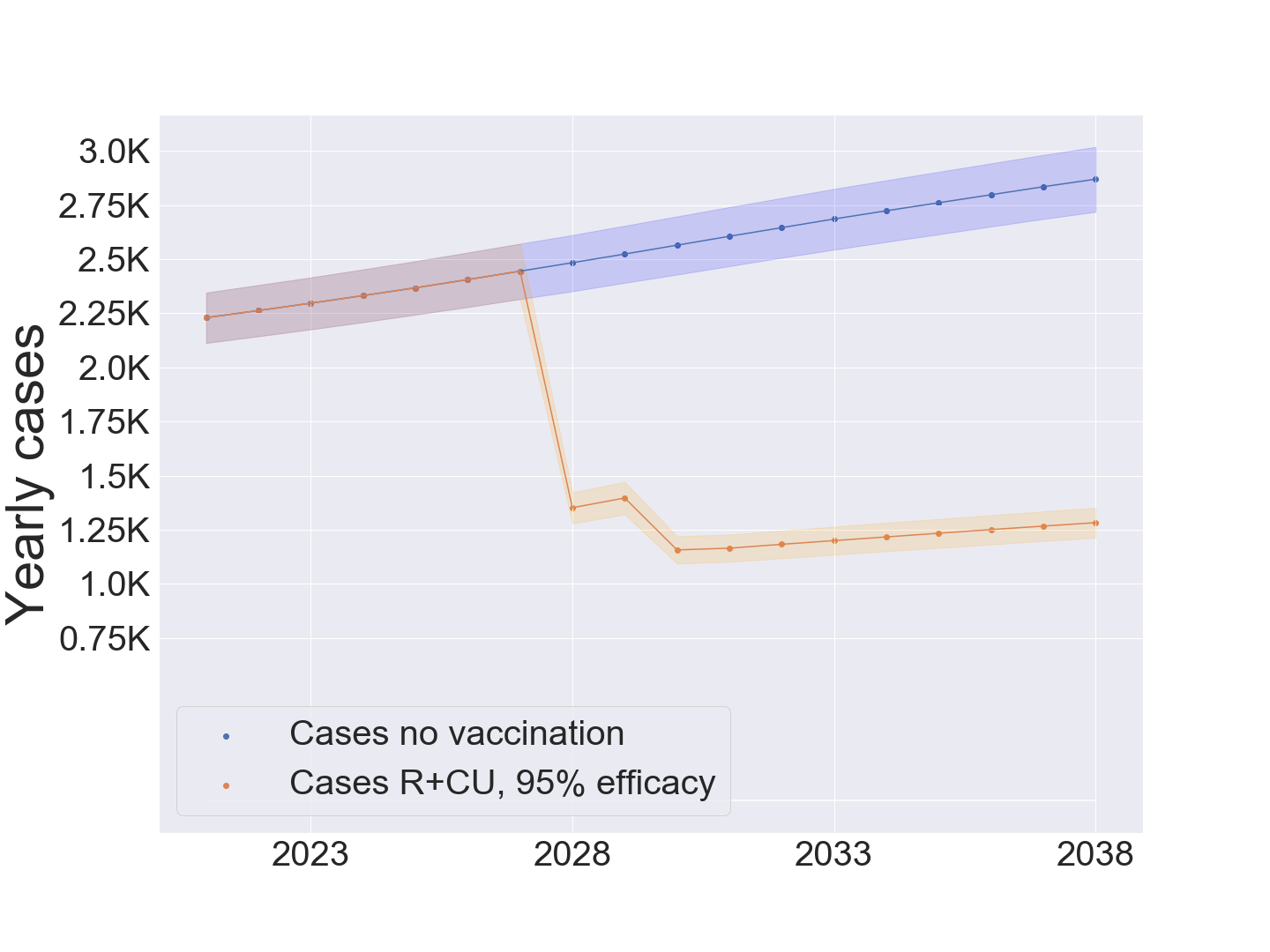}
\end{subfigure}
\caption{Mauritania cumulative (top)  and yearly (bottom) iNTS cases under the status quo and routine + catch-up vaccination ($95\%$ efficacy) scenarios. Shaded areas show the 25th and 75th percentiles, line shows the median over 1000 experiments, samples drawn from uniform distributions over (0.00020,0.00024) for $\beta_{2,n}$ and (0.0080,0.0084) for $\beta_{4,n}$. }\label{fig:Mauritania}
\end{figure}

\begin{figure}[htbp]
\renewcommand{\thefigure}{\textbf{Supplementary Fig. 40 Mauritius cumulative and yearly iNTS cases}}
\begin{subfigure}[b]{\textwidth}
\centering
\includegraphics[width=0.7\linewidth]{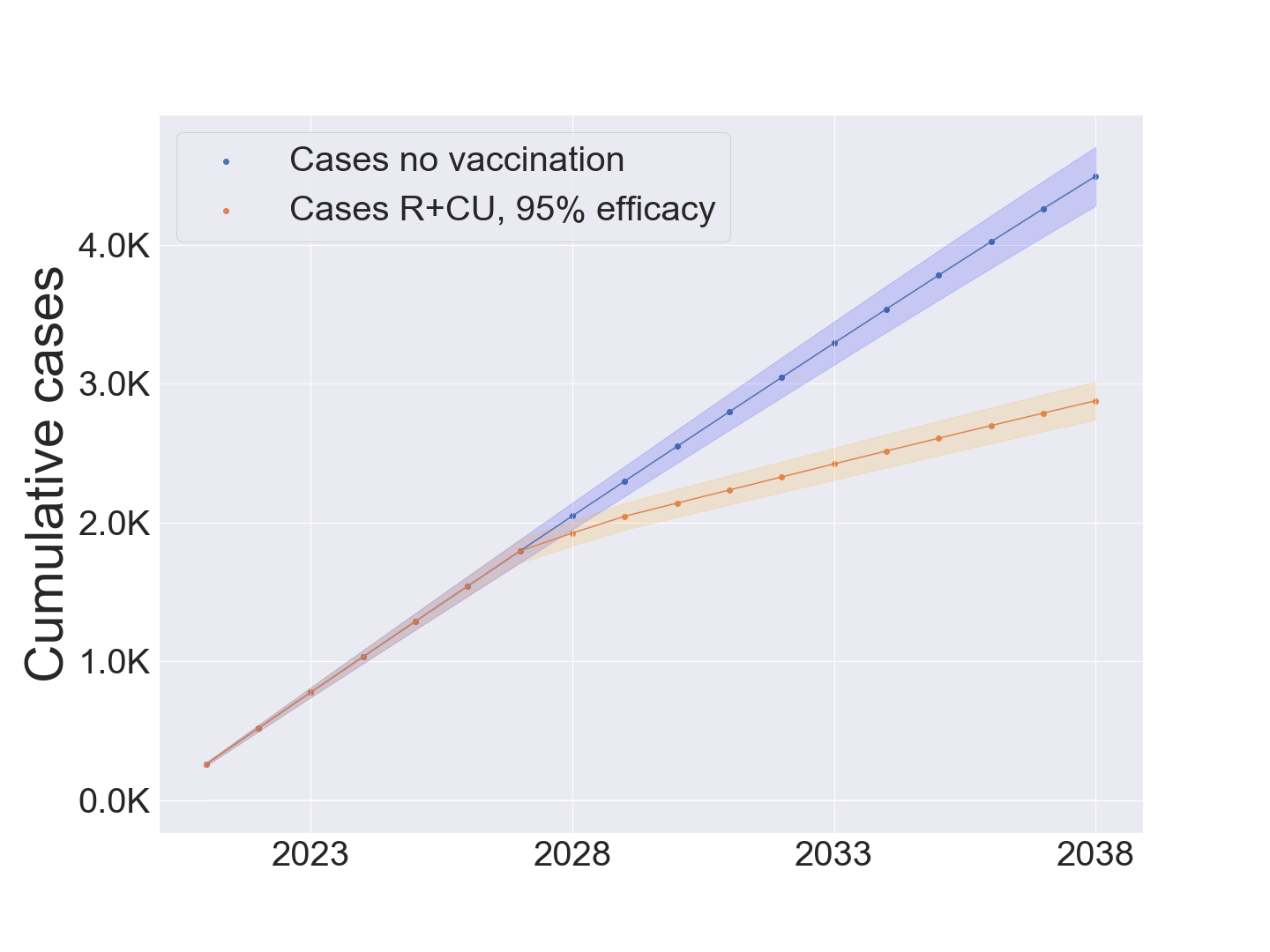}
\includegraphics[width=0.7\linewidth]{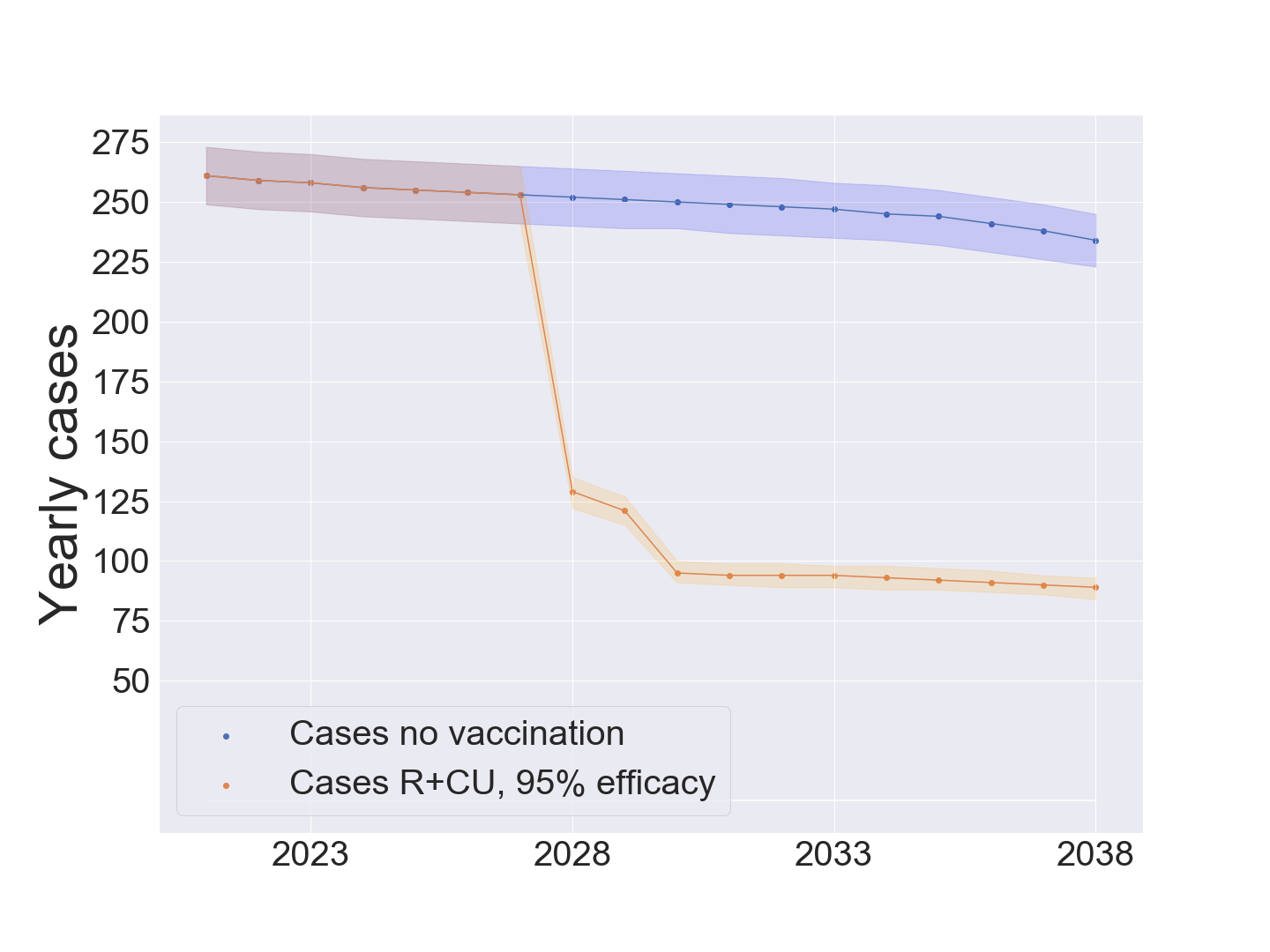}
\end{subfigure}
\caption{Mauritius cumulative (top)  and yearly (bottom) iNTS cases under the status quo and routine + catch-up vaccination ($95\%$ efficacy) scenarios. Shaded areas show the 25th and 75th percentiles, line shows the median over 1000 experiments, samples drawn from uniform distributions over (0.00020,0.00024) for $\beta_{2,n}$ and (0.0080,0.0084) for $\beta_{4,n}$. }\label{fig:Mauritius}
\end{figure}

\begin{figure}[htbp]
\renewcommand{\thefigure}{\textbf{Supplementary Fig. 41 Mayotte cumulative and yearly iNTS cases}}
\begin{subfigure}[b]{\textwidth}
\centering
\includegraphics[width=0.7\linewidth]{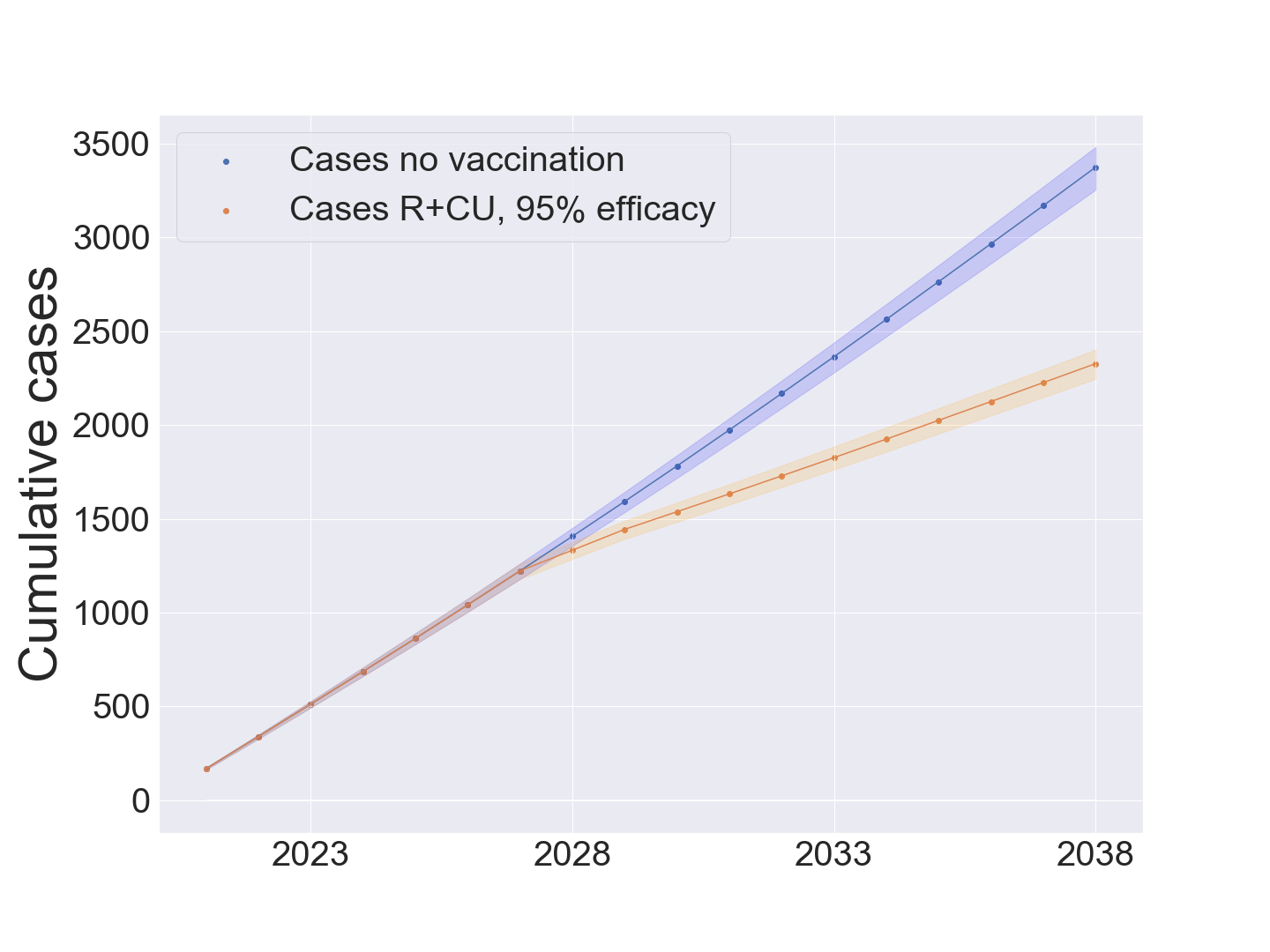}
\includegraphics[width=0.7\linewidth]{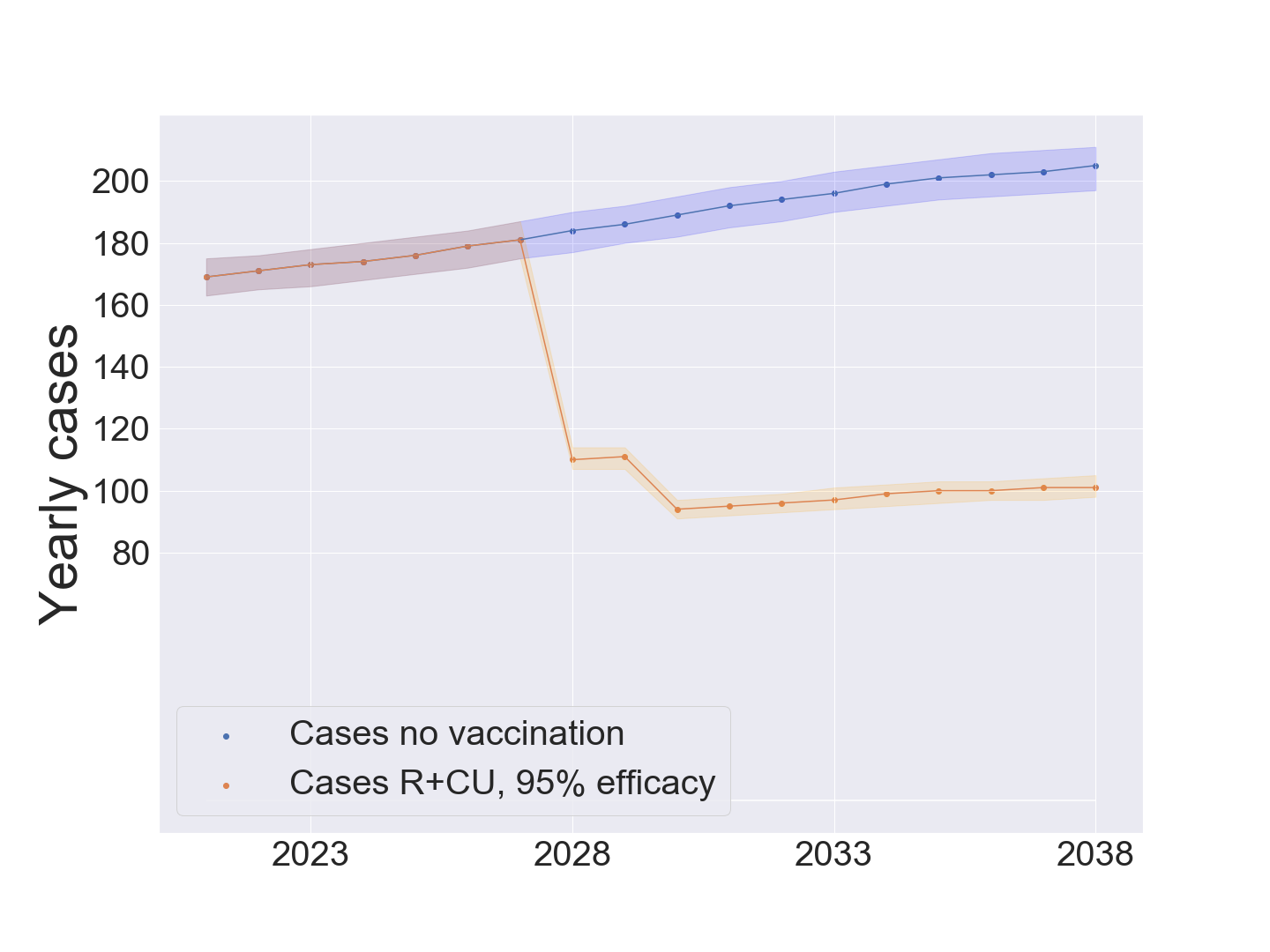}
\end{subfigure}
\caption{Mayotte cumulative (top)  and yearly (bottom) iNTS cases under the status quo and routine + catch-up vaccination ($95\%$ efficacy) scenarios. Shaded areas show the 25th and 75th percentiles, line shows the median over 1000 experiments, samples drawn from uniform distributions over (0.00020,0.00024) for $\beta_{2,n}$ and (0.0080,0.0084) for $\beta_{4,n}$. }\label{fig:Mayotte}
\end{figure}

\begin{figure}[htbp]
\renewcommand{\thefigure}{\textbf{Supplementary Fig. 42 Mozambique cumulative and yearly iNTS cases}}
\begin{subfigure}[b]{\textwidth}
\centering
\includegraphics[width=0.7\linewidth]{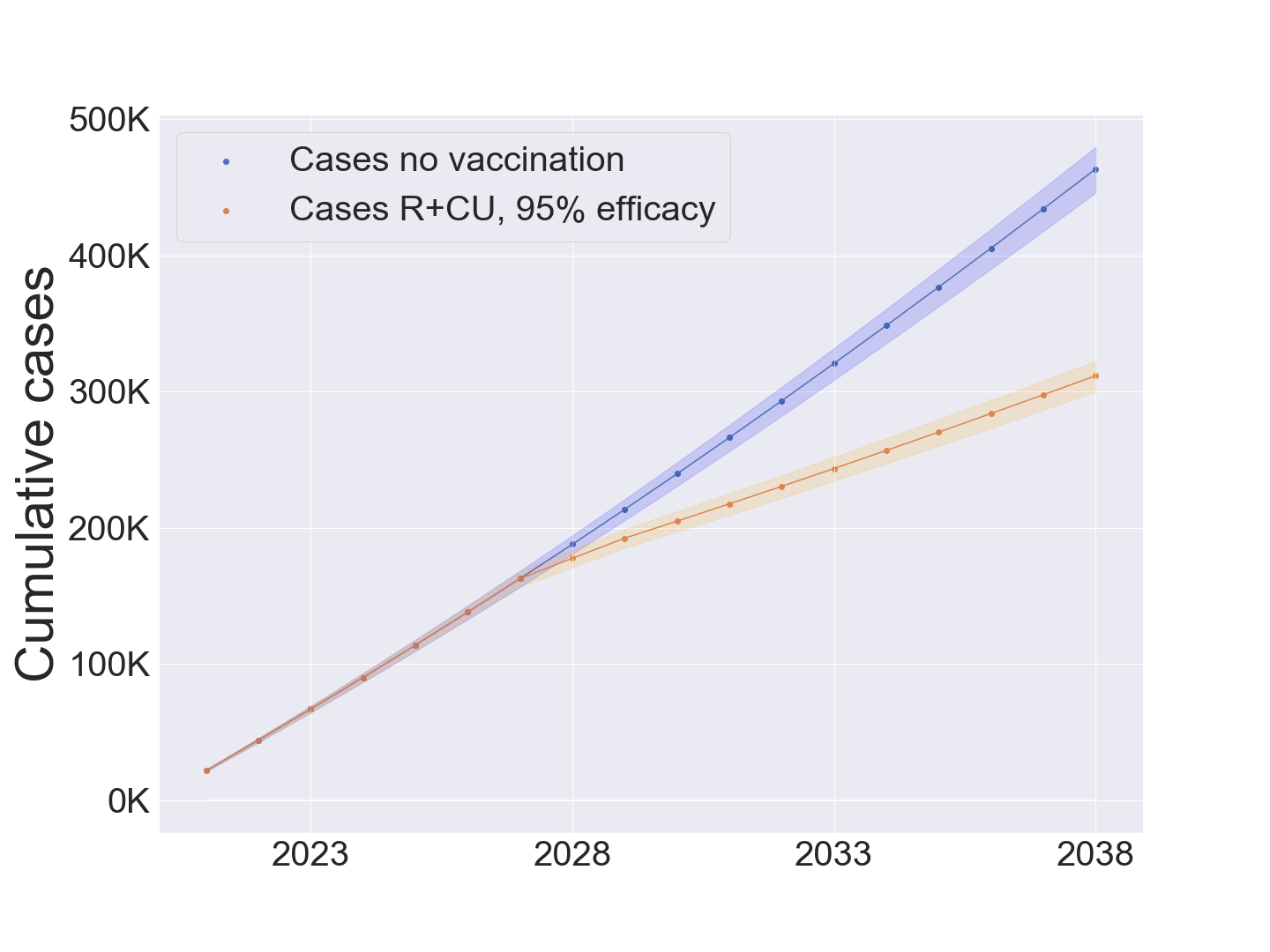}
\includegraphics[width=0.7\linewidth]{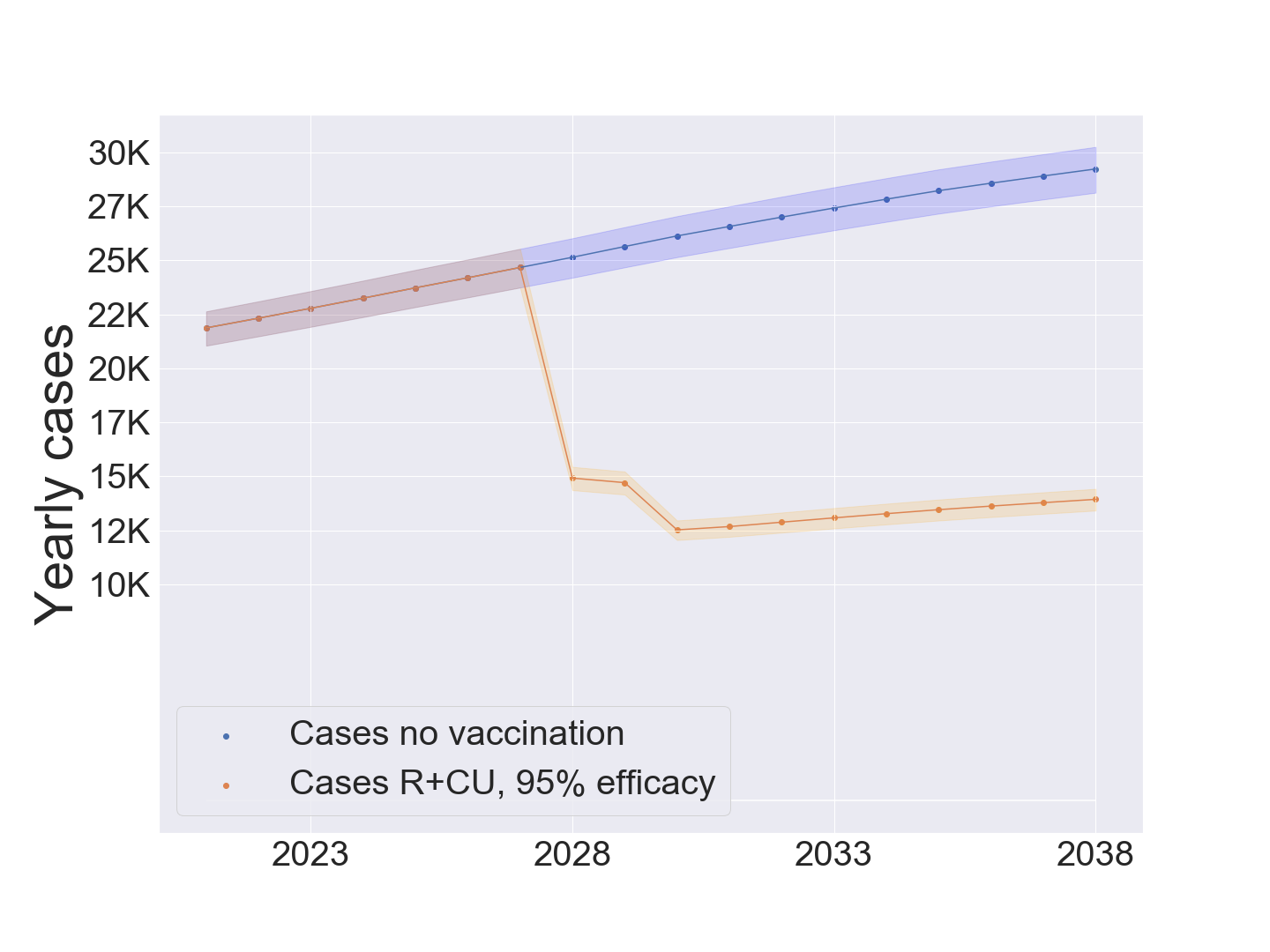}
\end{subfigure}
\caption{Mozambique cumulative (top)  and yearly (bottom) iNTS cases under the status quo and routine + catch-up vaccination ($95\%$ efficacy) scenarios. Shaded areas show the 25th and 75th percentiles, line shows the median over 1000 experiments, samples drawn from uniform distributions over (0.00020,0.00024) for $\beta_{2,n}$ and (0.0080,0.0084) for $\beta_{4,n}$. }\label{fig:Mozambique}
\end{figure}

\begin{figure}[htbp]
\renewcommand{\thefigure}{\textbf{Supplementary Fig. 43 Namibia cumulative and yearly iNTS cases}}
\begin{subfigure}[b]{\textwidth}
\centering
\includegraphics[width=0.7\linewidth]{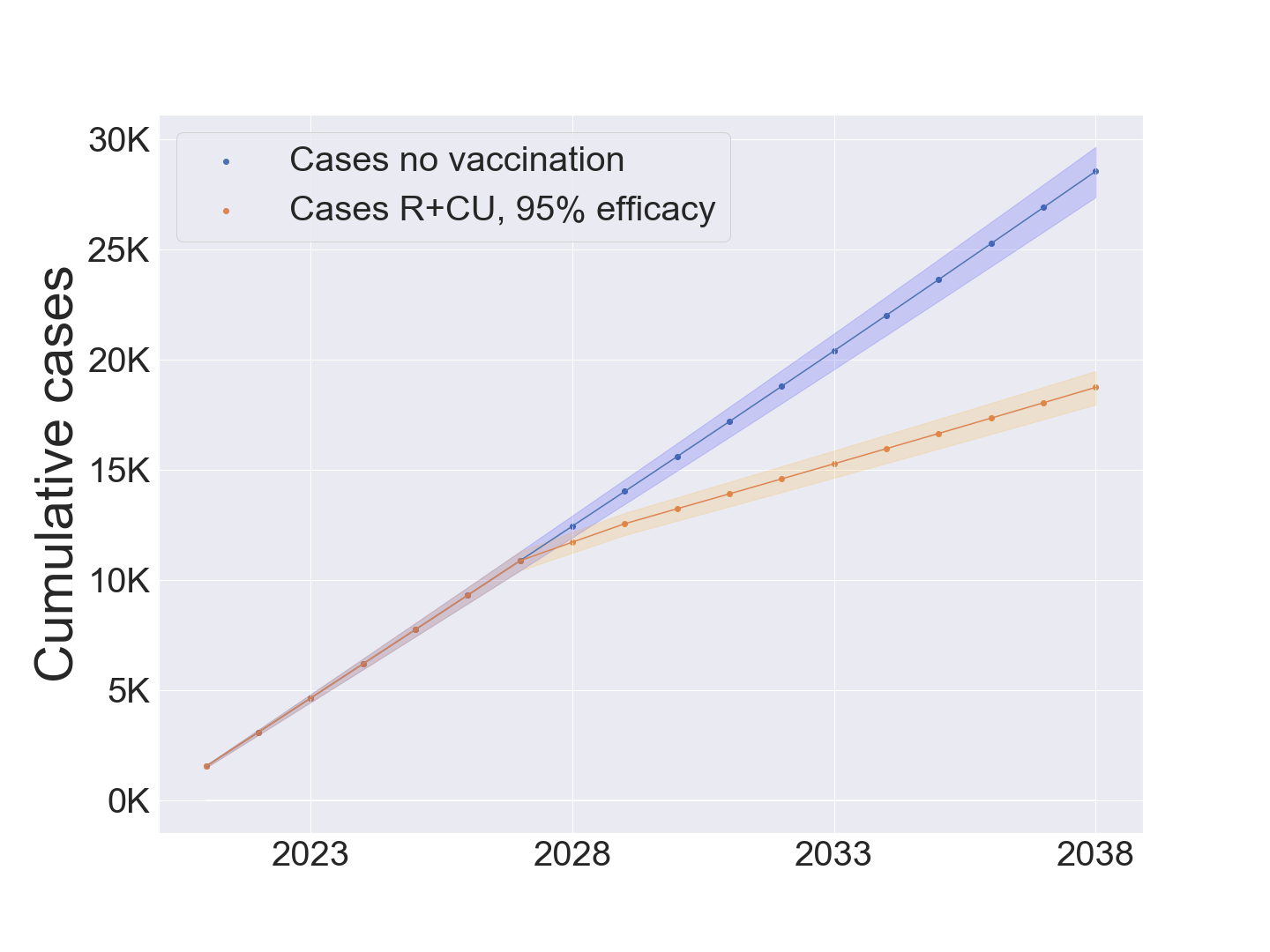}
\includegraphics[width=0.7\linewidth]{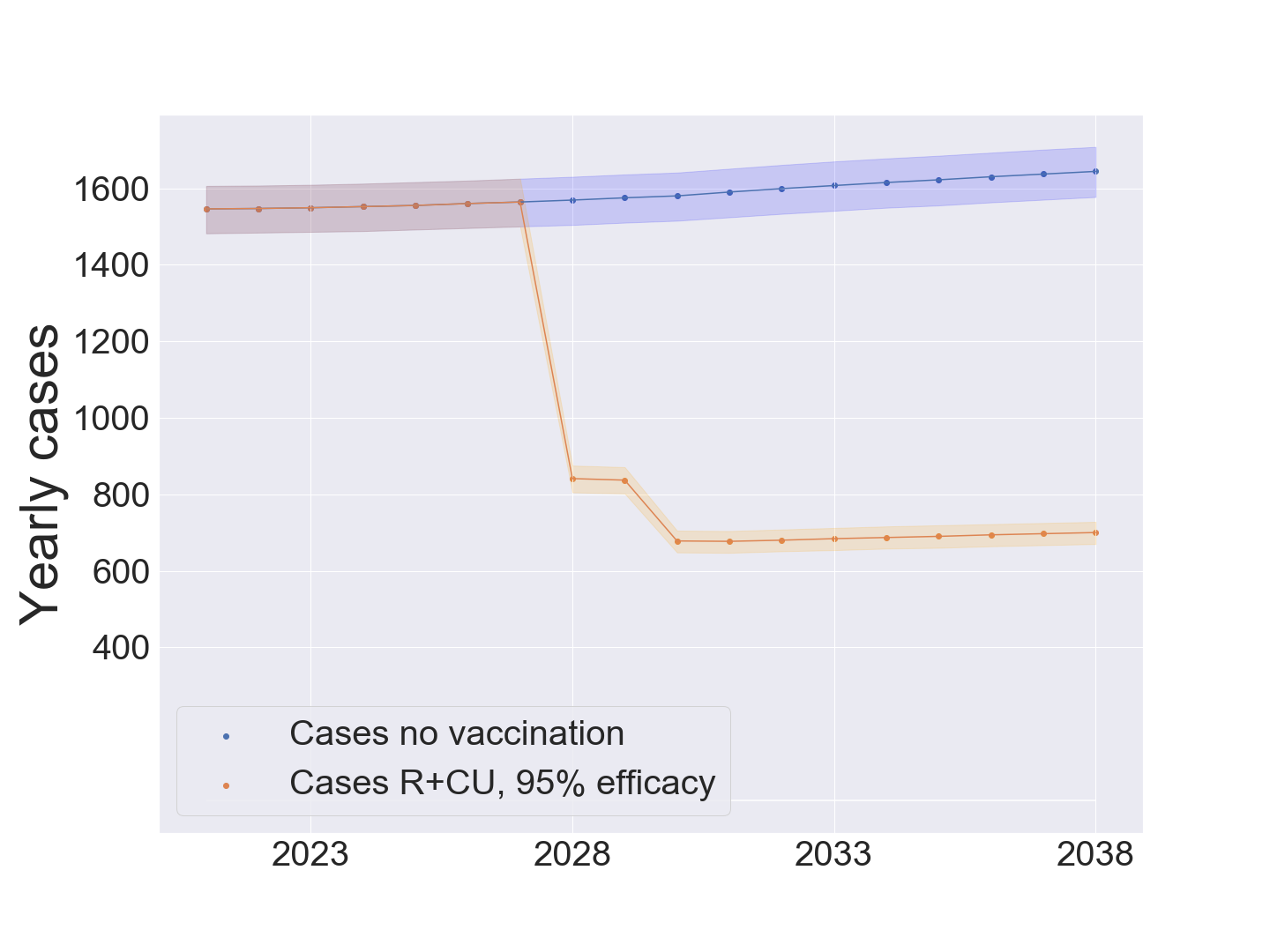}
\end{subfigure}
\caption{Namibia cumulative (top)  and yearly (bottom) iNTS cases under the status quo and routine + catch-up vaccination ($95\%$ efficacy) scenarios. Shaded areas show the 25th and 75th percentiles, line shows the median over 1000 experiments, samples drawn from uniform distributions over (0.00020,0.00024) for $\beta_{2,n}$ and (0.0080,0.0084) for $\beta_{4,n}$. }\label{fig:Namibia}
\end{figure}

\begin{figure}[htbp]
\renewcommand{\thefigure}{\textbf{Supplementary Fig. 44 Niger cumulative and yearly iNTS cases}}
\begin{subfigure}[b]{\textwidth}
\centering
\includegraphics[width=0.7\linewidth]{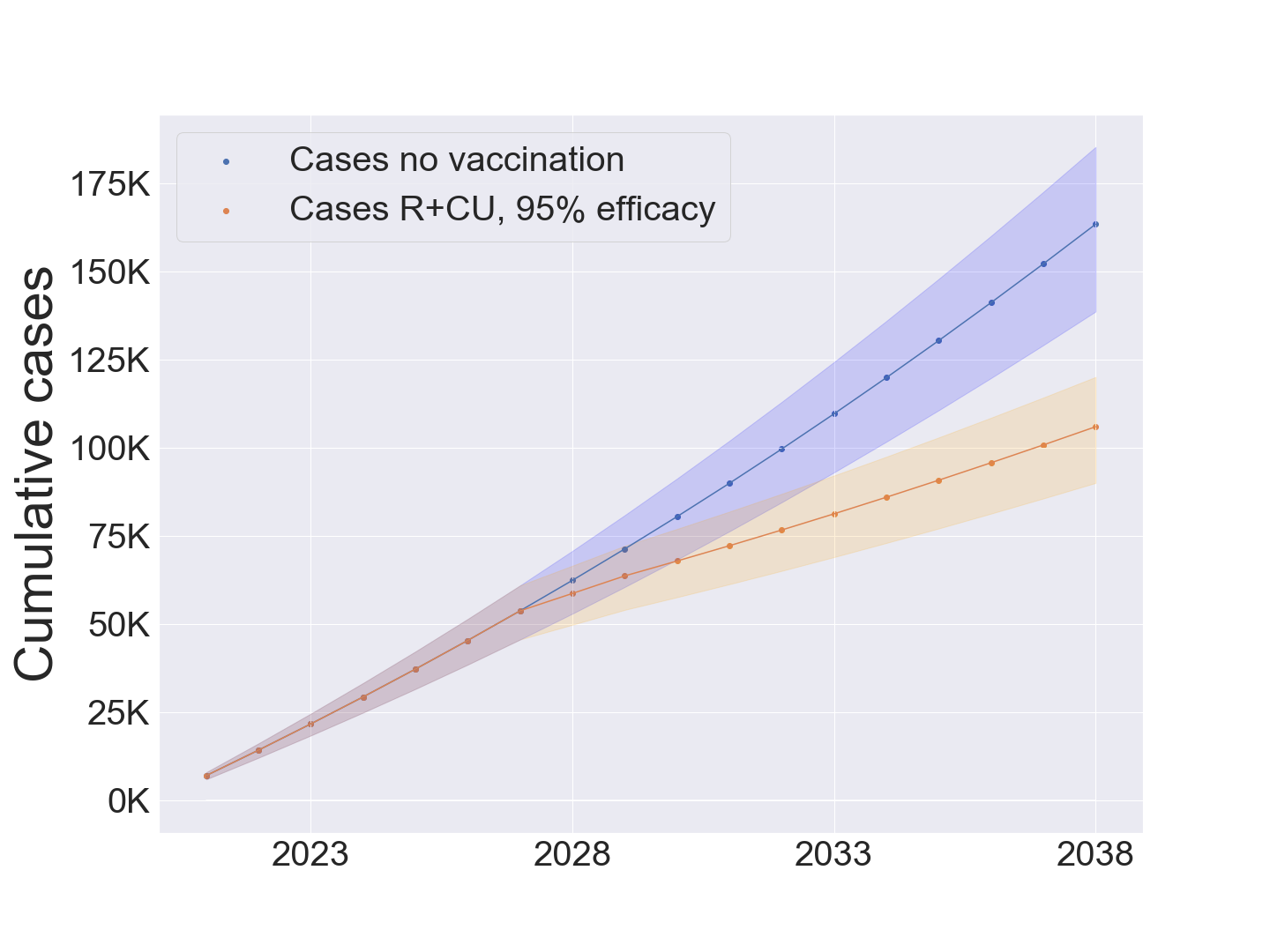}
\includegraphics[width=0.7\linewidth]{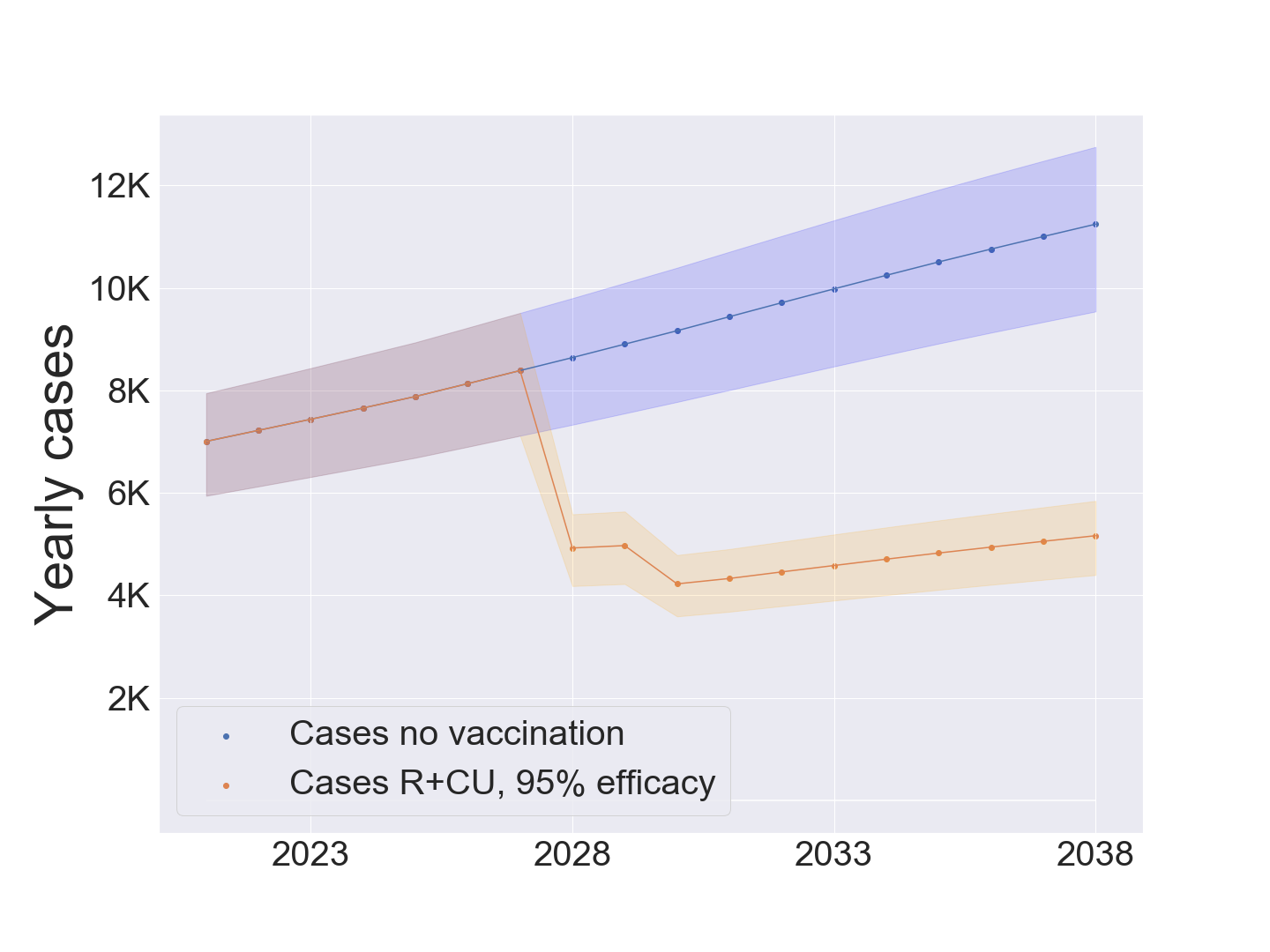}
\end{subfigure}
\caption{Niger cumulative (top)  and yearly (bottom) iNTS cases under the status quo and routine + catch-up vaccination ($95\%$ efficacy) scenarios. Shaded areas show the 25th and 75th percentiles, line shows the median over 1000 experiments, samples drawn from uniform distributions over (0.00020,0.00024) for $\beta_{2,n}$ and (0.0080,0.0084) for $\beta_{4,n}$. }\label{fig:Niger}
\end{figure}

\begin{figure}[htbp]
\renewcommand{\thefigure}{\textbf{Supplementary Fig. 45 Nigeria cumulative and yearly iNTS cases}}
\begin{subfigure}[b]{\textwidth}
\centering
\includegraphics[width=0.7\linewidth]{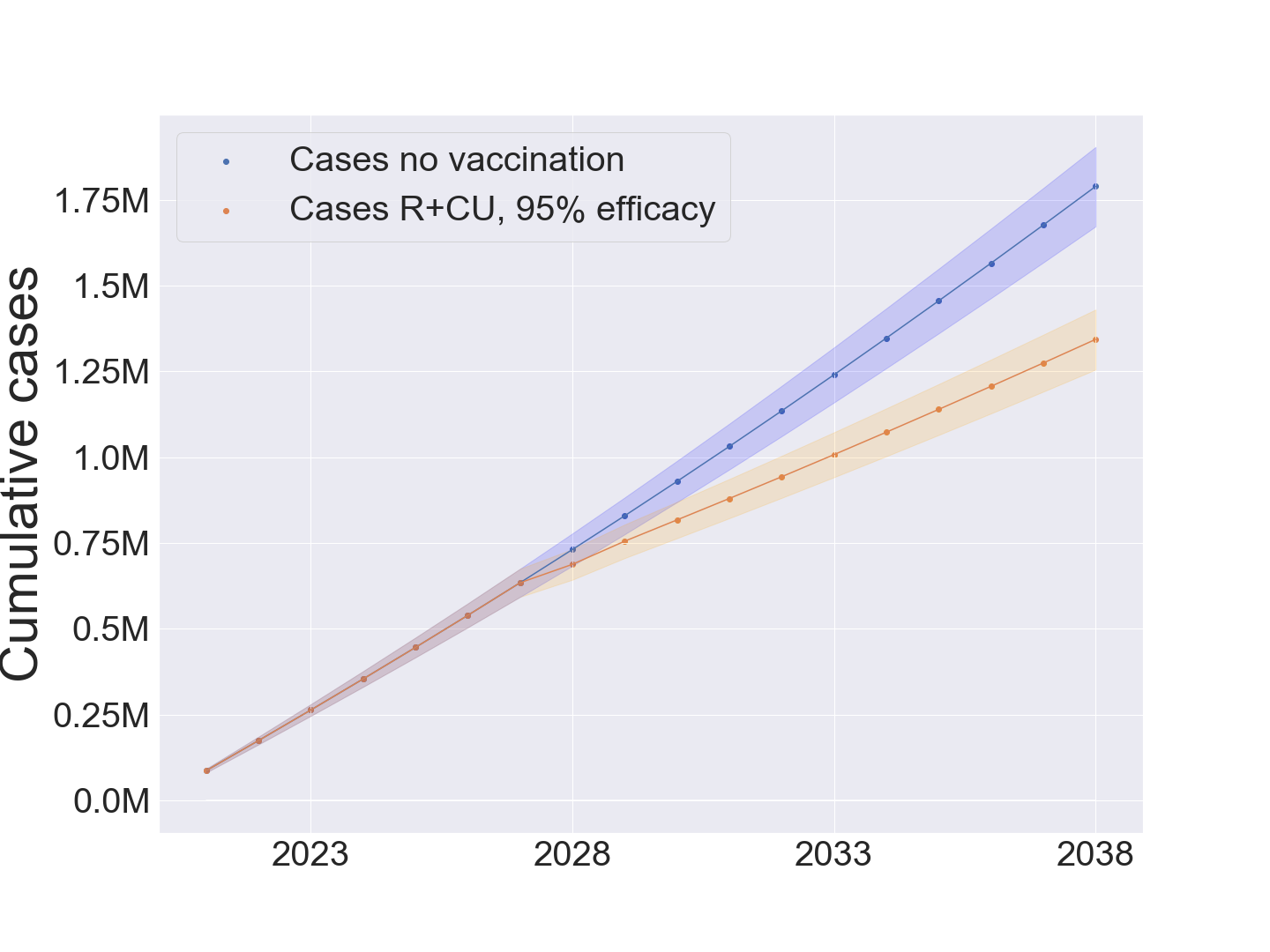}
\includegraphics[width=0.7\linewidth]{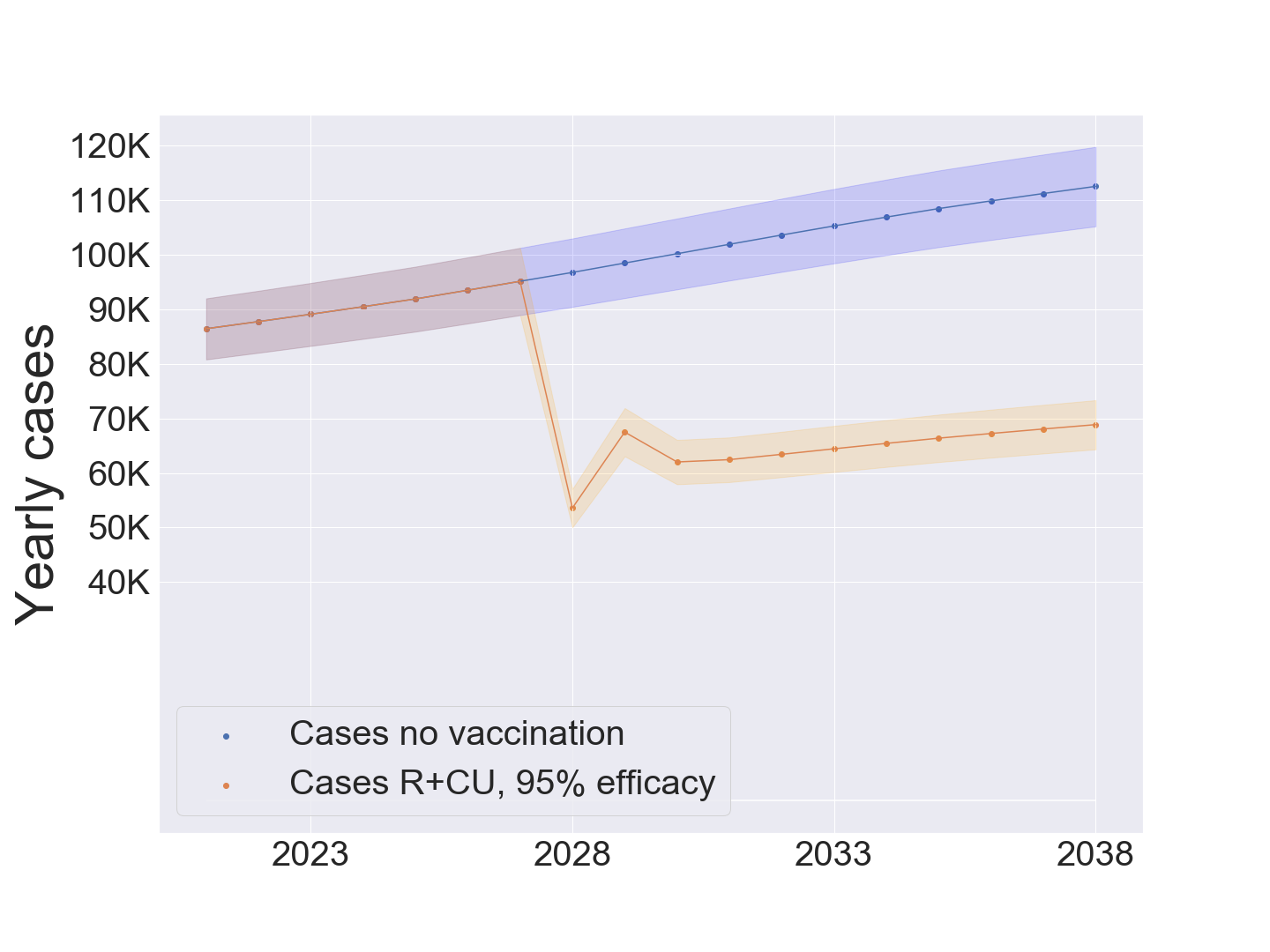}
\end{subfigure}
\caption{Nigeria cumulative (top)  and yearly (bottom) iNTS cases under the status quo and routine + catch-up vaccination ($95\%$ efficacy) scenarios. Shaded areas show the 25th and 75th percentiles, line shows the median over 1000 experiments, samples drawn from uniform distributions over (0.00020,0.00024) for $\beta_{2,n}$ and (0.0080,0.0084) for $\beta_{4,n}$. }\label{fig:Nigeria}
\end{figure}

\begin{figure}[htbp]
\renewcommand{\thefigure}{\textbf{Supplementary Fig. 46 Rwanda cumulative and yearly iNTS cases}}
\begin{subfigure}[b]{\textwidth}
\centering
\includegraphics[width=0.7\linewidth]{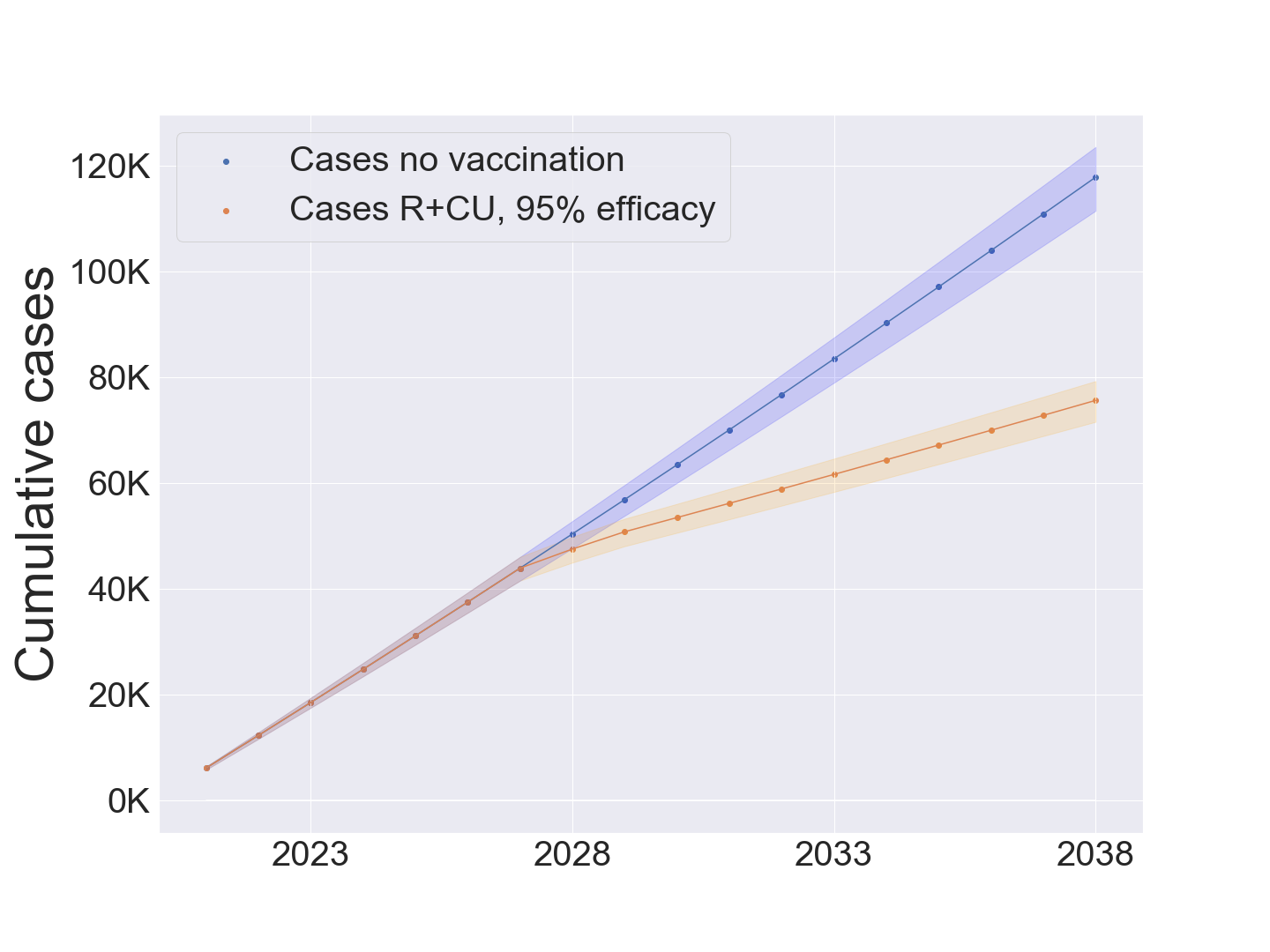}
\includegraphics[width=0.7\linewidth]{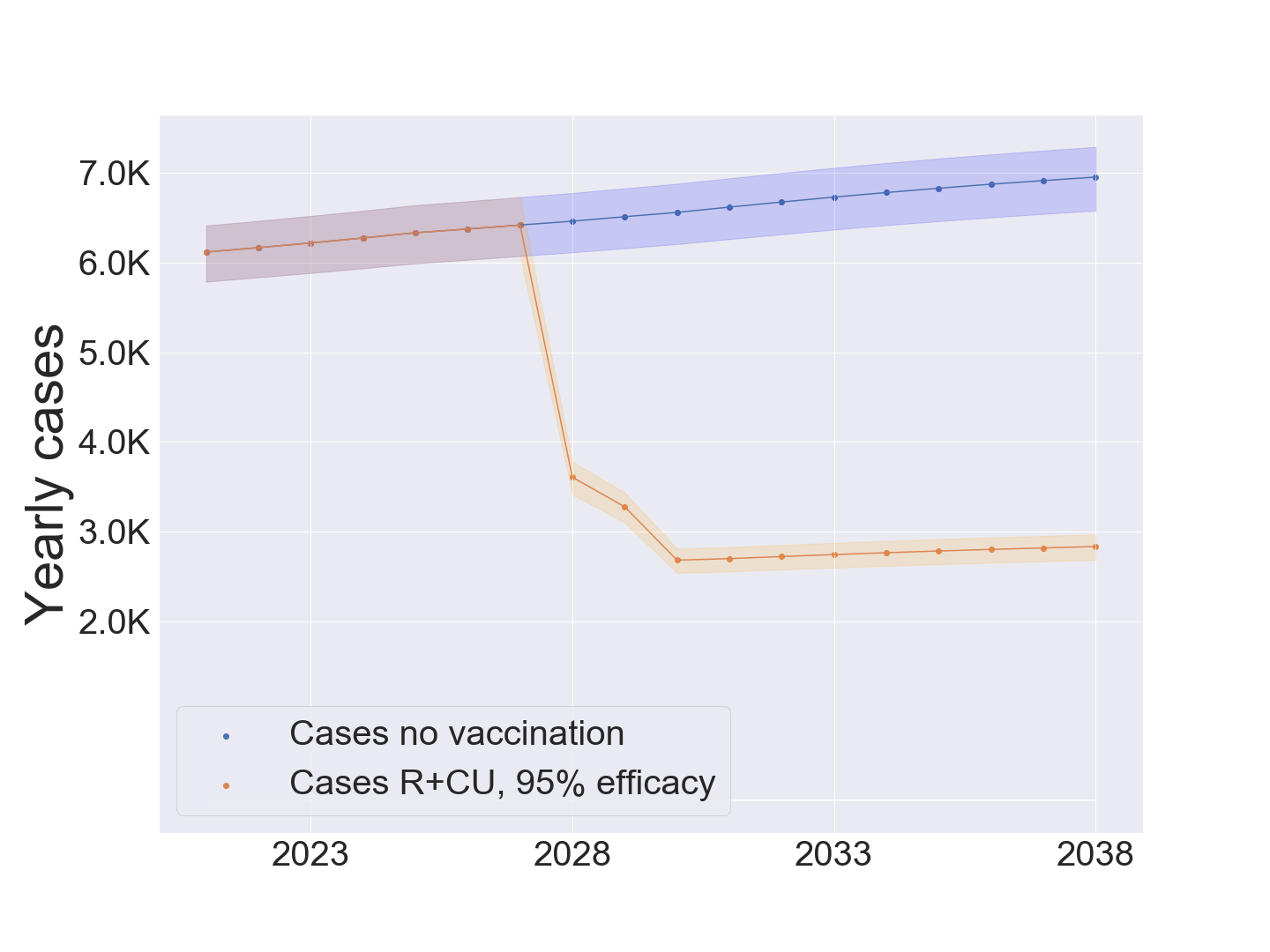}
\end{subfigure}
\caption{Rwanda cumulative (top)  and yearly (bottom) iNTS cases under the status quo and routine + catch-up vaccination ($95\%$ efficacy) scenarios. Shaded areas show the 25th and 75th percentiles, line shows the median over 1000 experiments, samples drawn from uniform distributions over (0.00020,0.00024) for $\beta_{2,n}$ and (0.0080,0.0084) for $\beta_{4,n}$. }\label{fig:Rwanda}
\end{figure}

\begin{figure}[htbp]
\renewcommand{\thefigure}{\textbf{Supplementary Fig. 47 R\'eunion cumulative and yearly iNTS cases}}
\begin{subfigure}[b]{\textwidth}
\centering
\includegraphics[width=0.7\linewidth]{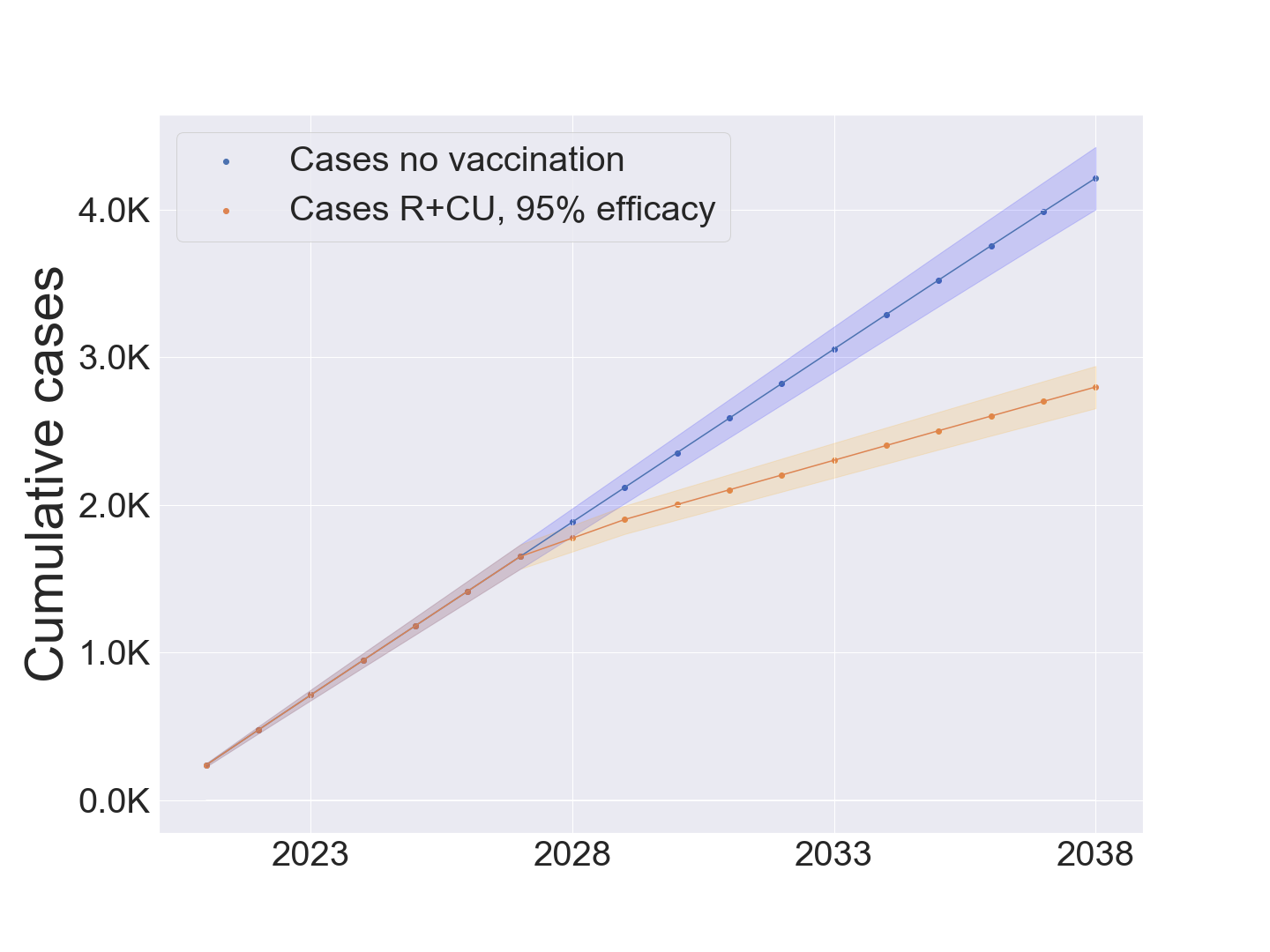}
\includegraphics[width=0.7\linewidth]{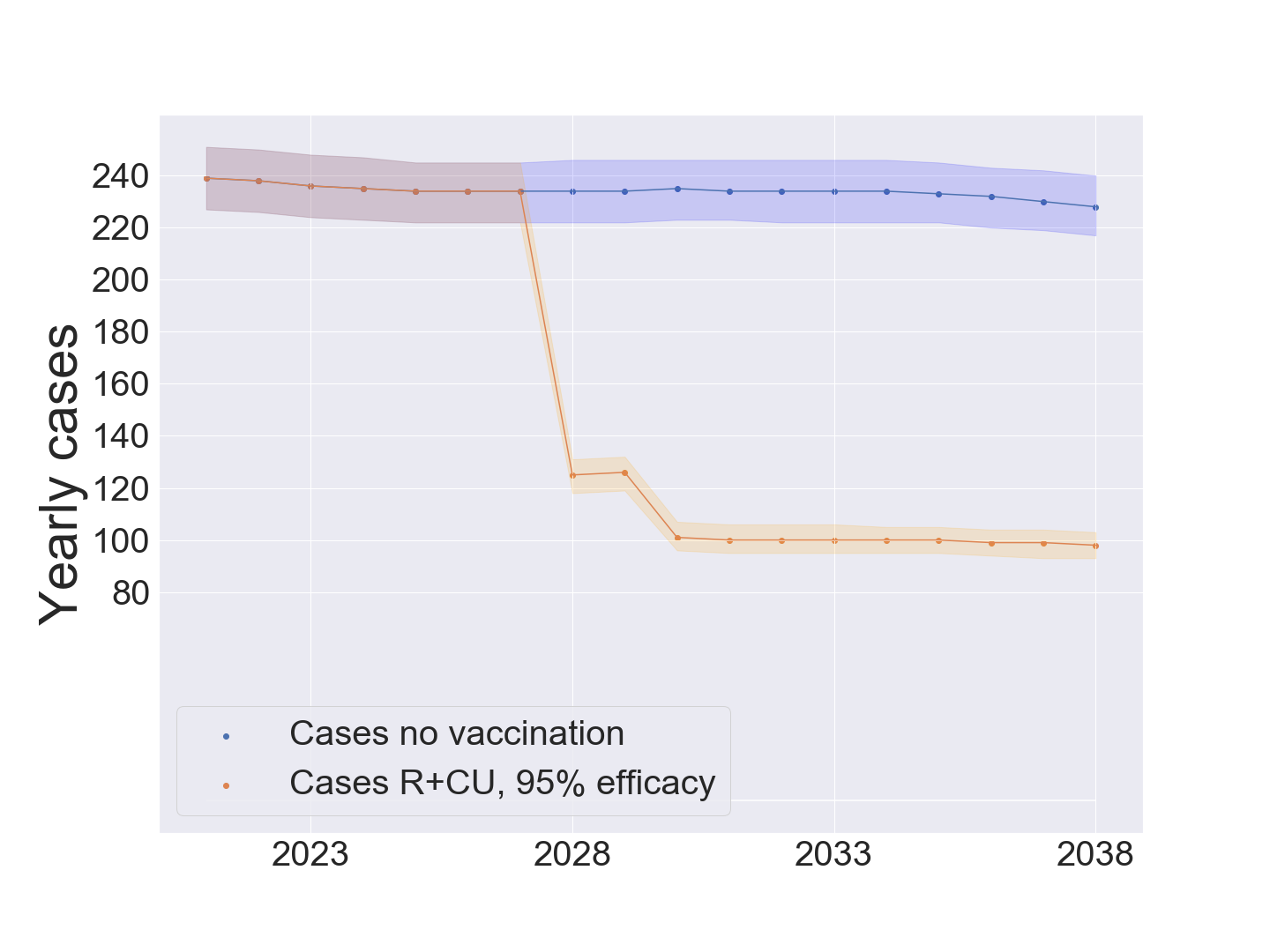}
\end{subfigure}
\caption{R\'eunion cumulative (top)  and yearly (bottom) iNTS cases under the status quo and routine + catch-up vaccination ($95\%$ efficacy) scenarios. Shaded areas show the 25th and 75th percentiles, line shows the median over 1000 experiments, samples drawn from uniform distributions over (0.00020,0.00024) for $\beta_{2,n}$ and (0.0080,0.0084) for $\beta_{4,n}$. }\label{fig:Réunion}
\end{figure}

\begin{figure}[htbp]
\renewcommand{\thefigure}{\textbf{Supplementary Fig. 48 Sao Tome and Principe cumulative and yearly iNTS cases}}
\begin{subfigure}[b]{\textwidth}
\centering
\includegraphics[width=0.7\linewidth]{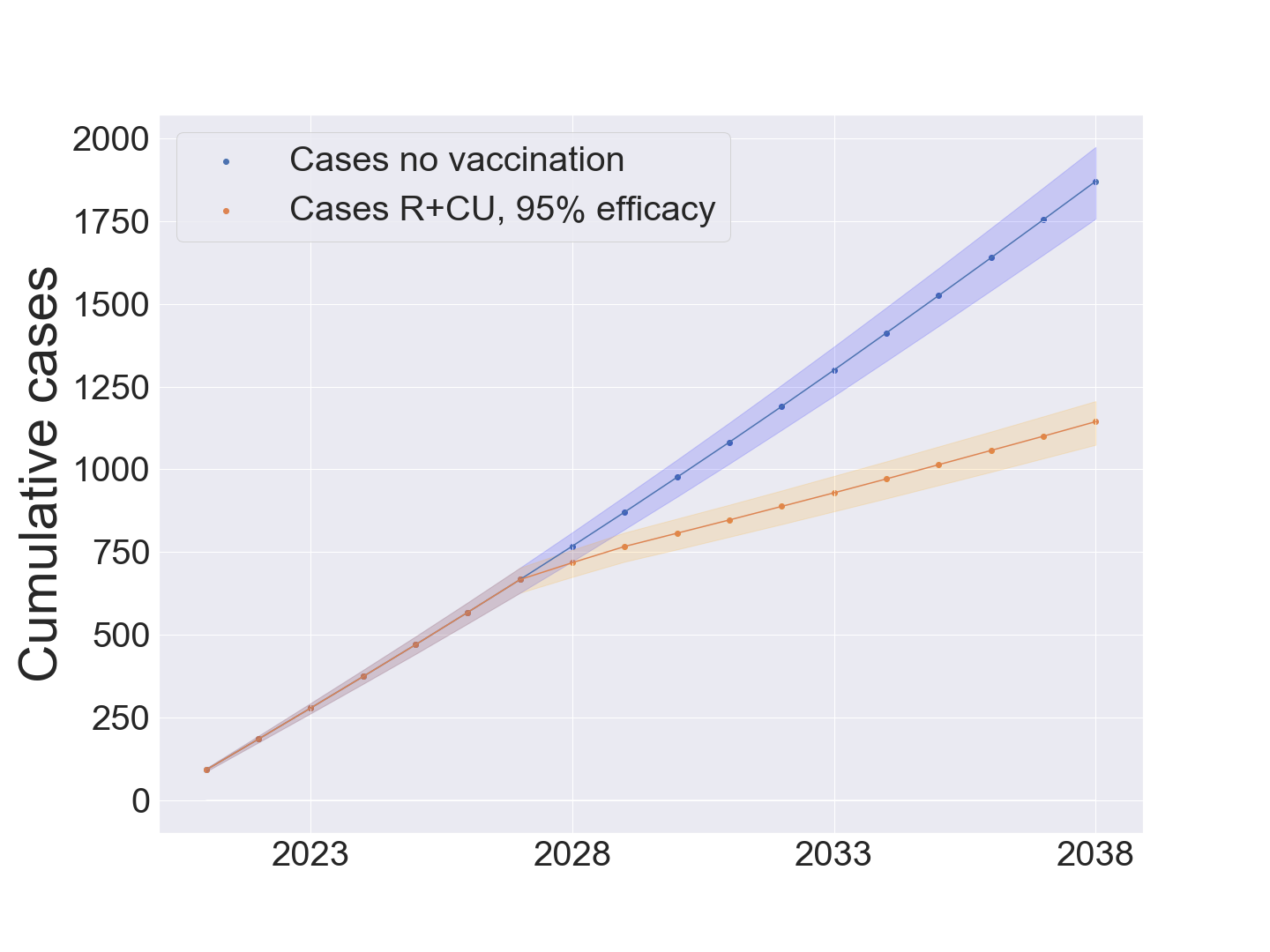}
\includegraphics[width=0.7\linewidth]{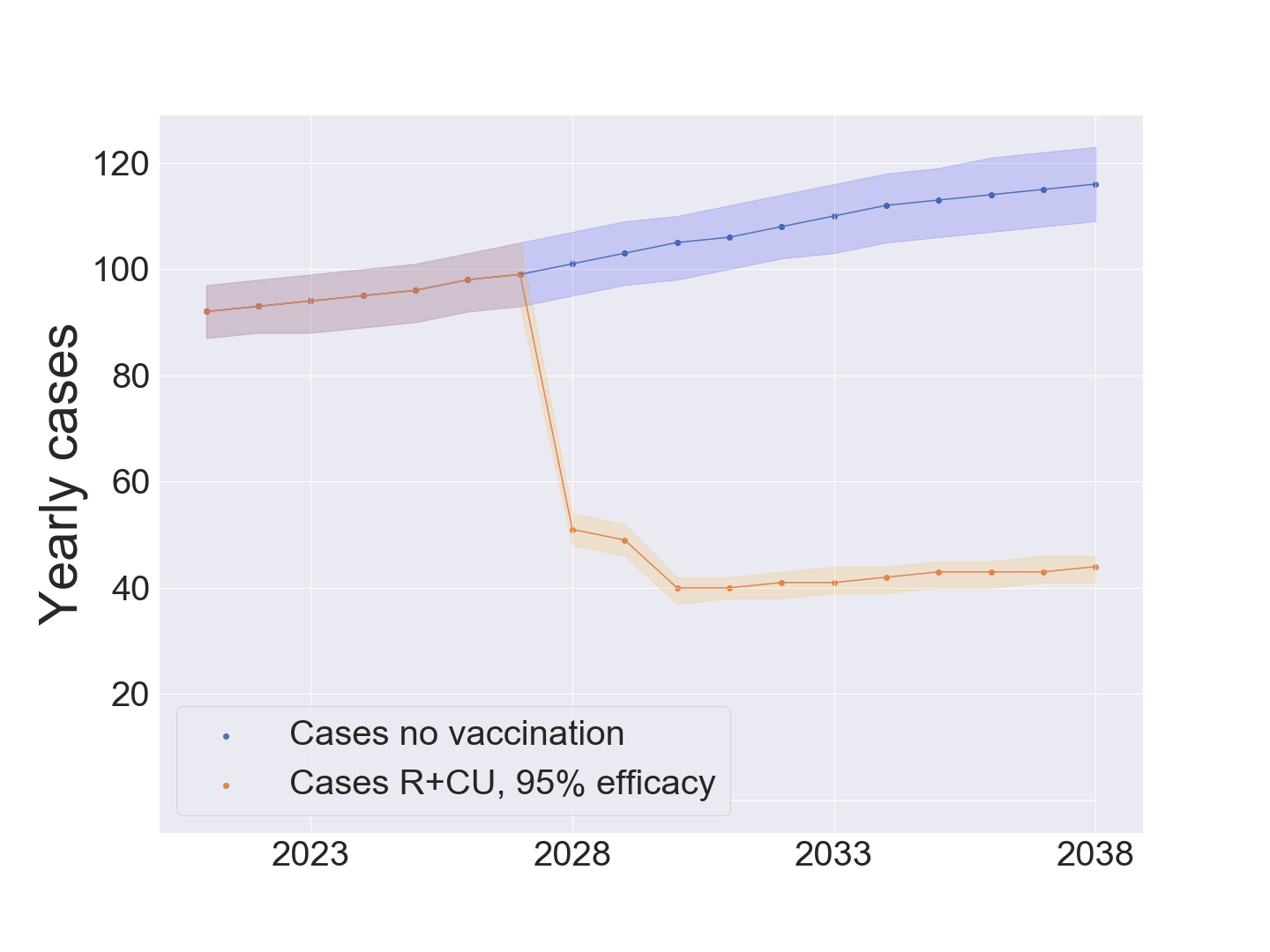}
\end{subfigure}
\caption{Sao Tome and Principe cumulative (top)  and yearly (bottom) iNTS cases under the status quo and routine + catch-up vaccination ($95\%$ efficacy) scenarios. Shaded areas show the 25th and 75th percentiles, line shows the median over 1000 experiments, samples drawn from uniform distributions over (0.00020,0.00024) for $\beta_{2,n}$ and (0.0080,0.0084) for $\beta_{4,n}$. }\label{fig:SaoTome}
\end{figure}

\begin{figure}[htbp]
\renewcommand{\thefigure}{\textbf{Supplementary Fig. 49 Senegal cumulative and yearly iNTS cases}}
\begin{subfigure}[b]{\textwidth}
\centering
\includegraphics[width=0.7\linewidth]{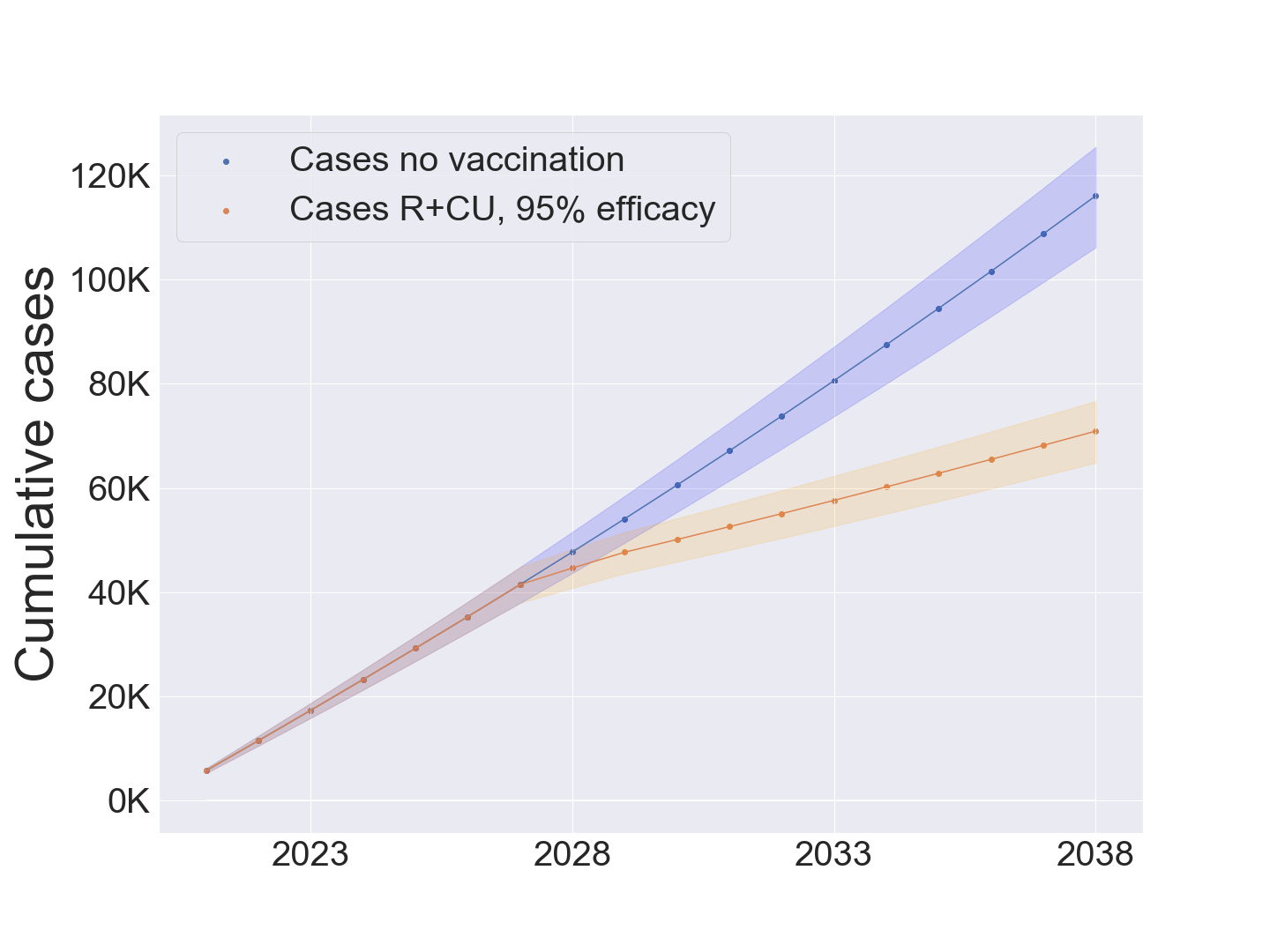}
\includegraphics[width=0.7\linewidth]{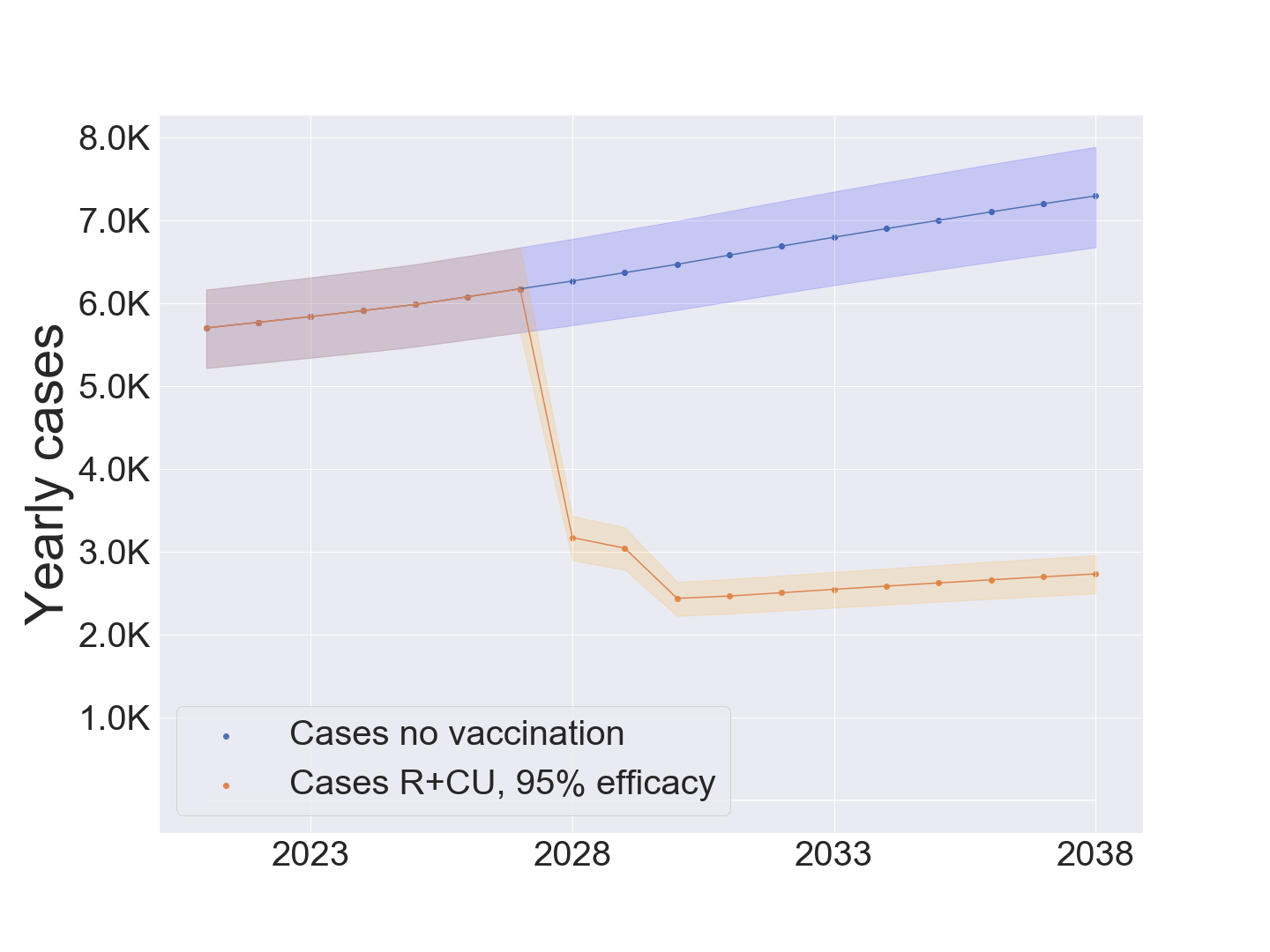}
\end{subfigure}
\caption{Senegal cumulative (top)  and yearly (bottom) iNTS cases under the status quo and routine + catch-up vaccination ($95\%$ efficacy) scenarios. Shaded areas show the 25th and 75th percentiles, line shows the median over 1000 experiments, samples drawn from uniform distributions over (0.00020,0.00024) for $\beta_{2,n}$ and (0.0080,0.0084) for $\beta_{4,n}$. }\label{fig:Senegal}
\end{figure}

\begin{figure}[htbp]
\renewcommand{\thefigure}{\textbf{Supplementary Fig. 50 Sierra Leone cumulative and yearly iNTS cases}}
\begin{subfigure}[b]{\textwidth}
\centering
\includegraphics[width=0.7\linewidth]{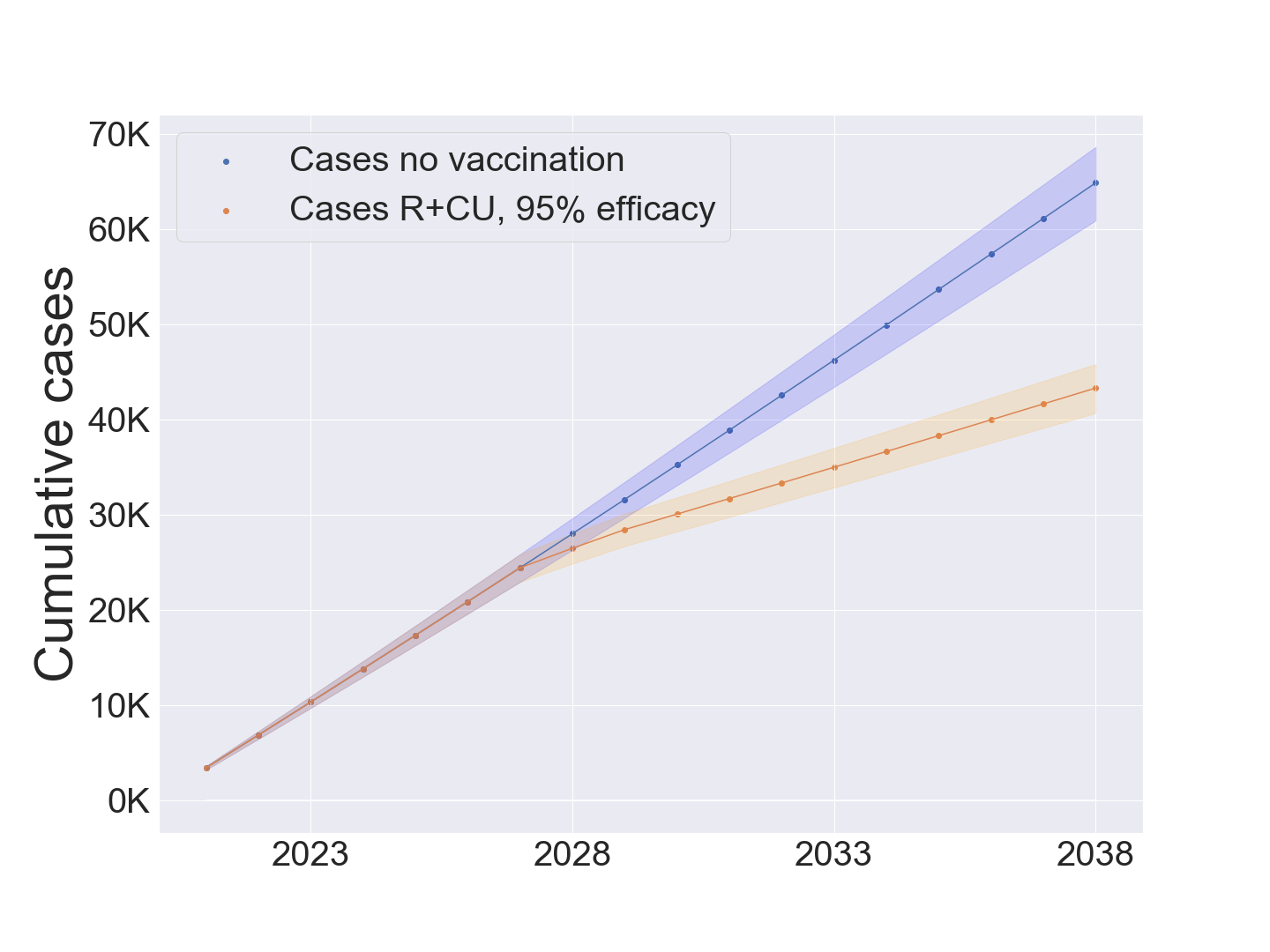}
\includegraphics[width=0.7\linewidth]{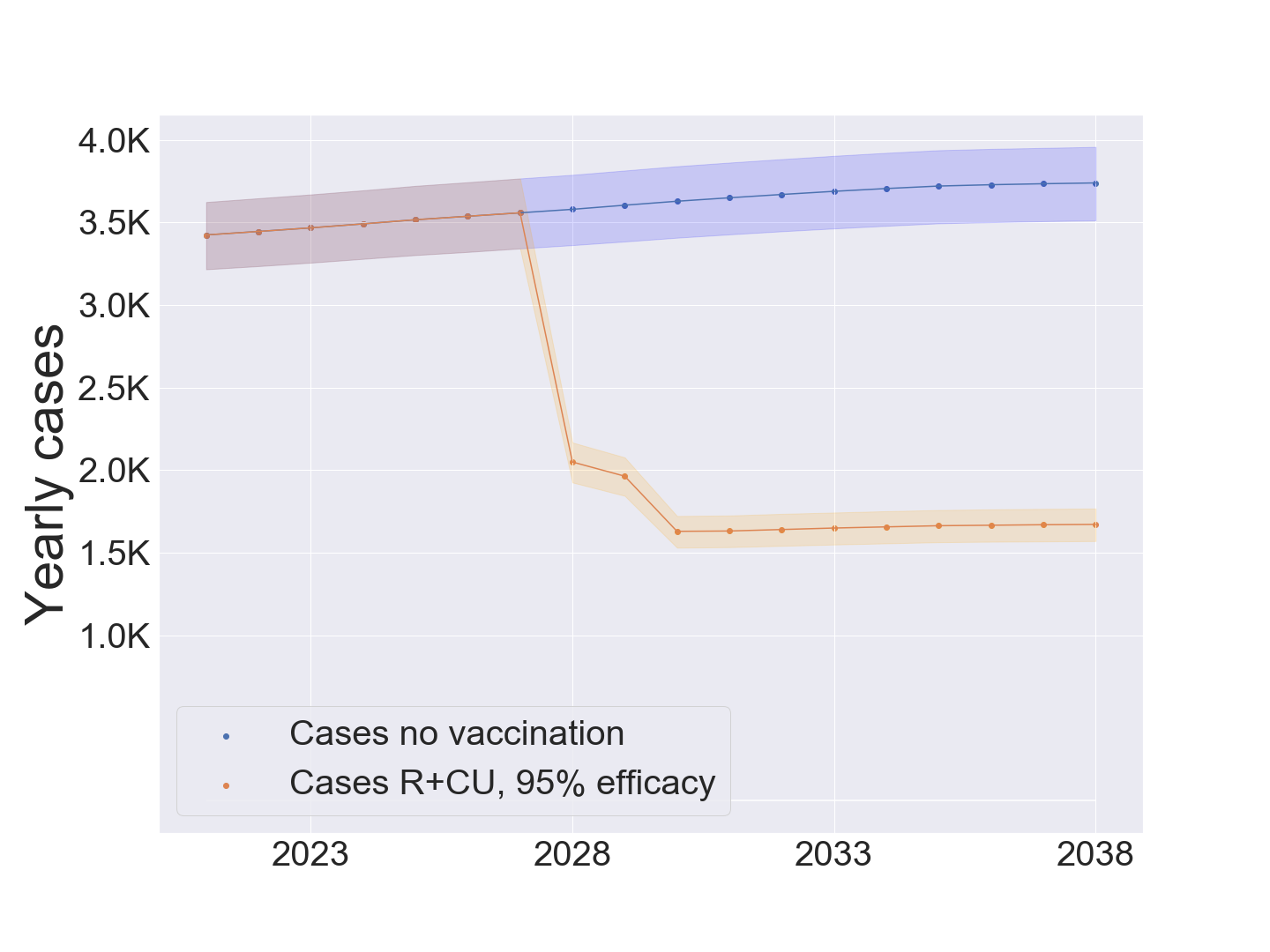}
\end{subfigure}
\caption{Sierra Leone cumulative (top)  and yearly (bottom) iNTS cases under the status quo and routine + catch-up vaccination ($95\%$ efficacy) scenarios. Shaded areas show the 25th and 75th percentiles, line shows the median over 1000 experiments, samples drawn from uniform distributions over (0.00020,0.00024) for $\beta_{2,n}$ and (0.0080,0.0084) for $\beta_{4,n}$. }\label{fig:SierraLeone}
\end{figure}

\begin{figure}[htbp]
\renewcommand{\thefigure}{\textbf{Supplementary Fig. 51 Somalia cumulative and yearly iNTS cases}}
\begin{subfigure}[b]{\textwidth}
\centering
\includegraphics[width=0.7\linewidth]{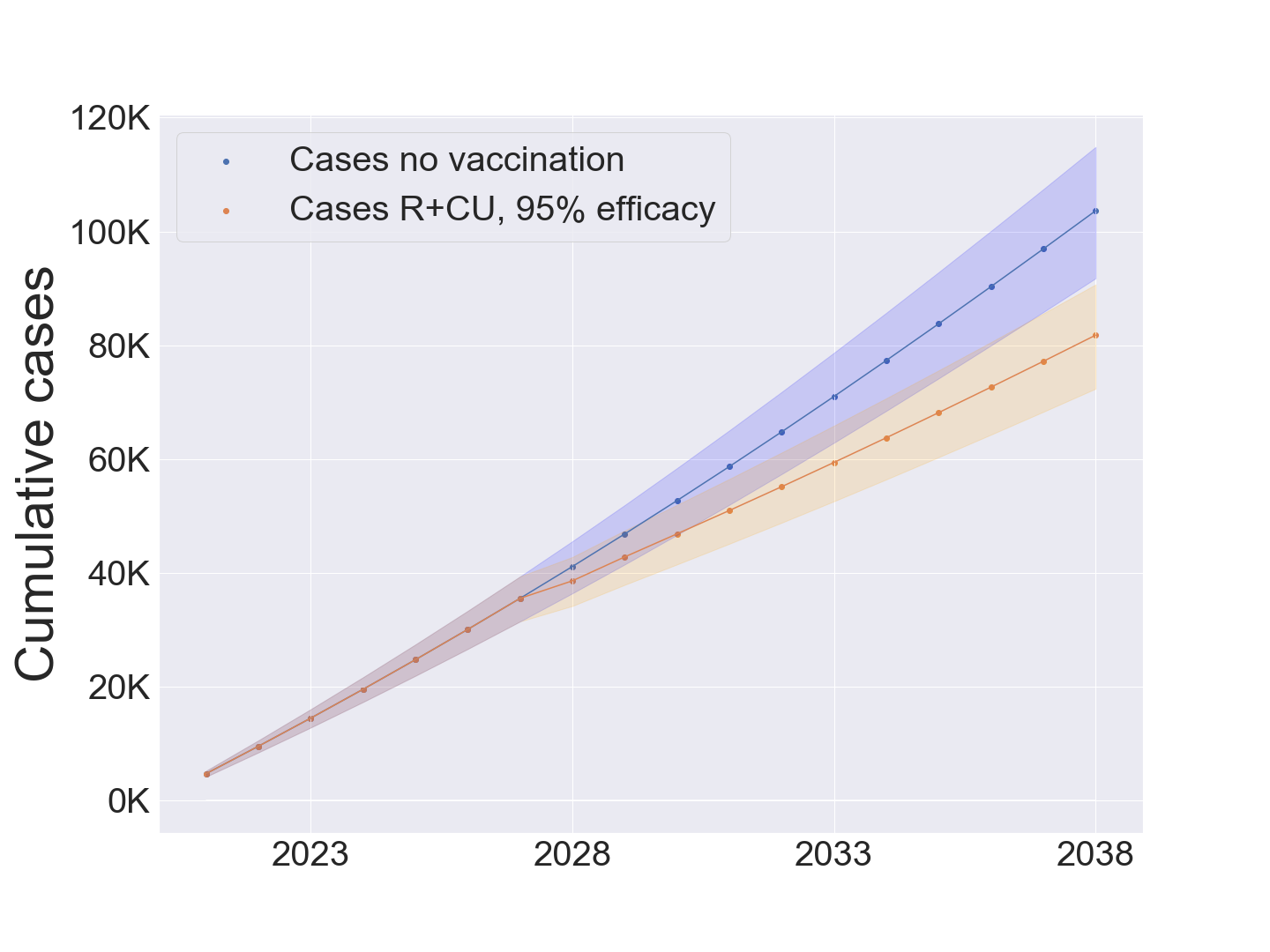}
\includegraphics[width=0.7\linewidth]{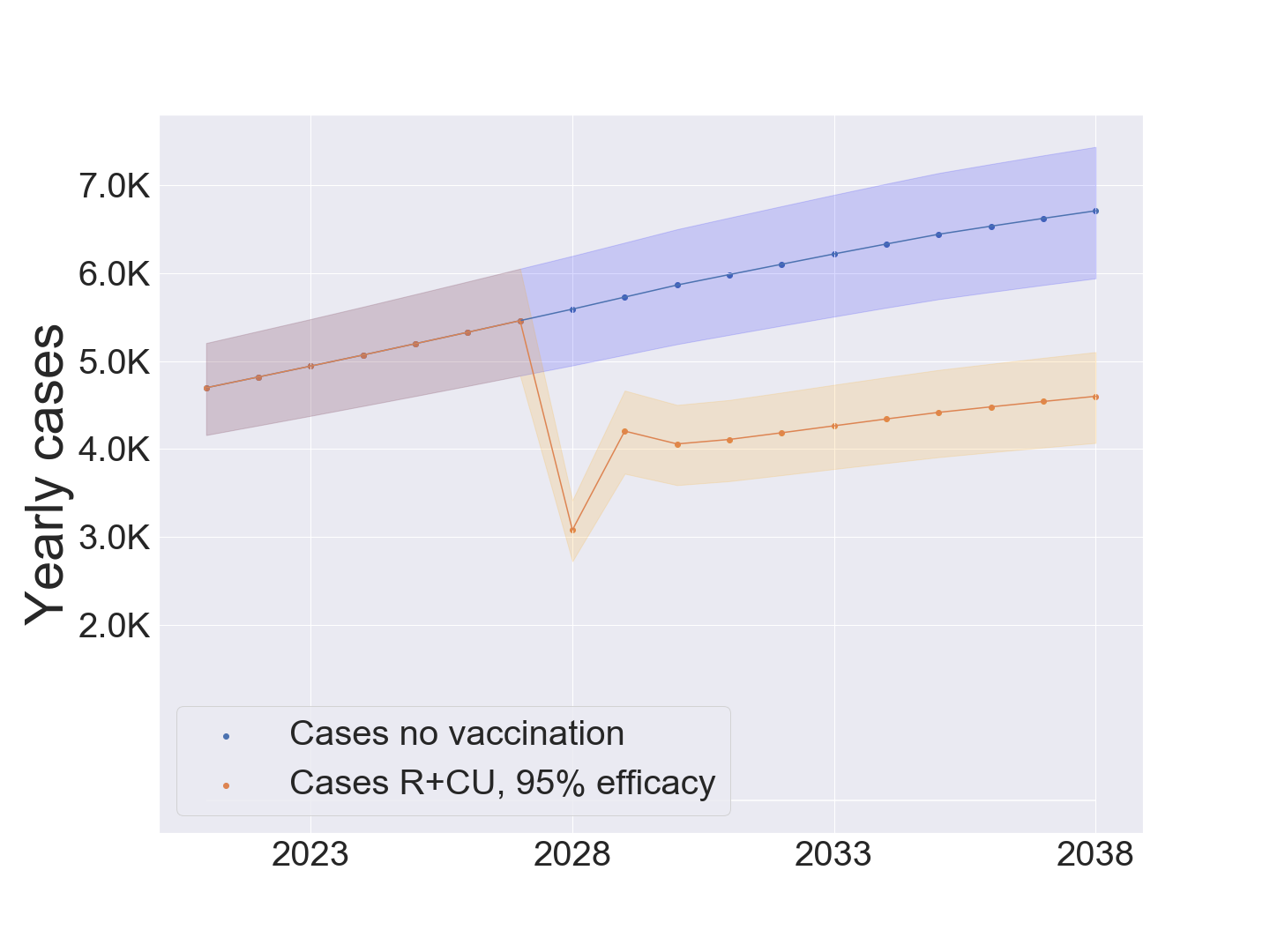}
\end{subfigure}
\caption{Somalia cumulative (top)  and yearly (bottom) iNTS cases under the status quo and routine + catch-up vaccination ($95\%$ efficacy) scenarios. Shaded areas show the 25th and 75th percentiles, line shows the median over 1000 experiments, samples drawn from uniform distributions over (0.00020,0.00024) for $\beta_{2,n}$ and (0.0080,0.0084) for $\beta_{4,n}$. }\label{fig:Somalia}
\end{figure}

\begin{figure}[htbp]
\renewcommand{\thefigure}{\textbf{Supplementary Fig. 52 South Africa cumulative and yearly iNTS cases}}
\begin{subfigure}[b]{\textwidth}
\centering
\includegraphics[width=0.7\linewidth]{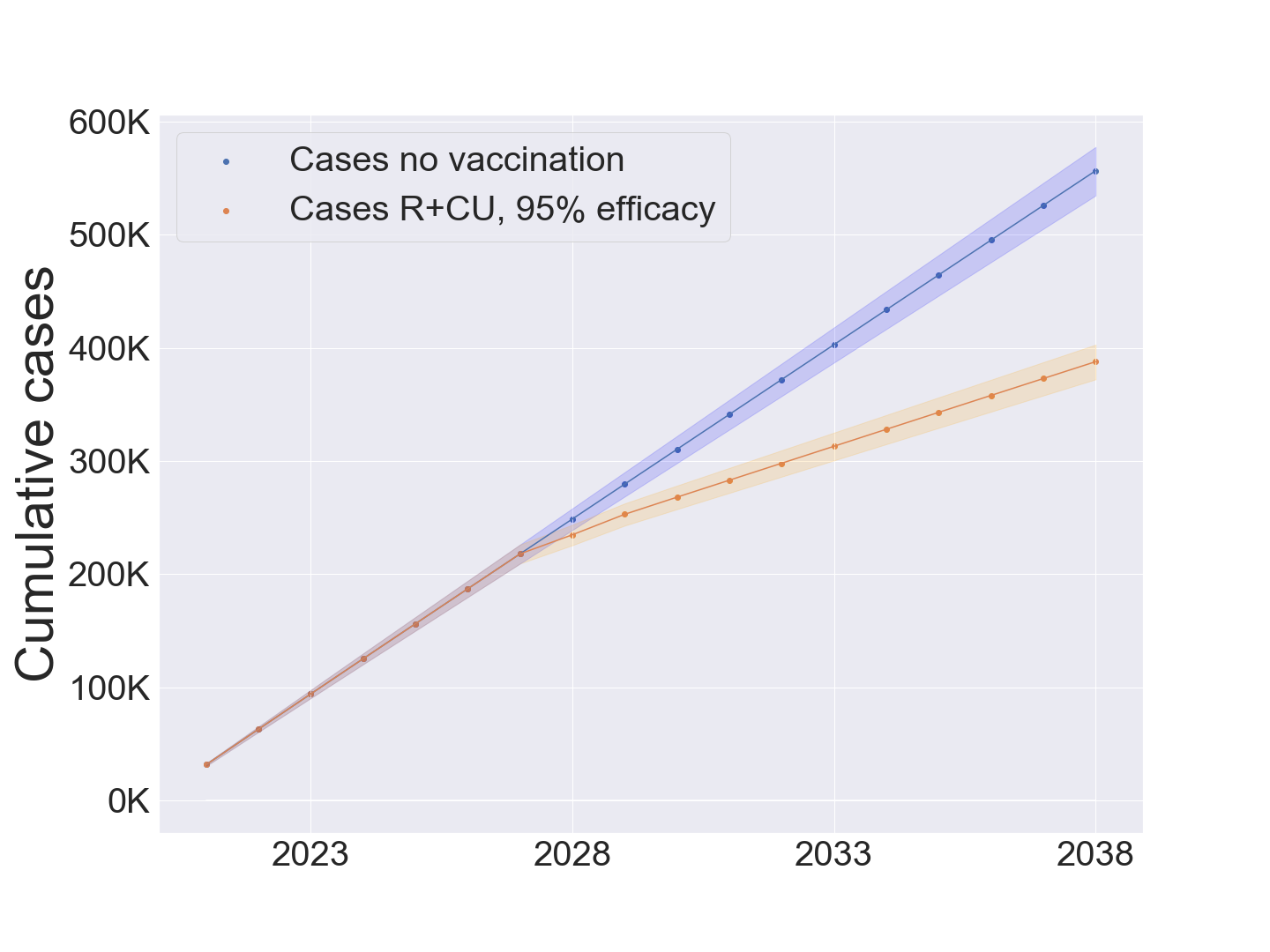}
\includegraphics[width=0.7\linewidth]{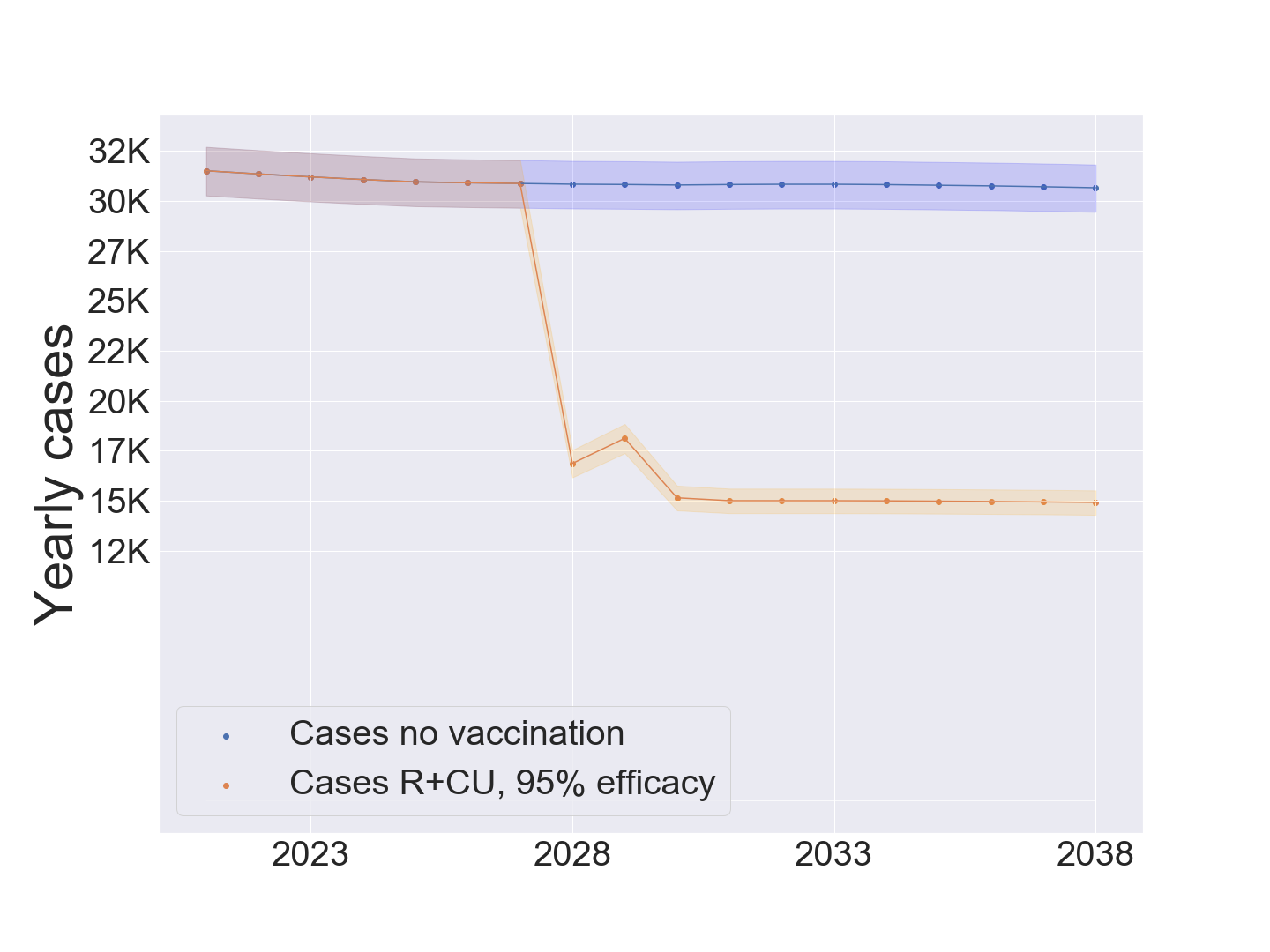}
\end{subfigure}
\caption{South Africa cumulative (top)  and yearly (bottom) iNTS cases under the status quo and routine + catch-up vaccination ($95\%$ efficacy) scenarios. Shaded areas show the 25th and 75th percentiles, line shows the median over 1000 experiments, samples drawn from uniform distributions over (0.00020,0.00024) for $\beta_{2,n}$ and (0.0080,0.0084) for $\beta_{4,n}$. }\label{fig:SouthAfrica}
\end{figure}

\begin{figure}[htbp]
\renewcommand{\thefigure}{\textbf{Supplementary Fig. 53 South Sudan cumulative and yearly iNTS cases}}
\begin{subfigure}[b]{\textwidth}
\centering
\includegraphics[width=0.7\linewidth]{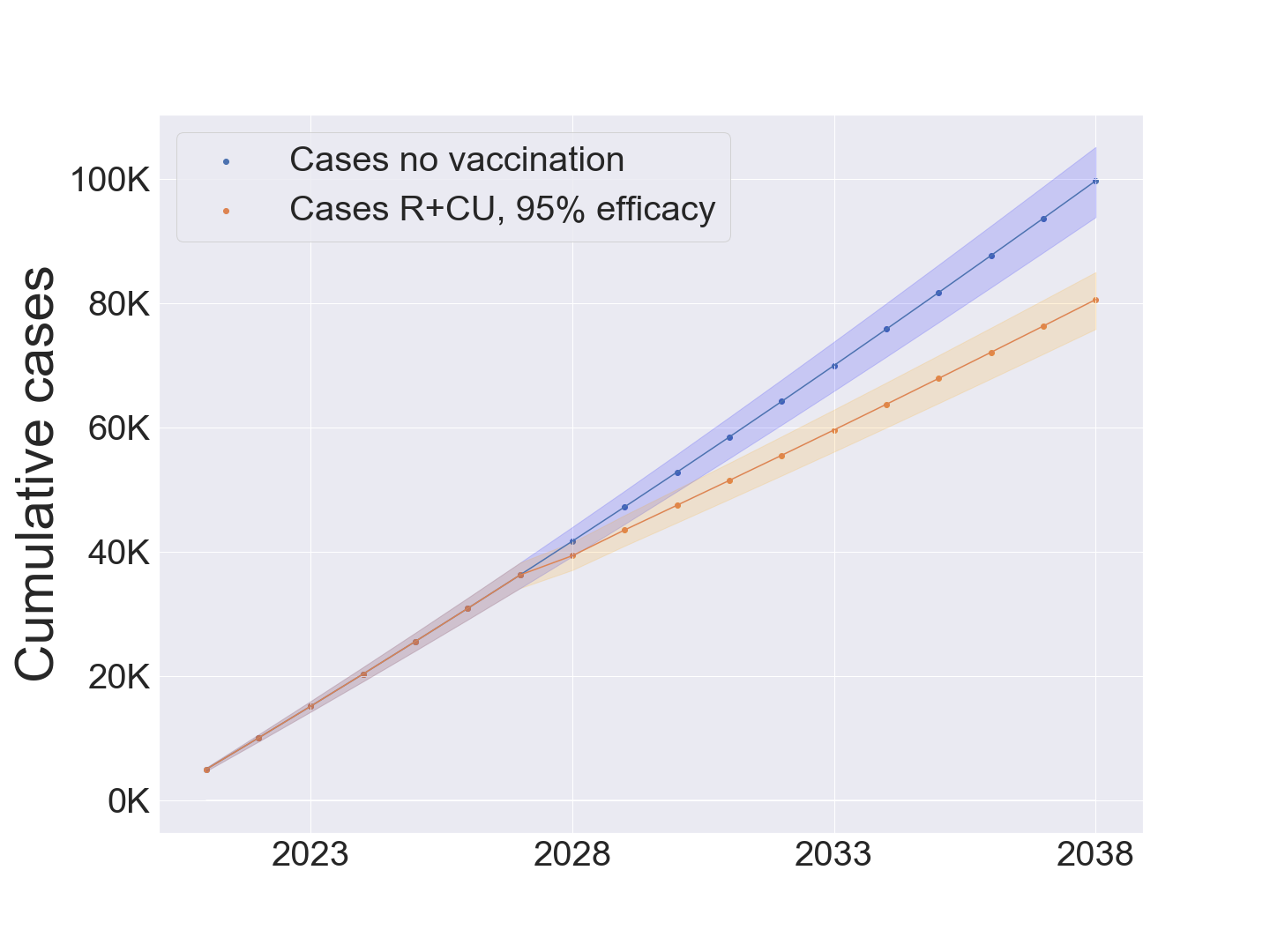}
\includegraphics[width=0.7\linewidth]{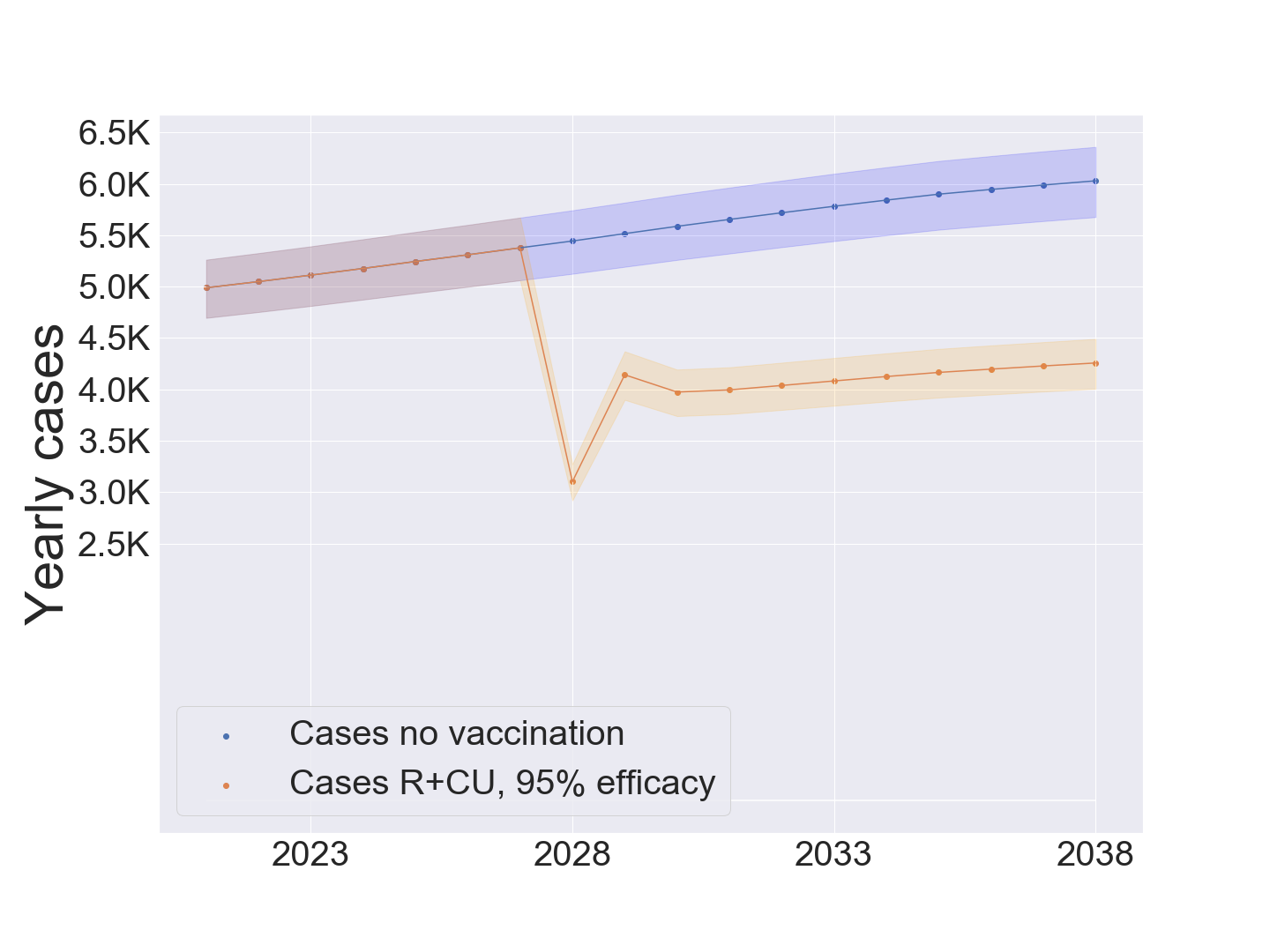}
\end{subfigure}
\caption{South Sudan cumulative (top)  and yearly (bottom) iNTS cases under the status quo and routine + catch-up vaccination ($95\%$ efficacy) scenarios. Shaded areas show the 25th and 75th percentiles, line shows the median over 1000 experiments, samples drawn from uniform distributions over (0.00020,0.00024) for $\beta_{2,n}$ and (0.0080,0.0084) for $\beta_{4,n}$. }\label{fig:SouthSudan}
\end{figure}

\begin{figure}[htbp]
\renewcommand{\thefigure}{\textbf{Supplementary Fig. 54 Tanzania cumulative and yearly iNTS cases}}
\begin{subfigure}[b]{\textwidth}
\centering
\includegraphics[width=0.7\linewidth]{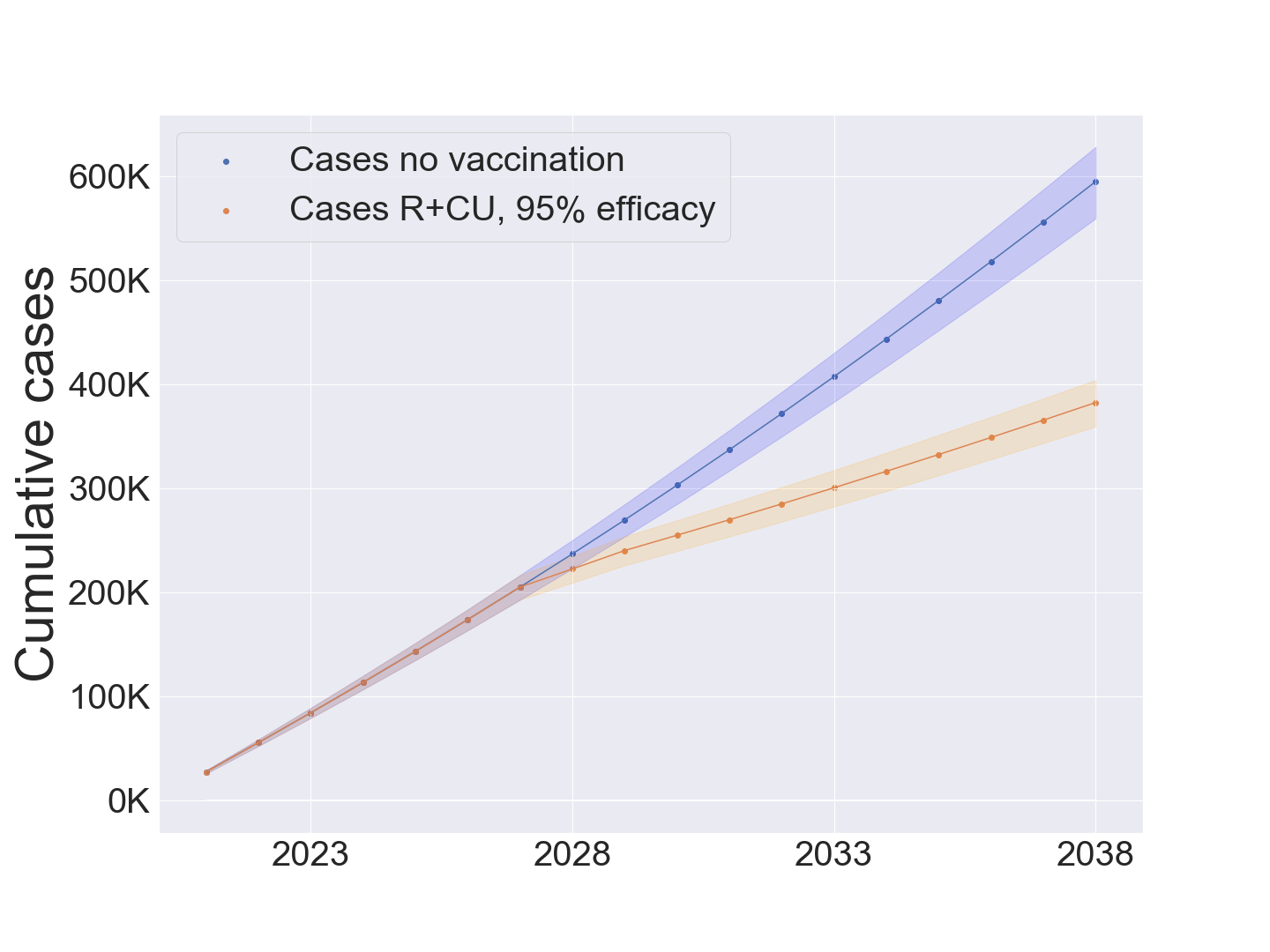}
\includegraphics[width=0.7\linewidth]{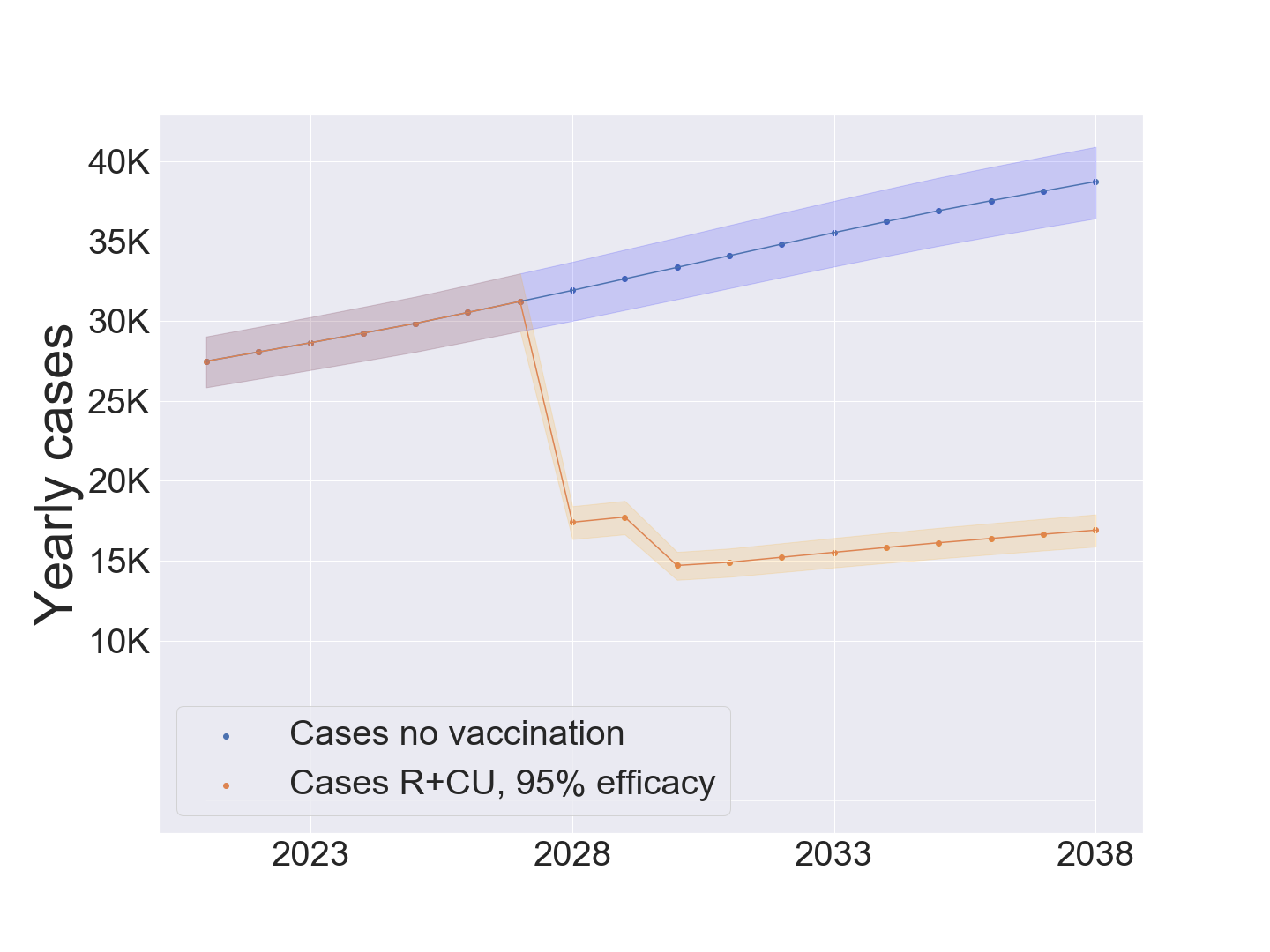}
\end{subfigure}
\caption{Tanzania (United Republic of) cumulative (top)  and yearly (bottom) iNTS cases under the status quo and routine + catch-up vaccination ($95\%$ efficacy) scenarios. Shaded areas show the 25th and 75th percentiles, line shows the median over 1000 experiments, samples drawn from uniform distributions over (0.00020,0.00024) for $\beta_{2,n}$ and (0.0080,0.0084) for $\beta_{4,n}$. }\label{fig:Tanzania}
\end{figure}

\begin{figure}[htbp]
\renewcommand{\thefigure}{\textbf{Supplementary Fig. 55 Togo cumulative and yearly iNTS cases}}
\begin{subfigure}[b]{\textwidth}
\centering
\includegraphics[width=0.7\linewidth]{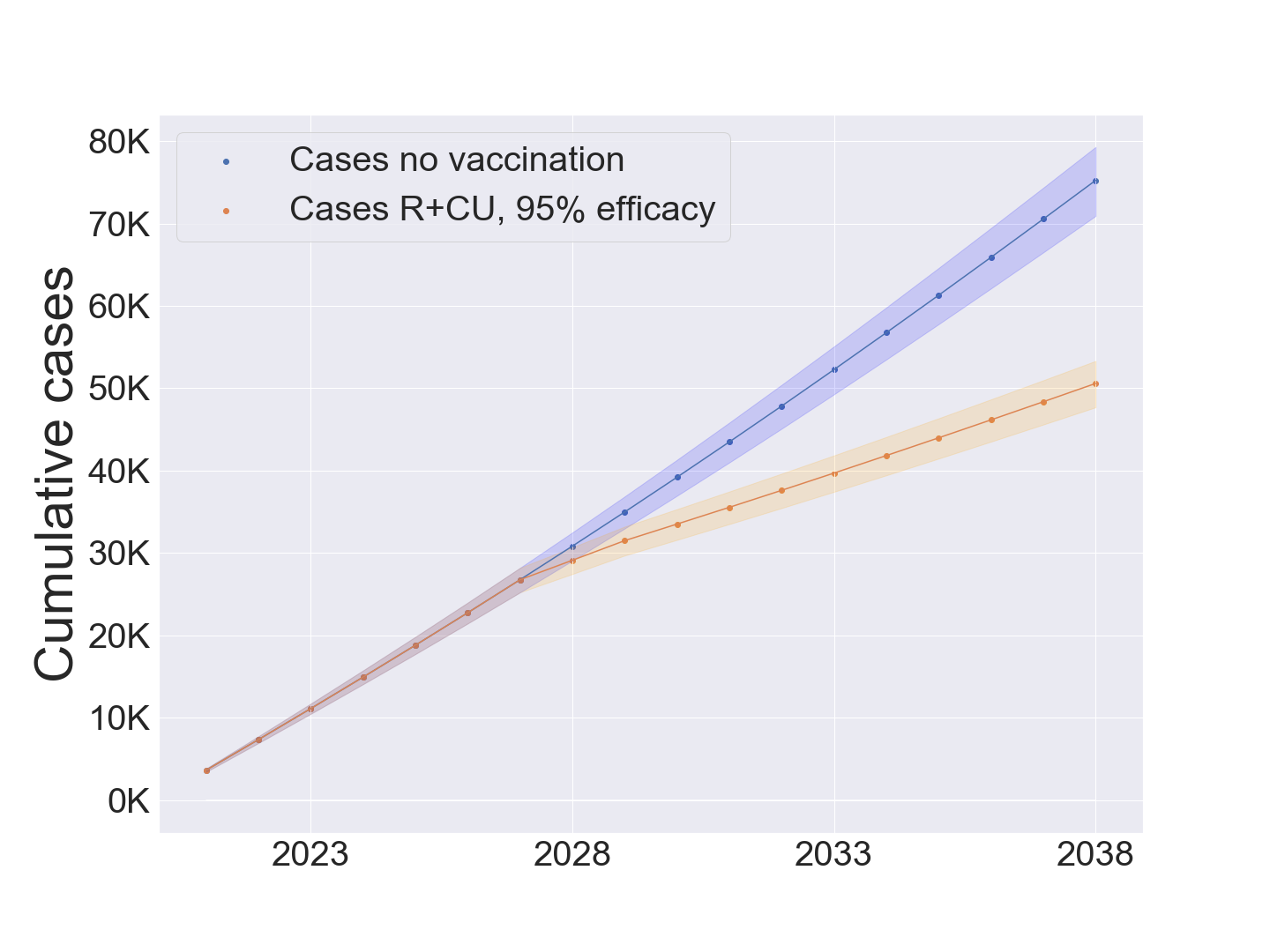}
\includegraphics[width=0.7\linewidth]{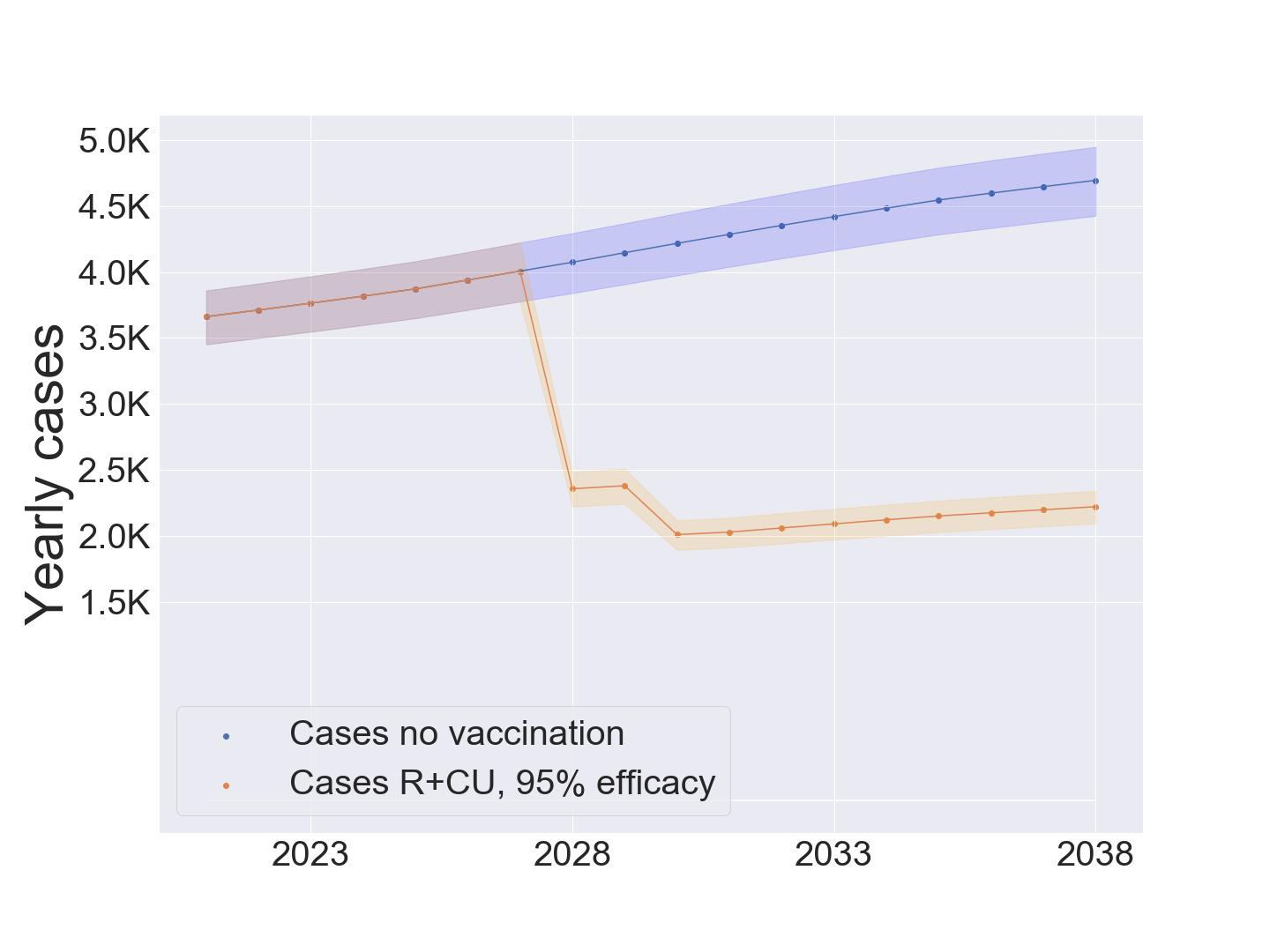}
\end{subfigure}
\caption{Togo cumulative (top)  and yearly (bottom) iNTS cases under the status quo and routine + catch-up vaccination ($95\%$ efficacy) scenarios. Shaded areas show the 25th and 75th percentiles, line shows the median over 1000 experiments, samples drawn from uniform distributions over (0.00020,0.00024) for $\beta_{2,n}$ and (0.0080,0.0084) for $\beta_{4,n}$. }\label{fig:Togo}
\end{figure}

\begin{figure}[htbp]
\renewcommand{\thefigure}{\textbf{Supplementary Fig. 56 Uganda cumulative and yearly iNTS cases}}
\begin{subfigure}[b]{\textwidth}
\centering
\includegraphics[width=0.7\linewidth]{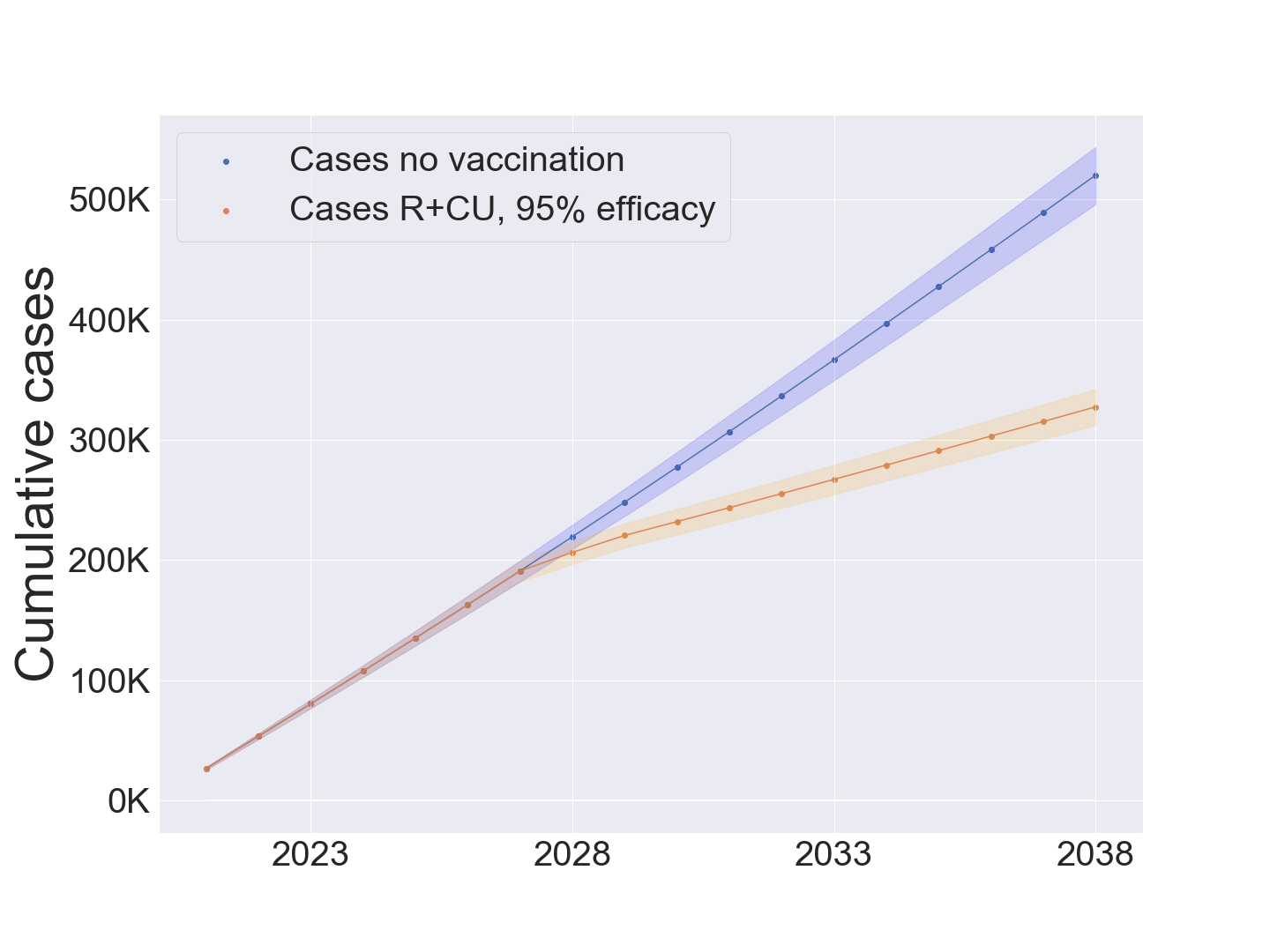}
\includegraphics[width=0.7\linewidth]{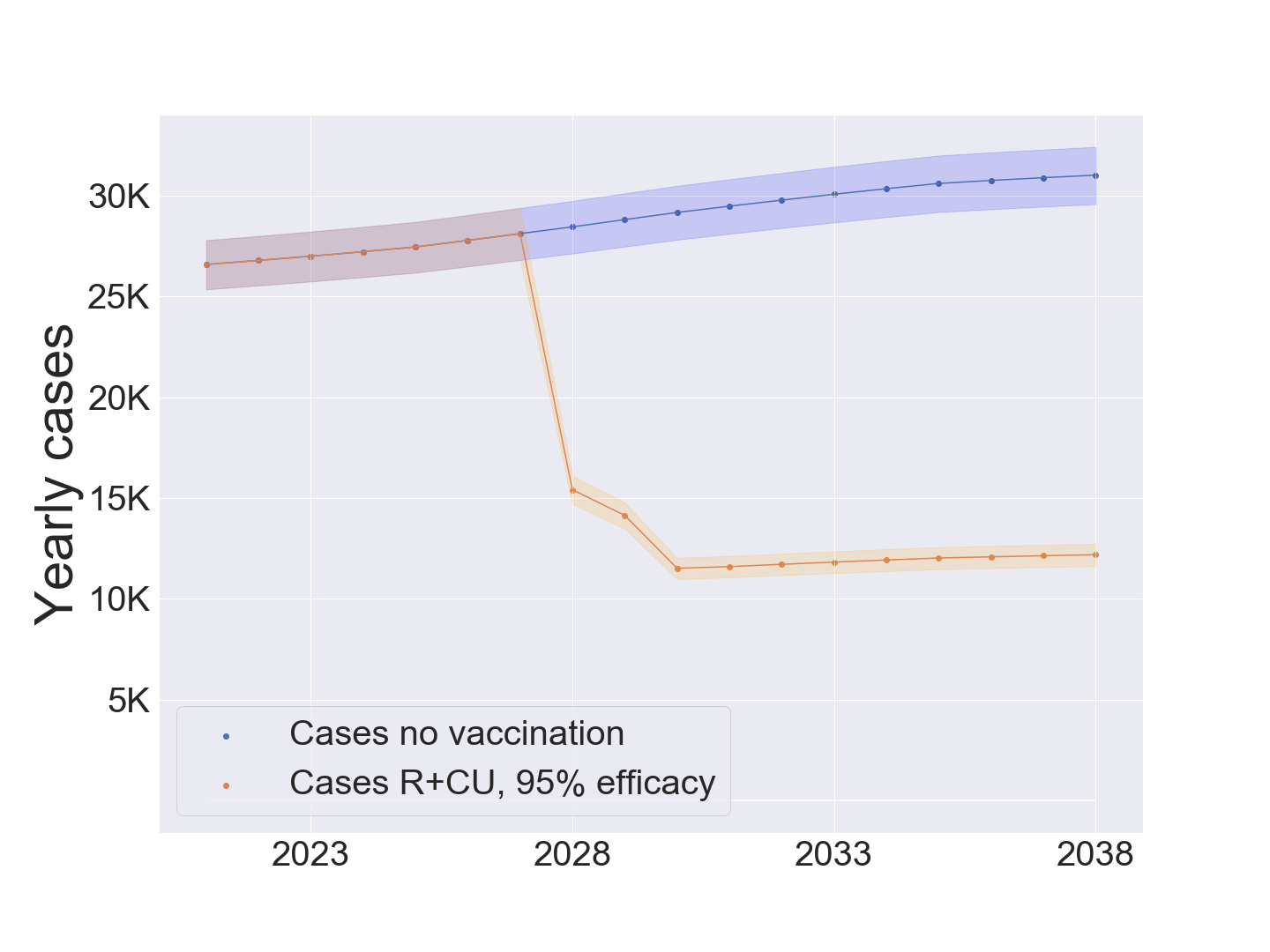}
\end{subfigure}
\caption{Uganda cumulative (top)  and yearly (bottom) iNTS cases under the status quo and routine + catch-up vaccination ($95\%$ efficacy) scenarios. Shaded areas show the 25th and 75th percentiles, line shows the median over 1000 experiments, samples drawn from uniform distributions over (0.00020,0.00024) for $\beta_{2,n}$ and (0.0080,0.0084) for $\beta_{4,n}$. }\label{fig:Uganda}
\end{figure}

\begin{figure}[htbp]
\renewcommand{\thefigure}{\textbf{Supplementary Fig. 57 Zambia cumulative and yearly iNTS cases}}
\begin{subfigure}[b]{\textwidth}
\centering
\includegraphics[width=0.7\linewidth]{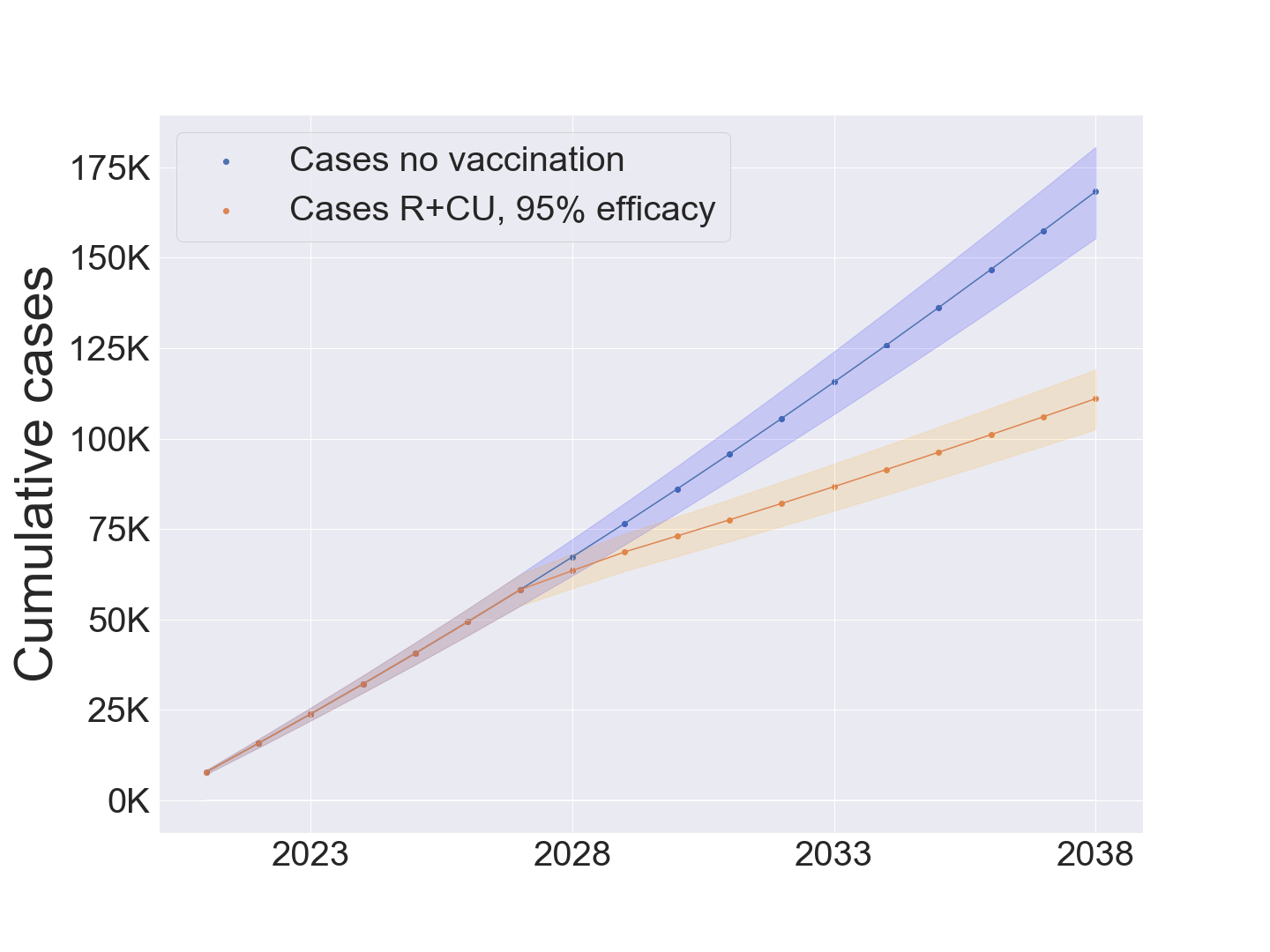}
\includegraphics[width=0.7\linewidth]{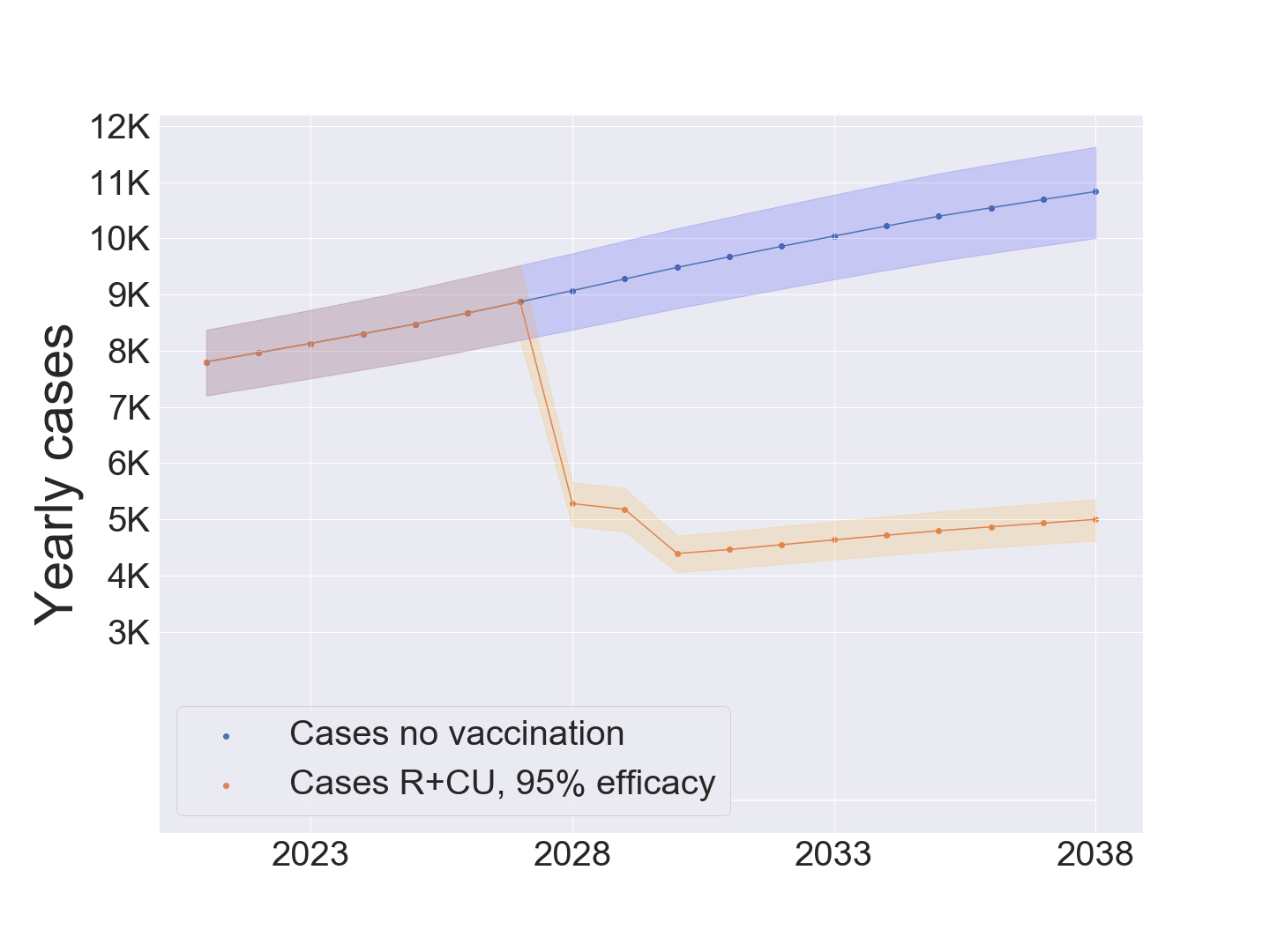}
\end{subfigure}
\caption{Zambia cumulative (top)  and yearly (bottom) iNTS cases under the status quo and routine + catch-up vaccination ($95\%$ efficacy) scenarios. Shaded areas show the 25th and 75th percentiles, line shows the median over 1000 experiments, samples drawn from uniform distributions over (0.00020,0.00024) for $\beta_{2,n}$ and (0.0080,0.0084) for $\beta_{4,n}$. }\label{fig:Zambia}
\end{figure}

\begin{figure}[htbp]
\renewcommand{\thefigure}{\textbf{Supplementary Fig. 58 Zimbabwe cumulative and yearly iNTS cases}}
\begin{subfigure}[b]{\textwidth}
\centering
\includegraphics[width=0.7\linewidth]{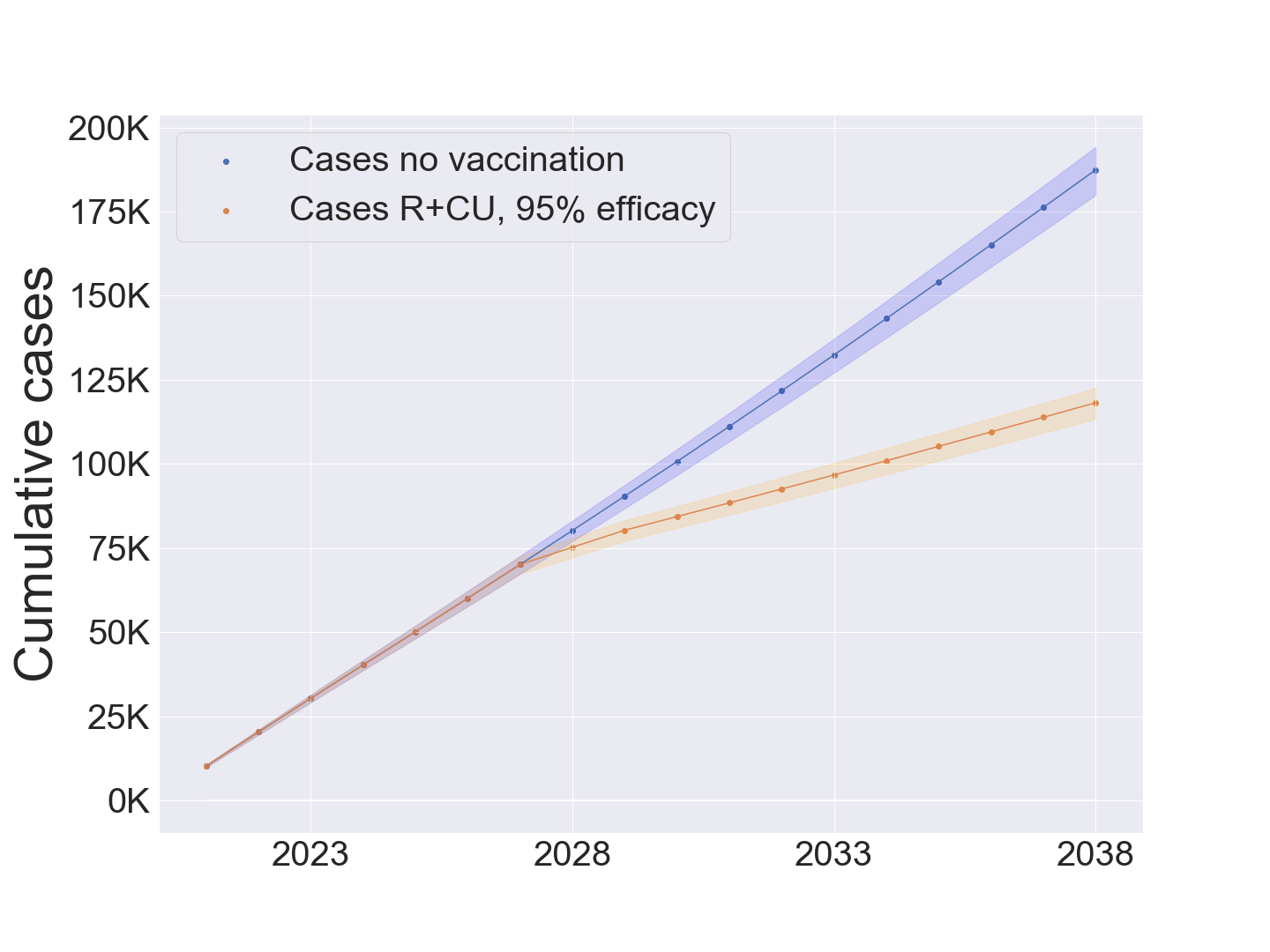}
\includegraphics[width=0.7\linewidth]{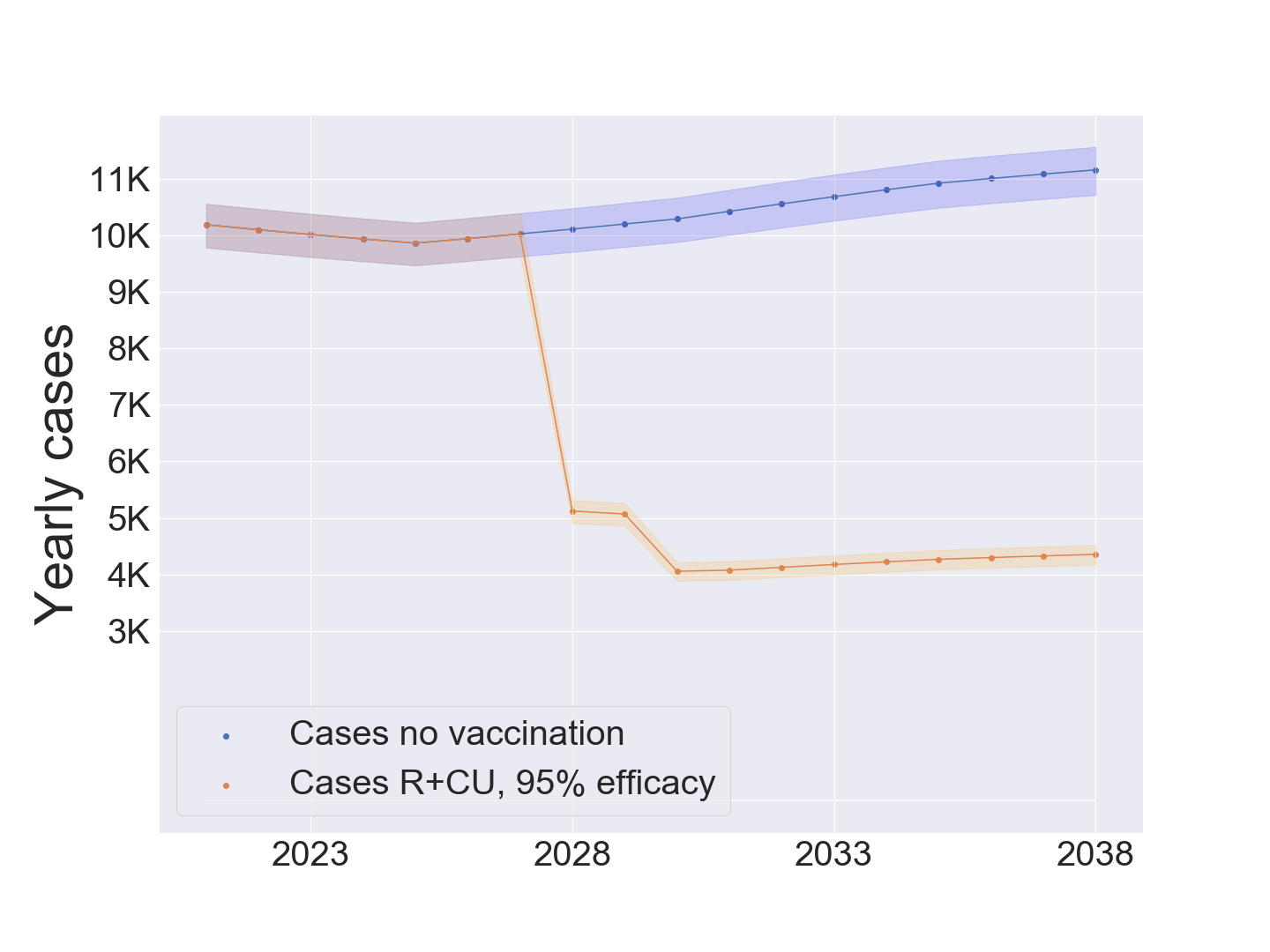}
\end{subfigure}
\caption{Zimbabwe cumulative (top)  and yearly (bottom) iNTS cases under the status quo and routine + catch-up vaccination ($95\%$ efficacy) scenarios. Shaded areas show the 25th and 75th percentiles, line shows the median over 1000 experiments, samples drawn from uniform distributions over (0.00020,0.00024) for $\beta_{2,n}$ and (0.0080,0.0084) for $\beta_{4,n}$. }\label{fig:Zimbabwe}
\end{figure}

\end{document}